\documentclass[a4paper,12pt]{report}

\usepackage[english]{babel}
\usepackage[utf8]{inputenc}
\usepackage[T1]{fontenc}

\usepackage{rotating}
\usepackage{lscape}
\usepackage{graphicx}
\usepackage{booktabs}
\usepackage[a4paper,top=3cm,bottom=3cm,left=3cm,right=3cm,marginparwidth=1.75cm]{geometry}
\usepackage{amsmath}
\usepackage{amssymb}
\usepackage{graphicx}
\usepackage[colorinlistoftodos]{todonotes}
\usepackage[colorlinks=true, allcolors=blue]{hyperref}

\usepackage{multirow}
\usepackage{setspace}
\newcommand*{\persspacing}{\setstretch{1}}
\usepackage[font=small,labelfont=bf,labelsep=period,aboveskip=6pt]{caption}
\let\oldcaption\caption
\renewcommand{\caption}{\persspacing\oldcaption}

\usepackage{caption}
\usepackage{subcaption}
\DeclareUnicodeCharacter{2212}{-}
\DeclareUnicodeCharacter{0300}{-} 
\DeclareUnicodeCharacter{00B0}{-}

\usepackage{pdfpages}
\usepackage[babel]{csquotes}
\usepackage{biblatex}
\title{Interacting Supernova Remnants: a population model for the Cherenkov Telescope Array }
\author{Silvia Crestan}

\begin{document}
\includepdf{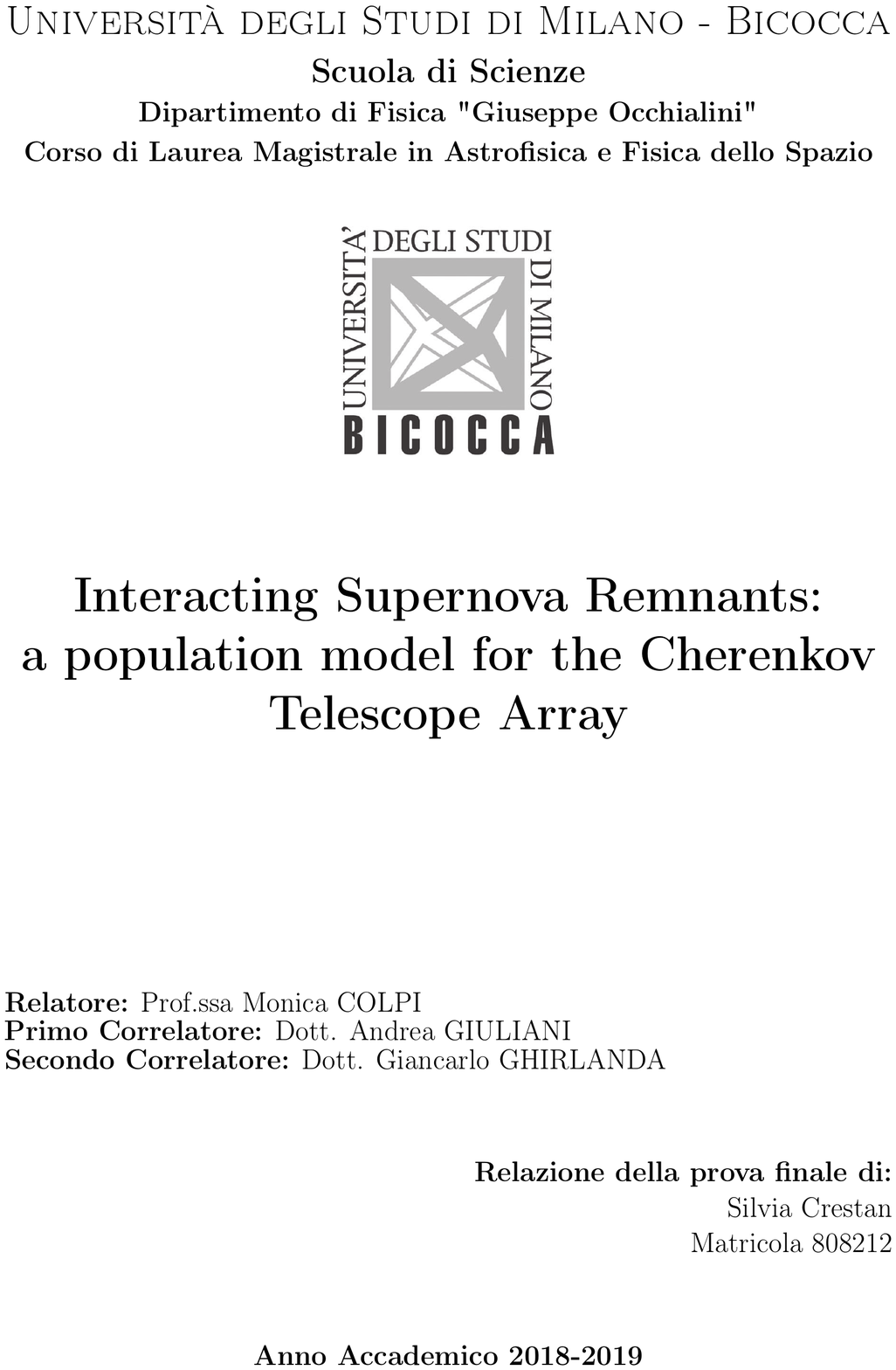}  

\newpage
\null
\thispagestyle{empty}
\newpage

\begin{abstract}
The work presented in this thesis is focused on the interacting supernova remnants (iSNRs), a class of gamma-ray emitting SNR where the radiation arise from the interaction of the SNR with a massive molecular cloud.  At the moment only 16 iSNR are known. Before this work, there was not any population study of these sources. 
Here is proposed a model for the Galactic population of iSNRs which can be used in order to predict the number of these systems in the Galaxy and how many of these will be detectable by the next generation of $\gamma$-ray instruments.  iSNRs are of particular interest for
the particle acceleration study because these objects have been proved to be sites of acceleration of protons up to high energies. Furthermore, high-energy $\gamma$-ray emission can pinpoint the presence of energetic leptons or ions and help to constraint the acceleration efficiency and maximum energy of accelerated particles

The model presented her was only achievable through the creation of a complete catalog of $\gamma$-ray (both GeV and TeV) supernova remnants that for each supernova remnants 
  gives collect the physical and spectral information available in litterature.
Simulating and analyzing the synthetic population, it was found that the Cherenkov Telescope Array (CTA) will be able to duplicate the number of detected interacting systems in its survey of the Galactic plane.
\end{abstract}

\chapter*{Summary}
Understanding the nature of galactic sources capable of accelerating cosmic rays (RC) is one of the main objectives of high energy astrophysics. In this context the supernova remnants (SNR) have a key role.
Since supernova remnants are capable to convert  $\sim 10\%$ of the kinetic energy of the explosion in protons acceleration up to energies higher than $10^{15}$ $eV$  $(PeV)$, they are considered the main sources of cosmic rays.
The current generation of Cherenkov telescopes (HESS, MAGIC, VERITAS) and $\gamma$-rays satellite (FERMI, AGILE) revealed $\gamma$-ray emission from 37 galactic supernova remnants compared to 295 SNR currently observed in the radio band\footnote{Green D. A., 2019, `A Catalogue of Galactic Supernova Remnants (2019 June version)' (available at "http://www.mrao.cam.ac.uk/surveys/snrs/")}.
Observations of supernova remnants at gamma-ray energies\footnote{Stefan Funk. Ground- and Space-Based Gamma-Ray Astronomy} showed that they can be divided into "Shell-like" sources and those that  are interacting with the dense molecular clouds in the surrounding environment .
"Interaction" refers to particles accelerated by supernova shocks that penetrate the molecular cloud.
For some of the interacting supernova remnants (iSNR) the hadronic origin of the $\gamma$-ray emission has been proven, so this sources are certainly cosmic rays accelerators potentially up to $10^{15} eV$. This makes interacting supernova remnants a very interesting class of sources to explain the cosmic rays spectrum at Pev energies.

\subsubsection{Purpose}
The thesis purpose is to give a population model for the galactic interacting supernova remnants to predict how many of these system will be observable by the next generation of $\gamma$-ray instruments.
The model I created will be used in the second data challenge (DC2) and in the simulation of the galactic plane survey (GPS) of the next  Cherenkov Telescope Array (CTA). 
Since interacting supernova remnants are systems composed of a supernova remnant and a molecular cloud, I modeled separately the population of molecular clouds and the spectrum of protons accelerated by the supernova explosions and then I associated a probability that these systems are close enough to interact.
Both synthetic population of molecular clouds and simulated spectra reflects the respective real populations since they are created starting from observations. 
In particular, for recover the simulated spectra, I created the first $\gamma$-ray emitting supernova remnants catalog.  
This catalog contains all information relating to physical and spectral parameters and those relating to the protons population responsible for the $\gamma$ ray emission.
This kind of informations are essential not only to classify SNR or to search for correlation between physical and spectral features, but also because are the starting point to perform any sort of simulation with the future high-energy instruments.
At the moment only 16 SNRs are classify as interacting in the main catalogs of $\gamma$-ray sources (TeVCaT, 1SNRCat\footnote{1st Fermi LAT Supernova Remnant Catalog}) compared to 37 $\gamma$-ray emitting SNRs. 
Another purpose is to estimate how many interacting systems CTA will detect in the Galaxy.
In conclusion, the goals are: \textbf{the creation of the first $\gamma$-ray emitting supernova remnants catalog}, \textbf{the creation of a population model for the interacting supernova remnants for the DC2 and GPS} and \textbf{estimate how many iSNR will be seen by CTA in the Galaxy}.

\subsubsection{Catalog of $\gamma$-ray Supernova Remnants}

For each SNR considered I carried out an extensive literature research of $\gamma$-ray spectra and physical characteristics.
To evaluate a possible classification based on spectral properties, I used a model-independent approach, fitting the available data with the hadronic emission model, that is, the relativistic proton-proton interaction followed by the $\pi^0$-decay in $E_{\gamma}>100MeV$ photons. 
I chose this radiative model because is the one expected for the interacting supernova remnants, object of interest of this work. 
I performed the spectral fit using the NAIMA python package: this software, through the maximum likelihood method, calculates the parameters of the particle distribution which reproduces the observations, based on the radiative model used. 
For each SNR I used a broken power law as proton distribution and $\pi^0$-decay as radiative model.
The main results obtained thanks to the catalog are: confirmation also in the $\gamma$-ray energy band that there are mainly two classes of SNRs, and the creation of the LogN-LogS curve of the interacting supernova remnants.
In principle, the slope logN-logS curve for a sources population distributed in the Galactic plane is equal to -1, more realistically, this curve has the expected slope only at high integrated fluxes, at lower fluxes the logN-logS curve deviates from -1 indicating that the sample is not complete.
The stretch of logN-logS with slope -1 will provide the calibration of the population model created, therefore the logN-logS of the synthetic iSNR must have the same slope in the completeness region.

\subsubsection{Interactiong SNR population model for the DC2 of CTA}
To create a synthetic population of interacting supernova remnant, as close as possible to what we observed in nature, I had to reproduce mainly three aspects: the population of molecular clouds 
, the spectrum of cosmic rays that interact with the nuclei of the cloud and the probability that the molecular clouds and SNRs are close enough to interact.
I realized 100 repetitions because the model is based on random extractions of spectral parameters and random associations between clouds and cosmic rays spectra, in order to not introduce any sort of effects due to  the extraction of a particular realization.

\subsubsection{Synthetic population of molecular clouds}
I created the synthetic population of molecular clouds with realistic features in terms of mass and distribution in the Galaxy.
For the spatial distributions I used the galaxy spiral model taken from Foucher et al 2006 and the galactocentric distribution function 
from Lorimer 2006 that is compatible with the young stellar distribution. This method gives a catalog of positions along the spiral arm of the Galaxy.
Clouds besides positions need an associate mass, for this reason I used the mass distribution function derived from the catalog of molecular clouds  by Rice et al.\footnote{A Uniform Catalog of Molecular Clouds in the Milky Way, Rice et al. 2016}, that is a power law with a spectral index equal to -1.72 for molecular clouds with $M_{cloud} > 10^5$ solar masses. 
I randomly associated the mass to the positions of the molecular clouds, assuming that I could extrapolate the mass distribution function of the real MC up to the current instrument sensitivity limit ($M_{cloud} \sim 10^3$).
The integral of the new mass function returns the number of molecular clouds that should be simulated, that is 13500.
\subsubsection{Synthetic spectra of cosmic rays injections}
Starting from the 16 spectra of the real interacting supernova remnant, I created the synthetic protons distributions that interact within the molecular clouds.
Thanks to the catalog realized, for each interacting supernova remnants, all parameters of the broken power law protons injection spectra (normalization, index before and after the break and the energy break) are available.
So for each parameters I created a Gaussian distribution with mean and standard deviation values calculated from the real sources and then I extracted randomly the spectral parameters values for each synthetic interacting supernova remnants.
I'm assuming that if there were more interacting supernova remnants their spectral parameters would be distributed following a Gaussian curve with mean and sigma value close to those I had found. 
Generate the synthetic interacting supernova remnants spectra requires besides the particle distribution, the density of the target protons. As the other this parameter must be randomly extracted from the distribution generated starting from the observed values of the molecular clouds taken from Rice et al..
Using the extracted values of the protons distributions parameters and of the clouds density, I created 13500 spectra and I randomly associated them with molecular clouds.

\subsubsection{The molecular clouds - SNR association}
Certainly not all the molecular clouds have a SNR nearby, therefore I introduced a probability function that in principle depend on the clouds mass. The only function that reproduces the real iSNR logN-logS curve is a flat probability at 1.5\%

\subsubsection{Results}
Using the interactive supernova remnant population model that I created, I estimated how many of these systems are potentially visible using the future Cherenkov Telescope Array. I simulated and analyzed with the CTA simulation software\footnote{http://cta.irap.omp.eu/ctools/index.html} some of the realization that I've made using 50h of observation for each source from 0.1 to 100 TeV .
For example in a realization I simulated 199 sources, 39 of these have TS $>$ 5$\sigma$. As expected, the logN-logS of the detected iSNR follow the real iSNR curves, but at the CTA sensitivity limit at 100 GeV-10 TeV it bends toward a greater number of sources. Depending on realization CTA detect between 30 and 50 synthetic iSNR.
Taking into account that simulations and analysis were done under ideal conditions (only one source for each simulation and only with instrumental background) this means doubling the number of SNR / MC systems already known.


\newpage
\null
\thispagestyle{empty}
\newpage

\tableofcontents

\chapter{Supernova remnants}
    \section{Observation of supernova remnants}
    Supernova remnants are visible in different energy bands across  the electromagnetic spectrum, from the radio band up to TeV $\gamma$-rays. About 300 supernova remnants have been discovered in the radio band and about one tenth of these are visible in $\gamma$-rays.
    The non-thermal radio spectra of supernova remnants are interpreted as synchrotron emission from relativistic electrons accelerated at the shock of the supernova remnant. The polarization of the radio synchrotron emission gives information about the magnetic field structure in the remnant. The observed X-ray emission consists of non-thermal and thermal emission. The thermal component stems from the heated gas within the shock, whereas the non-thermal emission is produced by electrons in the energy range (10 - 100) TeV. Both the non-thermal x-ray and radio emission are of leptonic origin. The observed GeV and TeV emission can be produced either by leptons due to Bremsstrahlung and inverse Compton scattering or by hadrons via pion decay. According to the theory of adiabatic shocks particles are accelerated during the free expansion phase and the Sedov phase as described in \ref{evol}.

\section{Evolution of supernova remnants}\label{evol}
The fate of a star depends on its mass. Stars with  $M \geq 8M_{dot}$ typically end in a supernova explosion caused by the collapse of the iron core after the star ends all the nuclear energy and the gravitational pressure surpasses the radiation pressure. The implosion following the collapse of the core leads to a neutron star or a stellar black hole as leftover. These supernovae are the so called core-collapse supernovae. Also less heavy stars can explode in a supernova if they have already reached the state of a white dwarf and they accrete material from a massive companion star until reaching the Chandrasekhar mass limit of about 1.3 $M_{sun}$. The typical energy released $E_{SN}$ during the supernova explosion, although it come from two different mechanisms,  is in the order of $10^{51} \, erg$.
After the explosion the stellar ejecta expand into the ambient medium and form the so-called supernova remnant. Typical velocities of the ejected material are $v_{eject} \approx 10^{4} km s^{-1}$ and due to the supersonic speed of the ejecta a shock front is formed. During the expansion the shock front sweeps up the ambient medium  and the shock velocity of the forward shock decreases. As a consequence a reverse shock, is formed. Observations of supernova remnants in different energy bands from radio to $\gamma$-rays show that particles get accelerated at the shock fronts during different stages in the evolution of an supernova remnant.

The evolution of a core-collapse supernova remnant can be simplified in 4 stages, as discussed in \cite{ev} and \cite{book}. \\
\textbf{Phase I: free expansion phase}: in this phase, the mass of the supernova ejecta, $M_{ej}$, is greater than the swept-up mass, $M_{sw}$. This phase last until $M_{ej}\sim M_{sw} $  The initial velocity $v_{0}$ of the ejected mass $M_{ej}$ after the explosion is given by:

\begin{equation}
    v_{0}=\sqrt{\frac{2E_{SN}}{M_{ej}}} = 10^9 \, cm\, s^{-1} \, \bigg( \frac{E_{SN}}{10^{51}\,erg} \bigg)^{1/2}\bigg( \frac{M_{ej}}{1M_{sun}}\bigg)^{-1/2}
    \end{equation}

 is larger than the speed of sound in the medium. Therefore, a shock front is formed. The timescale
  and the radius of the remnant at the end of this stage until the beginning of the following Sedov phase depend on the density of the ambient medium $\rho_{ISM}$. The free expansion phase ends at:
 \begin{equation}
    t\simeq 200 \, yr \, \bigg( \frac{E_{SN}}{10^{51} \, erg} \bigg)^{-1/2}\bigg( \frac{\rho_{ISM}}{10^{-24}g\,cm{^-3}}\bigg)^{-1/3}
 \end{equation}
    
 when the radius of the supernova remnant is:
 
 \begin{equation}
    r =\bigg( \frac{3M_{ej}}{4\pi\rho_{ISM}}\bigg)^{1/3}=2.1 \, pc \, \bigg( \frac{M_{ej}}{1M_{sun}} \bigg)\bigg( \frac{n_{0}}{1cm{^-3}}\bigg)^{-1/3}
 \end{equation}   
   where $n_{0}$ is the number density of the ambient medium.\\
\textbf{Phase II: Sedov Phase}: Radiative losses $dE/dt$ are still negligible since  $\int(dE/dt)dt \ll E_{SN}$. In this phase kinetic and thermal energies are conserved. The radius of the supernova remnant evolves as

\begin{equation}
    r\simeq 0.3 pc \, \bigg(\frac{E_{SN}}{10^{51}\,erg} \bigg)^{1/5}
    \bigg( \frac{n_{0}}{1cm{^-3}}\bigg)^{-1/5}\bigg( \frac{t}{yr}\bigg)^{2/5}
\end{equation}
The shock velocity $v_{shock}$ decreases with increasing radius as:

\begin{equation*}
    v_{shock} \simeq 5000 \, km \, s^{-1}\, \bigg(\frac{r}{2\,pc} \bigg)^{-3} \bigg(\frac{E_{SN}}{10^{51}\,erg} \bigg) \bigg(\frac{n_{0}}{1cm{^-3}}\bigg)^{-1}
\end{equation*}
and the heated gas reaches temperatures of about $\sim 10^6 K$, then emits thermal radiation visible in x-rays.
At the end of this phase the shock velocity is slowed down to $\sim 200 km s^{-1}$ and the radiative losses dominate. The outer shell of the shock decelerates first, allowing the inner material to catch up. During the deceleration process a shock wave at the inner edge is formed that leads to a heating of the matter in the outer shell. The transition between this phase and the following radiation phase occurs after at:
\begin{equation*}
    t_{rad}=3\cdot 10^4  \, yr \bigg(\frac{T}{10^6\, K}\bigg)^{-5/6} \bigg(\frac{E_{SN}}{10^{51}\,erg} \bigg)^{1/3} \bigg(\frac{n_{0}}{1cm{^-3}}\bigg)^{-1/3}
\end{equation*}

\textbf{Phase III: radiation phase} ($t > t_{rad}$): During this stage the temperature of the matter behind the shock decreases because of the radiation cooling is dominant . The radial momentum of the shell is nearly constant and the expanding shock piles up interstellar gas.  The radius increases with $R(t) \varpropto t^{1/4}$.\\
\textbf{Phase IV}: In this last phase the shock velocity drops to ($ 10 - 100  km  s^{-1}$ ) in about ($1 - 5) \cdot 10^5 \, yr$. The remnant loses its structure and the accumulated matter disperses into the interstellar medium.
\\
Typical values for the phase duration, radius of the remnant, shock velocity and swept-up material in this simplified view of the evolution of a core-collapse supernova remnant in a homogenous environment are summarized in Table \ref{tab:evo}.

\begin{table}
\begin{center}
\begin{tabular}{|c|cccc|}
\hline
Phases & t (yr)                      & r (pc) & v (km $s^{-1}$) & $M_{swp}$ ($M_{sun}$) \\ \hline
I-II   & 90                              & 0.9        & $10^4$              & 0.2                       \\
II-III & 22 $\cdot 10^3$  & 11         & 200                 & 180                       \\
III-IV & 750 $\cdot 10^3$ & 30         & 10                  & 3600                      \\ \hline
\end{tabular}
\caption{Characteristic values for the phase duration, radius r, shock velocity v and swept-up mass $M_{swp}$ for the transition between the different phases of the evolution of a supernova remnant taken from \cite{ev}}
\label{tab:evo}
\end{center}

\end{table}

\section{Particle acceleration}
Detection of non-thermal emission from supernova remnants from radio to TeV energy range suggests them as sites of particle acceleration. It is widely accepted that the acceleration mechanism is diffusive shock acceleration at the shock front of the supernova remnant. The general concept was developed contemporaneously in the late 1970s by Krymskii \cite{acc77}, Axford \cite{ax}, Bell \cite{bell1} \cite{bell2} and Blandford \& Ostriker \cite{bo} as first-order Fermi acceleration. The idea is that particles cross the shock of the supernova remnant many times, bouncing on  magnetic turbulences in the up- and down-stream regions of the shock gaining energy. Fermi (1949) originally proposed that the acceleration of cosmic rays is due to collisions with randomly moving magnetic fields in the galaxy.  For historical reasons, in Section \ref{2oa} the so-called “second-order” Fermi acceleration is discussed. The diffusive shock acceleration and “first-order” Fermi acceleration occurring at strong shocks are introduced in Section \ref{1oa}.

\subsection{Second order Fermi acceleration}\label{2oa}
The discussion of Fermi’s approach presented is based on Longair (2011) \cite{long}. In the second order Fermi mechanism, particles are assumed to be accelerated thanks to stochastic collisions with the clouds (and in particular with the magnetic
field irregularities inside them, also called magnetic mirrors) moving isotropically in the interstellar medium with a typical velocity ${v_M}$. Fermi showed that the particles gain energy stochastically in these reflections and if the particles remain within the acceleration region for some characteristic time $\tau_{esc}$, a power law distribution of particles energies is obtained. The mirror mass is much larger than the particle’s mass and simply assumed to be infinite for the calculation of the collision between a particle and a moving mirror so ${v_M}$ is unchanged in the collision. The energy $E'$ of the particle in the frame of the moving cloud is
\begin{equation}
    E' = \gamma(E + pv_{M} \cos\theta),
\end{equation}
where $\gamma = \sqrt{1/ 1 − v^2_M/c^2}$ is the Lorentz factor of the mirror, c is the speed of light and $\theta$ is the angle between the particle trajectory and the velocity direction of the mirror. The x-component of the particle’s momentum $p$ in the mirror frame is
\begin{equation}
    p'_{x} = p' \cos \theta' = \gamma (p\cos \theta + \frac{v_M \cdot E}{c^2})
\end{equation}
In this frame, the particle's energy is conserved, $E'_{before}= E'_{after}$, and its momentum along the x-direction is reversed, so $p'_x=-p'_x$. The energy of the particle after the collision in the observer frame is
\begin{equation}
    E''=\gamma(E'+v_M \cdot p'_x)=\gamma \cdot E \bigg[ 1+ \frac{2\cdot v_{M} \cdot v}{c^2} +\frac{v_M}{c}\bigg]
\end{equation}

with the particle’s velocity $v$. The average energy gain per collision for the particle after expanding Eq. 1.7 to second order in $v_M/c$ and averaging over $\theta$ for all angles between 0 and $\pi$ reads
\begin{equation}
    \bigg \langle \frac{\Delta E}{E}\bigg \rangle =\frac{8}{3} \bigg( \frac{v_M}{c}\bigg)^2
\end{equation}
This leads to an energy gain of second order in  $v_M/c$ and to a power-law energy spectrum of accelerated particles which is observed from non-thermal sources. But Fermi’s proposed acceleration mechanism is not efficient for random velocities of interstellar clouds with  $v_M/c \leq 10^{-4}$. Furthermore the mean free path of cosmic rays in the interstellar medium is $\approx 0.1$ pc. This results in only a few collisions per year and, therefore, the energy gain is too slow to explain the cosmic-ray energies observed.
Through this process, the spectral index  depends on the characteristics of the acceleration region. On the contrary, as it will be shown in next section, first-order Fermi mechanism predicts a power law spectrum with spectral index equal to −2, very close to the value required by the cosmic ray observations.

\subsection{First order Fermi acceleration }\label{1oa}
\begin{figure}
    \centering
    \includegraphics[width=0.9\textwidth]{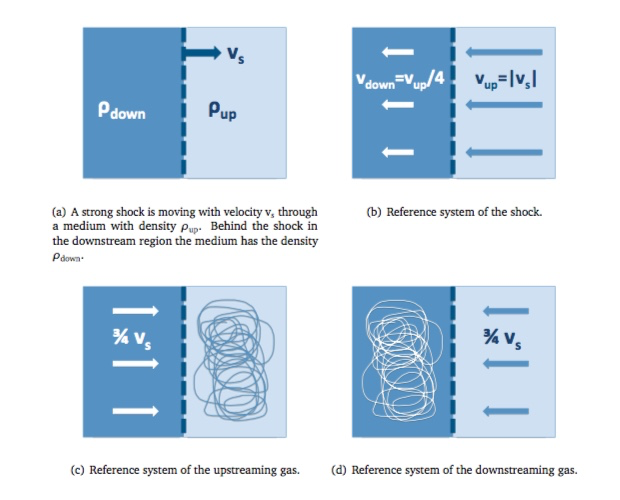}
    \caption{The dynamics of high-energy particles in the vicinity of a strong shock. The figure is adapted from Longair (2011).}
    \label{fig:fermi}
\end{figure}

The diffusive acceleration in strong shock waves,
results in an energy gain first order of the shock velocity $v_s$ and gives an energy spectrum with power-law form with a spectral index $\approx −2$.
The discussion in the following paragraphs is based on Longair (2011) \cite{long} and illustrated in figure \ref{fig:fermi}. The acceleration of particles takes place at the shock front that is formed in a supernova explosion. The shock is represented by the dashed line in figure \ref{fig:fermi}.  The fluid velocities before and after a hydrodynamical shock, moving with velocity $v_{s}$, are different. The particles are assumed to propagate with a speed much larger than the shock velocity $v_{P} >>  v_s$ and take no notice of the shock itself because their gyro radius is much larger than the shock thickness. The upstream fluid velocity is $v_{up} \approx v_s$ in the frame of the shock and the downstream velocity is $v_{down} = v_s/4$ for strong shocks as illustrated in figure \ref{fig:fermi}(b). This dependency follows from the continuity equation
\begin{equation}
    \rho_{up}\cdot v_{up}=\rho_{up}\cdot v_{s}=\rho_{down}\cdot v_{down}
\end{equation}
$\rho_{up}$ and $\rho_{down}$ are the up- and downstream densities and, in the case of strong shock, $\rho_{up}/\rho_{down} = (\tau + 1)/(\tau − 1)$ with $\tau$ denoting the ratio of the specific heat capacities. For a fully ionized or monoatomic gas is $\tau = 5/3$ and, hence, $v_{down} = vs/4$.
In the reference system of the upstream gas (figure \ref{fig:fermi}(c) ), the particle velocity distribution is isotropic and the downstream gas is moving with $3v_s/4$. Particles cross the shock and gain energy by scattering head-on off the downstream fluid. After the scattering the velocity distribution is now isotropic in the frame of the downstream fluid and the upstream fluid has a relative velocity of $3v_s/4$ in the system of the downstream fluid as illustrated in figure \ref{fig:fermi}(d). After crossing the shock the particles bounce off the upstream fluid and again gain energy. Therefore, the particles gain an amount of energy $\varpropto v_s/c$ with each crossing of the shock. Performing a Lorentz transformation between the different systems and averaging over all angles gives a fractional energy increase per shock crossing of
\begin{equation}
    \bigg \langle \frac{\Delta E}{E}\bigg \rangle =\frac{4}{3} \bigg( \frac{v_s}{c}\bigg)
\end{equation}
The approach discussed above shows that particles gain energy while crossing the shock and scattering off magnetic turbulences in the up- and downstream region. Bell in \cite{bell1} proposed the back-reaction of the accelerated charged cosmic rays as source for the magnetic turbulences by resonant Alfven waves. Alfven waves are generated by particles with velocities
larger than the critical speed $v_A \approx  B/\sqrt{4\pi\rho}$ in a medium with density $\rho$ and a magnetic field B.
For the definition of the Alfven speed and a more in-depth discussion of plasma physics see e.g. Kulsrud (2005).
Both Fermi acceleration processes, first and second order, naturally result in a spectral energy distribution following a power law. If the energy of a particle after one collision is $E = \beta E_0$ and there is a probability $P$ that the particle remains in the shock after the collision then there are $N=N_0 \cdot P^k$ particles left in the shock after k collisions with energies $E=\beta^kE_0$. Eliminating k gives
\begin{equation}
    \frac{ln(N/N_0)}{ln(E/E_0)}=\frac{ln(P)}{ln(\beta)}
\end{equation}
and therefore
\begin{equation}
    \frac{N}{N_0}=\bigg( \frac{E}{E_0} \bigg)^{LnP/ln\beta}
\end{equation}
and the energy distribution follows a power law
\begin{equation}
\label{pl}
    N(E)dE = constant \cdot E^{−1+(lnP/ln\beta)}
\end{equation}
The spectral index of the power law  depends on the probability P that particles remain in the shock and on the energy gain $\beta$. For diffusive shock acceleration the argumentation in \cite{bell1} gives $P = 1 − v_s/c$ and $\beta = E/E0 = 1 + (4v_s)/(3c)$. Inserting these values into equation \ref{pl} results in an energy spectrum with power-law form and spectral index of −2
\begin{equation}
    N(E)dE \varpropto E^{−2}dE.
\end{equation}
In conclusion, first-order Fermi acceleration mechanism is much more efficient than
the second-order mechanism, since the energy gain is linear in the velocity of the
shock. Furthermore, the predicted spectrum is a simple power-law with unique
value for the spectral index, which is also close to the value required to account
for the observed cosmic ray spectrum. The only requirements are the presence of
strong shock waves and that the velocity vectors of the high energy particles are
randomized on either side of the shock. For this reason, supernova remnants are
very good candidates to be acceleration sites for cosmic rays

\section{Gamma-ray emitting processes}
\begin{figure}
 \centering
    \includegraphics[width=0.7\textwidth]{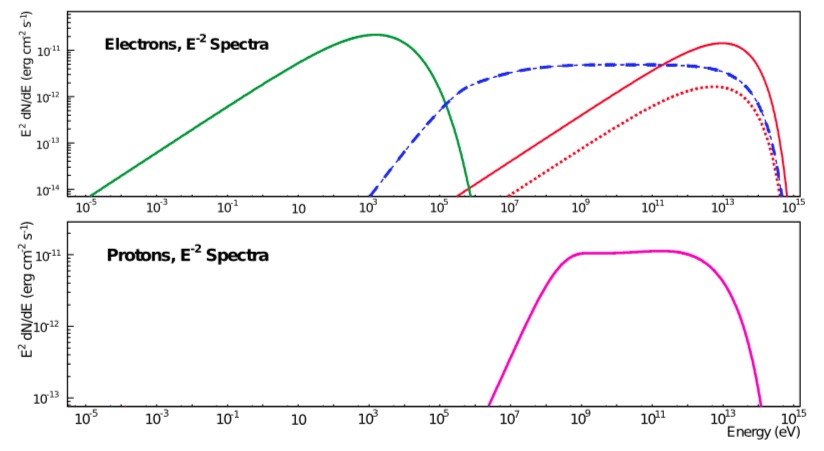}
    \caption{Photon spectra from leptonic and hadronic emission processes as a function of energy. Top panel: photon emission by electrons with a spectral form of a power law with a spectral index equal to $-2$ and an exponential cut off at 100 TeV. Synchrotron radiation (green), Bremsstrahlung (blue) and inverse Compton scattering on the cosmic microwave background and infrared photon fields (red dotted) and on starlight (red solid) for electrons.
Bottom Panel: $\gamma$-rays from proton-proton collisions via $\pi^0$-decay (magenta). Protons are distributed as a power-law  with index equal to $-2$ and an exponential cut-off at 100 TeV.
This figure is adapted from \cite{hh}.}
    \label{emis}
\end{figure}

The study of SNRs can give detailed information about the shock propagating in the interstellar medium, which in principle is able to accelerate particles up to very high energies through diffusive shock acceleration, as already discussed
in section \ref{1oa}. Since accelerated charged particles are deflected by the Galactic magnetic field during their propagation to Earth, direct observation at the source cannot be performed. However, charged particles interact with the SNR environment and produce photons from radio to TeV energy range, which can be directly detected. Different emission processes can be responsible for the emission of high energy photons  as illustrated in figure. \ref{emis}
In this example the $\gamma$-ray emission is modeled for electrons and hadrons with the spectral form of a power law with spectral index $\Gamma = −2$ and an exponential cutoff at an energy of 100 TeV. The emission in the radio and X-ray regime is produced by leptons via synchrotron radiation and is displayed in green on figure \ref{emis}. The blue curve shows the photon spectrum produced by the Bremsstrahlung mechanism that depends on the density of the medium. Inverse Compton scattering from high-energy electrons on photon fields, i.e. the cosmic microwave background and starlight (red dotted and red solid curves in figure \ref{emis} ), leads to emission peaking in the TeV energy band and is produced by the same lepton population as the synchrotron emission. TeV $\gamma$-ray emission can also be produced by high-energy protons colliding with atoms of the interstellar medium or molecular clouds producing pions (both neutral and charged) and in the subsequent neutral pion decay in $\gamma$-rays as illustrated by the magenta line on the lower panel of figure \ref{emis}. The hadronic and leptonic radiation processes are discussed in the following Sections
\subsection{Hadronic scenario}
Protons (p) produce $\gamma$-ray emission  colliding with interstellar material, mainly hydrogen, and subsequent decay of secondary particles. The main contribution originates from the decay of the neutral $\pi^0$ meson into two $\gamma$ rays while charged pions decay in muons and neutrinos (with a mean life of $2.6 \cdot 10^{−8} s$): 

\begin{equation}
    p+p \rightarrow \pi + X, 
\end{equation}
\begin{equation}
\pi^0 \rightarrow  \gamma + \gamma \, \, \, \, (98.8\%)
\end{equation}
\begin{equation}
\pi^+ \rightarrow \mu^+ + \nu_{\mu} \, \, (99.9\%), \, \, \, \, \mu^+ \rightarrow e^+ + \nu_e +\overline{\nu}_{\mu}
\end{equation}
\begin{equation}
\pi^- \rightarrow \mu^- +\overline{\nu}_{\mu} \, \, (99.9\%), \, \, \, \, \mu^- \rightarrow e^-  +\overline{\nu}_{e} + \nu_{\mu}
\end{equation}

The $\pi^0$ has a mass of $m_{\pi^0} \approx 135 MeV/c^2$ and the mean lifetime is about $8.4\cdot 10^{−17} s$.
In proton-proton collision also $\eta$ mesons are produced, those likewise have decay channels producing $\gamma$ rays, but as demonstrated by Kelner et al. \cite{kelner}, their contribution to the total number of $\gamma$ rays is at most 25\% depending on the photon energy.
The resulting $\gamma$-ray flux from pion-decay in an energy interval [$E_{\gamma}$, $E_{\gamma}$ +d$E_{\gamma}$] can be calculated as:

\begin{equation}
\frac{dN_{\gamma}}{dE_{\gamma}}= c n_H \int_{E_{\gamma}}^{\infty} \, \, \sigma_{inel}(E_p)J_p(E_p)F_{\gamma}(\frac{E_{\gamma}}{E_p},E_p)\frac{dE_p}{E_p} 
\end{equation}
where $c$ is the speed of light, $n_H$ the density of the medium and $J_p(E_p)$ is the proton flux at a certain proton energy $E_p$. $\sigma_{inel} $ is the inelastic cross section of proton-proton interactions and can be parametrized as:
\begin{equation}
\sigma_{inel}(E_p) = (34.3 + 1.88L + 0.25L^2) \cdot [ 1 − (\frac{E_{th}}{E_p})^4]^2 \,\, mb
\end{equation}
$L$ is defined as $L = ln(E_p/1TeV)$ and the threshold energy $E_{th}$ for protons to produce $\pi^0$ mesons is $1.2 \cdot 10^{-3} TeV$. The function $F_{\gamma}(\frac{E_{\gamma}}{E_p},E_p)$ is the spectrum of $\gamma$-rays from the $\pi^0$- and $\eta$-mesons decay channels for a certain proton energy $E_p$:
\begin{equation*}
F_{\gamma}(x,E_p) = B_{\gamma} \frac{ln(x)}{x} ( \frac{1-x^{\beta_{\gamma}}}{1+k_{\gamma}x^{\beta_{\gamma}}(1-x^{\beta_{\gamma}})} )^4 [ \frac{1}{ln(x)}-\frac{4\beta_{\gamma}x^{\beta_{\gamma}}}{1-x^{\beta_{\gamma}}}-\frac{4k_{\gamma}\beta_{\gamma}x^{\beta_{\gamma}}(1-sx^{\beta_{\gamma}})}{1+k_{\gamma}x^{\beta_{\gamma}}(1-x^{\beta_{\gamma}})}]
\end{equation*}
with $x=E_{\gamma}/E_p$. The parameters $B_{\gamma}$, $\beta_{\gamma}$ and $k_{\gamma}$, derived in \cite{kerner}, depend on L:
\begin{equation}
B_{\gamma}=1.30 + 0.14 L + 0.011 L^2
\end{equation}
\begin{equation}
\beta_{\gamma}=\frac{1}{1.70+0.11L+0.008 L^2}
\end{equation}
\begin{equation}
k_{\gamma}=\frac{1}{0.801+0.049L+0.014 L^2}
\end{equation}
In general, the spectral form of the proton population is conserved in the $\gamma$-ray emission. The produced photons receive about one tenth of the primary proton energy. For the total photon flux as described above a factor of 1.45 is multiplied to account for the contribution by nuclei heavier than helium present in the interstellar medium or in the molecular clouds.

\subsection{Leptonic scenario}
$\gamma$ rays are emitted by electrons (also positrons are included) losing energy in magnetic fields, dense medium or scattering on photon fields. Only photons emitted by Bremsstrahlung or those having gained energy via inverse Compton scattering can be observed in the TeV energy range, whereas the synchrotron radiation of electrons is typically emitted in radio and x-ray energy bands. The same electron population is responsible for the emission from synchrotron and inverse Compton emission, thus, synchrotron emission is also discussed shortly.

\subsubsection{Bremsstrahlung}
The term Bremsstrahlung refers to electromagnetic radiation produced by the deceleration of a charged particle when deflected by another charged particle, typically an electron by an atomic nucleus. The moving particle loses kinetic energy, which is converted into radiation (i.e., a photon), thus satisfying the law of conservation of energy. In astrophysics the Bremsstrahlung from relativistic electrons plays an important role, especially in the GeV energy range. Following the discussion in Blumenthal \& Gould \cite{brems} the energy spectrum of the emitted photons depends on the density $n_Z$ and  atomic number Z of the medium. The photon spectrum $dN /dE_{\gamma}$ for a given electron energy distribution is given by:

\begin{equation}
    \frac{dN}{dE_{\gamma}}=\int dE_i N_e(E_i)\cdot \frac{dN_i}{dE_{\gamma}}
\end{equation}
where $E_i$ is the electrons initial energy.
The emission spectrum from a single electron can be written, under the assumption that electrons mainly scatter off the Coloumb field of hydrogen atoms, as
\begin{equation}
    \frac{dN}{dE_{\gamma}}=c\cdot \rho_i\frac{d\sigma}{dE_i}
\end{equation}
$\rho_i$ is the hydrogen density of the medium the electron propagates through. The expression for the differential cross section $d\sigma/dE_{\gamma}$ for the Bremsstrahlung process is:

\begin{equation}
    \frac{d\sigma}{dE_i}=\alpha \cdot r_0^2\frac{f (E_{\gamma}/E_e)}{E_{\gamma}}
\end{equation}
in which $\alpha$ is the fine structure constant, $r_0$ is the classical electron radius and $f (E_{\gamma}/E_e)$
 is a energy-dependent factor.
 Under the assumption that hydrogen plasma is weakly shielding $f(\epsilon)$ is given by:
\begin{equation}
    f(\epsilon)=\Phi_{weak}\cdot \bigg[ 1 + (1 − \epsilon)^2 + \frac{2}{3} (1 − \epsilon)\bigg]
\end{equation}
with $\epsilon = E_{\gamma}/E_e$ and $\Phi_{weak}$  defined as
\begin{equation*}
   \Phi_{weak}=4(Z^2+Z_{el})  \left \{  ln \left [   \frac{2 E_{e} (E_{i}-E_{\gamma})}{E_{\gamma}} \right ] -\frac{1}{2}
 \right \}
\end{equation*} 
Z is the atomic number of the element the medium consists of and $Z_{el}$ is the number of electrons.
In general, the spectral form of the Bremsstrahlung emitted by relativistic electrons in a dense medium mainly depends on the underlying electron spectrum. Since the same electron population produced TeV Bremsstrahlung and x-ray synchrotron radiation of a source, these two emission mechanism are strongly related.

\subsubsection{Synchrotron radiation}
Synchrotron radiation is emitted when charged particles are accelerated radially, for example electrons gyrating in a magnetic field.  Only electrons will be discussed, since the energy loss of the particle and, therefore, the emitted radiation is suppressed for protons and ions because of much larger particle mass. This emission process dominates the non-thermal radiation in the radio to x-ray regime. A charged particle moving in a magnetic field with a strength B loses the following amount of energy per time unit
\begin{equation}\label{pot}
    \frac{dE}{dt}=\frac{4}{3}\sigma_T \gamma^2 \beta^2 c U_B
\end{equation}
$U_B$ is the energy density of the magnetic field, $\beta = v/c$ the ratio of the particle’s velocity and the speed of light. $\sigma_T$ is the Thomson cross section $\sigma_T = 8r_0^2/3 = 6.653 · 10^{−29} m^2$. Depending on the energy of the charged particle the Thomson cross section has to be replaced by the Klein-Nishina cross section, that can be write as $\sigma_{K-N}=8\pi /3 r_0^2 (1-2x)=\sigma_T(1-2x)$ (where $x=h\nu/m_ec^2$) for low energy photons.
According to \cite{rey} the synchrotron emission of an electron with energy E rises continuously up to a maximum energy
\begin{equation}
    h\nu_m = 1.93(E/100TeV)^2(B/10\mu G)keV
\end{equation}
and drops exponentially beyond this energy.
The energy of the emitted photons in a frequency range $\nu$ to $\nu + d\nu$ radiated by a distribution of electrons in the energy range E to E + dE can be expressed as
\begin{equation}
    J(\nu)d\nu= \bigg( -\frac{dE}{dt} N(E)dE \bigg)
\end{equation}
Following \cite{long} and substituting $E = \gamma m_ec^2 =  \sqrt{\nu/\nu_g}m_ec^2$ with $\nu_g = e · B/(2\pi m_e)$, differentiating for dE and replacing dE/dt with \ref{pot} gives:
\begin{equation}
    J(\nu) \varpropto B^{(p+1)/2}\nu^{−(p−1)/2}
\end{equation}
with p is the slope of the electron energy spectrum.
The electron energy ${E_e}$ required for x-ray radiation at a specific energy is, according to \cite{rey},
\begin{equation}
    E_e = 72 \, \bigg( \frac{h\nu_{x-ray}}{1keV} \bigg)^{1/2} \bigg( \frac{B}{10\mu G}\bigg) \, TeV
\end{equation}

The observation of non-thermal X-ray emission in supernova remnants thus demonstrates the presence of high-energy electrons accelerated up to energies of $\approx$ 100 TeV.

\subsubsection{Inverse Compton effect}
Inverse Compton scattering consists of the interaction of a high energy electron with a low energy photon; the electron transfers part of its energy to the seed photon, creating a gamma-ray. The energy loss of the electrons can be written similar to equation \ref{pot} as:
\begin{equation}
    \frac{dE}{dt}=\frac{4}{3}\sigma_T \gamma^2 \beta^2 c u_rad
\end{equation}
with the energy density of the photon field $u_{rad}$ replacing the magnetic energy density. In the Thomson regime only a small fraction of the electron energy is transferred to the photon. The spectral emissivity $I(\nu)$, under the assumption of an isotropic photon field in the ultra-relativistic regime,  according to Blumenthal \& Gould \cite{brems} calculation is: 
\begin{equation}
    I(\nu)d\nu= \frac{3\sigma_Tc}{16\gamma^4}\nu \bigg[  2\nu \cdot ln \bigg(\frac{\nu}{4\gamma^2 \nu_0}+\nu+4\gamma^2 \nu_0 -\frac{\nu^2}{2\gamma^2 \nu_0}  \bigg) \bigg] d\nu.
\end{equation}
The maximal energy a photon can gain by head-on collision with an electron is given by
\begin{equation}
    h\nu_{max} \approx 4\gamma^2h\nu
\end{equation}
Relevant photon fields in astrophysics are the cosmic microwave background (CMB), infrared light emitted by dust and optical starlight. Inverse Compton scattering of electrons on these photon fields are the standard scenario to explain, for example, the observed TeV $\gamma$-ray radiation from some supernova remnants.

\chapter{Interacting supernova remnants}
Current Cherenkov telescopes such as H.E.S.S., MAGIC and VERITAS detected $\gamma$-ray emission in the TeV energy range from more than twenty-five supernova remnants, considering also the $\gamma$-ray satellite that works in GeV energy range, $\gamma$-ray emission is detected from more than thirty SNR. For most of the supernova remnants it is not clear whether the observed $\gamma$-ray emission is mainly produced by accelerated leptons or hadrons. Molecular clouds near supernova remnants provide an appropriate environment to probe the protons acceleration by the supernova shocks, as $\gamma$-ray emission coincident with molecular clouds proves the hadronic origin of the emission. Leptonic emission processes, except possibly Bremsstrahlung in the Galactic centre region, are negligible regarding the $\gamma$-ray emission of clouds in the TeV energy band. The accelerated proton spectrum is obviously the most interesting in order to have hints of the acceleration of cosmic rays in SNRs and therefore, the $\gamma$-ray spectrum acquires a key role in the search for evidences of the so called "SNR paradigm" for cosmic rays. One of the most convincing way of proving the hadronic origin of the $\gamma$-ray flux is the observation of the shape of the spectrum at energies below 100 MeV, where the $\pi^0$-decay spectrum presents a break due to the threshold energy of the pion production, often called pion bump. At this energy there are two key instruments (Fermi-LAT and AGILE) to proof the hadronic origin of the gamma ray emission of some supernova remnants.
Two different scenarios for $\gamma$-ray emission from molecular clouds are reported. Section 2.2 describes the scenario of “passive molecular clouds”, clouds only embedded in the cosmic ray sea without a individual source nearby. Molecular clouds near cosmic ray sources such as molecular clouds interacting with the shock of a supernova remnant are introduced in Section \ref{mci}. First the general properties of molecular clouds are outlined in Section \ref{gp}. 

\section{General properties of molecular clouds}
Molecular clouds are dense accumulations of molecular gas within the interstellar medium, located mostly in the spiral arms. Molecular clouds are known since the early 1980s to harbour star forming regions \cite{tadd}, they  can be observed directly using molecular line emission at radio wavelengths. The main constituent is molecular hydrogen ($H_2$). Since $H_2$ does not posses a permanent dipole moment, thus, no transitions are excited at low temperatures as usually found in the environment of molecular clouds. For $H_2$ line emission to be produced, temperatures of about 150 K are required much higher than commonly found in molecular clouds. Hence, other molecules are used as tracers for $H_{2}$ such as the carbon monoxide molecule (CO). The most commonly used emission line, according to \cite{kawa}, is the rotational transition $^{12}C ^{16}O(J = 1 − 0)$ with a frequency of 115.27120 GHz \footnote{http://www.cv.nrao.edu/php/splat/}.
$^{12}C ^{16}O(J = 1 − 0)$ has an excitation energy of $\approx$ 5 K and a critical density of $\approx 1000 cm^{-3}$. Simplified, the critical density is the density at which the line emission gets observable. Line emissions of other transitions or molecules have different excitation energies and critical densities, thus the density distribution of the interstellar medium can be probed using different molecules or different transitions as tracers.
The $H_2$ column density and the cloud mass can be estimated from the intensity of the CO line emission W($^{12}C ^{16}O(J = 1 − 0)$) via the conversion factor $X_{CO}$ as discussed in \cite{bolatto}:
\begin{equation*}
    N_{H_2}=X_{CO}\cdot W (^{12}C ^{16}O(J = 1 − 0))
\end{equation*}
With the exception of the Galactic centre, in the galactic plane $X_{CO}$ has a typical value of $ \approx 2 \cdot 10^{20}(K km s^{−1})^{−1}$ and can be determined via measurements of the virial mass of the cloud, dust emission, extinction and from GeV $\gamma$-ray emission along the Galactic plane. 
The first observation of CO emission was reported by Wilson et al. (1970) \cite{wilson} observed in the galactic interstellar medium from the Orion nebula. Subsequent radio surveys in the 1970s and ’80s established molecular clouds as major constituents of the Galactic interstellar medium \cite{kawa}. CO emission was also detected in other galaxies \cite{comb}. The first complete CO survey of the Galactic plane was published by Dame \cite{dame}. Since then Galactic CO emission has been observed not only from molecular clouds and giant molecular clouds (GMCs; molecular clouds with masses of about $\geq 10^5 M_{sun}$ and sizes around 50 pc) but also as cold CO “bridges” between the spiral arms and extending beyond the galactic disk. 
\begin{figure}
    \centering
    \includegraphics[width=1\textwidth]{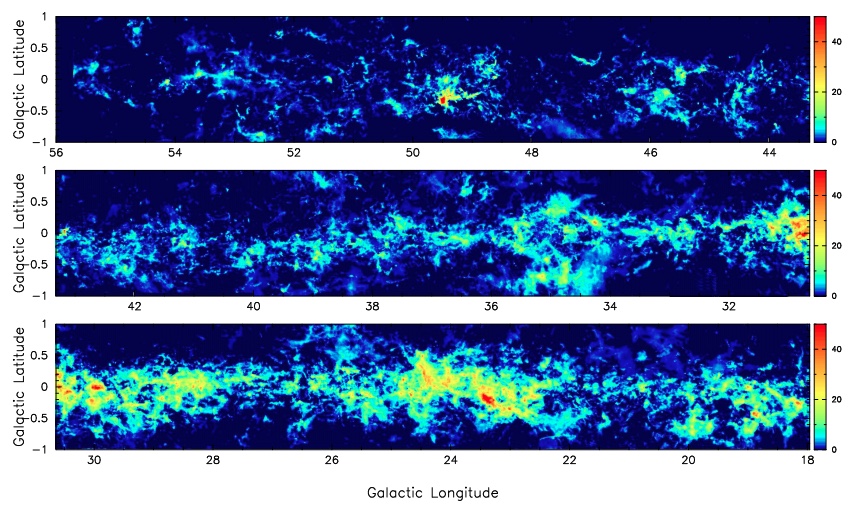}
    \caption{Integrated intensity image of GRS (Galactic Ring Survey) $^{13}CO$ line emission, integrated over all velocities   (from $V_{LSR} = −5$ to 135 $km s^{−1}$ for Galactic longitudes $l \leq 40^{\circ}$ and  $V_{LSR} =$ −5 to 85 $km s^{−1}$
for Galactic longitudes $l > 40^{\circ}$. The image shows that most of the emission is confined to $b \sim 0^{\circ}$, with concentrations at $ l \sim 23^{\circ}$ and $l \sim 31^{\circ}$. A striking aspect of the image is the abundance of filamentary and linear structures and the complex morphology of individual clouds. The image is in units of Kkm $s^{−1}$. Image taken from \cite{jack} )}
    \label{13co}
\end{figure}

Molecular clouds have no homogeneous density distribution as it is clearly visible in \ref{13co}, but a rather clotted structure with dense cores, often called clumps. This figure shows $^{13}CO$ emission from the Galactic Ring Survey \cite{jack} integrated over velocities between -5 $kms^{−1}$ and 135 $kms^{−1}$. The magnetic field within molecular clouds can be estimated, for example by observing the polarized emission of dust or using the Zeeman effect in HI, OH and CN line emission. The maximum interstellar magnetic field strength is $\approx 10 \mu G$ up to densities of $300 cm^{−3}$ and increases at higher densities following a power law with index $\approx 2/3$. Discussion about magntic field within molecular clouds and different methods for the magnetic field estimation can be found in \cite{mf}.

\begin{figure}
    \centering
    \includegraphics[width=0.7\textwidth]{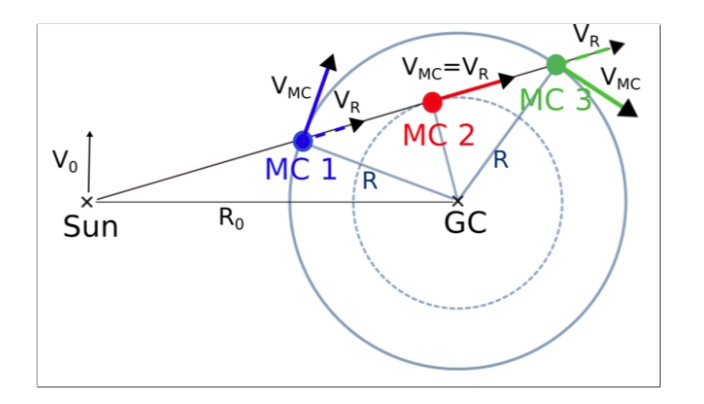}
    \caption{Three molecular clouds have different distances but the same line-of-sight velocity $V_R$. The cloud 2 is located at the tangent point where the orbital velocity $V_{MC}$ is equal to the line-of-sight velocity $V_R$. The illustration is adapted from Roman-Duval \cite{rom}}
    \label{dnubi}
\end{figure}
The distances of molecular clouds can be determined via the so-called kinematic distance method. The disk of the Milky Way is rotating and therefore the Doppler shift of the observed molecular line gives the relative line-of-sight velocity $V_R$ of the cloud. This is the projection of the orbital velocity $V_{MC}$ of the molecular cloud onto the line of sight as shown for the three molecular clouds on figure \ref{dnubi}. The galactocentric radius R, (the distance of the cloud to the Galactic centre), can be determined with rotation curves of the galaxy \cite{brand}. The distance d of the molecular cloud to the sun can be calculated, according to \cite{rom}, as
\begin{equation*}
    d=R_{0}\cdot cos(l)\pm \sqrt{R^2-R^2_{0}\cdot sin^2(l)}
\end{equation*}

with the longitude position $l$ of the cloud and the galactocentric radius $R_0$ of the sun. The molecular clouds 1 and 3 on figure \ref{dnubi} have different distances but the same line-of-sight velocity. This is represent the “near-far ambiguity” and can only be resolved with additional information, for example those obtained by looking for absorption features in HI emission as done by \cite{rom}. 
Calculate the molecular clouds mass is more delicate, since the major contributor to the cloud mass, namely H2, is practically invisible.  In order to estimate the mass, it is necessary to
rely on a tracers CO is the most used tracer of molecular hydrogen. Assuming to know precisely the conversion factor between the tracer and $H_2$ column density, then derive the mass is pretty easy. Usually it is possible to obtain the
brightness temperature vs. velocity profile $T_b = T_b(\nu)$ of the tracer from observations. The
brightness temperature is the temperature of an object that has the same brightness (e.g.
energy emitted per unit time, area and solid angle) than a black body of that temperature:
\begin{equation}
I_{\nu}=\frac{2h\nu^3}{c^2}\frac{1}{\exp(\frac{h\nu}{kT})-1}
\end{equation}
where $h$ is the Planck constant, $\nu$ is the frequency, $k$ is the Boltzmann constant and c the
speed of light. On the other hand velocity is related to physical distance as explained in the
former section. Once the temperature profile is measured, the abundance of the tracer can be
derived from the integral emissivity (W) and, under some assumptions, this quantity can be
related to the $H_2$ column density $N(H_2)$, namely the number of protons per unit area. Then
having $N(H_2)$ the mass M can be derived integrating over the physical size of the MC:
\begin{equation}
M(H_2)=2m_p \int N(H_2)dA
\end{equation}
where $m_p=1.67\cdot10^{-27}kg $ is the proton mass, and $d$A is the surface element. The factor two takes into account that the molecule is made of two protons.

\section{Passive molecular clouds}\label{gp}
Cosmic rays at GeV energies remain in the Galaxy for about $\sim 10^7$ yr \cite{ptu}. Particles got accelerated by individual sources, but lose the information of their origin and contribute to the “cosmic ray sea”. Molecular clouds are embedded in this cosmic ray sea and provide “passive” targets for cosmic ray hadrons to interact and produce $\gamma$-ray emission via neutral pion decay. The integral $\gamma$-ray flux $F_{\gamma}$ above an energy threshold $E$ depends on the cosmic ray flux, on the molecular cloud mass $M_{MC}$ and on the distance $D$ according to \cite{a}:
\begin{equation*}
    F_{\gamma}( \geq E)= 1.5\cdot 10^{-13} \bigg( \frac{E}{1TeV}\bigg)^{-1.75} \bigg( \frac{M_5}{D^2_{kpc}}\bigg) cm^{-2} s^{-1}
\end{equation*}
$M_5$ is the mass of the molecular cloud in units of $10^5 M_{sun}$, $D_{kpc}$ is the distance of the cloud in kpc and this approximation is only valid for energies much larger than 1 GeV. $\gamma$-ray emission from local molecular clouds is observed in the GeV energy range and can be used to probe the level of the cosmic ray sea at different locations of the Galaxy \cite{issa}. Passive molecular clouds have not been detected yet in the TeV energy range because of the low expected fluxes and their extension, but the future experiment CTA observatory should be capable of observing nearby clouds as explained in \cite{peda}.

\section{Molecular clouds interacting with shocks of supernova remnants}\label{mci}
Associations of supernova remnants with star-forming regions are established $\gamma$-ray sources in GeV astronomy since the 1970s \cite{mont}. These regions are embedded in molecular clouds and therefore the ambient medium provides an ideal dense target material for hadrons accelerated in the supernova remnant and $\gamma$-ray emission due to pion decay is observed. In those regions either the supernova explodes within the dense medium, where particle acceleration cannot take place efficiently due to the low shock speed, or the remnant expands into a nearby cloud \cite{aha94}. Supernova remnants that interact with molecular clouds are bright sources detected with the Fermi-LAT and AGILE satellites in the GeV energy range and are also strong in the TeV energy band where they are observed by the major ground-based Cherenkov instruments. The observed $\gamma$-ray spectrum reflects the spectrum of accelerated hadrons in the shock of a supernova remnant.
As will be discuss in \ref{intera} the interaction of the shock of a supernova remnant with a molecular cloud is usually established with observation of OH 1720MHz maser emission within the extent of the remnant \cite{frail}. This type of emission is defined by compact narrow emission lines and high effective temperatures and is produced by OH molecules colliding with $H_2$ molecules. This process takes place in the compressed and heated material behind the shock. Following Jiang et al. \cite{cav} other evidence for supernova remnant/molecular cloud interactions includes: line broadening in molecular emission or asymmetric profile, line emission with large high-to-low excitation line ratio, near infrared emission due to shock excitation and morphological agreement of the molecular emission with supernova remnant features. 
A prominent example for a supernova remnant/molecular cloud association observed both in the TeV and GeV bands \cite{28} \cite{28ag} is the supernova remnant W28. The CO emission from this region is shown in figure \ref{}. The TeV $\gamma$-ray emission observed with H.E.S.S. is indicated by the green contour lines \cite{w28radio}. The H.E.S.S. source HESS J1801–233 is coincident with a molecular cloud clearly visible in CO emission. The black circle indicates the radio size and position of W28. The $\gamma$-ray emission is spatially coincident with parts of the shell of the supernova remnant and the molecular cloud. Also the three sources HESS J1800–240 A,B and C are coincident with dense molecular material and the $\gamma$-ray emission observed there is interpreted as pion decay from accelerated particles escaped from the remnant and propagated through the interstellar medium illuminating the molecular clouds.
\begin{figure}
    \centering
    \includegraphics[width=0.7\textwidth]{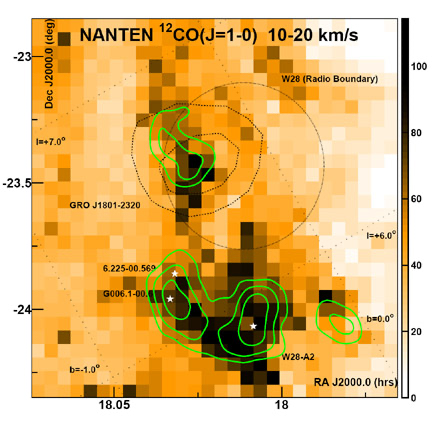}
    \caption{The W28 region. The $^{12}CO$ emission is overlaid with the 4,5,6$\sigma$ green contours of the TeV $\gamma$-ray emission from H.E.S.S. \cite{w28radio} }
    \label{dnubi}
\end{figure}

\section{Illuminated molecular clouds}\label{ill}
As said in previous sections, $\gamma$-ray production from relativistic particles can proceed through inverseCompton (IC) scattering of ambient photons by the energetic electrons, non-thermal
Bremsstrahlung (NTB) from collisions between relativistic electrons and ambient material, and the decay of neutral pions formed after the collisions of energetic protons with ambient nuclei.
\begin{figure}
    \centering
    \includegraphics[width=0.5\textwidth]{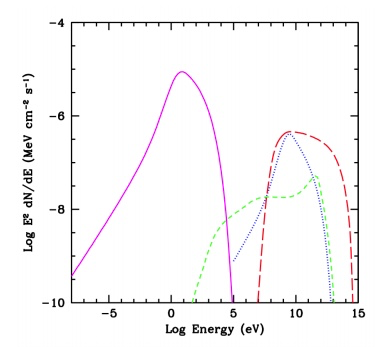}
    \caption{Simulated broadband spectrum from SNR undergoing efficient diffusive shock acceleration of electrons and protons. The solid magenta curve represents synchrotron emission, the dotted blue curve is IC emission, the dashed
green curve corresponds to NTB, and the long-dashed curve represents the $\pi^0$-decay emission.}
    \label{spettro_intro}
\end{figure}

Figure \ref{spettro_intro} presents a simulation of the broadband spectrum produced by an SNR undergoing efficient acceleration of electrons and ions. The magenta curve represents synchrotron emission from the relativistic electrons, and the dotted blue curve corresponds to IC emission produced from that same electron population up-scattering photons from the cosmic microwave background (CMB). The dashed green curve represents NTB from the relativistic electrons interacting with ambient material, and the long-dashed red curve corresponds to $\gamma$-rays from the decay of neutral pions produced by collisions of the relativistic proton component with ambient nuclei. The emission from both NTB and $\pi^0$-decay scales with the ambient density $n_0$. As a result, in high density environments such as those encountered in SNR/MC interactions, significant $\gamma$-ray emission is expected if the SNR has been an active
particle accelerator. For electron-to-proton ratio for the injected particles equal to $K_{ep} \sim 10^{−2}$, as measured locally, the $\pi^0$-decay emission dominated, making such $\gamma$-ray emission an important probe of the hadronic component of the particles accelerated by the SNR.
An important additional consideration for $\gamma$-ray production is the local photon energy density. Contributions from starlight as well as infrared (IR) emission from local dust can increase the IC emission and change the spectral shape of this component due to the different effective temperatures of these photon components. These contributions are highly dependent upon galactocentric radius. Moreover, IR emission produced by the SNRs themselves can contribute significantly to the IC $\gamma$-ray emission. Modeling of the broadband spectra from such SNRs, in order to ascertain the nature of the $\gamma$-ray emission, is complicated and has led to mixed interpretations, making the evidence for ion acceleration controversial in some cases. However, in a growing number of cases, $\gamma$-ray emission from some SNRs known to be interacting with MCs seems to clearly require a significant component from pion decay. 

\section{Signatures of SNR/MC interaction}\label{intera}
Resolved studies of SNRs and their associated clouds are most easily carried out
for systems within the Milky Way, though with the increasing capacity of new observational facilities, the studies are rapidly extending to our neighbor galaxies, as shown by the observation of 20 shocked clouds in the supernova remnant N132D in the Large Magellanic Cloud (LMC) \cite{lmd} . To date, studies have identified $\gamma$-ray emission from 37 galactic SNRs, about half of which are interacting with MCs. Of these, the evidence for energetic hadrons as the source of these $\gamma$-rays is compelling for more than 50\% of these based on energetic arguments and/or broadband spectral modeling. 
Unambiguously establish whether an SNR is physically associated
with a MC, is not trivial.  Several distinct criteria can be used to demonstrate a possible SNR/MC
interaction as discuss in depth by Slane et al. \cite{slane}. Morphological traces along the periphery of the SNRs such as arcs of gas surrounding parts of the SNR or indentations in the SNR outer border encircling
dense gas concentrations can be observed in images of SNRs.
Usually such features indicate that a dense external cloud is disturbing an otherwise
spherically symmetric shock expansion. These initial signatures need to be confirmed
with more convincing, though more rare, features like broadenings, wings, or asymmetries in the the molecular line spectra, high ratios between lines of different excitation state, detection of near infrared
$H_2$ or [Fe II] lines, peculiar infrared colors, or the presence of OH (1720) MHz masers, the most powerful tool to diagnose SNR/MC interactions. Once the association between an SNR and a molecular feature is  established, it serves to provide an independent estimate for the distance to the SNR
through the observed Doppler shift of the line and by applying a circular
rotation model for the Galaxy.
The basic tool used to investigate cases of SNR/MC interactions is the survey of
the interstellar medium in a field around the SNR using different spectral lines, is: atomic hydrogen emitting at $\lambda$ 21 cm to the dense shielded regions of molecular hydrogen emitting in the millimeter and infrared ranges. The most widely used proxy to track molecular gas is CO. This molecule has a nonzero dipole moment, and radiates much more efficiently than the abundant $H_2$ (with no dipole moment) and can be
detected easily.
\subsubsection{HI emission}
The $\lambda$ 21 cm line of atomic hydrogen is the basic resource to observe the environs of supernova remnants searching for candidate structures with which the SNRs may be interacting. However, because of the high abundance and ubiquity, confusion with unrelated gas along the line of sight represent the major problem of this method, so the detection of an HI candidate structure needs to be confirmed with other indicators that reinforce the hypothesis of association. The
HI can be studied either in emission or in absorption, the latter being a very effective tool for constraining the distance to the SNR. Moreover, the HI mapping of large
fields around SNRs is an excellent tool to explore the history of the precursor star, for example by detecting large wind-blown bubbles around the SNRs, and to estimate the gas density of the medium where the blast wave expands.

\subsubsection{Molecular emission}

\begin{figure}
    \centering
    \includegraphics[width=1\textwidth]{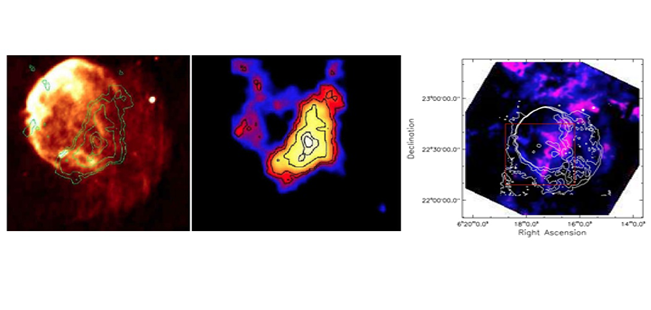}
    \caption{The SNR IC443: left: radio-continuum at 330 MHz (in grey) with
$\gamma$−ray emission from VERITAS (contours), middle: VERITAS image of very-high-energy $\gamma$-ray emission. Image taken from \cite{radiosnr} Right: $^{12}CO$ (J=1–0; blue) and $^{13}CO$ (J=1–0; red) intensity maps in the −10 km $s^{−1}$ to 10 km $s^{−1}$, overlaid with the 1.4 GHz radio continuum emission contours \cite{443co}.}
    \label{ic443comb}
\end{figure}

As mentioned above CO studies are the most widely used tool to analyze distribution and kinematics of cold, dense clouds with high molecular content. $^{12}CO$ and $^{13}CO$ lines in their different excitation states have been surveyed over most of the sky “The Milky Way in Molecular Clouds”\footnote{ http://www.cfa.harvard.edu/mmw/MilkyWayinMolClouds.html} by Dame et al., the “FCRAO CO Survey of the Outer Galaxy”\footnote{http://www.astro.umass.edu/ fcrao/telescope/2quad.html} by
Heyer et al., the “Galactic Ring Survey”\footnote{ http://www.bu.edu/galacticring/} by Jackson et al
, or the more recent “MOPRA Southern Galactic Plane CO Survey”\footnote{www.mopra.org › data} by Burton et al..
Additionally, dedicated studies using different facilities and in different molecular transitions have been conducted towards many Galactic SNRs, such as the case of IC 443. The SNR IC443 because of its location in a relatively confusion
free region of the outer Galaxy, is a text book case to analyze shock chemistry, and
as such it has been thoroughly studied in many molecular transitions The morphology consisting of two semi-circular shells with
different radii is an indication of expansion in an environment with a marked density
contrast. IC443 is also an excellent example where the interaction of the SNR with
a molecular cloud almost surely gave origin to the $\gamma$−ray emission through a hadronic
mechanism. In effect, there is an excellent concordance between VERITAS very high energy $\gamma$−ray radiation \cite{443ver} and the $^{12}CO$ J=1-0 cloud, see figure \ref{ic443comb}

\subsubsection{Masers}

\begin{figure}
    \centering
    \includegraphics[width=0.4\textwidth]{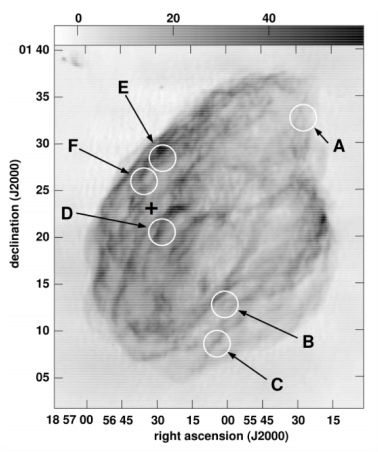}
   \includegraphics[width=0.367\textwidth]{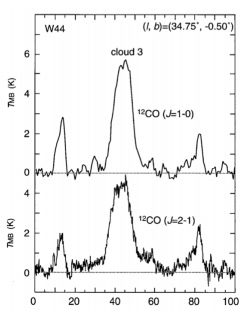}
   \caption{Left: VLA image of W44. Circles indicate the positions of observed OH maser emission, and cross
indicates the position from which the CO profile at the right is taken. Right: CO line profile from Cloud
C in field of W44, illustrating broad wings produced by an interaction between the cloud and the SNR. Image taken from \cite{slane} }
    \label{44comb}
\end{figure}

OH (1720 MHz) masers have long been recognized as signposts for supernova remnants molecular clouds interactions.
The conditions required for the maser formation are rather narrow, with temperatures in the range 50 − 125 K, densities between
n = $10^3$ − $10^5$ $cm^{−3}$, and OH column densities of $10^{16}$ − $10^{17}$ $cm^{−2}$. As the physical conditions needed to pump this maser are so strict, it has to be noted that their presence is sufficient to demonstrate interaction, but its absence does not rule it out.
According to \cite{brogan} , OH (1720 MHz) masers have been found in $\sim$  10\% of
the known SNRs in our Galaxy. 
The location of masers show, in general, very good coincidence with density/shock tracers. In the case of W44 there is a strong correlation between the morphology of the molecular gas and the relativistic gas traced by synchrotron emission at centimeter wavelengths, see figure \ref{44comb}.
In addition to providing a clear indication of SNR/MC interactions, OH(1720
MHz) masers also permit an independent estimate for the kinematic distances to
the clouds, and thus for the remnants.

\chapter{Catalog of $\gamma$-ray Supernova Remnants}
At the moment there is no one complete catalog of $\gamma$-ray (both GeV and TeV) supernova remnants that for each object gives the main physical (age, distance, positions, dimensions) and spectral information. This kind of information are the essential to classify supernova remnants or to relate the $\gamma$-rays emission mechanism with their features, but also because is the starting point to generate the spectra of synthetic population of interacting supernova remnants,  (see chapter \ref{spectra} for details). Thanks to an extensive research 37 $\gamma$-rays emitting supernova remnants have been found. The catalog was made recovering all physical data in catalogs already existing (A Catalogue of Galactic Supernova Remnants by D.A Green \cite{green}, High Energy Observations of Galactic Supernova Remnants maintained by the SNR group of the University of Manitoba \cite{unimanitoba} , TeVcat by S. Wakely and D. Horan \cite{tevcat}) and all available spectral data in the most recent literature of each instruments collaboration considered (Fermi-LAT, AGILE, HESS, MAGIC, VERITAS) and in the Fermi Large Area Telescope Third Source Catalog (3FGL, \cite{3fgl}) or in the Third Catalog of Hard Fermi-LAT Sources (3FHL, \cite{3fhl}) and in every publications containing $\gamma$-ray spectral data of supernova remnants. 

Excluding data from fermi catalogs discuss in the next paragraph, the other spectral data present in this work are taken directly from the publication using PlotDigitizer software which allows to take a scanned image of a plot and quickly digitize values off the plot. I chose to work in $\nu F \nu $ flux units so all available data was conveerted in $erg\cdot cm^{-2}\cdot s^{-1}$ and in $eV$ for energies. Table \ref{tab:strum} contains all data sets found divided by instruments, symbol "$\ast$" marks those used in the spectral fitting procedure as explained in section \ref{fit}.

\subsubsection{Fermi catalogs}\label{fermicat}
The 3FGL catalog refers to  high-energy $\gamma$-ray sources detected in the first four years of the Fermi Gamma-ray Space Telescope mission by the Large Area Telescope (LAT) in the 100 MeV–300 GeV energy range. The 3FHL catalog instead is constructed from the
first 7 years of data in the 10 GeV - 2 TeV energy range.

In the Fermi-LAT catalogs all the objects have a classification. In the Fermi catalogs there is a distinguish between associated and identified sources, with associations depending primarily on close positional correspondence and identifications requiring measurement of correlated variability at other wavelengths or
characterization of the 3FGL or 3FHL source by its angular extent (\cite{3fhl}).
Catalogs cited above were recovered from the data access section of the Fermi mission site\footnote{https://fermi.gsfc.nasa.gov/ssc/data/access/} and then  from the catalogs.fits file were extracted only the sources with the "SNR" or "snr" classifications. The "SNR" keywords refers to identified sources, instead "snr" refers to associated sources. In the 3FGL catalog the source photon fluxes are reported in five energy bands (100 to 300 MeV; 300 MeV to 1 GeV; 1 to 3 GeV; 3 to 10 GeV; 10 to 100 GeV) in $erg\cdot cm^{-2}\cdot s^{-1}$ instead of the flux uncertainties are reported in $photons\cdot cm^{-2}\cdot s^{-1} $. While in the 3FHL catalog energy limits were set to 10, 20, 50, 150, 500 GeV and 2 TeV in $photons\cdot cm^{-2}\cdot s^{-1} $ units like the flux uncertainties.

\begin{table}[ht!]

\begin{center}
\begin{tabular}{|c|ccccc|}
\hline
 SNR Name  & FERMI-LAT & AGILE & HESS & MAGIC & VERITAS \\
\hline
\hline
G359.1-0.5  &  (a) - \cite{g359fermi}$^{\ast}$ &  ---  & \cite{g359hess}$^{\ast}$ & ---  & ---  \\ 
HESS J1731-347  &  \cite{J1731fermi15} - \cite{j1731fermi18}$^{\ast}$  & ---   & \cite{j1731hess}$^{\ast}$  & --- & ---  \\ 
CTB 37B  & (b) - \cite{37bfermi}$^{\ast}$  &---& \cite{37bhess}$^{\ast}$  & --- & ---  \\ 
CTB 37A  & (a,b$^\ast$) - \cite{37afermi08} - \cite{37afermi13}$^{\ast}$  & ---   & \cite{37ahess}$^{\ast}$ & ---  & ---\\ 
RX J1713.7-3946  & (a,b) - \cite{1713fermi}$^{\ast}$  &  ---  & \cite{1713hess}$^{\ast}$ & --- & --- \\ 
SN 1006 (NE)  & \cite{sn16} - \cite{sn17}$^{\ast}$  & ---   & \cite{sn10}$^{\ast}$ & --- & --- \\ 
SN 1006 (SW)  & \cite{sn16}  - \cite{sn17}$^{\ast}$ &  ---  & \cite{sn10}$^{\ast}$ & --- & --- \\ 
G318.2+0.1  & ---   & ---   & --- & --- & ---  \\ 
RCW 86  & ($b^{\ast}$)  \cite{86fermi}$^{\ast}$  &    & \cite{86hess}$^{\ast}$ & --- & --- \\ 
G298.6-0.0  & ($a^{\ast},b^{\ast}$)   &  ---  & --- &---  & --- \\ 
Vela Jr.  & ($a^{\ast},b^{\ast}$) - \cite{velefermi}  &  ---  & \cite{velahess}$^{\ast}$  & --- & --- \\ 
Puppis A  & (a,$b^{\ast}$) - \cite{puppfermi}$^{\ast}$ &   --- & \cite{pupphess}$^{\ast}$ & --- & --- \\ 
IC 443  & (a,b) - \cite{443fer}$^{\ast}$  & \cite{443ag}$^{\ast}$  & ---  & \cite{443mag}$^{\ast}$ & \cite{443ver}$^{\ast}$ \\ 
Tycho  & (a,b) - \cite{ty}$^{\ast}$  & ---   & --- & --- & \cite{ty}$^{\ast}$ \\ 
Cas A  & (a,b) - \cite{casamag}$^{\ast}$  & ---   & --- & \cite{casamag}$^{\ast}$ & \cite{casaver}$^{\ast}$  \\ 
Gamma Cygni  & (a,b) - \cite{gcfermi}$^{\ast}$  & ---    &  ---& --- & \cite{gcver}$^{\ast}$   \\ 
Cygnus Loop  & ($a^{\ast},b^{\ast}$) -  \cite{cl}$^{\ast}$    &  ---  & ---  &---  &---  \\ 
W51 C  &  (a,b) - \cite{51f}$^{\ast}$  & ---   & --- & \cite{51m}$^{\ast}$ & ---  \\ 
W49 B  & (a,b) -  \cite{49}$^{\ast}$ &  ---  &  \cite{49}$^{\ast}$ & --- & --- \\ 
W44  & (a,$b^{\ast}$) - \cite{443fer}$^{\ast}$  &  \cite{w44-agile}  & ---  & --- & ---  \\ 
W41  & ($a^{\ast}$) - \cite{41} & ---   & \cite{41}$^{\ast}$ & \cite{41mag}$^{\ast}$ & ---  \\ 
W28 north  & \cite{28}$^{\ast}$  & \cite{28ag}   & \cite{28}$^{\ast}$  & --- & --- \\ 
W28 A & \cite{28}$^{\ast}$  & ---   & \cite{28}$^{\ast}$  & --- & --- \\ 
W28 B  & \cite{28}$^{\ast}$  & ---   & \cite{28}$^{\ast}$  & --- & ---\\ 
W28 C  & \cite{28}$^{\ast}$  & ---   & \cite{28}$^{\ast}$  & --- & --- \\ 
G349.7+0.2  & (a,$b^{\ast}$) - \cite{349}$^{\ast}$   & ---  & \cite{349}$^{\ast}$ & --- & --- \\ 
HESS J1912+101  & ---  & ---   & \cite{j}$^{\ast}$ & ---  & ---  \\ 
HESS J1534-571  & \cite{1534}$^{\ast}$  & ---   &  \cite{1534}$^{\ast}$ & --- & --- \\ 
MSH 17-39  & ($a^{\ast}$) - \cite{1739}   & ---   & --- &---  & --- \\ 
HB 21  & ($a^{\ast}$) - \cite{21}  &  ---  & --- & --- & --- \\ 
HESS J1614-518  & ---  &    ----&  \cite{j}$^{\ast}$  & ---  & --- \\ 
W30  & ($a^{\ast}$,b) - \cite{30f}  &  ---  & \cite{30h}$^{\ast}$ & --- & --- \\ 
3C 391  & (a,$b^{\ast}$) - \cite{391f}$^{\ast}$  & ---   &  ---& ---  & ---  \\ 
CTB 109  & ($a^{\ast},b^{\ast}$) - \cite{109}   & ---   &  ---& ---& --- \\ 
G337.0-0.1  & (a,$b^{\ast}$) - \cite{1739}$^{\ast}$   & ---   & --- &  ---& --- \\ 
S147  &  (a,$b^{\ast}$) - \cite{147}$^{\ast}$ & ---   & --- & --- & --- \\ 
Kes17  & ($a^{\ast},b^{\ast}$) - \cite{k17}  & ---   & --- & --- & --- \\ \hline

\end{tabular}
\caption{All data and catalogs data considered in this work, (a) 3FGL, (b) 3FHL.  Data actually used in the fitting procedure are marked with "$\ast$" symbol } 
\label{tab:strum}
\end{center}
\end{table}

\section{Distance and Age of supernova remnants}
Reliable distances for the Galactic supernova remnants are crucial in order to obtain the other basic parameters such as for example size, age and the explosion energy. Rarely for the supernova remnants there are unique information about distance and age, especially because of different measurement method. Distances and ages used in this work are taken from literature, whose results are reported in table \ref{da}. For each object there is an indication of the minimum and maximum distance and age and the exact value used in this work.  
In some cases the the value chosen is not the mean value, this is because there is strong evidence or a more recent observation which favors a certain measure in the possible range of values.

In sections \ref{distance} and \ref{age} are described the main ways to calculate distances and ages of these sources.

\subsection{Distance}  \label{distance}
\subsubsection{The surface brightness - diameter relation }
There is evidence that there should be a relationship between the surface brightness ($\Sigma$) and the diameters of a supernova remnants ($D$). This relation can be written as $\Sigma=aD^{\beta}$. However, the validity of this relation has not yet been universally accepted. 
This method is not very precisely in fact as reported in \cite{case} the associated errors can reach 40\% of the measure. The reasons are related to:
\begin{itemize}
    \item The wide range of supernova explosion energies
    \item The presence of interstellar medium around the remnant, normally higher densities led to higher $\Sigma$
    \item The presence of a neutron star within the remnant
    \item Calibration uncertainties of distance and diameters
\end{itemize}
The analysis of 36 Galactic shell SNRs by Case et al. suggest that \cite{case} $\beta = -2.38$, but more recent analysis by Pavlovic et al. \cite{pablo} set $\beta = -4.8$,  significantly steeper than those in previous works. 

\subsubsection{The kinematic distance}
\begin{figure}
    \centering
    \includegraphics[width=.7\textwidth]{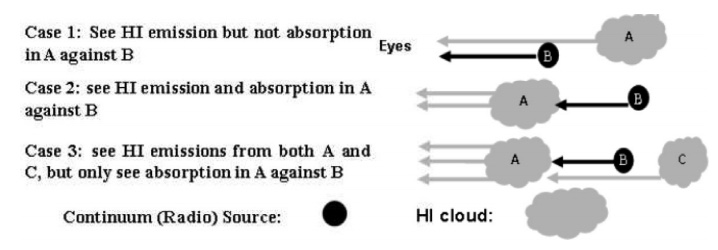}
    \caption{Three cases of relative positions between continuum source and HI clouds}
    \label{kin}
\end{figure}

Kinematic distance measurement is based on rotation curve of the Milk Way. For objects that follow the rotation curve well, we can derive their distances from the observed radial velocity (e.g. HI clouds). 21 cm line observation toward SNRs is used to build HI absorption spectra. Cold gases
between observer and SNRs will lead to absorption, gases behind an SNR will not be seen
in this SNR’s absorption spectrum (see figure \ref{kin} ). Therefore, the HI absorption features
are used to give a lower distance restriction and the emissions without absorptions are
used to derive an upper distance restriction. For SNRs showing continuum emission
and absorption features, their positions are given by the velocity at which the continuous
absorption stops.

\subsubsection{Other methods}
Kassim et al. \cite{kassim} found that X-ray observations could be used to calculated the Sedov distance of shell type SNRs which are in the adiabatic expansion phase using:
\begin{equation}
D_s = 8.7 \cdot 10^6 \epsilon^{0.4}_0 P(\Delta E,T)^{0.2} \theta^{-0.6} F^{-0.2}_{x0}T^{-0.4}
\end{equation} 
Here $\epsilon_0$ is the initial energy of the SNR explosion in unit of $10^{51}$ ergs,$\theta $ is the observed angular diameter of the SNR shell in arcminutes, $F^{-0.2}_{x0}$ is the measured x-ray flux corrected for interstellar absorption in units of $ergs s^{-1} cm^{-2}$, T is the measured thermal temperature of the x-ray emitting gas in K and function $P(\Delta E,T)$ described the power emitting by hot electrons in a low-density plasma via Bremsstrhalung emission.
The uncertainties in determining $D_s$ are dominated by the following assumption:
\begin{itemize}
\item the SNR is in te adiabatic expansion phase
\item the measured x-ray temperature gives a reliable estimate of the SN shock velocity
\item $\epsilon_0$ = 1 is approximately corrected for all the supernova 
\end{itemize}
Another possibility is that some SNRs are associated with distance-known objects, such as pulsar, HII region, CO cloud.
Errors in this method are caused by not only the distance’s errors of the associated objects, but also the uncertainty of such association.
\subsection{Age} \label{age}
There are mainly five ways to estimate the age of an SNR:
\begin{itemize}
    \item 1. If a supernova is associated with a neutron star, it's age can be estimated by using the pulsar characteristic age derived from: $t=P/2\Dot{P}$, where P is the rotation period and $\Dot{P}$ is the rate of change of period.
    \item 2. The age of a SNR can be estimated using the total explosion energy (E) and the energy loss rate $(\Dot{E})$ according to $t=e/\Dot{E}$.
    \item 3. Knowing the diameter (D) or equally the radius (R) of a supernova remnants and the expansion velocity ($v_s$), the dynamical time can be calculated as $t=CR/v_s$, where C is a constant that depends on the evolution phase in which the remnant is (2/5 for the adiabatic expansion phase, 2/7 for the radiative remnant and 1 for a freely expansion remnant) and on the expanding velocity of the shock wave.
    \item 4. For SNRs with a known radius (R) and thermal temperature (T) measured from X-ray observation, the age can be obtained by
    $t=3.8 \cdot 10^2 R_{pc} (kT)^{-1/2}_{keV}$ yr derived from the Sedov Relation $R=(2.026E/ \rho)^{1/5}t^{2/5}$.
    \item 5. For SNRs that already have in their radio spectra the characteristic break at frequency $\nu_d$ due to synchrotron losses in a magnetic field B, the age can be derived using the relation $t=40000B^{-1.5}\nu_d^{-0.5}.$  

\end{itemize}
The ages estimated with these methods have uncertainties up to an order of magnitude. There are a few supernova remnants with associated pulsars, so the first method can be use for a very limited number of sources. For some SNR information about R and $v_s$ are available so the age estimations have a good accuracy. When this parameters are not available, the last two method can give a plausible indication or limits on the age.

\begin{table}[ht!]
\begin{center}
\begin{tabular}{|cccccc|}
\hline

 \multirow{2}*{SNR Name} & $D_{min}$ - $D_{max}$ & $D$ & $A_{min}$ - $A_{max}$ & $A$ & Ref.   \\ 
   & (kpc) & (kpc) & (kyr) & (kyr) & \\         
\hline
G359.1-0.5  & 5 - 8-5 & 7.6   & -  & > 10   & \cite{yus} \cite{aha08} \\ 
HESS J1731-347  & 3.2 - 5.2  &  3.2   & 2 - 27 & 14.5 &\cite{fuku} \cite{guo}\\ 
CTB 37B  & 8.3 - 11.3  & 13.2  & 0.6- 6 &  5 &\cite{blu} \\ 
CTB 37A  & 6.3 - 9.5   &  9  & 1 - 100 &  16  & \cite{blu}\\ 
RX J1713.7-3946  &  -  & 1 & 1.6 - 2  & 1.6 & \cite{fukui} \cite{tsuji} \\ 
SN 1006   &   -   & 1.6 & - & 1 & \cite{wink} \\ 
G318.2+0.1  &  -   & 9.2 & - & 8 & \cite{hofv}  \\ 
RCW 86  & 1.2 - 2.5  &  2.5   & 1.8 - 10 & 1.8 & \cite{bocc} \cite{ajello} \cite{shan} \\ 
G298.6-0.0  &   5 - 10 & 5 & - & 1 & \cite{bamba} \\
Vela Jr.  &  0.5 - 1   & 1 & 2.4- 5.1 & 3& \cite{allen}\\ 
Puppis A  & 1.3 - 2.2  & 2.2 &  3.7 - 5.2  &  4.4 & \cite{rey} \cite{becker} \\ 
IC 443  &   0.7 - 2  & 1.5 & 3 - 30& 10&  \cite{yu} \cite{lee} \\ 
Tycho  &  2 - 4.5   & 4.5  & -  &  0.4 & \cite{yato} \\ 
Cas A  &   3.3 - 3.9  & 3.4 &  & 0.3 & \cite{ala} \cite{fe06}\\ 
Gamma Cygni  & 1.5 - 2.6      & 1.5 & 6- 7& 6.6 & \cite{clump} \cite{Uchiyama} \\ 
Cygnus Loop  &   0.8 - 10  & 0.8 & 10 - 20 & 15 & \cite{Floop} \\ 
W51 C  & 4 - 6  &  5.6 & 16 - 30 & 30 & \cite{lea} \cite{rana} \\ 
W49 B  & 8 - 14  &  8   & 1 - 4  & 2 &\cite{Z49} \cite{lea} \\ 
W44  & 1.9 - 3.3  &   3  & - & 20 & \cite{w44-agile} \cite{K} \\ 
W41  &  3.9 - 4.8 &  4.8    & 60 - 200 & 100 &\cite{rana} \cite{tia} \\ 
W28  &  1.8 - 3.3  &  1.9 & 35- 150 & 35 & \cite{clump} \cite{28ag}\\ 
G349.7+0.2  &  18.3 - 22.4      & 11.5  &  & 1.8 & \cite{ti} \cite{frail}  \\ 
HESS J1912+101  &  -  & 4.1& 70 - 200 & 100 & \cite{su}\\ 
HESS J1534-571  & 3.5 - 6  & 3.5 & - & 10 &\cite{maxted} \cite{raya} \cite{sa} \\ 
MHS 17-39  & -  & 11.8 & -   & - & \cite{brog} \\ 
HB 21  & 0.8 - 1.7   & 1.7   & 5.6 - 45 &  15 &\cite{ambrogio} \cite{kooo} \\ 
HESS J1614-518  & 1.2 - 5.5 &  1.5    & - & 30 & \cite{sakai}\\
W30  & 3.2 - 6   &  4.6   &  15 - 28  & 20 & \cite{k30} \cite{aje} \\ 
3C 391  & 5.4 - 7.2  &  7.1& 4 - 9  & 9 & \cite{lea} \cite{pat} \\ 
CTB 109  &  2.8 - 7.5 &  3.2  & 8.8 - 24 & 14 &\cite{Ko} \cite{cru}  \\ 
G337.0-0.1  &  - &  11  & - & 5 & \cite{frail} \cite{corbel}\\ 
S147  & 0.9 - 1.6  & 1.39  & 30- 100   & 30 &\cite{karl} \cite{D} \\ 
Kes17  & -  &  9.7  & 2 - 40& 14 &\cite{G17} \cite{2G17} \\ 
\hline

\end{tabular}
\caption{Distances and ages of all the $\gamma$-ray emitting SNR.} 
\label{da}
\end{center}
\end{table}

\begin{figure}
    \centering
    \includegraphics[width=.4\textwidth]{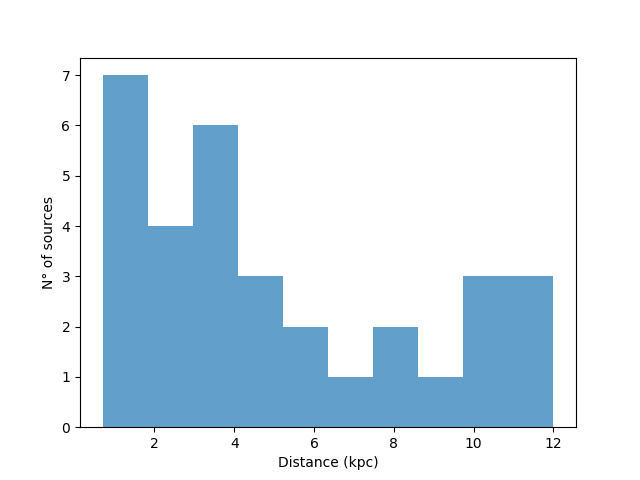}
    \includegraphics[width=.4\textwidth]{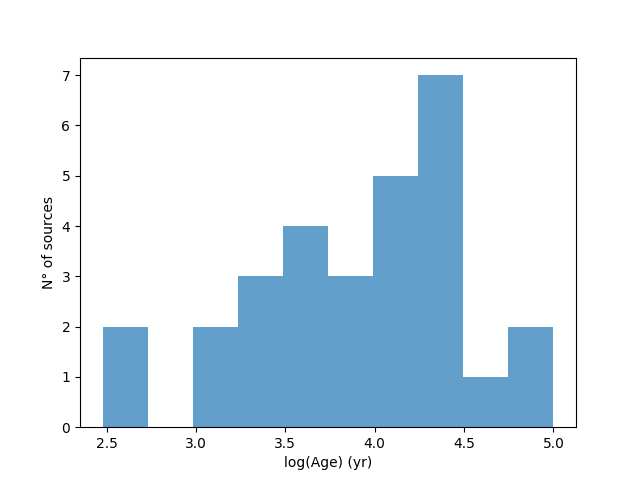}
    \caption{Histogram of distance (left) ad age (right) of all the SNR.}
    \label{fig:hist_ad}
\end{figure}

In figure \ref{fig:hist_ad} are plotted distribution of distance and age.

\section{Spectral fitting} \label{fit}
Supernova remnants spectra are obtained joined together all the available data. To eventually introduce a classification based on some spectral  properties, was used a model independent approach  for the emission mechanism, meaning use a unique model for all the supernova remnants.
Some of the known interacting supernova remnants (like W44 \cite{agile}, W51 C \cite{w51c}, IC443 \cite{w44-fermi} ) have in their $\gamma$-ray spectrum the so called "pion bump", so their emission is certainly hadronic. Reasonably 
also the other interacting supernova remnants emits $\gamma$-rays in the same way, therefore the $\gamma$-ray emission was fitted using an hadronic model for all the SNRs considered.
Obviously the hadronic model isn't a good model for all the supernova remnants but, as said this choice excludes the possibility of introduce a artificial classification depending on the emission model.
 
In order to pass from spectral points to the pion decay spectrum Naima \cite{naima} was used, a Python package developed with the aim of computing the non-thermal radiative output of relativistic particle distributions (protons in this case).
The systematic method discuss above, means, in this case, a broken power law (BPL) distributions for the protons populations and the Pion decay radiative model for all sources.
Naima can derive the best-fit and uncertainty distributions of spectral model parameters through Markov Chain Monte Carlo (MCMC) sampling of their likelihood distributions \cite{emcee}. The measurements and uncertainties in the provided spectrum are assumed to be correct, Gaussian, and independent Under this assumption, the likelihood of observed data given the spectral model $S(\Vec{p},E_i)$, for a parameter vector $\Vec{p}$, is:

\begin{equation}
\mathcal{L}= \prod_{i=1}^N\frac{1}{\sqrt{2\pi\sigma_i^2}}\exp{(-\frac{(S(\Vec{p},E_i)-F_i)^2}{2\sigma_i^2})}
\end{equation}
where ($F_{i}$,$\sigma_i$) are the flux measurement and uncertainty at an energy $E_i$ over $N$ spectral measurements. The $\ln{\mathcal{L}}$ function in this assumption can be related to the $\chi^2$ parameter as $\chi^2=−2\ln{\mathcal{L}}$, so that maximization of the log-likelihood is equivalent to a minimization of $\chi^2$.
Since for some sources there is more than one data set referred to the same energy range ( 100 MeV-10 GeV, 10 GeV-1 TeV, > 1 TeV bands), to be sure to consider data points with their appropriated weight was choosen one set for energy band, in order to not influenced the maximum likelihood fitting procedure. 
Below is reported an example of the fitting procedure used for all the supernova remnants starting from the choice of the data set.

\subsection{Example of W44} \label{w44}
W44 is a middle-aged supernova remnant ($\sim 2 \cdot 10^4 $ years old) located at $\sim 3 kpc$ from Earth in the Galactic Plane. Since this remnant is one of the more studied there are multiwavelenght observations that revealed interesting features. In the Radio band it is characterized by an elliptical shell due to expansion in an inhomogeneous ISM. The OH maser emission revealed that W44 is interacting with a molecular cloud in the south-est region.

\begin{figure}[ht!]
\centering
      	\label{dist}
        \includegraphics[width=0.7\textwidth]{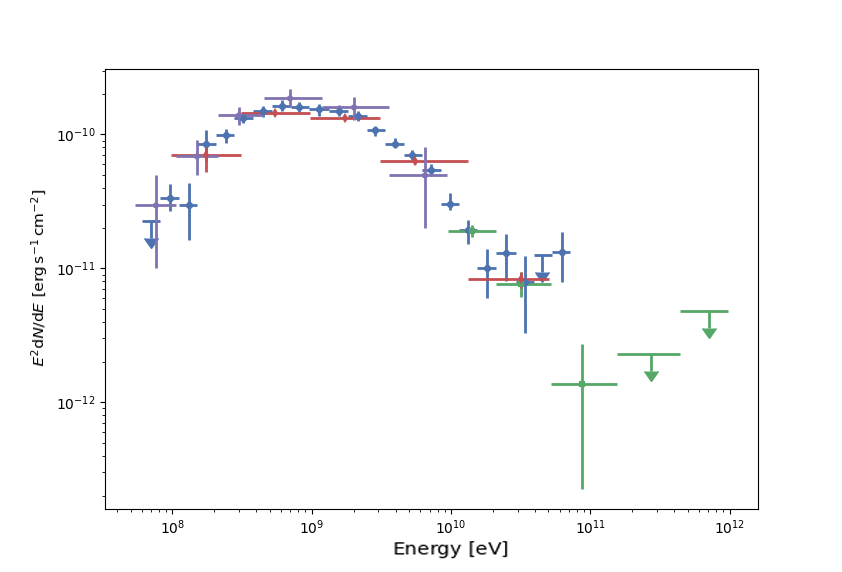}
        \caption{W44 spectral data. Blue points refers to 4 years of observations with the Fermi-LAT telescope (4 August 2008 to 16 July 2012)  \cite{w44-fermi}, violet is for AGILE \cite{w44-agile} instead red and green points are for the Fermi 3FGL and 3FHL data.}
        \label{fig:w44_data}
\end{figure}
The figure \ref{fig:w44_data} show all the data set eventually available, 3 data set are almost in the same energy band. I used Fermi data extract in the analysis of Ackermann et al. \cite{w44-fermi} (blue points) because of better band sampling and moreover the data has a smaller error bars associated than those from the systematic analysis of the 3FGL Fermi catalog. The 10 GeV-1 TeV band is sampled only in the 3FHL catalog, no other very-high-energy instrument has detected W44 above 100 GeV up to now.
As initialization of Naima, I set the broken power law(BPL) as protons distribution and the $\pi^0$-decay as radiative model.
In addition, it is required the prior function, i.e., a function that encodes any (eventually) previous knowledge about the parameters, limit the amplitude and the energy break to be positive, and spectral indices to be
between -1 and 5. In appendix are are reported the models and the prior function definitions code sections

\begin{figure}
\centering
\includegraphics[width=.7\textwidth]{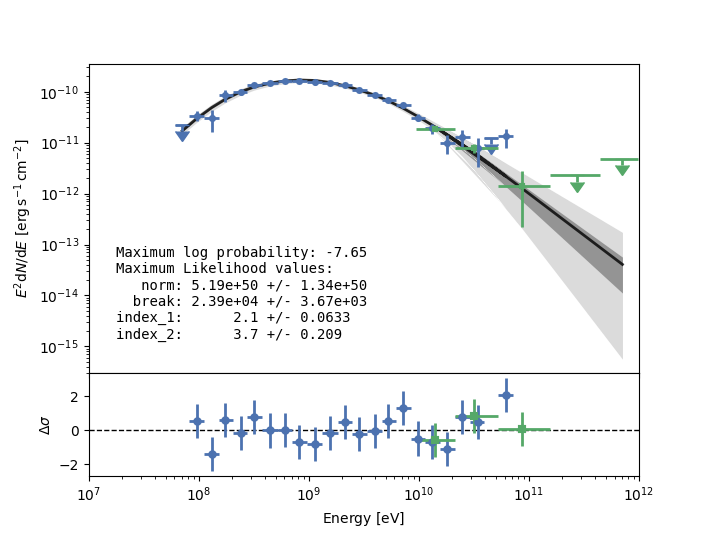}
\caption{W44 fitted spectrum}
\label{a}
\end{figure}

\begin{figure}
\centering
\includegraphics[width=.7\textwidth]{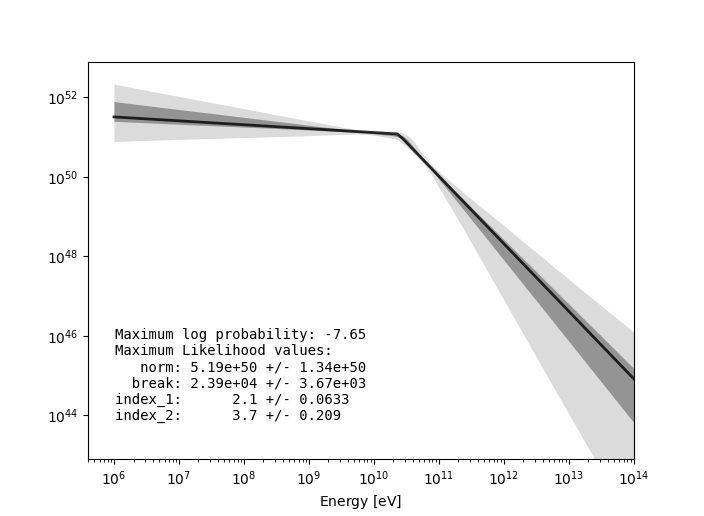}
\caption{Protons population of W44}
\label{b}
\end{figure}

After define the array with the initial parameters estimates, the MCMC procedure start. The maximum likelihood optimization takes about 15 minutes (for each SNR), it depends mainly on how much the initial BPL parameters are close to the "real" one, and on the $nwalkers$, $nburn$, $nrun$ parameters described in detail in \cite{emcee}.
Fitting results can be viewed both through plots that illustrate also the stability of the procedure and through a table (in ecsv format) that contains all the run information, the parameters array of the protons population of the best fit and the maximum value of the negative log-likelihood.
Results for W44 are reported in \ref{a} and \ref{b}.


\section{Catalog}
\begin{figure}[ht!]
    \centering
    \includegraphics[width=\textwidth]{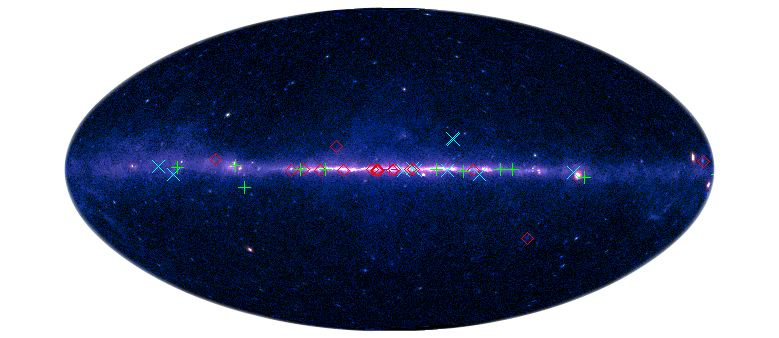}
    \caption{Full sky view of the catalog SNRs. Red rhomb is for interacting supernova remnant, cyan cross is for shell-like SNR and green cross are for undefined type. Image was realized using Aladin software and Fermi-LAT data up to 2016 \cite{skymap}.}
    \label{fig:sky}
\end{figure}

All the $\gamma$-ray supernova remnants considered and their position in the Galaxy are shown in figure \ref{fig:sky}, red rhomb marks the interacting supernova remnants, cyan cross is for the shell-like SNR, and the green plus is for the undefined SNR. The procedure in section \ref{w44} was reproduced for all the supernova remnants obtaining the parameters of the protons distribution that is responsible (together with the proton target density fixed at 1 $proton/cm^{-3}$ ) for the neutral pion decay emission.
Having all spectra allows to get the integrated flux over the entire $\gamma$-ray band (from 100 MeV to 10 TeV ) as over any other band, for my purpose, two important band are 0.1-10 TeV and 1-10 TeV. 
Catalog of gamma emitting supernova remnants is reported in table in appendix. Below is reported a brief catalog columns description.

\begin{enumerate}
    \item Name: Supernova remnant common name
    \item Type: INT is reserved for supernova remnants that in the TeVCat or in the Fermi SNR catalog \cite{snrcat} are considered interacting with a molecular cloud. Instead SHELL is the tag for shell like supernova remnants according to the TeVCat classification. No tag means that the classification is no obvious.
    \item Other Names: Other known names
    \item TeVCat: SNR name (eventually) stores in the TeVCat
    \item l: Galactic Longitude in degree unit
    \item b: Galactic Latitude in degree unit
    \item RA: Right Ascension, J200 in degree unit
    \item DEC: Declination, J200 in degree unit
    \item d:  Distance of SNR in kpc
    \item Age: SNR age in years
    \item MC MASS: The molecular cloud mass if the supernova is interacting, in solar mass unit
    \item Flux100MeV: Integrated flux from 100 MeV to 10 TeV, in $\mathrm{erg\,s^{-1}\,cm^{-2}}$
    \item Flux100GeV: Integrated flux from 100 GeV to 10 TeV, in $\mathrm{erg\,s^{-1}\,cm^{-2}}$
    \item Flux1TeV: Integrated flux from 1 TeV to 10 TeV, in $\mathrm{erg\,s^{-1}\,cm^{-2}}$
\end{enumerate}

\section{Some catalog results}
\begin{figure}[ht!]
    \centering
    \includegraphics[width=0.7\textwidth]{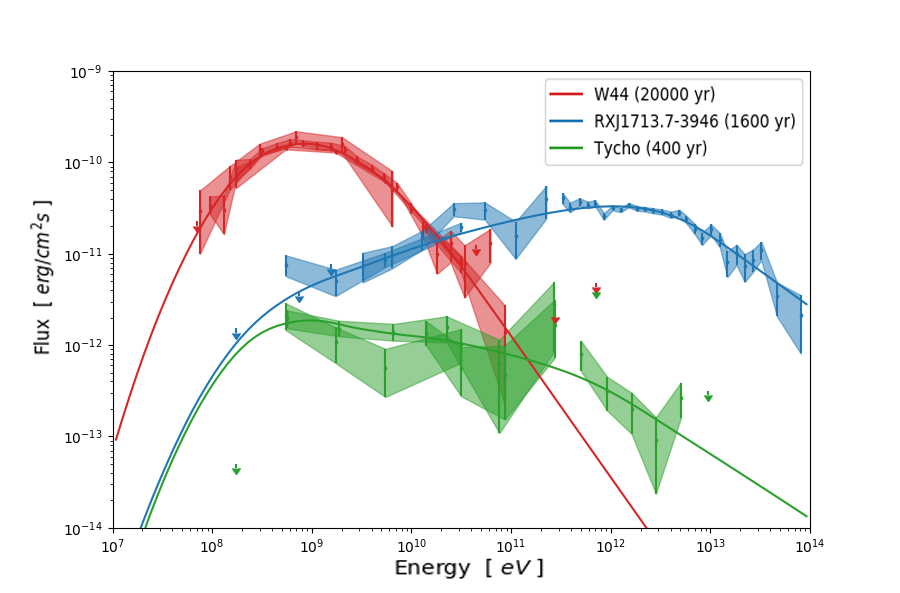}
    \caption{Spectra of SNRs W44 (red), RXJ1713.7-3946 (blue) and Tycho (green). All points available are plotted see table \ref{tab:strum}, shades indicate the $1\sigma$ error region.}
    \label{fig:funk}
\end{figure}

This $\gamma$-ray  supernova remnants catalog contains all the relevant data to compare the real population of SNR with the synthetic one results from the model presented in chapter \ref{pop}. 
Furthermore I could verify the theory according to which supernova remnants are divided into "Shell-like" SNR that are typically young object ($\sim 10^2-10^3 yr$), expanding in a relatively low density medium, with $\gamma$-ray emission morphology correlated with the radio and X-ray shell, and the interacting SNRs, older ($\sim 10^3-10^4 yr$) mixed morphology sources, that interact with giant molecular clouds and with a emission morphology
that correlates better with MC than with the supernova shell. This classification must be evident in some 
aspects of the SNR population, for example in the age  distributions, in the spectral shape and in the integrated flux distributions.
In my catalog there are all the useful elements to check this classification. In figure \ref{fig:funk} the spectral shape of the main exponent of the two classes is shown, any other object which belongs to this classes follows the  behavior of the example reported in figure \ref{fig:funk}. The W44 supernova remnant is one of the brightest SNRs on the GeV sky, the characteristic low-energy cutoff in the energy spectrum (the pion bump) has been detected. This observation demonstrates that the $\gamma$-ray emission in the GeV band is dominated by pion decay. The shell-like class comprises young SNRs that are typically less luminous at GeV energies, have harder spectra, and are often also detected at TeV energies. The most prominent shell-like SNR that have
been detected both at GeV and TeV energies is RXJ J1713.7-3946. The absence of thermal
X-ray emission suggests that in this region the is too low density target material for interaction of protons potentially
accelerated by the SN shock to produce neutral pions. Furthermore, the rather hard photon index in the GeV band, strongly indicates that the emission of this SNR has a leptonic origin. 
The Tycho SNR instead belong to a class of supernova remnant that haven't a clear classification.

\begin{figure}[ht!]
     \centering
     \begin{subfigure}[b]{0.4\textwidth}
         \centering
         \includegraphics[width=\textwidth]{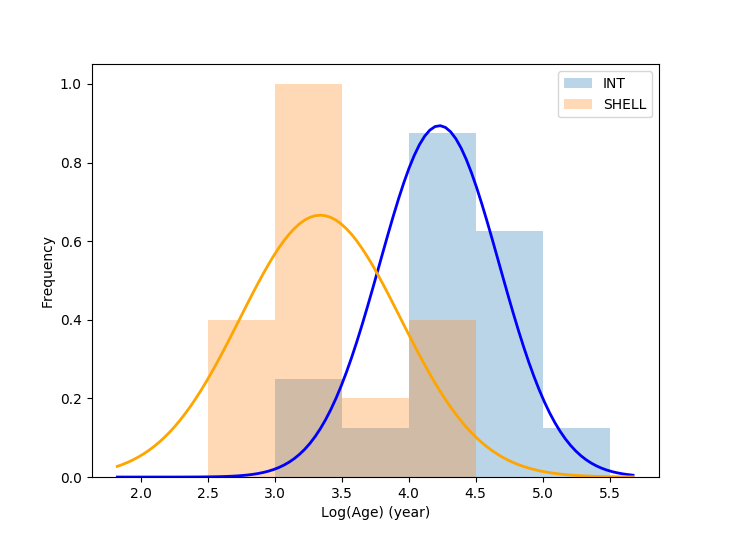}
         \caption{}
         \label{fig:age}
     \end{subfigure}
     \hfill
     \begin{subfigure}[b]{0.4\textwidth}
         \centering
         \includegraphics[width=\textwidth]{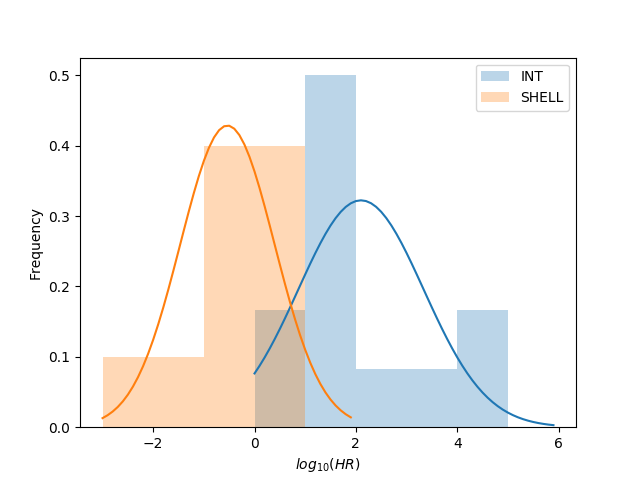}
         \caption{}
         \label{fig:hr}
     \end{subfigure}
     \hfill
        \caption{Histogram of the age of the supernova remnants with Gaussian fit superimposed. Orange and Blue color represent shell-like and interacting SNR respectively.}
        \label{fig:resw44}
\end{figure}

I could test if the division between interacting and shell-like supernova remnants based mainly on other wavelengths observation is reflected also in some other features. For example  that interacting supernova remnants are older than the shell like because of accelerated protons takes some time (it depends on how far is the MC and on the acceleration of the particles) to reach the molecular cloud target protons. The shell-like supernova remnants emission instead is due to electrons inverse compton on CMB local photons.
The age distributions of the two classes are reported in figure \ref{age}.
Further indication that the two classes have a different $\gamma$-ray emission mechanism came from the ratio between the flux value at 1 GeV and 1 TeV (Hardness ratio  HR ). As on can see in figure \ref{fig:hr}, it's hard to explain this hardness ratio distribution (that reflects the spectral shape) with only one emission mechanism. As before the   the two classes are not compatible at 10\% significance level.

\begin{figure}[ht!]
    \centering
    
    \includegraphics[width=1\textwidth]{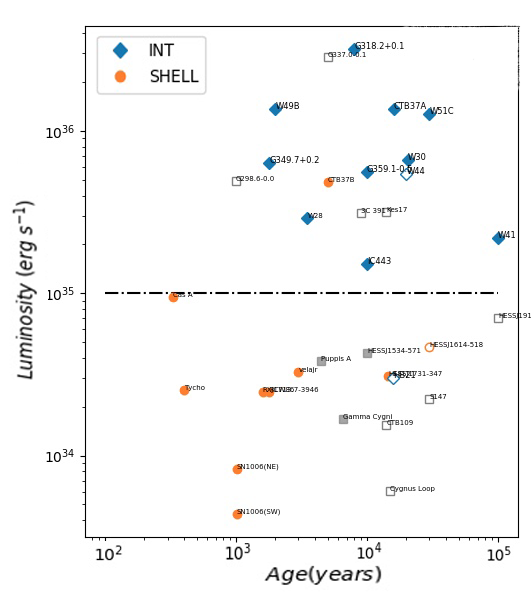}
    \caption{Luminosity (100 MeV-100 TeV) in units of $erg \, s^{-1}$ plotted as function of the age of SNRs (left) and of the hardness ratio (right).
    Blue color is for interacting supernova remnants, orange points are shell-like SNR and gray square are the undefined ones. Solid color meaning that available data cover almost all the energy band.}
    \label{fig:lumage}
\end{figure}

In figure \ref{fig:lumage} are investigated correlations between  the $\gamma$-ray  luminosity $L_{\gamma}= f_{100MeV-100TeV} \cdot 4 \pi D^2$ and the age and between the luminosity and the hardness ratio (HR). Solid points represents object that have a data coverage both on the GeV and on the TeV energy band. This plot show that all the interacting supernova remnants have luminosity above $10^{35} \, erg\, s^{-1} $ and are in general older. The only  exception is HB 21 that has $L_{\gamma}\sim 3\cdot 10^{34} \, erg \,s^{-1} $, but this discrepancy is probably due to the lack of TeV data, that will results in underestimate the $\gamma$ luminosity.
Another interesting features is that all the iSNRs have HR $>1$, so this objects have a spectrum with the dominant contribution to the flux at GeV energy band, like W44. The Shell-like supernova remnants with HR $>1$ are those have a "flat" spectrum similar to the Tycho SNR spectrum (see figure \ref{fig:funk}). 
Note that for the W28 supernova remnants in figure \ref{fig:lumage} (left) was considered the luminosity of the whole source, instead in \ref{fig:lumage} (right)  the source was divided in W28 north and W28 south . The HR value of W28 south (source A+B+C) is the ratio of the sum of the flux of the three sources at 1 GeV and the sum at 1 TeV.

\begin{figure}[ht!]
    \centering
    
    \includegraphics[width=0.5\textwidth]{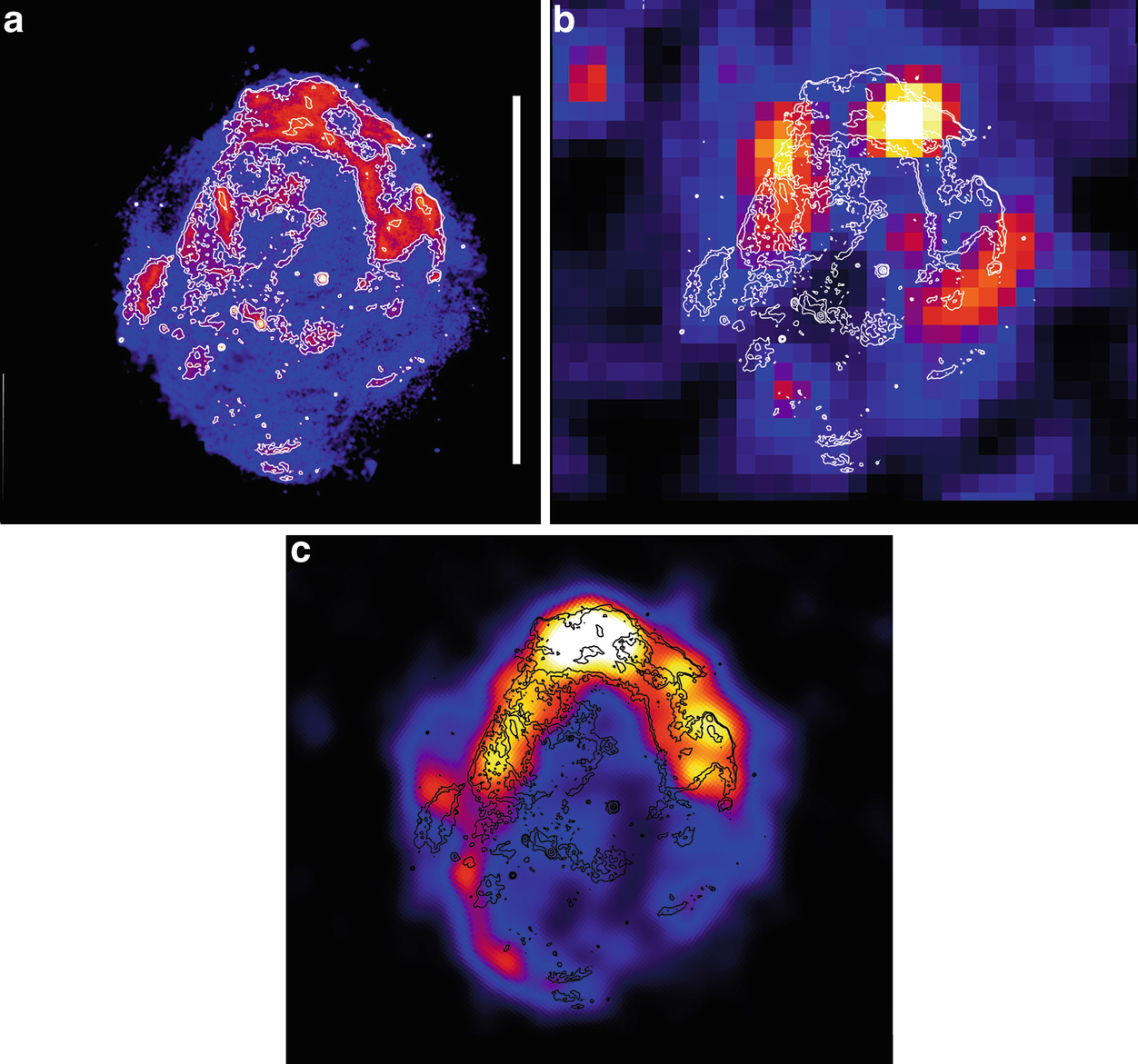}
    \label{fig:rxj1713}
    \includegraphics[width=0.4\textwidth]{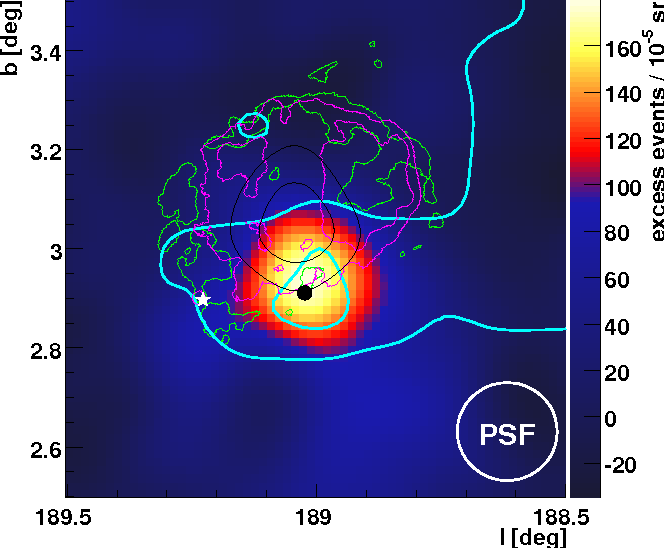}
    
    \caption{Left: Multiwavelength data for supernova remnant RX J1713.7-3946 in Galactic coordinates. The white bar is 1$\deg$ long. Data are taken from (a) XMM-Newton observations, (b) GeV $\gamma$-rays observed by Fermi-LAT (source class, 6.5 years of data), and (c) H.E.S.S. observations. Right: IC 443 MAGIc $\gamma$-ray map. Overlayed are 12CO emission contours, that trace the molecular cloud (cyan), contours of 20 cm VLA radio (green), X-ray contours from Rosat (purple) and  $\gamma$-ray contours from EGRET (black) .
    Image credits: \cite{funk} (left) and \cite{magic} (right).}
    \label{fig:ic443}
\end{figure}

All the features presents in this section confirm the scenario that there supernova remnants are divided in two classes. The shell like supernova remnants that are younger and have a low intrinsic luminosity furthermore the spectra don't have the pion bump features (like RXJ 1713.7-3946). The $\gamma$ ray and radio map of this kind of object indicates that the emission is rather symmetric and came from the ejected material. All this considerations indicates that the emission mechanism is leptonic. 
The interacting supernova remnants, object of interest of this work, instead have a spectrum that can be well explain with the hadronic emission, moreover in some case the presents of the pion bump is clearly evident. The evidence that this object are brighter and older and the fact that the hadronic mechanism needs target protons can be explain with the explosion of a supernova close to a molecular cloud. The shock accelerated protons (Fermi mechanism), takes some time to reach the MC that have density of hundreds $proton \cdot cm^{-3}$. This results in $\gamma$-ray brighter sources with a delayed emission compares to the shell-like supernova remnants.
This scenario is confirmed looking at the $\gamma$-ray emission map of the sources, infact the $\gamma$-ray emission  is well correlated with the molecular cloud (figure \ref{fig:ic443} right) instead of with the supernova shell.


\section{The LogN-LogS of real interacting supernova remnants}

\begin{figure}[ht!]
    \centering
    \includegraphics[width=0.8\textwidth]{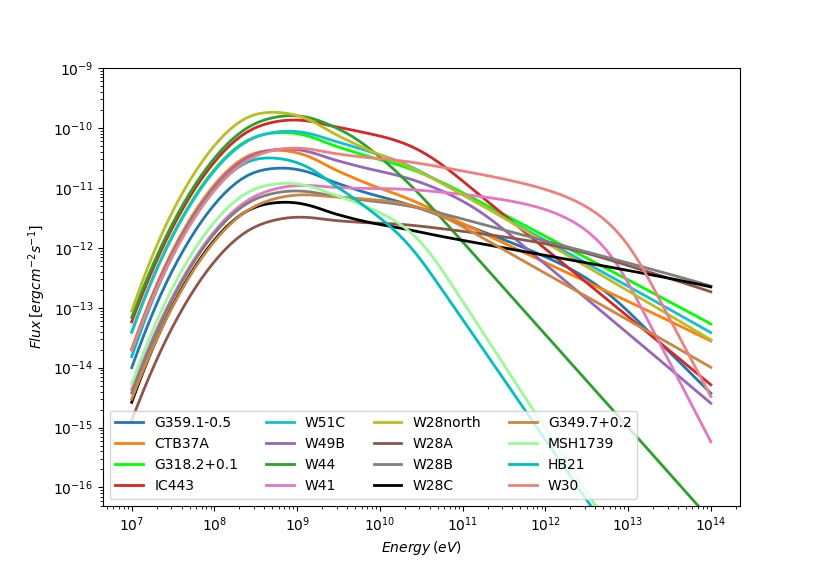}
    \includegraphics[width=0.8\textwidth]{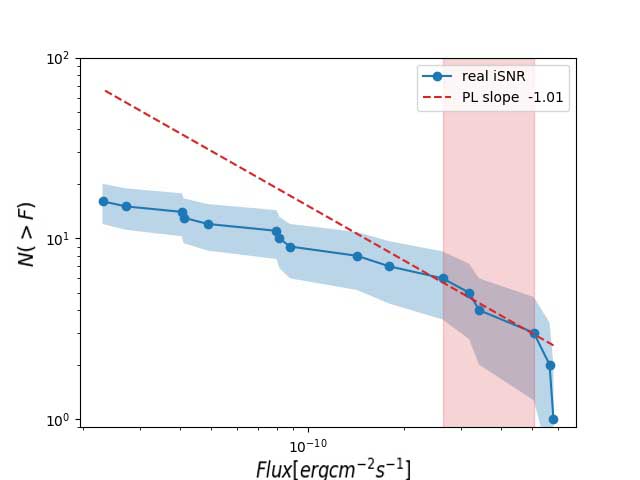}
    \caption{Top: spectrum of all the interacting supernova remnants. Bottom: the LogN-LogS curve of the real interacting supernova remnants. Shade blue represent the $\pm \sqrt{n}$ error region.
    Red region identify the  completeness region.}
    \label{fig:lognlogs}
    \end{figure}

The main target of this work is to create a population model for the interacting supernova remnants and give an estimate of how many of this sources CTA can be identify. 
The strategy is to realize the LogN-LogS of the real interacting supernova remnants and compare it with those obtained from the population model presented in \ref{pop}, and then extrapolate the iSNR number at the CTA flux sensibility.
The way to check if the resulting model, is consistent with what observed by the major experiments is to compare the LogN-LogS curve of the real SNRs (figure \ref{fig:lognlogs}) with that of the synthetic population. 
The LogN-LogS curve was obtained by integrating all the interacting SNRs spectra above 100 MeV. The associated error is the Poisson standard deviation, $\pm \sqrt{n}$. The features of the LogN-LogS that the population model of interacting supernova remnants have to reproduce is the slope in the region in which the sample is complete (red region in figure \ref{fig:lognlogs}).
Assuming a population uniformly distributed in the galactic plane (thin disk), thanks to a Monte Carlo simulation was found that the slope of the LogN-LogS is equal to -1. Therefore where the slope of the real LogN-LogS curve deviates from -1, the sample is incomplete.
To calculate where the sample is complete, I fixed one real iSNR and I fitted with a power law the LogN-LogS between any other iSNRs, repeating for each point was obtained that the best accordance is $-1.00998$. The red region in figure \ref{lognlogs} indicates the resulting completeness region.

\chapter{CTA - Cherenkov Telescope Array}
The Earth’s atmosphere is opaque for high-energy cosmic radiation. A direct observations of high-energy photons is possible only in the open space, by making use of satellites. The first gamma-ray space missions, such as SAS-2 and COS-B, were launched in 1970th. The major improvement of the sensitivity was achieved with the EGRET instrument, and then with the AGILE and the Fermi-LAT facilities, which are still in operation observing objects at energies up to 50 and 300 GeV, respectively. However, the steep very-high-energy spectra of the majority of the $\gamma$-ray emitting sources make the space TeV observations essentially ineffective. Satellites have to deal with width limitations (($\sim 1 m^2$ for the Fermi-LAT) that prevent them to observe photons more energetic than some hundreds of GeV. Ground-based observatories instead can perform deeper, even though indirect, measurements of the high-energy particles, by observing their secondaries particles, with Extensive Air Shower (EAS) detectors, or their Cherenkov light, by means of Imaging Atmospheric Cherenkov Telescopes (IACT). The first ones are designed for particles and
are basically silicon scintillators, located at high altitudes, to encounter the shower at its
maximum development. IACTs instead measure the radiation produced in these cascades, this technique is based on the phenomenon of the Cherenkov radiation discovered by P. Cherenkov in 1934. When a charged particle moves in a dielectric medium with a speed higher than the phase velocity of light in that medium, the wave front of blue Cherenkov emission
appears. In 1948, Blackett \cite{chere} found out that there is a detectable contribution to the night sky emission (Cherenkov radiation) from air showers of particles, produced as a consequence of the $\gamma$-ray photons and/or cosmic rays interactions with the upper layers of the Earth’s atmosphere. It is the Cherenkov
light  that can be observed from the ground using the imaging atmospheric Cherenkov technique. A few notable
experiments that implement the IACT, the High-Energy Stereoscopic System (H.E.S.S.), the Major Atmospheric Gamma-ray Imaging Cherenkov (MAGIC) and the Very Energetic Radiation Imaging Telescope Array System (VERITAS) have helped to establish this field of ground-based astronomy. The discoveries made by these ground-based $\gamma$-ray experiments have furthered our understanding of the technique itself as well as the violent objects in our universe that produce very high-energy gamma-rays.

\section{Detecting Cherenkov light}
\begin{figure}
    \centering 
    \includegraphics[width=.6\textwidth]{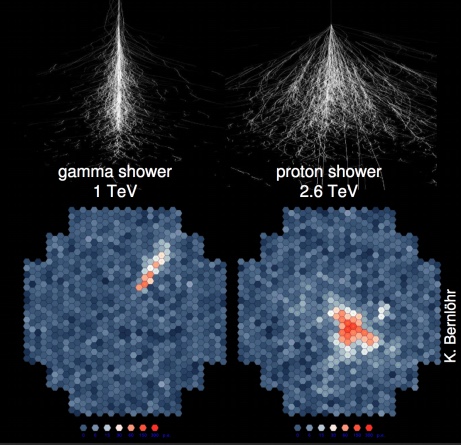}
    \caption{Difference between electromagnetic and hadronic showers. }
    \label{shower}
\end{figure}

Passing through the atmosphere, a $\gamma$-ray (primary) interacts with its upper layers at an altitude of $\sim$ 25 km and produces a particle shower. The energy of the primary
is so high that produced particles (mainly electrons and positrons) are moving with velocities higher than the speed of light in the atmosphere and, thus, emit the Cherenkov radiation. The number of particles in the air shower reaches the maximum at an altitude of $\sim$ 10 km and
then it continuously decreases as the shower goes deeper into the atmosphere and looses its energy. The Cherenkov radiation from the air shower forms a cone of blue visible light (light pool) around the primary gamma ray. Such flash of Cherenkov light lasts
for about 10 ns. A typical size of a light pool on the ground is about 250 m in diameter with a density of $\sim$ 50 Cherenkov photons per $m^2$ for a 1 TeV $\gamma$-ray primary. An
optical telescope located inside the Cherenkov light cone will be able to detect it, if its dish is large enough for such low fluxes and its camera is fast for such short signals. An effective collecting area of a single Cherenkov telescope can be approximated by the area of the Cherenkov light pool on the ground ($\sim 5 \times 10^4 m^2$). This provides a remarkable advantage to IACTs comparing to the space $\gamma$-ray telescopes. 
\begin{figure}
    \centering
    
    \includegraphics[width=.6\textwidth]{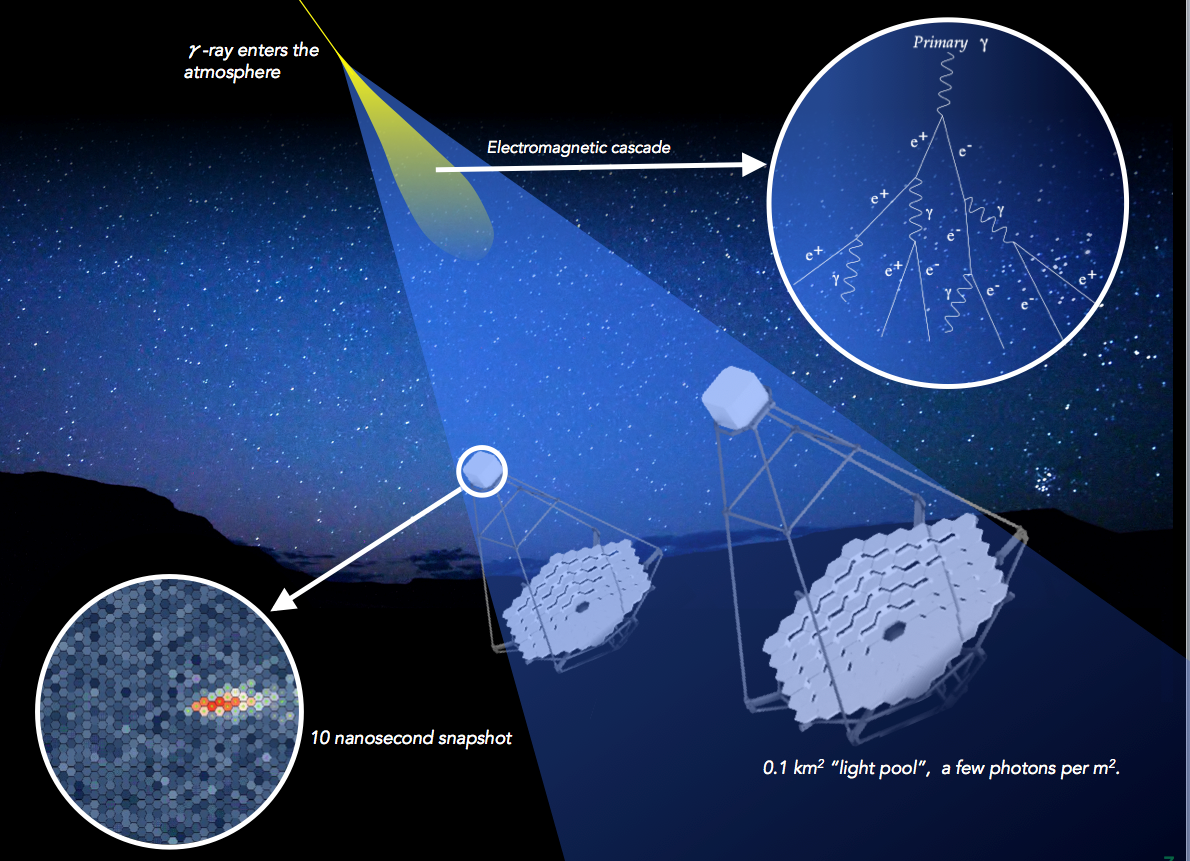}
    \caption{The schematic view of the IACTs technique}
    \label{IACT}
\end{figure}
By means of the analysis of a signal detected with the IACT, it is possible to identify properties of the
primary. The size and the brightness of obtained elliptical image corresponds to the energy of the initial $\gamma$-ray photon, while its orientation allows us to reconstruct
the direction of the primary or, in other words, the position of emitting source. A schematic view of the IACT technique is shown in figure \ref{IACT} $\gamma$-rays are not the only particles, which produce air showers: a $\gamma$-ray photon accounts for $\sim 10^4$
cosmic-ray hadrons. Interacting with atoms and molecules of the Earth’s atmosphere, cosmic rays also produce energetic showers of pions, nuclei and products of their decays. Showers originated from cosmic rays are much more extended than the electromagnetic ones, formed by $\gamma$-ray primaries
(see figure \ref{shower} ). 
The differences between the structures of the $\gamma$-ray and hadronic showers are imprinted on their images seen with Cherenkov telescopes. As shown in figure \ref{shower}, an image of an electromagnetic shower has an elliptical shape on the camera, whereas that of an hadronic event is irregular. Analyzing obtained images of Cherenkov pools, we can discriminate signal events ($\gamma$-ray) from the
background (hadrons) with $\sim$99\% efficiency (see gamma-hadron separation method in \cite{hillas}). The energy threshold of Cherenkov telescopes is defined basically by a $\gamma$-hadron
separation efficiency. An image of an electromagnetic shower produced by a low-energy gamma ray (< 100 GeV) is faint and it can be misinterpreted as a hadronic event with an
irregular shape. Using IACTs with large mirrors, we can obtain images of a shower of a better quality. This leads to more efficient $\gamma$-hadron discrimination and, therefore,
to lower values of the energy threshold attainable with this telescope. 
Using several Cherenkov telescopes as a single $\gamma$-ray instrument provides a number
of advantages. Observing with a number of IACT-units spaced by a distance of $\sim$ 100 m
can enlarge the collective area and increase the sensitivity of the whole instrument. In
addition, analyzing multiple images of the same shower (stereoscopic observations) it is possible to improve the reconstruction of the primary gamma ray, i.e. to
improve the determining of its energy and initial direction. This provides better angular
and energy resolutions of a Cherenkov instrument. Stereoscopic
observations also improve the gamma-hadron separation efficiency: e.g. by accepting
only events, which simultaneously trigger all IACTs under the light pool.

\subsection{Status of current IACTs}
A great improvement of ground-based Cherenkov instruments was made since a detection
of the first very-high-energy $\gamma$-ray source, the Crab Nebula, with the Whipple 10-meter IACT
in 1989 \cite{crab}. Using the legacy and experience of the first-generation IACT projects
(e.g. Whipple telescope, HEGRA, CAT, CANGAROO), the currently
operating Cherenkov telescopes such as MAGIC, VERITAS and H.E.S.S. have been developed. An order of
magnitude improvement of the sensitivity was achieved with these instruments. The \textbf{MAGIC} (Major Atmospheric Gamma Imaging Cherenkov Telescopes) instrument
consists of two identical 17-meter Cherenkov telescopes spaced by 85 meters, which are
situated at the Roque de los Muchachos Observatory (La Palma, Spain). The first
telescope operates since 2004, the second unit joined in 2009. This IACT performs observations in the energy range from 50 GeV up to 30 TeV. \textbf{VERITAS} is the Very Energetic
Radiation Imaging Telescope Array System of four 12-meter telescopes separated by a
100-meter distance, working at energies from 85 GeV to $\sim$ 50 TeV. This facility is
in operation in Mount Hopkins (Arizona, USA) since 2007. \textbf{H.E.S.S.} (the High Energy Stereoscopic System) was initially developed as an array of four 12-meter telescopes
at a distance of $\sim$ 120 meters, observing the very-high-energy $\gamma$-ray sky in the 100 GeV–100
TeV energy band since 2002. The second stage of the H.E.S.S. project started in
2013 after a construction and commissioning of the fifth large 28-meter telescope in the
center of the array, this telescope enlarged the energy range
of the H.E.S.S. instrument down to $\sim$ 20 GeV. In contrast to MAGIC and VERITAS,
H.E.S.S. operates in the southern hemisphere (Khomas Highland, Namibia). Taking
into account a large field of view of small H.E.S.S. telescopes (5$^{\circ}$), this array can be
considered as the best currently operating IACT machine for the Galactic Plane Survey. A wide range of astrophysical objects is being investigated.
 A large amount of important investigations, which expanded the frontier of our knowledge concerning the VHE sources, was already performed with the current generation of Cherenkov instruments.
However, the potential of the IACT technique has not been fully realized yet.

\section{CTA Telescopes}
The Cherenkov Telescope Array (CTA) is a modern project of the next-generation
ground-based very-high-energy $\gamma$-ray facility, which aims to use all the advantages of the
IACT technique. The CTA observatory will comprise two arrays, one in
each hemisphere, with hundred IACT telescopes of 3 different types: Large Size
Telescopes (LSTs, 23-meter diameter), Medium Size Telescopes (MSTs with diameters
of 10–12 meters) and Small Size Telescopes (SSTs, 4-meter diameter). The northern site will be in the Canary island of La Palma, the southern one at Paranal, in Chile. The two sites will have a different configuration due to the different geographical conformations: the northern one will be more limited in size because of the features of the site, and will not have SSTs. In table \ref{tab:layout} is reported a scheme of the different configuration of the two sites. However this configuration, known
as baseline configuration is not granted, the CTA community for now guarantees only for a threshold configuration, with less telescopes as reported in table \ref{tab:layout}.\\
\begin{figure}[ht!]
    \centering
    \includegraphics[width=1\textwidth]{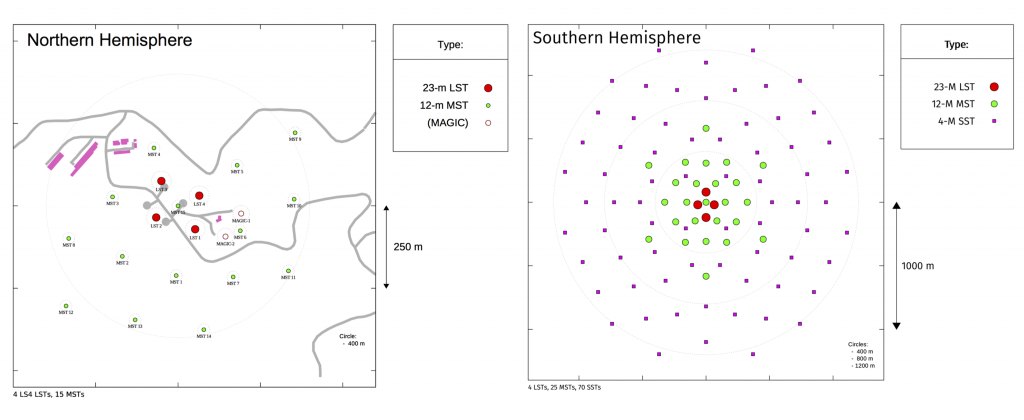}
    \caption{Layouts designed for the northern and southern array of CTA. Image from https://www.cta-observatory.org/about/how-cta-works/}
    \label{fig:layout}
\end{figure}

\begin{table}[ht]
    \centering
    \begin{tabular}{|c|c|c|c|}
    \hline
    &     & Baseline & Threshold  \\ \hline
    &   Energy range  & South   North &    South   North \\ \hline
    LST & 20GeV-200GeV & 4    4 & -       4  \\ \hline 
    MST & 100GeV-10TeV & 25   15 & 15      5  \\ \hline 
    SST & 1TeV-300TeV & 70   - & 50       -  \\ \hline 
    \end{tabular}
    \caption{CTA telescopes specifications}
    \label{tab:layout}
\end{table}

Large sized telescopes (LST) with their 23-m diameter reflector have the lowest energy threshold among the three kinds of telescoped. The LST is designed to be an alt-azimuth telescope
a parabolic reflective surface. Although the LST will stand 45 m tall and weigh around 100
tonnes, it will be extremely nimble, with the goal to be able to re-position within 20 seconds.
Both the re-positioning speed and the low energy threshold provided by the LSTs are critical
for CTA studies of Galactic transient phenomena, high red-shift active galactic nuclei and
gamma ray bursts. Besides the LSTs will stand the MSTs that will cover the central energy range
from 100 GeV to 10 TeV. They will have a modified Davies-Cotton (e.g. sub-mirrors are disposed in a spherical configuration) mirror of 12 m diameter, with polar mount connected to
a camera of 8$^{\circ}$of field of view. Finally SSTs will be built and spread into a large area. Due to
the power law behavior of the spectrum in fact, the high energy photons flux is lower. With
the contribution of these SSTs CTA will extend its energy domain up to 300 TeV. SSTs will
have a diameter of 4m and a field of view of 9$^{\circ}$.


\section{CTA performance}\label{perf}

For any kind of measurement one has to deal with the instrument capability. As in equation \ref{e} the detected signal $e(E'
, \Vec{p'})$ is the result of the convolution between the flux $F(\Vec{p})$ of the source and the instrument response functions $R(E', \Vec{p'}|E, \Vec{p})$. Where $E'$ and $\Vec{p'}$
label the detected event, and $E$ and $\Vec{p}$ the physical one.
\begin{equation}
    e(E',\Vec{p'})= \int F(\Vec{p}) \times R(E',\Vec{p'}|E, \Vec{p})d\Vec{p}dE
    \label{e}
\end{equation}
The instrument response functions for CTA comprise the effective area $A_{eff}(p,E,t)$, the point spread function $PSF(p'|p,E,t)$ and the energy dispersion $E_{disp}(E',p,E,t)$:
\begin{equation*}
R(p',E',t'|p,E,t)=A_{eff}(p,E,t)\times PSF(p'|p,E,t) \times E_{disp}(E'|p,E,t)    
\end{equation*}
where are indicated the dependences on the physical properties of the original event $(p,E,t)$
and of the detected one $(p',E',t')$. Each of these quantity is computed by simulating showers
with Monte Carlo methods. The CTA consortium on their website provides the results of
their simulations. [Ambrogi et al., 2016] in their paper modeled the resulting points and
derived an analytical parametrization in the energy range from 50 GeV to 100 TeV for a
point like source positioned at the center of the field of view (FoV). The plot of the fitted
formula for each quantity are reported in 4.4.
\subsection{Effective area}
For a single telescope the effective area is determined by the radius of the Cherenkov light
pool at ground, while for a multi-telescope system is determined essentially by the total geometrical area. CTA bigger array, when completed, will cover an area of $\sim 300 km^2$. Particles
of different energies create cascades of different area (more energetic particles generate wider showers and vice-versa) that causes an energy dependency of $A_{eff}$ . The parametrization of
[Ambrogi et al., 2016] for the CTA southern observatory effective area yields:
\begin{equation*}
    A_{eff}(x)=\frac{A}{1+Bexp(-x/C)}
\end{equation*}
where $x= \log(E/1TeV)$, $A=4.36\times10^6 m^2$ is the saturation value of the effective area, whereas $B=6.05$ and $C=3.99\times 10^{-1}$ define the rate of change of $A_{eff}$ with respect to energy.

\begin{figure}
    \centering
    \includegraphics[width=0.45\textwidth]{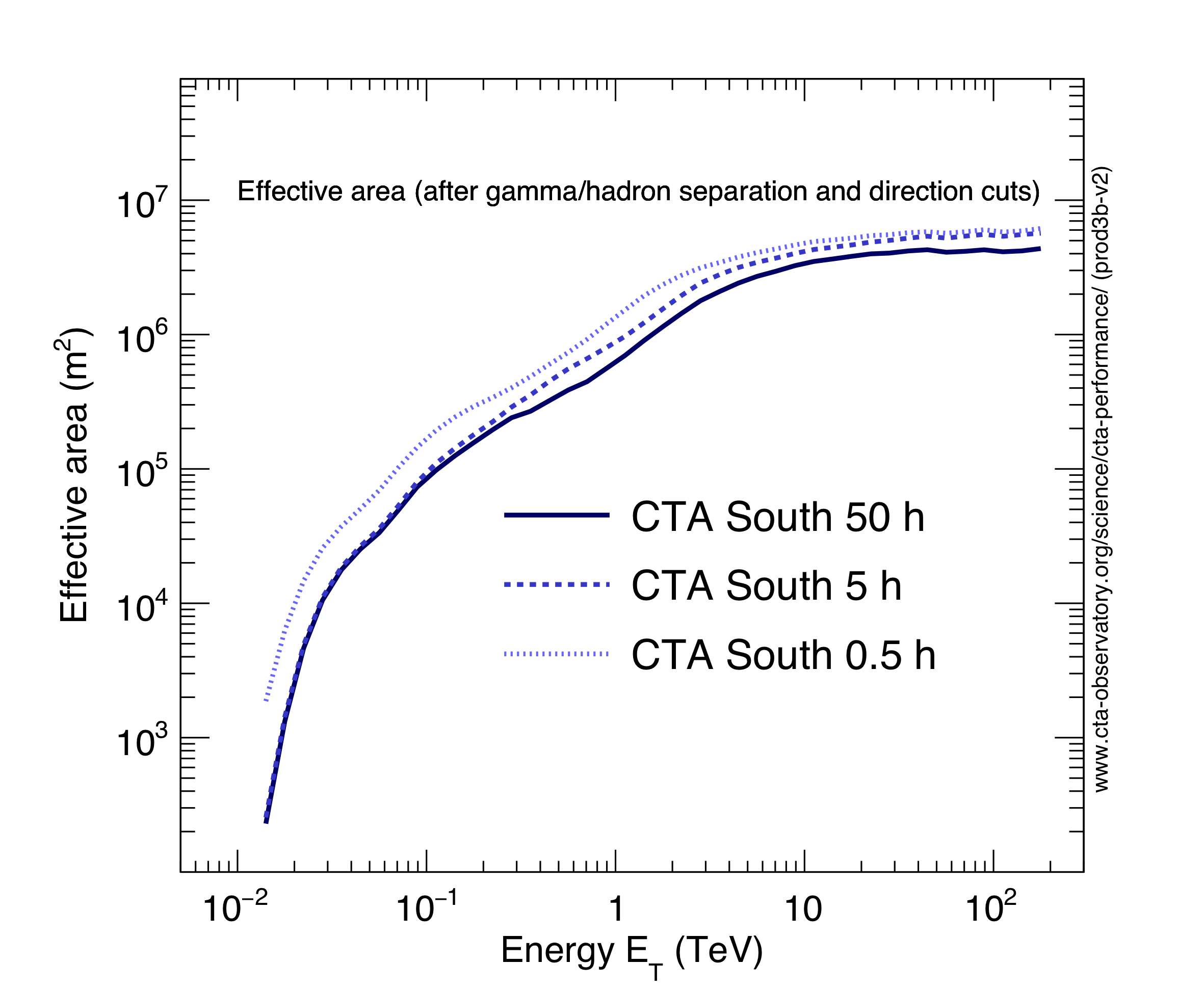}
    \includegraphics[width=0.45\textwidth]{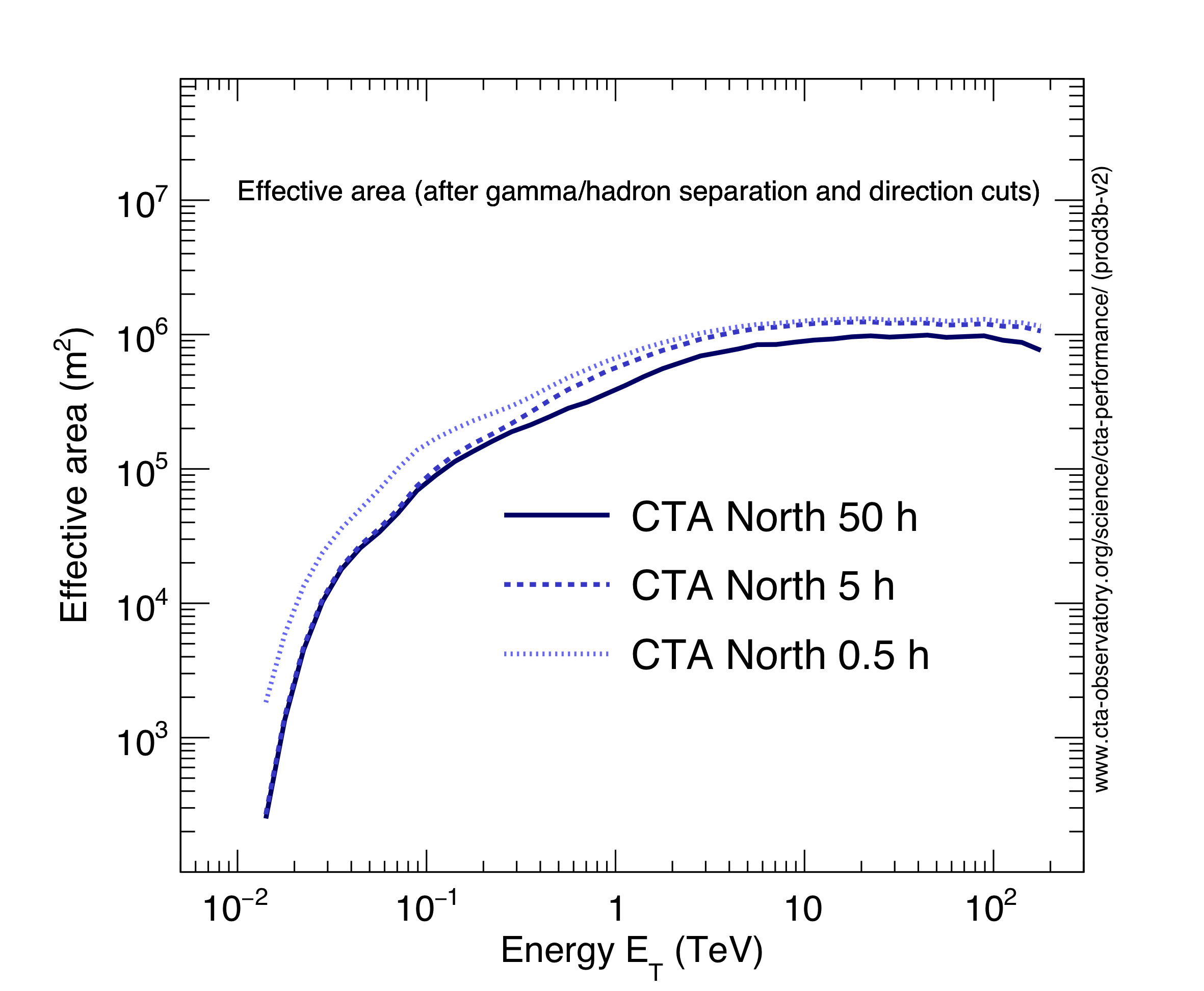}
    \caption{The effective collection area for gamma rays from a point-like source as a function of energy optimised for 0.5-, 5- and 50-h observations.}
    \label{fig:my_label}
\end{figure}

\subsection{Angular resolution}
\begin{figure}
    \centering
    \includegraphics[width=0.4\textwidth]{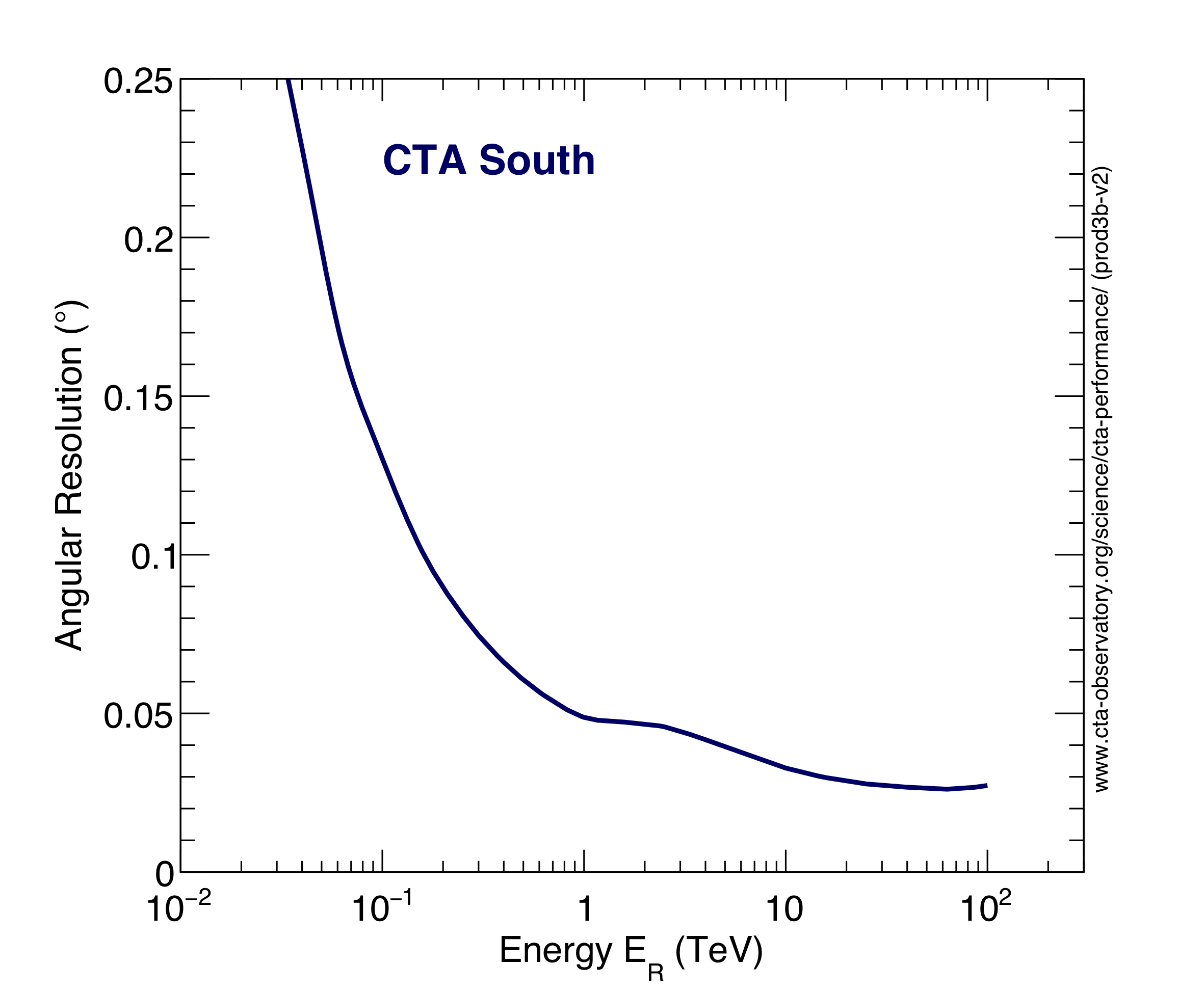}
    \includegraphics[width=0.4\textwidth]{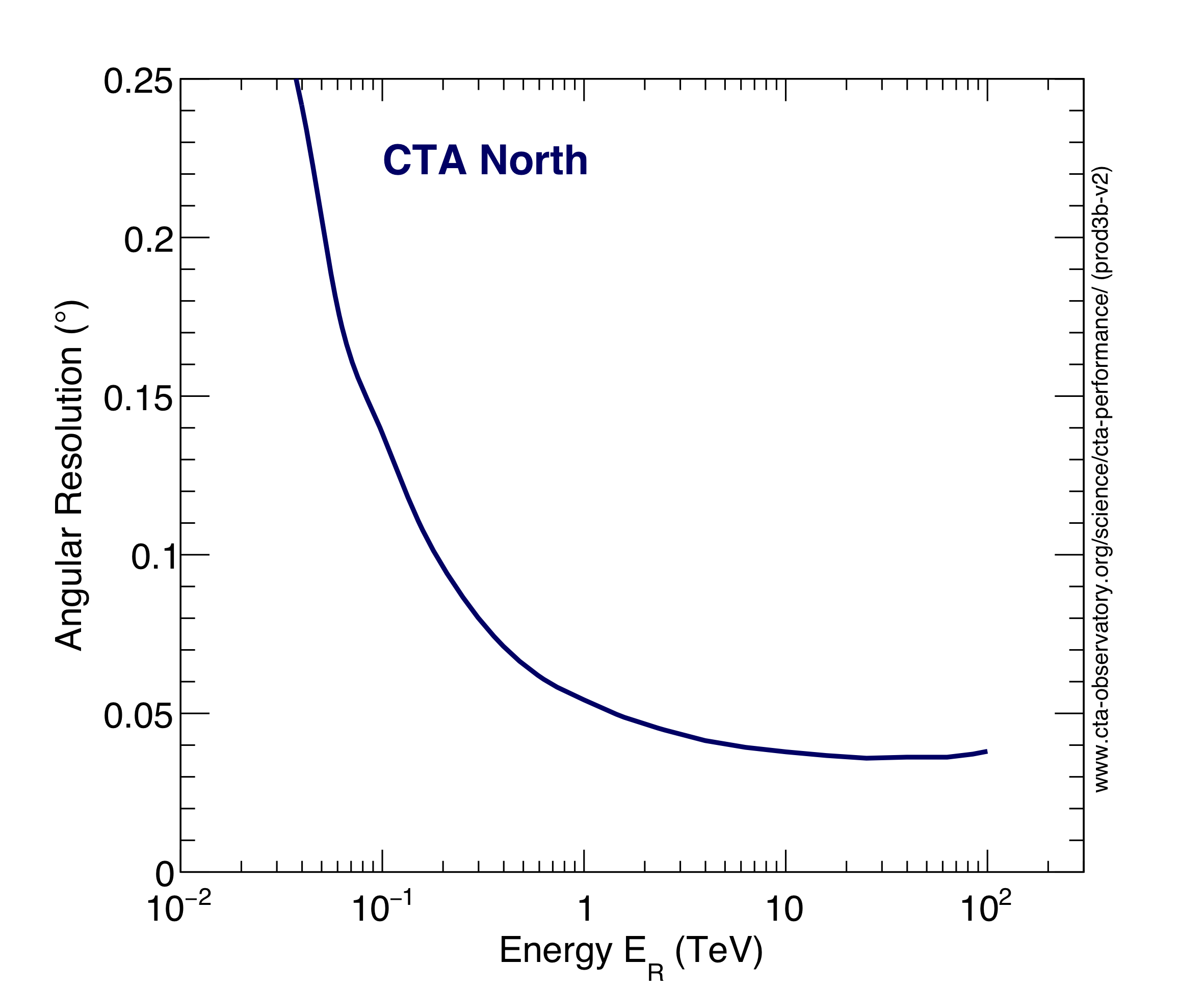}
    \includegraphics[width=0.4\textwidth]{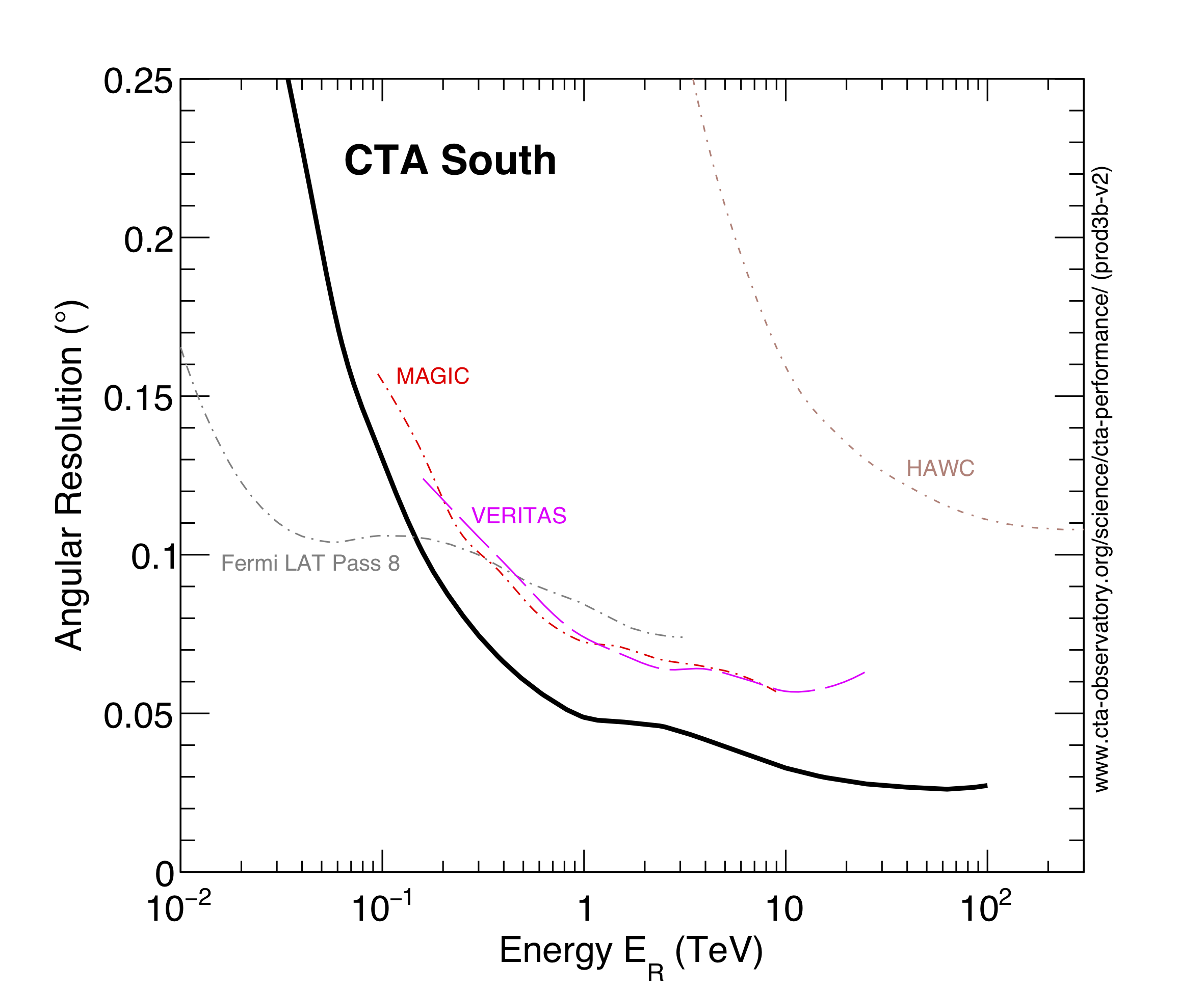}
    \caption{Angular resolution as a function of energy of CTA south, CTA north and a  comparison with other instruments respectively. }
    \label{fig:ar}
\end{figure}

A point like source, when detected from any camera appears not as a point but as a circle
with a certain extension. The distribution of the detected photons depends on the instrument, and it is known as point spread function (PSF). In first approximation the PSF is a two
dimensional Gaussian:
\begin{equation*}
   PSF = exp ( \frac{x_{f}^{2}+y_{f}^{2}}{2\sigma_{PSF}^{2} } )
\end{equation*}
where $x_f$ and $y_f$ are the coordinates in the camera focal plane.
The angular resolution in this context is defined as the standard deviation of the gamma ray PSF. The angular resolution depends on the energy, the higher the energy the smaller is the
PSF. For CTA a reference value is an angular resolution of 0.05$^{\circ}$ at 1 TeV, as shown in figure
\ref{fig:ar}.
The fit of the data points produced the function in \ref{2}
\begin{equation}
    \label{2}
    \sigma_{PSF}=A \cdot [ 1+exp (-\frac{x}{B} )  ]
\end{equation}
here $A=2.71 \cdot 10^2 $ deg is the best angular resolution achievable with the considered configuration and $B=7.90 \times 10^{-1}$ is a scaling factor that determines how fast $\sigma_{PSF}$ changes with energy.

\subsection{Energy resolution and energy dispersion}
The measure of the gamma photon energies is not direct, it is derived from the energy of the photons generated in the cascades. For this reason it is called reconstructed energy and it corresponds to the peak of a probability function, similar to the PSF, known as Point Dispersion Function. The energy resolution, quantifies the ability to distinguish between two energy values and it is defined as the full width at half maximum (FWHM) of the energy distribution centered on the reconstructed energy. CTA energy resolution should reach the level of $\sim$6\% above 1 TeV. The parametrization of [Ambrogi et al., 2016] for the energy resolution is
given by:
\begin{equation}
    \Delta E/E (x) = A \times [(x-B)^2+(x-B^4)]+C
\end{equation}
with $A=6.33 \times 10^{-3}, B=8.34 \times 10^{-1} $ and $C=6.24 \times 10^{-2}$.
\begin{figure}
    \centering
    \includegraphics[width=0.45\textwidth]{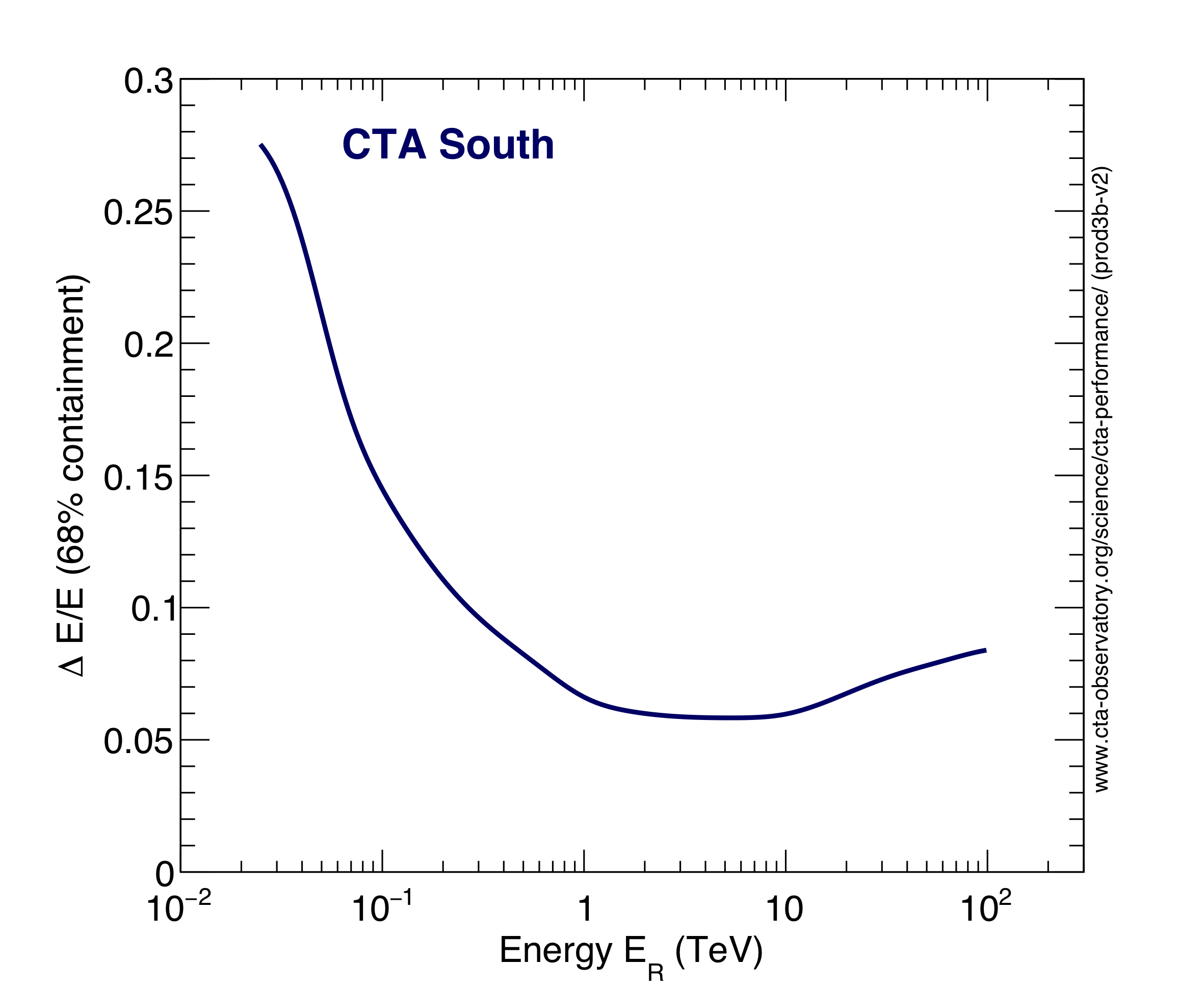}
    \includegraphics[width=0.45\textwidth]{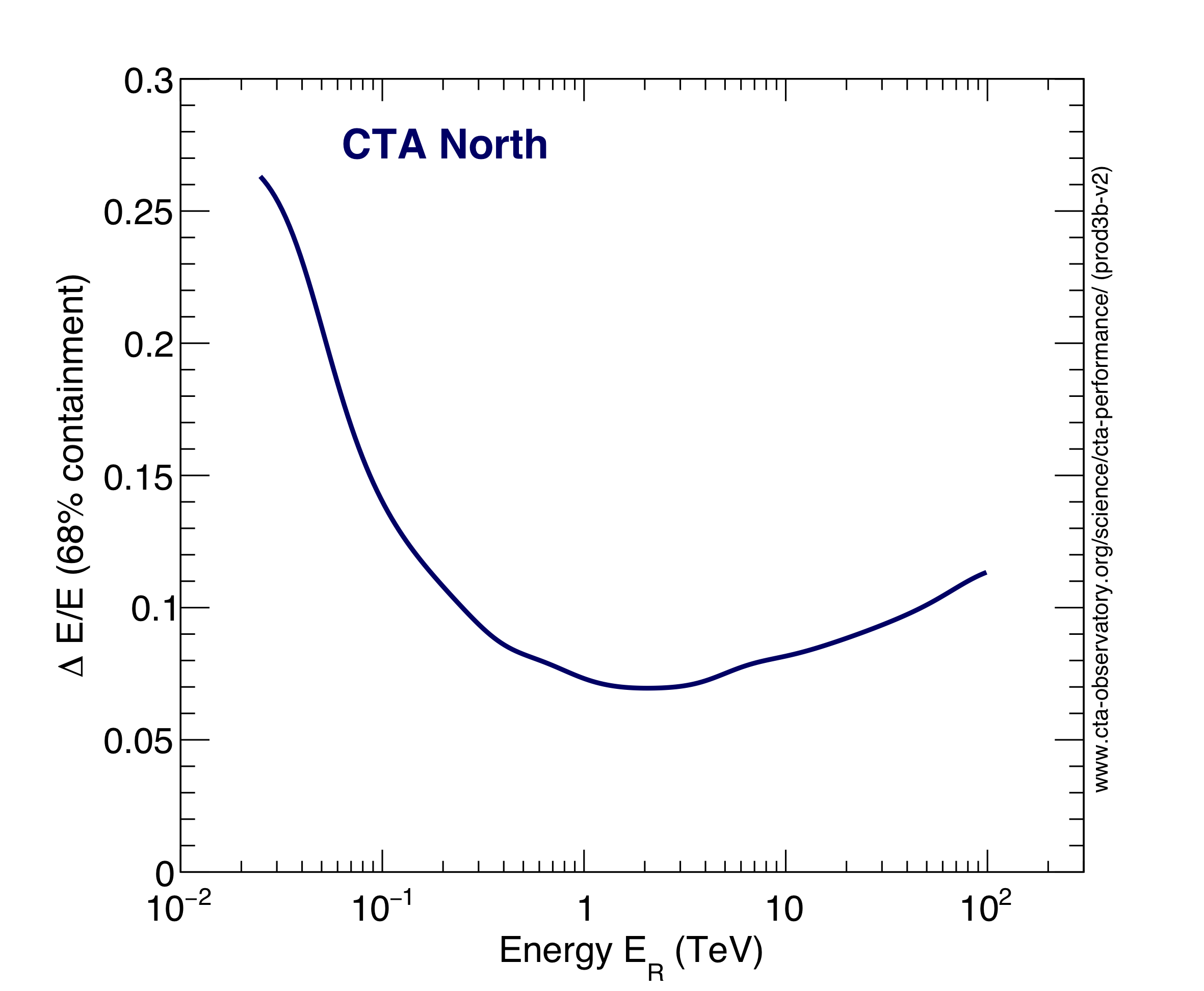}
    \caption{Energy resolution as a function of energy for CTA south and north respectively}
    \label{fig:my_label}
\end{figure}

\subsection{Background rate}
The main sources of background for IACT arrays are gammas generated in the atmosphere
from the impact of protons and heavier nuclei of the CR. The CTAconsortium simulates the arrival of these particles, assuming for them a power law spectrum consistent with the satellites measurements. This contribution is influenced by the zenith angle of the observation, since higher zenith angles correspond to more layers of atmosphere that the particles cross, enhancing the chance for interaction. Besides since the final detection products are visible photons, all the sources (both of astrophysical and human origin)
of contamination in the optical band have to be taken into consideration. Usually CTA background is evaluated for a dark sky, namely with no moon, and offset from the galactic plane.
The electronics noise must be considered as well.
\begin{figure}
    \centering
    \includegraphics[width=0.45\textwidth]{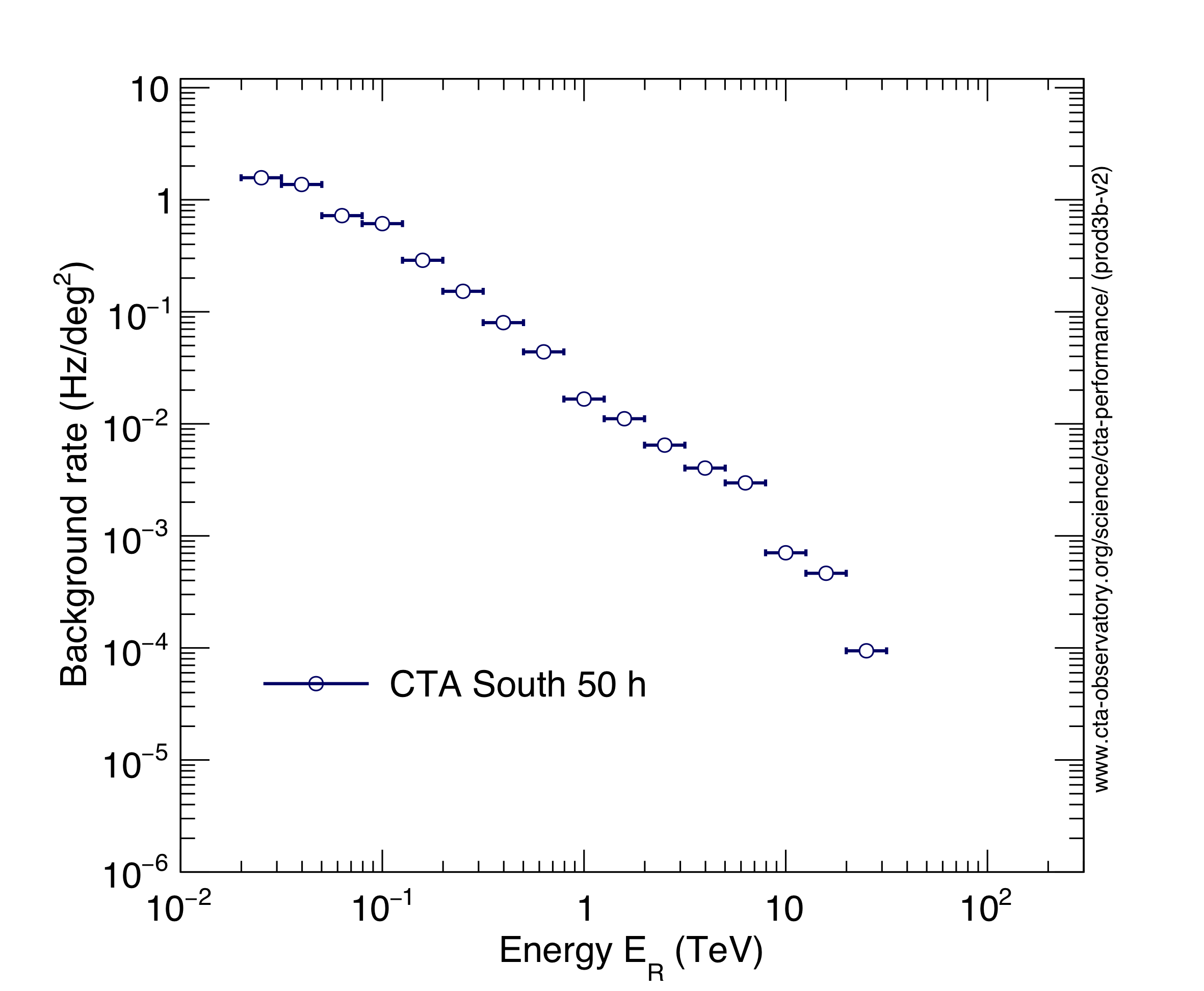}
    \includegraphics[width=0.45\textwidth]{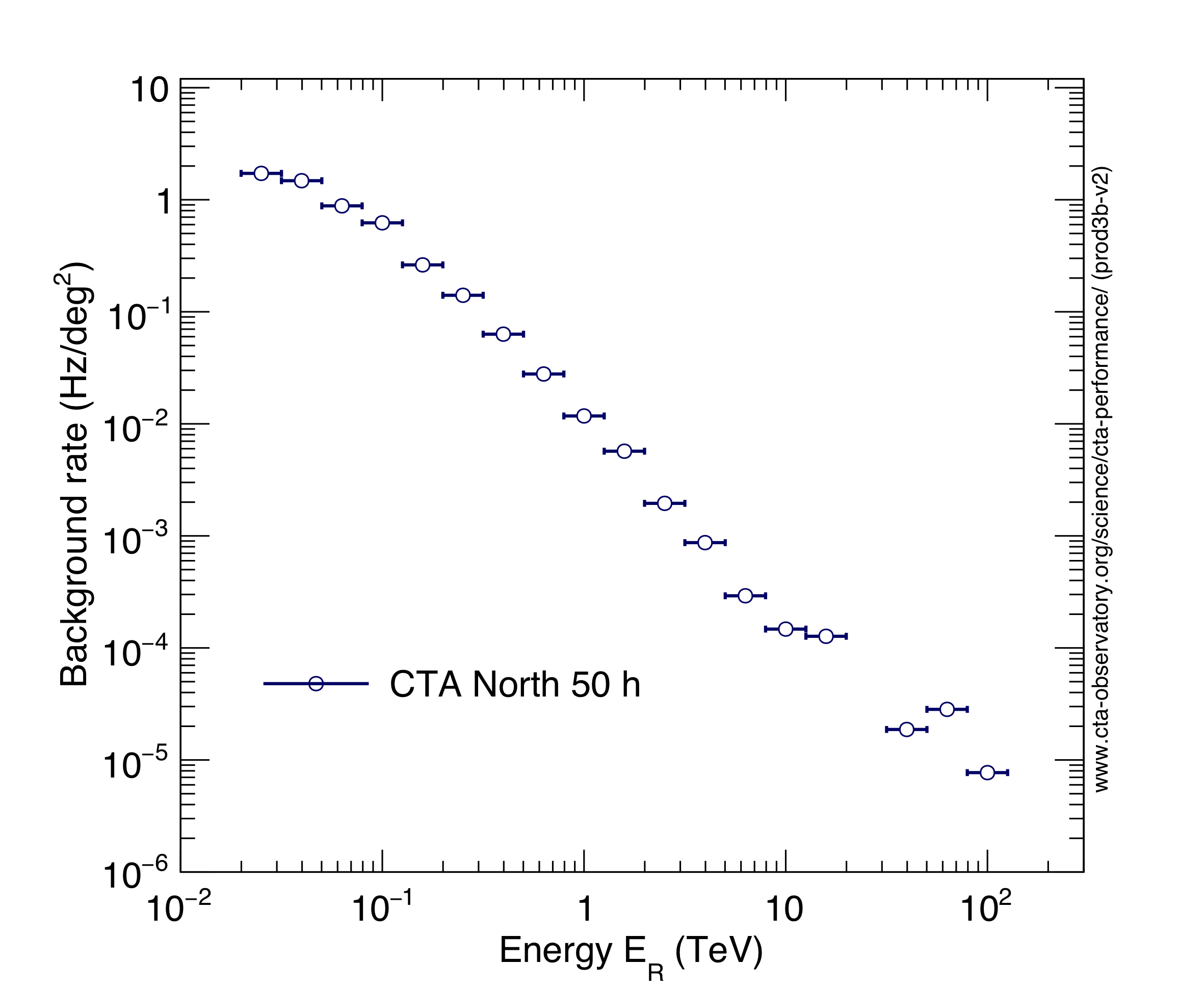}
    \caption{The (post-analysis) residual cosmic-ray background rate per square degree vs reconstructed gamma-ray energy for CTA south and north respectively.}
    \label{fig:my_label}
\end{figure}

\subsection{The sensitivity}
\begin{figure}
    \centering
    \includegraphics[width=0.8\textwidth]{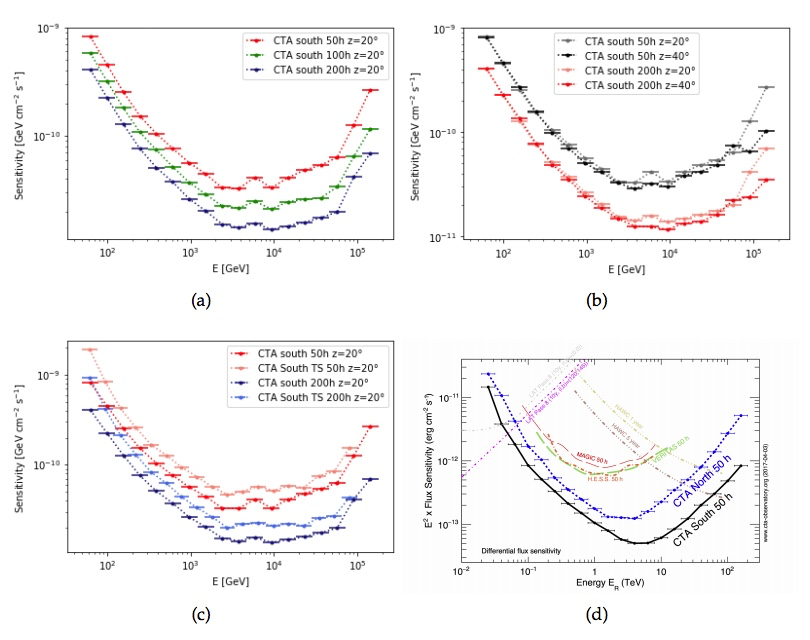}
    \caption{Comparison of CTA sensitivity for different cases. In
(a) sensitivities for the southern array in its definitive configuration for different observation time intervals. In (b) the sensitivity for observation at different zenith angles. In (c) we considered the provisional threshold configuration. Finally in (d) CTA
southern and northern arrays are compared with the main current
gamma ray detectors. This last figure is from https://www.ctaobservatory.org/science/cta-performance/}
    \label{fig:sens}
\end{figure}

The sensitivity of a telescope is the minimum flux detectable with a required statistical significance, e.g. with enough number of events to distinguish the signal from the background.
For CTA the requirements to be fulfilled for each energy bin are:
\begin{itemize}
    \item at least 5$\sigma$ significance level;
    \item the presence of at least 10 excess events;
    \item a signal of at least 5\% of the background systematic error.
\end{itemize}
The sensitivity is then calculated simulating the flux of a source; for most applications a
point source with the spectrum of the Crab nebula is used. Statistical limits are calculated
using a maximum likelihood approach, and the background systematics are assumed to
have an uncertainty of 1\%. The sensitivity of a gamma-ray detector is determined by the
effective collecting area, the residual background and the angular resolution and hence depends strongly on the energy. Furthermore different layouts of the array, will give different
effective areas, and consequently different sensitivities. The two arrays of CTA will hence
have different performances: the northern one will be approximately two times less sensitive
than the southern one.
Moreover, these sensitivity curves are background dependent, as one can see from the sensitivity definition itself and hence depends on the zenith angle and other factors that influence
the background.
The plots in figure \ref{fig:sens} show the computed differential sensitivity for a point-like source.
Different exposure times and zenith angles were assumed. Different configurations were
also tested: it was chosen to consider not only the differences between North and South, but
also the difference from the baseline configuration, namely the final designed configuration
and the threshold configuration (as in table \ref{tab:layout}).
For what concerns the extended sources, the question is more delicate. In this case the background rate increases proportionally to the factor $\sqrt{\theta^2+\sigma^2_{PSF}}$ [Funk et al., 2013], and therefore the sensitivity can be assumed to scale according to the relation:
\begin{equation*}
    sens_{ext}=sens_{point}\cdot \frac{\sqrt{\theta^2+\sigma^2_{PSF}}}{\sigma^2_{PSF}}
\end{equation*}

A reference value for CTA differential sensitivity is the sensitivity of a point source observed at low zenith angle ($z \leq 20^{\circ}$) at 1TeV, with an exposure time of 50 hrs, namely:
\begin{equation}
    sens(E=1TeV, z=0, t_{exp}=50h)=6.55 \times 10^{-14} \frac{TeV}{cm^2s}
\end{equation}

\subsection{Off-Axis sensitivity}
All performance parameters presented above are valid for a source located close to the centre of the CTA field of view (FoV). The differential sensitivity curves for a point-like source at increasing angular distances from the centre of the FoV are shown below. The radius of the FoV region in which the sensitivity is within a factor 2 of the one at the centre is around 2 degrees near the CTA threshold, and >3 degrees above a few 100 GeV.
Angular and energy resolution also degrade as one approaches the edge of the FoV. The provided IRFs contain the evolution of all performance parameters with off-axis angle.

\begin{figure}
    \centering
    \includegraphics[width=0.4\textwidth]{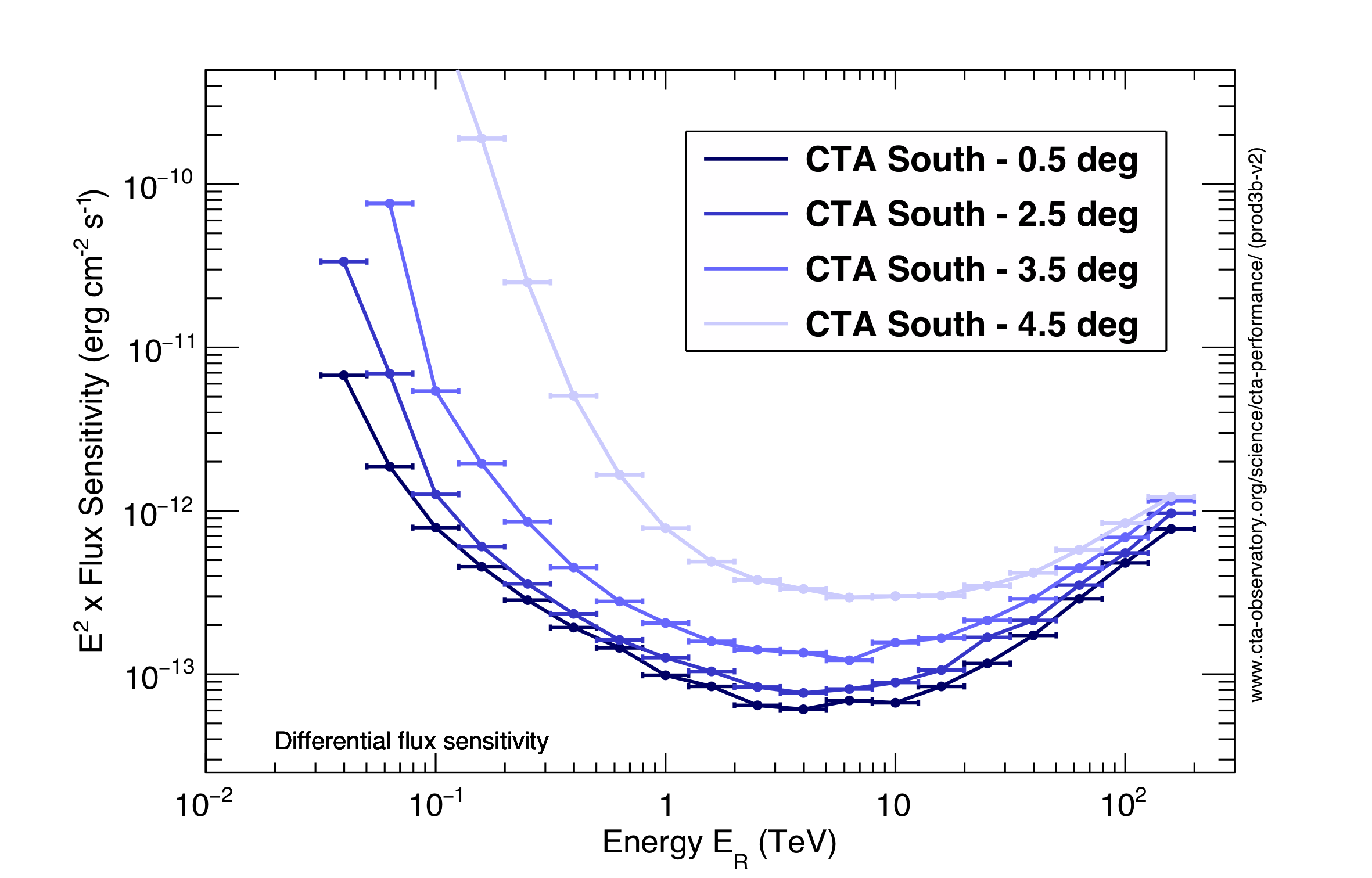}
    \includegraphics[width=0.4\textwidth]{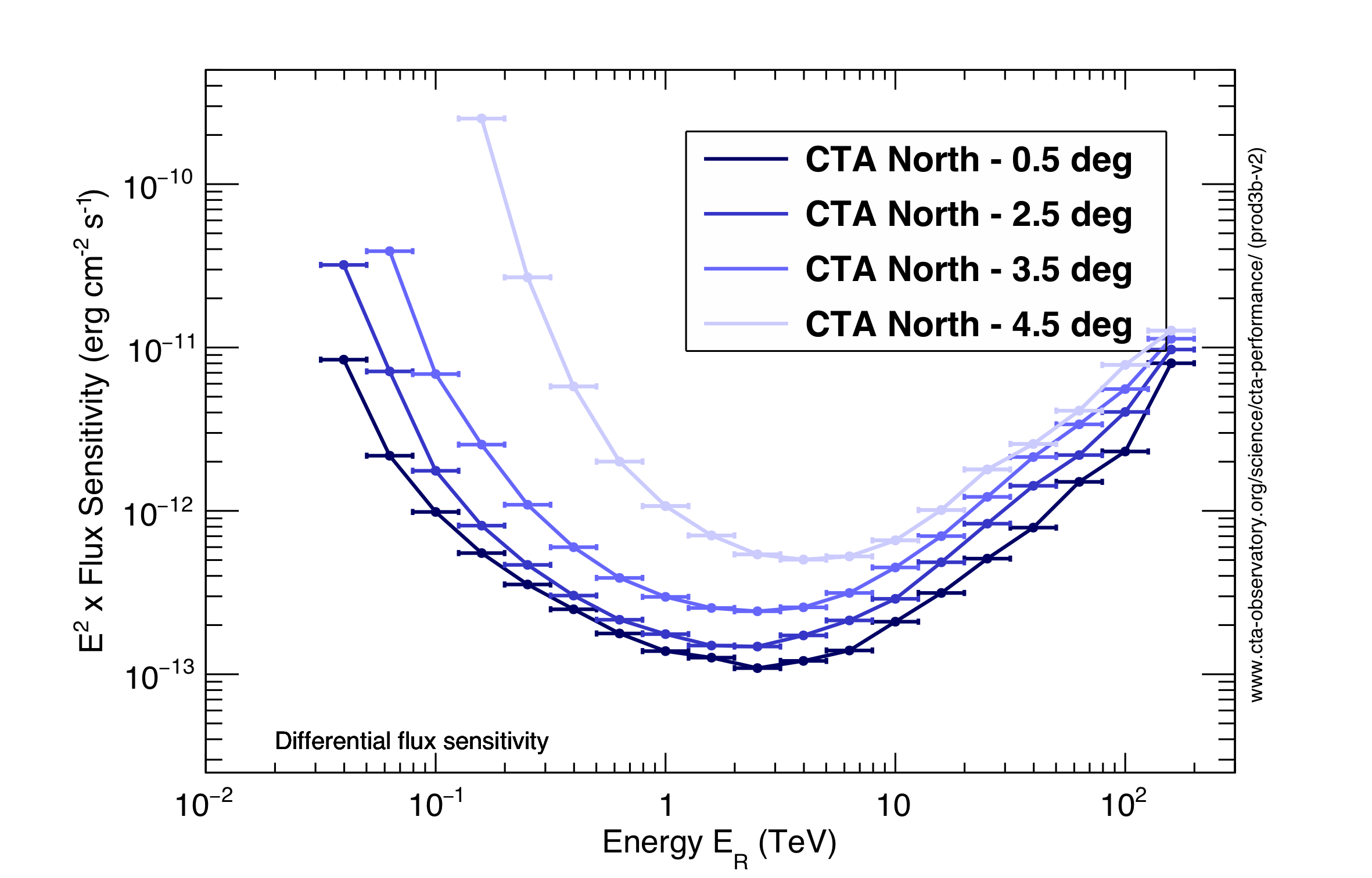}
    \caption{Off-axis sensitivity of CTA south and north respectively. }
    \label{fig:off}
\end{figure}

\section{CTA Data Challenges}
CTA data challenges (DC) are a large-scale semi-realistic test of the high-level analysis system, each with a different focus and an increasing level of realism in particular towards instrument imperfections and processing tools. At the moment only the first data challenge (DC-1) is complete, the second 
is scheduled for next year. In DC-1  the CTA collaboration has produced a set of simulated high-level science data (DL3) - i.e. the reconstructed properties (direction, energy, time) of gamma-ray like showers (real gamma rays and misclassified charged particles). In DC-1 the focus was on the validation of the core versions of the Science tools (currently ctools and gammapy).  A large number of different science cases was tested in the science working groups (SWGs). In addition, a systematic comparison of the two high-level analysis tools that
are currently in development within CTA was performed. In DC-1, 1980 hours were scheduled for the Southern array and 1815 hours for the Northern array
each under the assumption of a total of 1100 hours/year observation time at each site. These hours
were distributed amongst the different KSPs, see table \ref{cta}
\begin{table}[ht]
\centering

\begin{tabular}{|c|c|c|}
\hline 
KSP & Duration & Pointings \\ 
\hline 
Galactic Plane Survey & 1020 south, 600 north & 3270 \\ 

Galactic Center Survey & 825 south & 1671 \\ 

Extragalactic Survey & 200 south, 300 north & 1271 \\ 

AGN monitoring & 960 north & 1920 \\ 
\hline 
\end{tabular}
\caption{Distribution of observational time among the KPS}
\label{cta}
\end{table}

The science working groups provided the input to DC-1 in the form of xml files that describe the flux from
a certain sky direction. These models are consistent with existing TeV and Fermi-LAT data and in many
cases include an extrapolation to lower fluxes based on population models. The second data challenge is motivated by the requirement of a more realist models both of population of sources and of specific cases. In particular in the case of supernova remnants only few new SNRs have
actually been detected using a systematic analysis of the DC-1 data. Although this is probably partially due to the fact that the pipeline lacks a treatment of shell-type morphology, the model used predicts many large and faint
SNRs, difficult to disentangle from large-scaled diffuse emission. Infact, in DC-1 only the population of young SNR is modeled, instead a population model for the interacting supernova remnants is fully missing. The model presented in section \ref{pop} will be the population model of iSNR for the DC-2.

\section{Galactic Plane Survey}
\begin{landscape}
\begin{figure}[ht]
    \centering
    \includegraphics[scale=1.5]{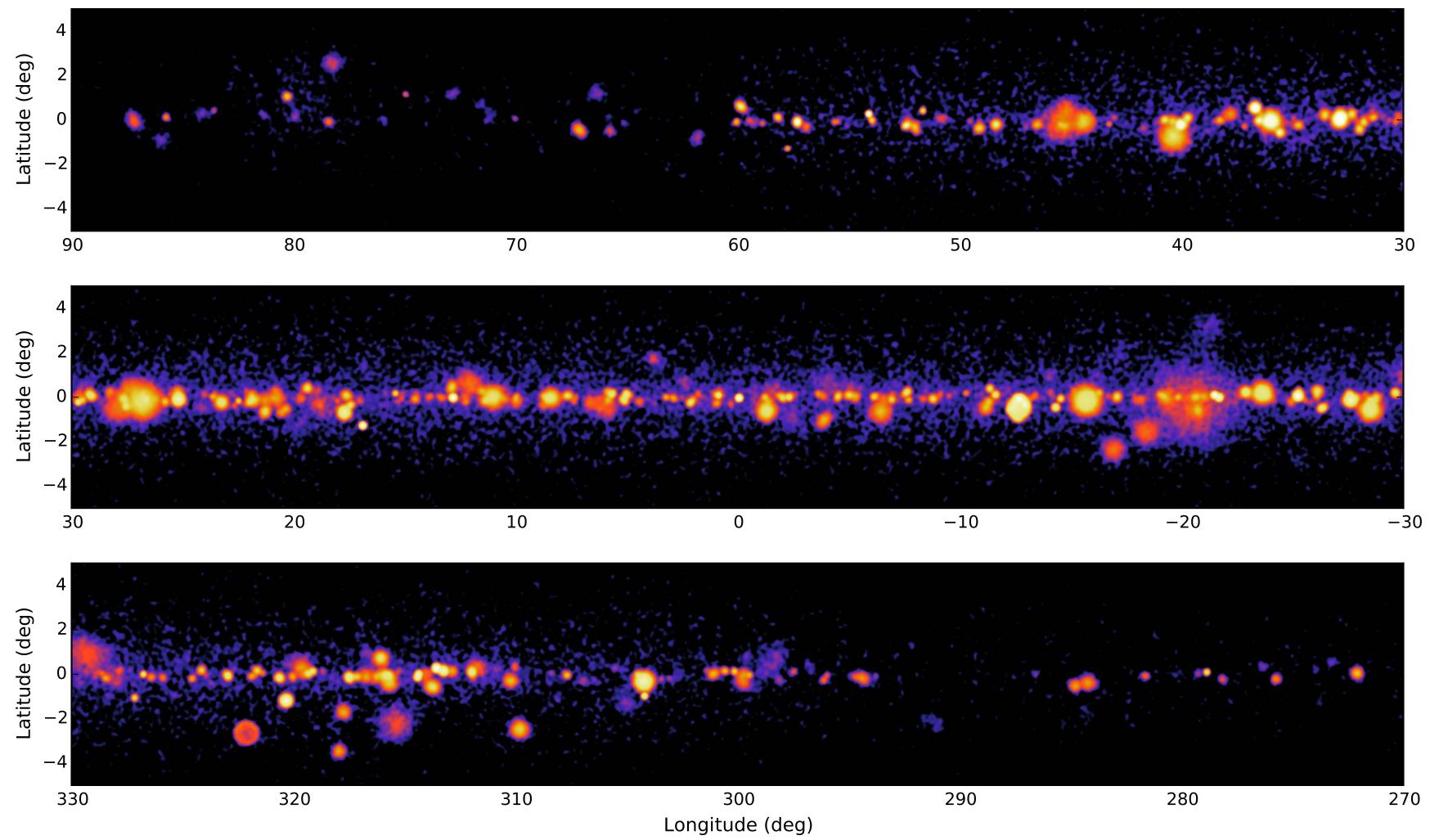}
    \caption{Simulated CTA image of the Galactic plane for the inner region −90$^{\circ}$ < $l$ < 90$^{\circ}$
, adopting the
actual proposed GPS observation strategy, a source model incorporating both supernova remnant and pulsar
wind nebula populations and diffuse emission.}
    \label{fig:gps}
\end{figure}
\end{landscape}

CTA is an observatory available for guest observer programs. In addition, the CTA consortium will be granted with a fixed amount of time that will be devoted to sky surveys. One of
these so called key science projects (KSP) is a Galactic Plane Survey (GPS). This survey will achieve a sensitivity better than 4.2 mCrab over the entire Galactic Plane and of 1.8 mCrab in the Galactic center and is expected to lead to the discovery of many new VHE objects. The GPS will be divided into two phases: a short term (Years 1-2) preliminary phase, and a long term (Years 3-10) one with a total observing time of 1020 hours with CTA South and
600 hours with CTA North.
The observations will proceed in a scan of the sky, composed of subsequent pointing with a double row strategy with a nominal pointing separation of $3^{\circ}$, like sketched in figure \ref{fig:gps}.
The observations will be performed at low zenith angles ($z \leq 45^{\circ}$) in order to preserve the
sensitivity. That will allow to cover the $|b| < 2^{\circ}$ region, where the majority of the known sources, like SNRs and PWNe, is located.
The major scientific objectives for the CTA GPS include the following:
\begin{itemize}
    
\item discovery of new and unexpected phenomena in the Galaxy. These would include completely new
source classes, new types of transient and variability behavior, or new forms of diffuse emission.
\item discovery of PeVatron candidates that are of key importance in our search for the origin of cosmic
rays. These candidates will likely require deeper, follow-up observations to confirm and characterize their PeVatron nature.
\item detection of many new VHE Galactic sources (of order 300 – 500), particularly
PWNe and SNRs, to increase the Galactic source count by a factor of five or more. The substantially increased statistics and more uniform sensitivity will allow more advanced population studies
to be performed. The ultimate goal is to significantly advance our understanding of the origin of
cosmic rays.
\item measurement of the large-scale diffuse VHE gamma-ray emission, to better understand its
origin in terms of inverse-Compton, $\pi^0$ decay, and unresolved source components.
\item discovery of new VHE gamma-ray binary systems, a unique class of objects with periodic emission on varying timescales, where physical processes are observed from different vantage points
depending on each system’s orbital inclination. Only five such systems are currently known in the
Galaxy.
\item  production of a multi-purpose legacy data set comprising sky images and source catalogues of
the complete Galactic plane at very high energies. This dataset will have long-lasting value to the
entire astronomical and astroparticle physics communities, far beyond the lifetime of CTA.
\item the GPS will produce and periodically release sky maps and catalogues
\end{itemize}

Figure \ref{fig:gps} shows a simulated image of what could result from a CTA survey of a portion of the Galactic
plane using a model that incorporates SNR and PWN source populations as well as diffuse emission.
The conclusion that can be drawn from these population estimates is that CTA can
expect the detection of many hundreds of Galactic sources that would follow from a survey of the plane
with a sensitivity at the level of a few mCrab.
\chapter{Population model for interacting supernova remnants}\label{pop}

The aim of this section is create a synthetic population model of Galactic SNR-MC systems emitting $\gamma$-rays.
I had then to reproduce three aspects of the real supernova remnants - molecular clouds systems, starting from the $\gamma$-SNR catalog and from the Rice et al. \cite{rice} catalog of galactic molecular clouds. First of all I had to create a synthetic population of molecular cloud that reflects the features of the molecular clouds in the Milky Way in terms of positions, mass and radial velocity. Then I had to reproduce the cosmic rays injection spectra in the molecular clouds. The last aspects which I had to consider is the real fraction of supernova remnants with a molecular cloud nearby, it means that I had to introduce a probability function that returns the number of interacting systems comparable to the observed one. 

\section{Synthetic population of molecular clouds}

\subsection{Catalog of molecular clouds}
\begin{figure}
\centering

\includegraphics[width=1\textwidth]{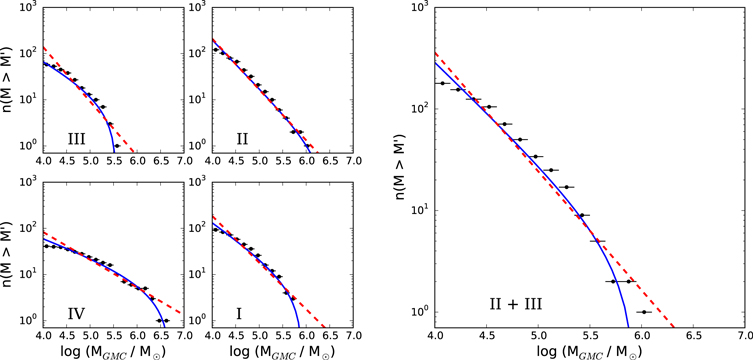}

\caption{Cumulative mass functions for clouds beyond the solar circle. In each panel, both a
truncated (solid blue) and non-truncated (dashed red) power law mass spectrum is fit. For the outer
Galaxy, the non-truncated, dashed red power law functional form is preferred}
\label{mass2}
\end{figure}

\begin{figure}
\centering
\includegraphics[width=1\textwidth]{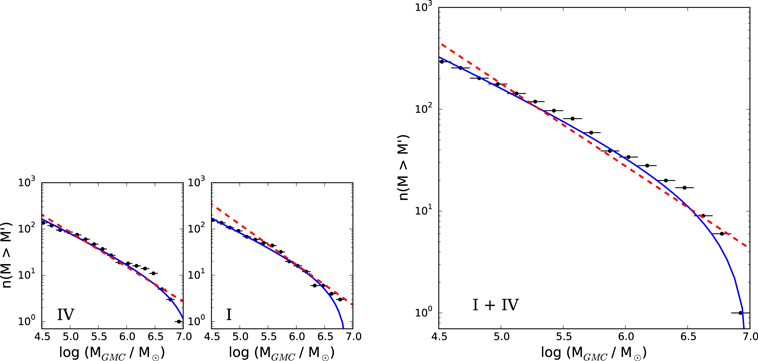}

\caption{Cumulative mass functions for clouds within the solar circle. In each panel, both
a truncated (solid blue) and non-truncated (dashed red) power law mass spectrum is fit.}
\label{mass}
\end{figure}

\begin{figure}
\centering
\includegraphics[width=1\textwidth]{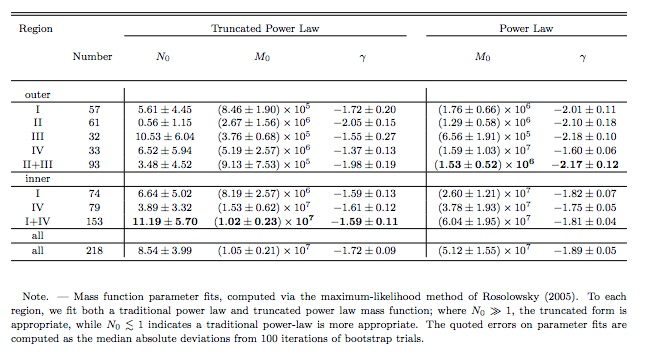}
\caption{}
\label{tablerice}
\end{figure}

Rice et al.\cite{rice} have created a catalog of 1064 massive molecular clouds throughout the Galactic plane, using the $^{12}CO$ CfA-Chile survey, which is by far the most uniform survey of molecular gas in the Galaxy. This catalog, with 1064 massive clouds totaling $2.5 \cdot 10^8 M_{sun}$, contains $25^{+10.7}_{−5.8}$\% of the molecular gas mass of the Galaxy, and trace several spiral arms, most notably the Sagittarius,Perseus, Outer, Carina, and Scutum-Centaurus arms.
Rice et al. derived a cloud mass spectrum, which describes the distribution by number of clouds of different masses expressed as a truncated power law:

\begin{equation}
N(M^{'}>M) = N_0[(\frac{M}{M_0})^{\gamma+1}-1]\\
\end{equation}

where $\gamma$ is an index describing how mass is distributed amongst clouds. $\gamma$ > −2 indicates that the majority of mass is contained in massive clouds; $\gamma$ $\approx$ −2 means that mass is roughly equally distributed in all mass bins; and $\gamma$ < −2 indicates that low-mass clouds contain the majority of mass.
To measure mass spectra in their study,  Rice et al. used the maximum-likelihood method and they carry out this measurement region-by-region, only considered clouds with $v_{rad}|>20$ \phantom{} km $s^{-1}$  located at least 20$^{\circ}$ from $l$ = 0$^{\circ}$ or $l$ = 180$^{\circ}$ in order to minimize kinematic distance errors, which have a substantial effect on estimated masses. Results are summarized in figure \ref{tablerice} and each region mass distribution function are in figure \ref{mass} and \ref{mass2}

The authors also suggest that their catalog is complete in the solar circle for mass above $10^5$ $M_{sun}$
With this restrictions the number of molecular clouds drops from 1064 to 153. 
A first test for my model is to simulate an appropriate number of molecular clouds so that by applying the same restriction roughly the same number of clouds is obtained.
\begin{figure}
\centering
\includegraphics[width=0.7\textwidth]{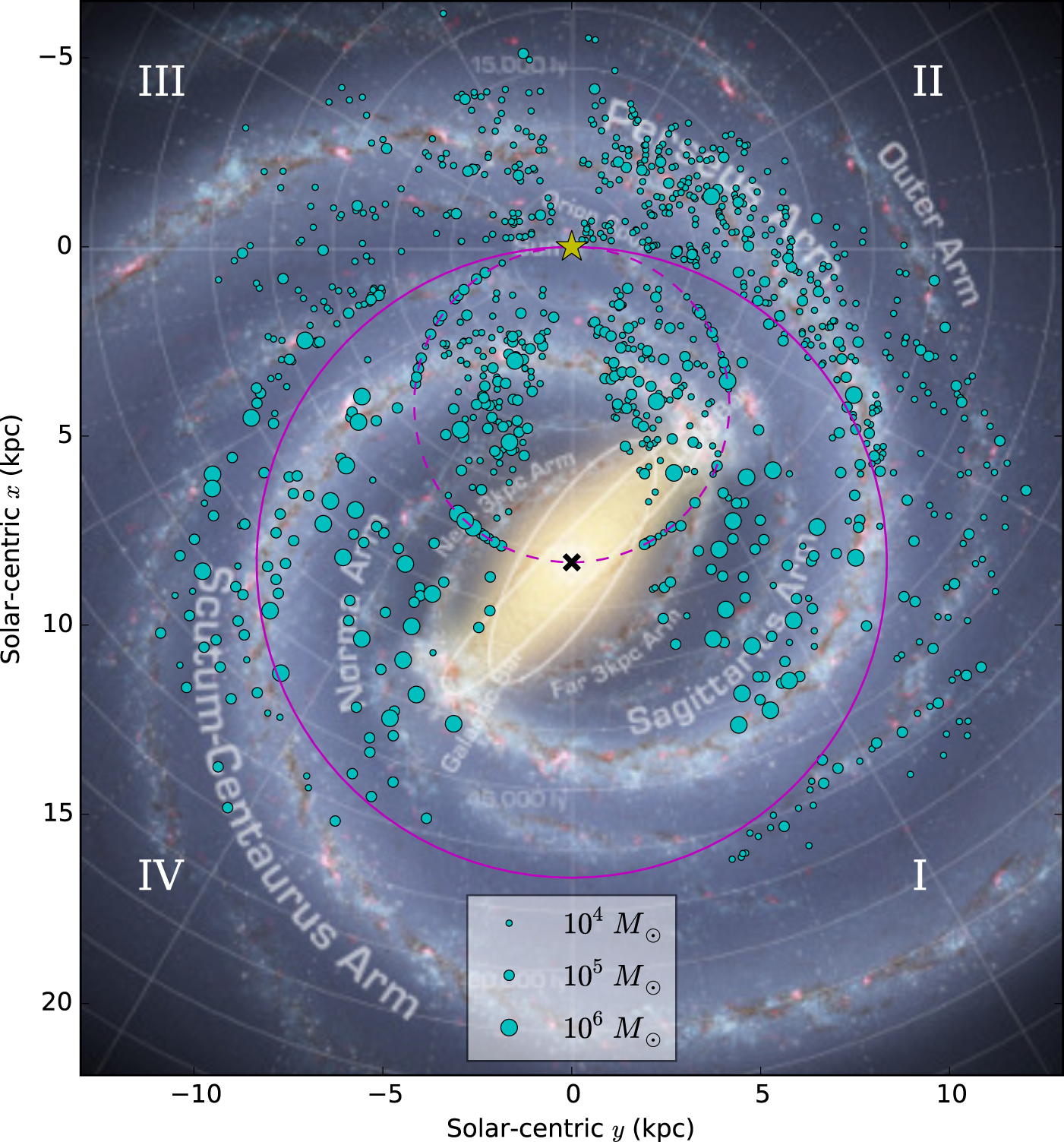}
\caption{Top-down Galactic map of all clouds with circle sizes indicating cloud masses. The solar circle is drawn in solid purple, and the tangent circle in the inner Galaxy is drawn in dashed purple. The Sun's location is marked with a yellow star and the Galactic Center is marked with a black "x."}
\end{figure}

\subsection{Synthetic population of molecular clouds}
Taking into account the Rice et al.'s studies, I had to reproduce the positions of the molecular clouds along the spiral arm, the molecular clouds mass distribution and their radial velocity distribution in order to create a realistic and normalized model.
\subsubsection{Spatial distribution}
The models for the spatial distribution of $\gamma$-ray sources in our galaxy are mainly based on
recent pulsar data. In general it is necessary to distinguish between evolved distributions,
that are obtained from observational data to describe the current population of objects
and birth distributions, that are used in studies to reproduce the current population of
objects. The spatial distribution is described via the surface density
of the objects, commonly it is radial symmetric and just varies with the distance to the Galactic center. The vertical distribution (height above the Galactic plane) is treated
separately.
\subparagraph{Radial distribution}\,\\

\textbf{Evolved distributions}: Based on the data of the current ATNF catalog Yusifov et al. analyzed the
radial distribution of pulsars in our galaxy. They found that the surface density could
be well described by a $\Gamma$-function:
\begin{equation}
\label{ev}
p(r) \propto (\frac{r}{R_{sun}})^a exp[-b \frac{r}{R_{sun}}]
\end{equation}
$R_{sun}$ denotes the position of the sun in the Galaxy. It was assumed that $R_{sun}$ = 8.5 kpc.
According to Case and Bhattacharya the distribution of SNRs in the galaxy can also be described by formula \ref{ev}. In
table \ref{tab:sp} all values of the parameters a and b are listed.

\begin{figure}
    \centering
    \includegraphics[width=0.6\textwidth]{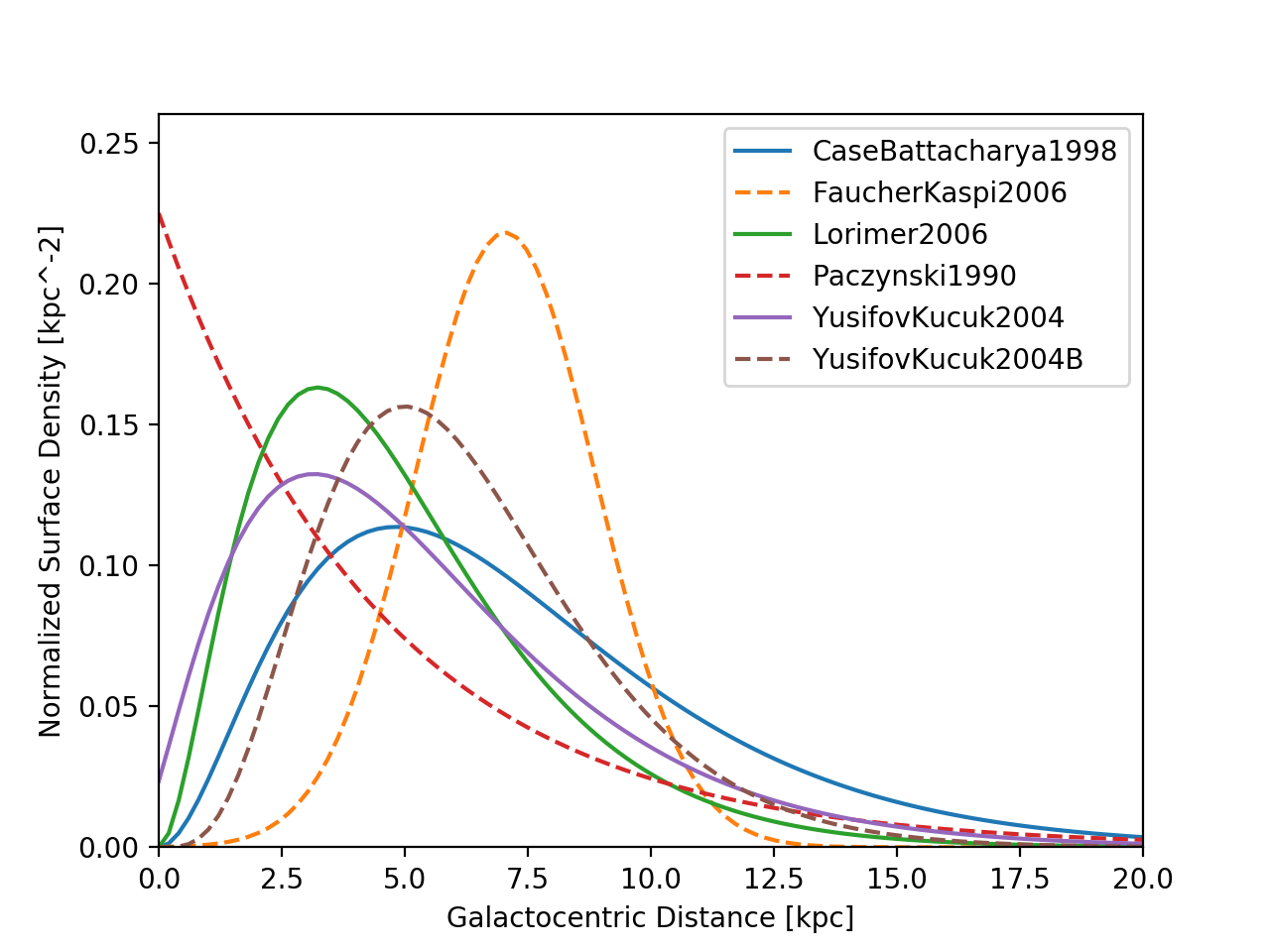}
    \caption{Comparison of radial distribution models. The dot dashed lines indicate
birth distributions, the solid lines evolved distributions. The area under the
curves is normalised to 1}
    \label{sp}
\end{figure}

\textbf{Birth distributions}:
In their optimal model Faucher et al. proposed a gaussian birth distribution  that led to:
\begin{equation}
p(r) \propto exp [ - \frac{(r-R_{\mu})^2}{2\sigma^2}]
\end{equation}
Paczynski (1990) made the assumption that the birth rate just varies exponentially with
distance from the Galactic center.
\begin{equation}
p(r) \propto e^{- r/r_0}
\end{equation}
Yusifov et al. instead proposed a birth distribution that follows the distribution of OB stars (which are considered as progenitor stars of neutron stars) in the Galaxy.
A comparison of all distribution models is shown in figure \ref{sp}. The solid curves represent
the current distributions, the dash-dotted lines the birth distributions. In general the peak of the birth distributions is positioned farther from the Galactic center. This holds
also for [CB98], as the distribution of SNRs can be considered as a birth distribution of
pulsars. The distribution of Paczynski is an exception, because it is not based on
current data.

\begin{table}[]
\centering
\begin{tabular}{|lllll|}
\hline
\multicolumn{1}{|l|}{Model}      & \multicolumn{1}{l|}{Abbrev.} & \multicolumn{1}{l|}{Parameters}  & \multicolumn{1}{l|}{Birth} & Evol. \\ \hline
Case and Bhattacharya (1998)     & {[}CB98{]}                       & a=2.0 b=3.53                     &                           $\surd$ &      $\surd$   \\
Faucher-Giguere and Kaspi (2006) & {[}F06{]}                        & $R_{\mu}$=7.0 $\sigma$=1.8kpc &                           $\surd$ &         \\
Paczynski (1990)                 & {[}P90{]}                        & $r_0$=4.5 kpc                    &                           $\surd$ &         \\
Yusinov and Kucuk (2004)         & {[}YK04{]}                       & a=1.64 b=4.01                    &                            &    $\surd$     \\
Yusinov and Kucuk (2004)         & {[}YK04B{]}                      & a=4.0 b=6.80                     &                           $\surd$ &         \\
Lorimer et al. (2006)            & {[}L06{]}                        & a=1.9 b=5.0                      &                            &     $\surd$    \\ \hline
\end{tabular}
\caption{Parameters of the radial distribution models.}
\label{tab:sp}
\end{table}

\subparagraph{Z-distribution}
It is commonly assumed that the current vertical distribution of pulsars follows an exponential law:
\begin{equation}
p(z)= e^{-z/z_0}
\end{equation}
Faucher-Giguere and Kaspi demonstrate that an initial scaleheight of $z_0 \approx 50$ pc reproduce the current distribution of pulsars.

\subparagraph{Spiralarm modelling}
Based on observations of other spirals it is known that massive stars are mostly born
in spiralarms. As it is not expected that they move far from where they were formed
during their lifetime, the spiralarm structure should be considered in the distribution
of birth positions of pulsars. A simple model can be taken from Faucher-Giguere and
Kaspi. The raw shape of each arm follows a logarithmic spiral that is defined by:
\begin{equation}
\Theta(r)=k ln (r/r_0)+\Theta_0
\end{equation}
The parameters of the four known spiralarms of the milky way are shown in table \ref{spiral}
\begin{table}
\centering
\begin{tabular}{|cccc|}

\hline 
Spiralarm & k[rad] & $r_0$[kpc] & $\Theta_0$ [rad] \\ 
\hline 
Norma & 4.25 & 3.48 & 1.57 \\ 

Carina-Sagittarius & 3.48 & 4.25 & 4.71 \\ 
 
Perseus & 4.89 & 4.90 & 4.09 \\ 

Crux-Scutum & 4.89 & 4.90 & 0.95 \\ 
\hline

\end{tabular} 

\caption{ Parameters of the four known spiralarms of the milky way}
\label{spiral}
\end{table}

For the spatial distribution in the Galaxy, the Lorimer et al. radial distribution [L06] was used, this model together with the Z-distribution and the spirlarm model are already implemented in the gammapy libraries \footnote{//docs.gammapy.org/0.6/astro/population/index.html}\footnote{http://docs.gammapy.org/0.6/astro/population/index.html}. 

\subsubsection{Velocity distribution}
\begin{figure}
\centering
\includegraphics[width=1\textwidth]{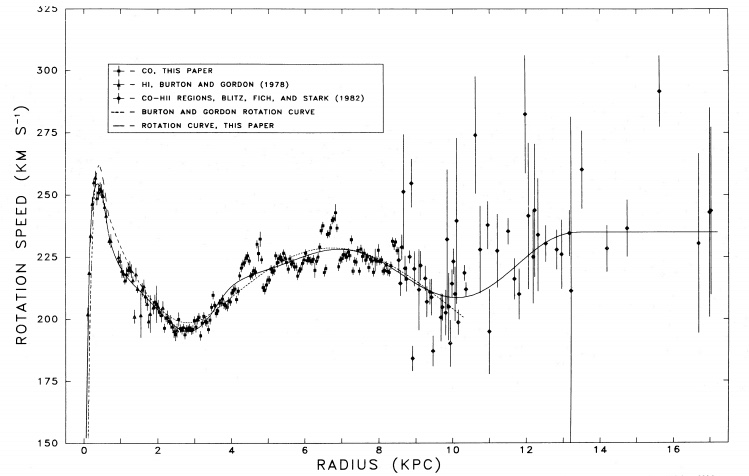}

\caption{Plot of the rotation speed versus galactocentric radius. The solid line correspods to the polynomial function}
\label{rot}
\end{figure}

Plot of rotation speed versus distance from the center of the Milky Way reveals that the stars do not orbit the center of the galaxy like planets orbiting the Sun (Keplerian rotation) but rather follow a much flatter rotation curve. This difference implies much more mass lying far from the Galactic Center than one would infer from stars and gas detectable at visible, IR, and radio wavelengths. This in turn implies that most of the mass ($\sim$90\%) of the galaxy is  dark matter. Other spiral galaxies have similar ratios of visible to invisible matter.
Rotation of the Milky Way can be used to determine distances of clouds along the line-of-sight. The Galaxy rotation curve (figure \ref{rot}) is well described by a 7th grade polynomial function \cite{} whose coefficients depends on the distance from center, it means that if we know the distance of the cloud, we compute its velocity. Note that the velocity calculated by this function is a rotational velocity so we need to convert in radial velocity taking into account the Sun's velocity.

\begin{figure}
\centering
\includegraphics[width=0.6\textwidth]{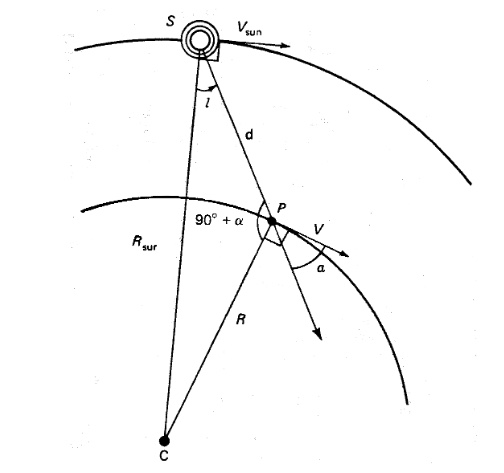}
\caption{Plot of the rotation speed versus galactocentric radius. The solid line correspods to the polynomial function}
\label{velrot}
\end{figure}
 
As one can see in figure \ref {velrot}, the observed radial velocity of the cloud will be the difference between the Sun's component of rotation and the cloud's along the line-of-sight:
\begin{equation}
V_{rad}=V \cos \alpha - V_{sun}\sin l
\end{equation}
where $V_{sun}$ is the sun's velocity around the galactic center (220 $km s^{-1}$, V is the rotational velocity and l is the galactic longitude.
After some trigonometry calculations the radial velocity can be write as:
\begin{equation}
V_{rad}=R_{sun}\cdot \sin(l) \cdot (\frac{V_{rot}}{r}-\frac{V_{sun}}{R_{sun}}) 
\end{equation}
where R is the distance from galactic center and $R_{sun}=8.5$ $kpc$

\subsubsection{Mass}

I had a catalog of Molecular cloud positions, without any information about the mass of the object. To associate a reasonable mass I used the mass distribution function derived in the Rice catalog. I considered only regions in the solar circle (I-IV) and mass above $10^5M_{sun}$, since the catalog is almost complete in these conditions. As one can see in figure \ref{tablerice} this function is a truncated power law with index $\gamma=-1.72$.
If the differential mass distribution function is extrapolated up to the current instrument sensitivity limit ($M_{cloud} \sim 10^3$) and then this function is integrated, the number of molecular clouds obtained is 13500. This is the number of the clouds that should be simulated. 
To associate randomly a mass to each cloud, the idea was to choose a set of 13500 values on the $N(M^{'}>M)$ axis that intersect the mass distribution function and get the corresponding mass value, then associated the mass values with the positions of the molecular clouds.\\

\begin{figure}
\centering
\includegraphics[width=0.7\textwidth]{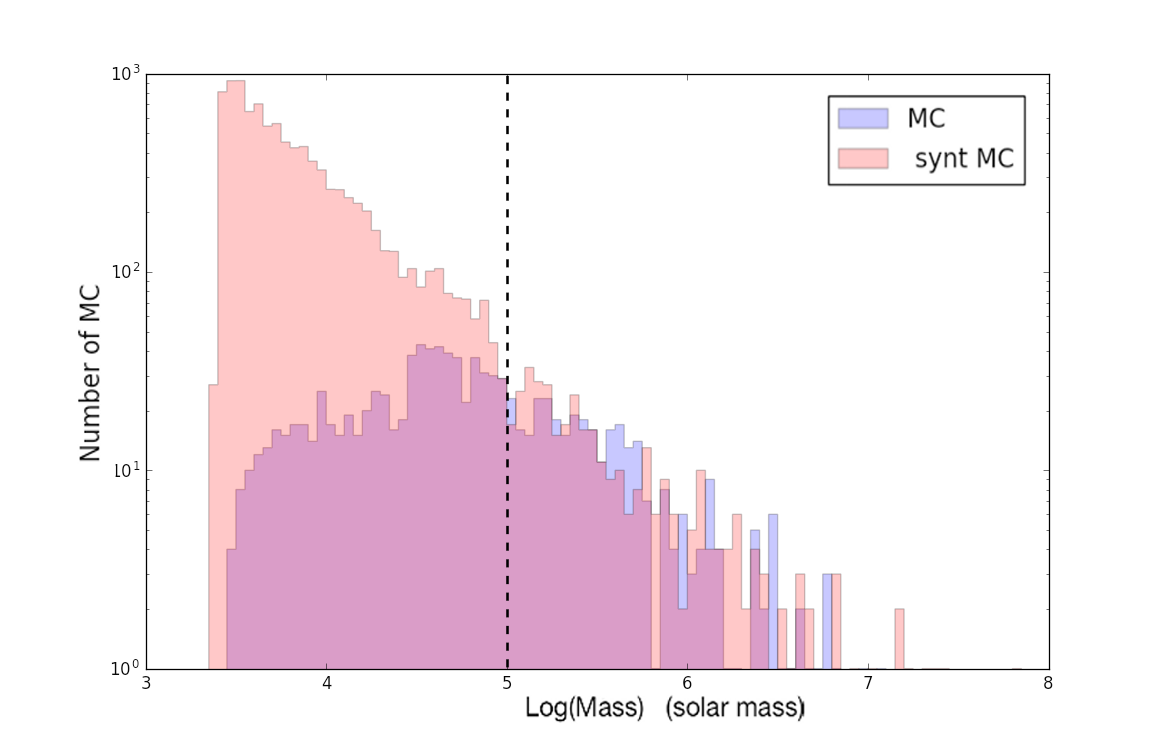}
\caption{Comparison between the mass distribution of real (blue) and synthetic (red) molecular clouds}
\label{cmpmass}
\end{figure}

\begin{figure}
\centering
\includegraphics[width=0.8\textwidth]{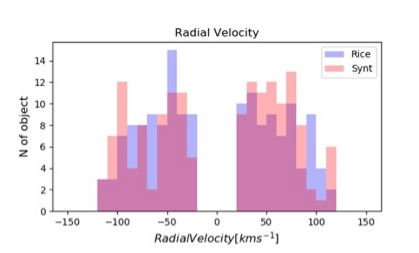}
\caption{Comparison between the radial velocity distribution of real (blue) and synthetic (red) molecular clouds}
\label{cmpvrad}
\end{figure}

\begin{figure}[ht!]
\centering
\includegraphics[width=1\textwidth]{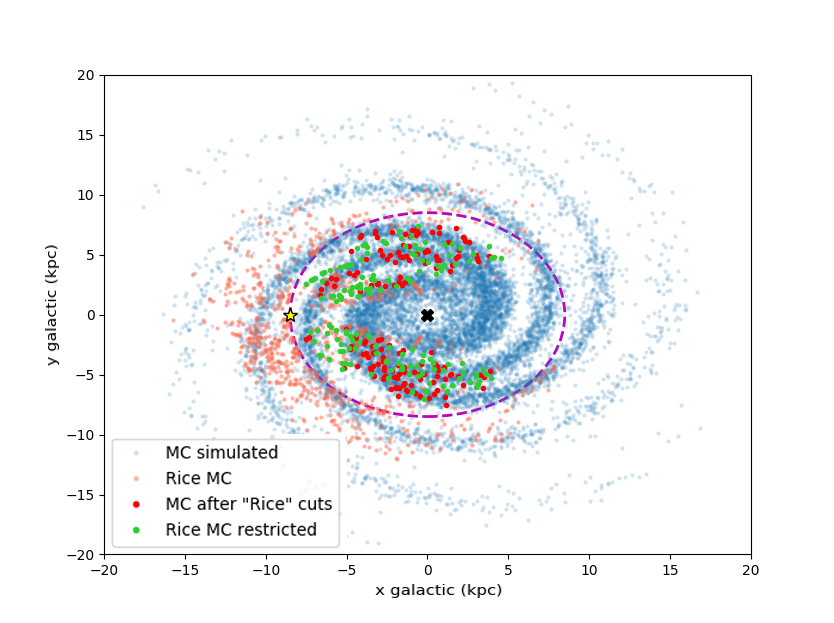}

 \caption{ Galactic map of all clouds, in blue are reported all the synthetic clouds, the pink point are for all the Rice MCs while the red and the green point are for synthetic and real (respectively) MCs after the "Rice cut". The Sun’s location is marked with a yellow star, and the Galactic Center is marked with a black “x”. The dashed purple line represent the solar circle}
\label{myrice}
\end{figure}

I had all the  physical parameters to compare the synthetic population of molecular clouds with the real one.
The aim of the comparison is to calibrate the model with observations, since the Rice et al. catalog is complete only within the restriction mentioned above, applying that restriction 156 molecular clouds were obtained, comparable with the 153 real molecular clouds in the Rice et al. catalog. In figure \ref{myrice}  the synthetic molecular cloud population and the MC of the Rice et al. catalog are superimpose.
Figure \ref{cmpmass} represent the  comparison between the mass distribution of real (blue) and synthetic (red) population of molecular clouds, the dashed black line is the completeness limit suggest by Rice et al.
In figure \ref{cmpvrad} are compared the radial velocity distributions of the two populations.
Executing a two sample Kolmogorov-Smirnov test on mass, positions and radial velocity distributions, the two population of molecular clouds (the synthetic and the real one) are compatible, this means that in principle the Rice et al. sample could be extract from the synthetic population.

\section{Spectra of Cosmic rays in Molecular clouds}\label{spectra}

The gamma emission from the interacting supernova remnants comes from the interaction between cosmic rays protons accelerated by the supernova shock and the nuclei of atoms present in the molecular cloud.
Reproduce the interaction between protons and nuclei means creating the protons injection spectra and estimate the synthetic molecular clouds density.
\subsubsection{Injection spectra}

Thanks to the catalog I've made, for each of the 16 interacting supernova remnants confirmed up to now, all the spectral parameters of the protons distribution are available. 
The idea was to create a gaussian distribution for each parameters ($\alpha$, $\beta$, Break Energy) starting from mean and standard deviation values of the real interacting supernova remnants (see table \ref{tab:param}).

\begin{table}[]
\centering
\begin{tabular}{|c|c|c|c|}

\hline
     & $\alpha$         & $\beta$  & Break Energy ($E_b$) \\
     &                  &          & (MeV)                \\ 
\hline
Mean   & 2.43  &  3.60   &  $10^{5.49}$                    \\ \hline
Std   & 0.21  &  0.78   &  $ 10^{1.11}$                    \\ \hline
\end{tabular}
\caption{Mean and standard deviation values of the proton power law spectra of the real SNR.}
\label{tab:param}

\end{table}

\begin{figure}[ht!]
\centering
  \includegraphics[width=0.45\textwidth]{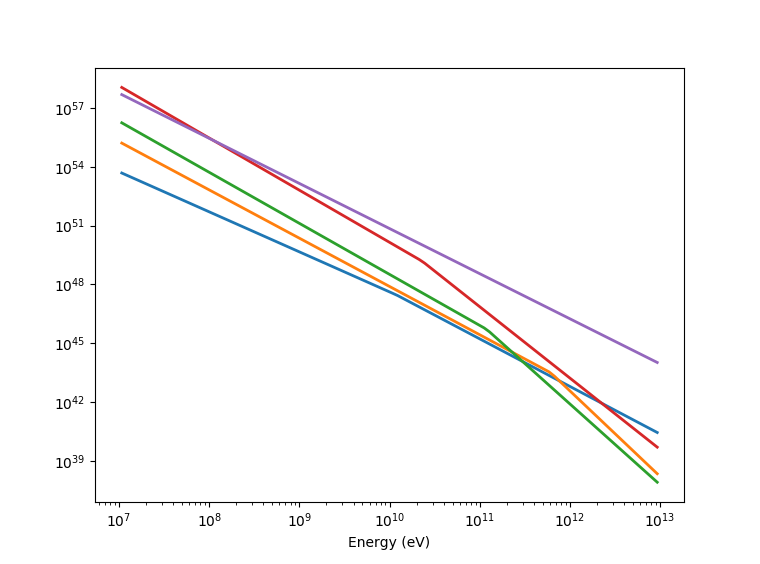}
  \includegraphics[width=0.45\textwidth]{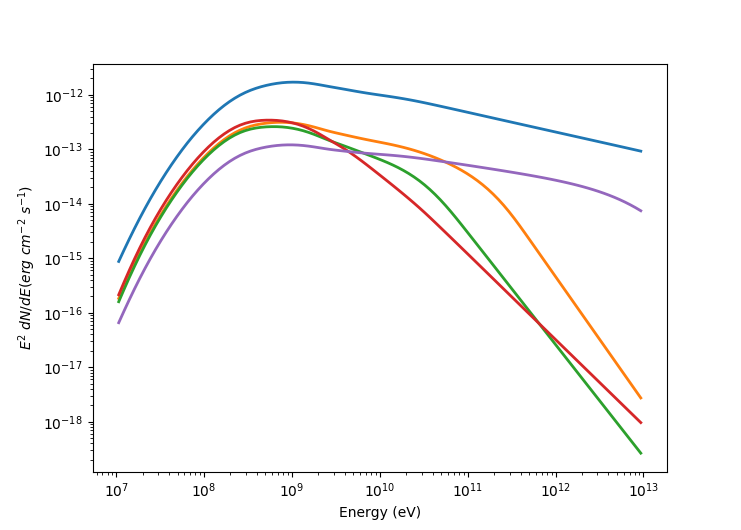}
  \caption{Example of Protons distributions and $\pi^0$-decay spectra resulting from the parameters extracted}
  \label{spe}
\end{figure}

Note that this values are not statistically very significant because of there are few objects, therefore I assumed that if we knew much more interacting supernova remnants their spectral parameters would be distributed following a Gaussian curve with mean and sigma value close to those are in table \ref{tab:param}. I extracted randomly the spectral parameters value for each synthetic interacting supernova remnants. These extracted parameters will be those of the broken power law proton distribution for the Naima pion decay spectra calculation, figure \ref{all} and \ref{spe} clarify the method.
I point out that the $\beta$ index must be greater than the $\alpha$ index. This problem was solved creating the Gaussian distribution of $\alpha - 2 $ instead of $\alpha$ and extracting only positive values.

\begin{figure}[ht!]
\centering
  \includegraphics[width=0.7\textwidth]{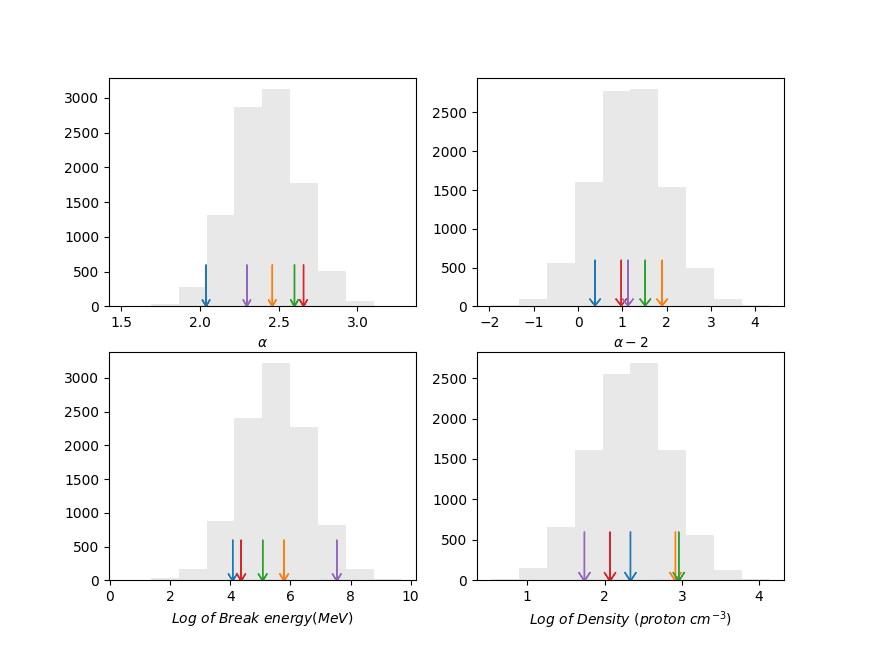}
  \caption{Example of the spectral parameters extraction}
  \label{all}
\end{figure}
\subsubsection{Molecular clouds density}
Generate the synthetic interacting supernova remnants spectra requires besides the particle distribution, the density of the target protons. As the other this parameter must be randomly extracted from the distribution generated starting from the observed values. Since I hadn't the information about the density of the target protons in the molecular clouds of real interacting supernova remnants, I used the Rice et al. catalog.
Some reasonable hypothesis must be done, the first assumption is that the molecular clouds have the same density everywhere, so the target protons density 
can be confused with the clouds density. Another approximation is that  molecular clouds are spherical.
Now using the mass and the size (diameter) of the real molecular clouds, I calculated the density and then the mean and standard deviation values of proton's density. 
Note that, to not introduce any bias, only molecular clouds with mass above $10^5 M_{sun}$ are considered in the density calculation.\\

\begin{figure}[ht!]
\centering
  \includegraphics[width=0.5\textwidth]{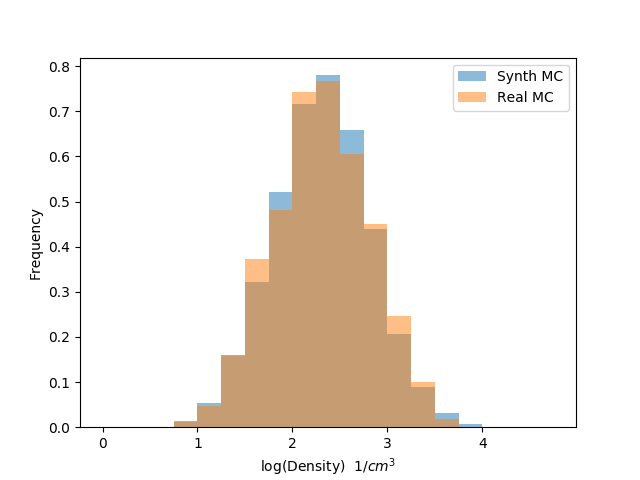}
  \caption{Density distributions of real (orange) and synthetic (blue) MCs}
  \label{all}
\end{figure}

For calculation efficiency, I associated an injection spectra to each synthetic molecular cloud, as if all molecular cloud have a supernova remnants nearby. Surely not true, but this allows me to obtain the LogN-LogS of the all synthetic interacting supernova remnants (red line in figure \ref{all_p0}) that is crucial to find a probability function that returns the correct number of synthetic iSNR.

\section{The molecular clouds - SNR association }
\begin{figure}[]
    \centering
    \includegraphics[width=0.8\textwidth]{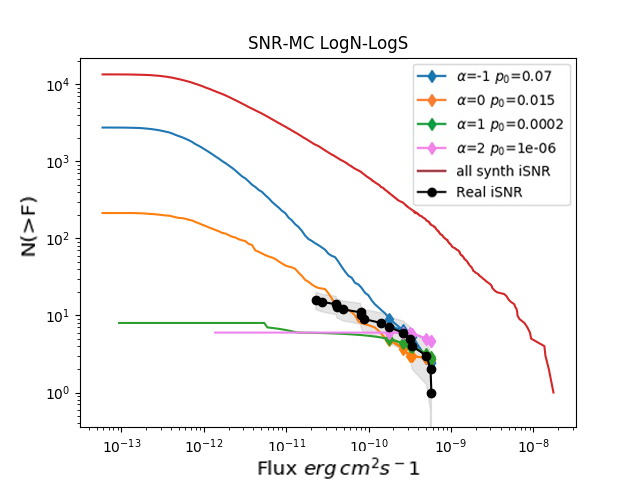}
    \caption{LogN-LogS curves of: all the synthetic iSNR in red color, black is for the real iSNR, $\alpha =-1, p_0=0.05$ in blue, $\alpha =0, p_0=0.015$ in orange, $\alpha =1, p_0=0.002$ in green and pink for $\alpha =2, p_0=10^{-6}$.  }
    \label{all_p0}
\end{figure}

The results is a catalog of 13500 interacting supernova remnants distributed in the Galaxy. Certainly not all the molecular clouds have a SNR nearby, therefore I had to estimate how many of supernova remnant-molecular cloud systems can form in the  Galaxy. I introduced a probability function that in principle depend on the clouds mass as

\begin{equation}
 P(M_{MC}) = p_0 \cdot (\frac{M_{MC}}{m_0})^{\alpha}
\end{equation}
in which $M_{MC}$ is the molecular cloud mass, $m_0$ is the reference mass fixed at $10^4 \, M_{sun}$. The free parameters are $p_0$ (probability parameter) and $\alpha$, by acting on these parameters the LogN-LogS curve of all the synthetic interacting systems (i.e red line in in figure \ref{all_p0}) changes towards the curve of the real interacting supernova remnants. Since in principle there are almost infinite pairs of parameters that reproduce the slope on the real iSNR LogN-LogS in the completeness region, I set the $\alpha$ exponent a priori equal to: -1, 0, 1, 2 and then for each exponent the best $p_0$ value was found, results are reported in table \ref{tab:a0_p0}. 
The best probability values was found by making 1000 extraction from the synthetic iSNR catalog, meaning 1000 LogN-LogS, keeping fixed the probability function exponent and calculated the average LogN-LogS curve.
The method  used to averaging 1000 curves in the completness region, is fix the flux of the real iSNR in that region  and interpolate the curves at this fixed fluxes.
The goodness of the obtained LogN-LogS is evaluated making a two sample Kolmogorov-Smirnov test between the real interacting supernova remnants LogN-LogS and the synthetic one in the significant region. 
Results reported in figure \ref{all_p0} clarify that the only possible probability function is the flat one (orange line, $\alpha = 0, \, p_0=0.015 $), because the linear and the quadratic option do not reproduce the real iSNR LogN-LogS curve at all, instead the curve with $\alpha = -1, \, p_0=0.07 $, even if fit the real iSNR behavior, provides  $\sim$ 3000 interacting SNR in the Galaxy, about 10 times all the supernova remnants seen in the Galaxy. For these reason I considered only the flat probability at 1.5\%, therefore, the number of the interacting system expected in the Galaxy in this model is about 200.
The next step is to simulate each of these synthetic interacting supernova remnants with CTA to estimate how many of these systems have flux above the CTA sensibility threshold.

\begin{table}[ht!]
    \centering
    \begin{tabular}{|c|c|c|c|c|}
       \hline
       $\alpha$ & -1 & 0 & 1 & 2 \\ \hline
         $p_0$  & 0.07 & 0.015 & 0.002 & $10^{-6}$ \\
         \hline
    \end{tabular}
    \caption{$p_0$ values found.}
    \label{tab:a0_p0}
\end{table}

\section{Repetition}
\begin{figure}[]
    \centering
    \includegraphics[width=0.7\textwidth]{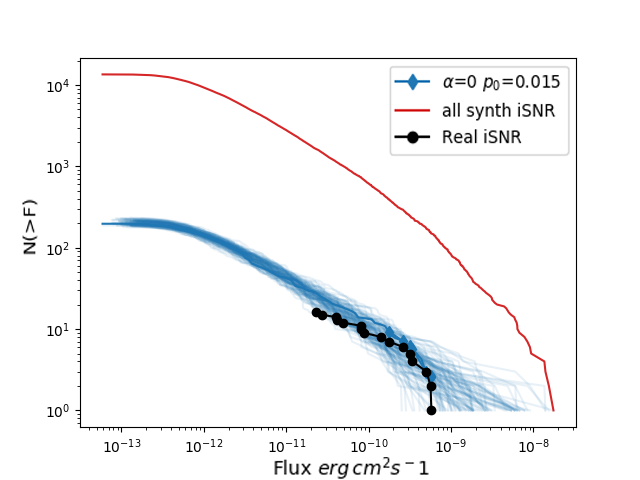}
    \caption{LogN-LogS curves of: all the synthetic iSNR in red color, black is for the real iSNR, 100 realization with $\alpha =1, p_0=0.002$ in blue color }
    \label{100rea}
\end{figure}

\begin{figure}[]
\centering
      
        \includegraphics[width=0.7\textwidth]{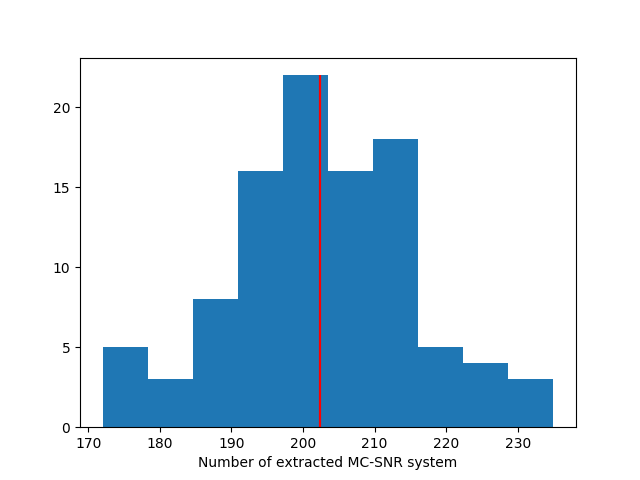}
        \caption{Distribution of the number of interacting system extracted, the red line is the average number of iSNR extracted.}
        \label{fig:dist}
\end{figure}

Since the number of the interacting systems extracted from the synthetic iSNR catalog depends on the realization, to exclude any possible selection effects  100 realizations were made keeping fixed probability function parameters (see figure \ref{100rea}). 
As one can see in figure \ref{fig:dist} the interacting system number goes from 172 to 235 and the mean value is 202 .
Basically each realization is a set of synthetic interacting supernova remnants positions, dimensions, spectra, integrated fluxes in 100 MeV-10TeV or 0.1-100 TeV energy bands.


\chapter{Simulation and analysis}
In general, simulating an astronomical observation consists into randomly generate events that will be produced by both background and sources, assuming a precise spatial and spectral model. The simulation algorithms are implemented in ctools and are based on Monte
Carlo methods. All the generated events are  weighted with the instrumental response
functions (IRF) in order to estimate the number of events that the instrument will effectively be
able to detect. The analysis tools then allow to fit this generated sample of data and to calculate the statistical significance of them. In this way is possible to predict the range
of confidence in which the  parameters  of our model will be constrained.

The likelihood \textit{L} is defined as the probability of obtaining a certain result starting from an
input model. The input model is defined in a ".xml" file and contains a description of the
gamma ray sources. Each source is divided into a spatial component, namely the celestial coordinates and the shape of the object and a spectral component which provides the energy distribution of the photons. In the input model we choose the
free parameters, i.e. the parameters to be fitted. The fit maximizes the likelihood.
The $\chi^2$ is related to the likelihood as : $\chi^2$= −2$log \textit{L}$ so maximize the likelihood is equivalent to minimize the $\chi^2$.
The input model $M(E, \vec{p}, t|\alpha)$ will be then the differential flux per unit area in the observed
region of the sky as a function of the true energy E, the true arrival direction $\vec{p}$ and the time
of observation t, that is important for variable phenomena. The term $\alpha$ labels the free parameters of the model. Once obtained the number of events, the flux is estimated according to:
\begin{equation}
\Phi(E', \vec{p},|\alpha) = \int dE d \vec{p}R(E', \vec{p'}|E, \vec{p})M(E, \vec{p}|\alpha)
\end{equation}
where $R(E', \vec{p'}|E, \vec{p})$ are the IRFs, as described in \ref{perf}.
The more the model resembles the reality, the more probable is to find results that are close
to that model. In other words \textit{L} is a test of
the goodness of the model.

The population model of interacting supernova remnants that I realized gives the possibility to predict how many of this systems will detectable by the Cherenkov Telescope Array and more in general by any future gamma ray instruments. 
The intent of the simulations and analysis reported in this chapter is to give an estimate of how many supernova remnants - molecular cloud systems CTA can detect. I simulated the sources individually, in the center of the field of view with only the CTA instrumental background, certainly since this is an ideal case, the resulting number of detected iSNR is overestimated but anyway it can provide a good indication on the possibilities of CTA regarding these extended systems.

\section{Simulation}
In order to avoid any possible selection effect and to increase the statistic, I simulated 10 realization randomly selected.  The CTA simulation tools in ctools (version 1.4.2) is ctobssim.
Ctobssim simulates events from the input model and the instrument response functions (IRF) using a numerical random number generator. Both the astrophysical sources and the background are simulated.
To perform the simulation basically are required the Instrument Response Function of CTA and the source model, both the spectral  and the spatial model. The general model is describe in ctools using a model definition XML file. Below is a simple example of such a file comprising one source and one background model. Source model is factorized into a spectral (tag <spectrum>), a spatial (tags <spatialModel>) instead the instrument response functions contain templates that describes the spatial and spectral distribution of the background.

\begin{equation}
M(x,y,E,t)=M_{spatial}(x,y|E) \times M_{spectral}(E) \times M_{temporal}(t)
\end{equation}

\begin{figure}
    \centering
    \includegraphics[width=1\textwidth]{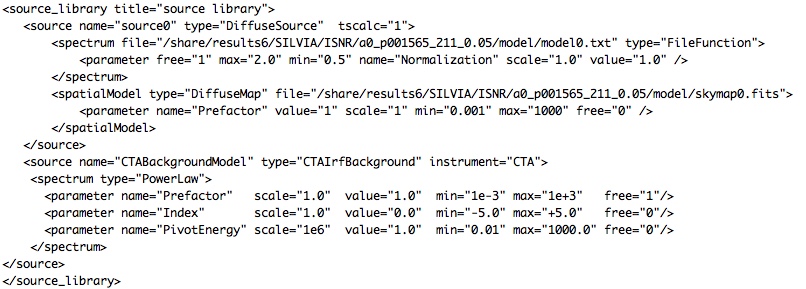}
	\caption{Example of a source model file }
    \label{fig:modelxml}
\end{figure}

In this specific example (figure \ref{fig:modelxml}), the source component describes a diffuse source with a spectral model imported from a file that contains the punctually defined photons spectra. The spatial model is a source map and the background component is the CTAIrfBackground, so only instrumental background is considered. 
Model parameters are specified by a <parameter> tag with a certain number of attributes. The name attribute gives the  parameter name that needs to be unique within a given model component. The scale attribute gives a scaling factor that will be multiplied by the value attribute to provide the real (physical) parameter value. The min and max attributes specify boundaries for the value term of the parameter. And the free attribute specifies whether a parameter should be fitted (free="1") or kept fixed (free="0") in a maximum likelihood analysis.

\begin{figure}
    \centering
    \includegraphics[width=1\textwidth]{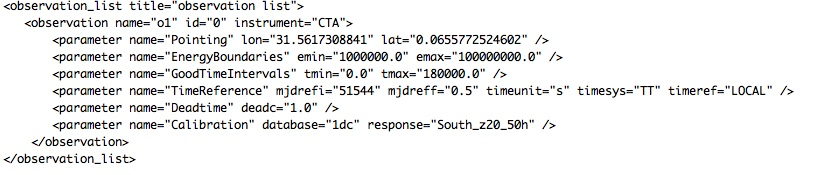}
\caption{Example of an observation file }
    \label{fig:obs}
\end{figure}

To start the simulation with ctobssim another file is required, namely the observation file. This file (see figure \ref{fig:obs}) contains the information about the parameters of the simulation position of the center of the FOV, energy boundaries, time, calibration database and irf to be used.

I randomly chose 10 realization, and I simulated each synthetic interacting supernova remnants as explained above. Ctobssim write two files in the working directory: events.fits and ctobssim.log. The first file contains the simulated events in FITS format. The FITS file contain three extensions: an empty primary image, a binary table named EVENTS that holds the events (one row per event), and a binary table named GTI holding the Good Time Intervals. The second file produced by ctobssim is a human readable log file that contains information about the job execution.

\begin{figure}    
\centering
    \includegraphics[width=1\textwidth]{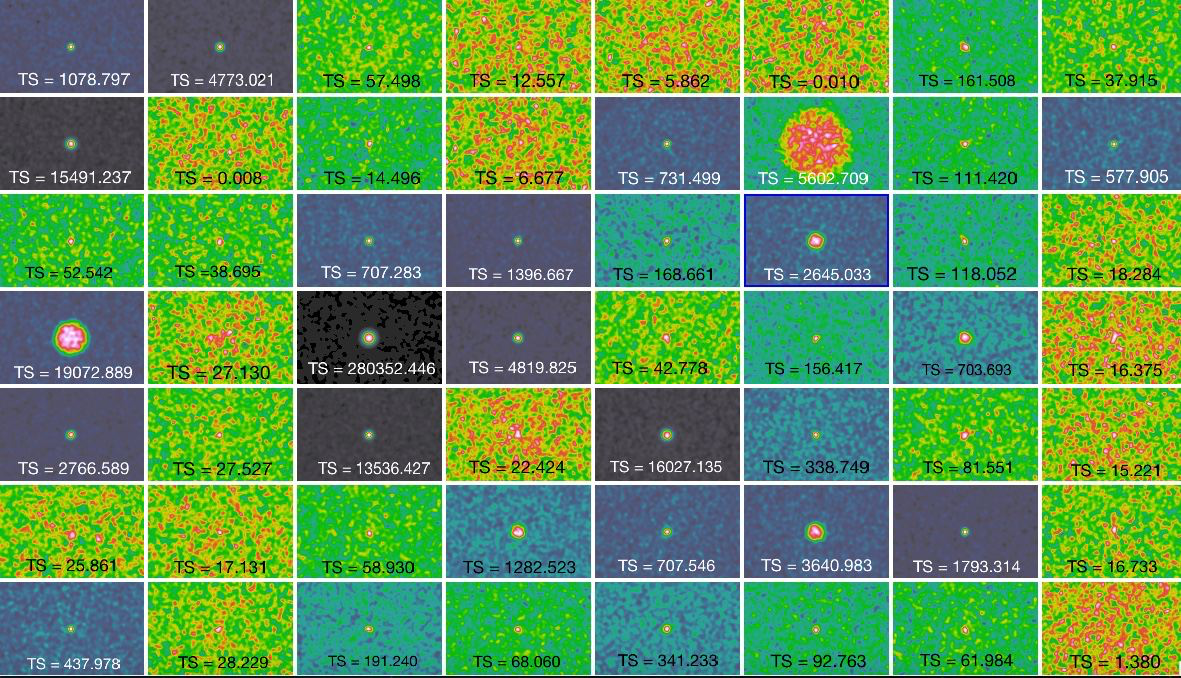}
    \caption{Sky map of some simulated interacting supernova remnants, the colorbar and amplitude are the same for all sources. }

    \label{map}
\end{figure}

I could look at the simulated sources map using ctskymap. The tool produces the file skymap.fits which contains a sky map of the events in FITS format (see figure \ref{map}). The sky map is centered on the location of the sources and consists of 200 x 200 spatial pixels of 0.02 x 0.02 degrees in size, covering an area of 4 deg x 4 deg.  
The figure \ref{map} shows a few simulated interacting supernova remnants, some of these clearly emerge from the background, instead the others have too weak flux at TeV energies to be detectable. 

\subsection{Spectral and spatial models}

\begin{figure}
\centering
    \includegraphics[width=\textwidth]{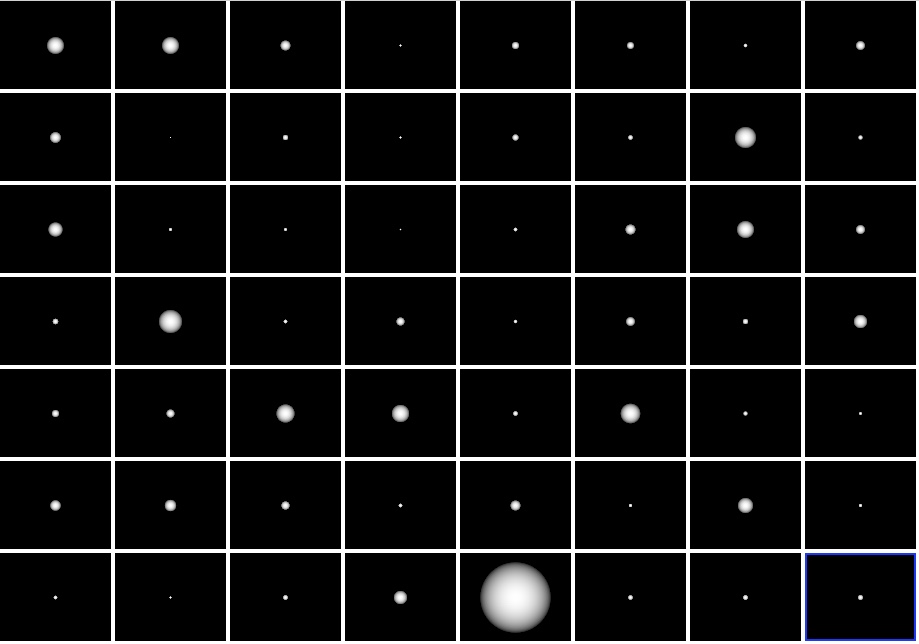}
    \caption{Some interacting supernova remnants spatial template. The size of  the maps is 4 $\times$ 4 degree.}

    \label{spatial}
\end{figure}

Spectra models are created as described in \ref{spectra} and converted in  $ph\cdot cm^{-2}s^{−1}MeV^{−1}$. Thus each simulated source has a photons spectra associated, example of simulated $\pi^0$-decay spectra are in figure \ref{spe} (right). 
This spectra are defined by a numerical function because it is the simplest way to simulate a source with spectral model not defined by a classical functional form. For the spatial model I created a source map for each simulated interacting supernova remnants, using a python code that fills a pixels matrix with values from 0 to 1. The filling criteria is the projection of a sphere on a plane perpendicular to the line of sight, results of some simulated map are in figure \ref{spatial}

\section{Analysis}

\begin{figure}[ht]
    \centering
    \includegraphics[width=\textwidth]{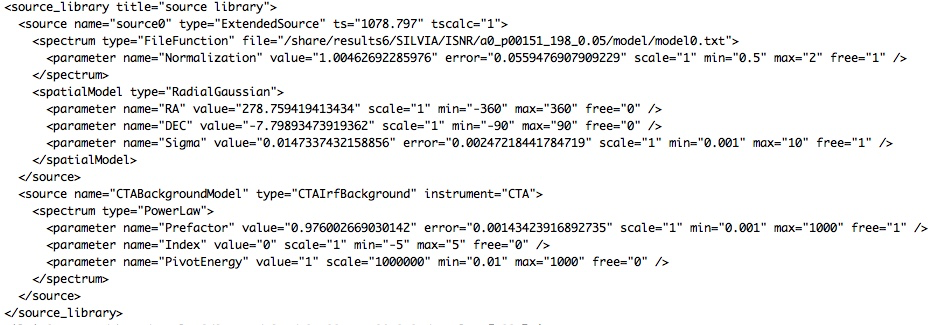}
\caption{Example of file resulting from the analysis with ctlike}
    \label{resu}
\end{figure}

The CTA analysis tools is ctlike that performs a maximum likelihood model fitting on the data and gives the estimation with the relative error of the needed parameters. It computes the detection significance, namely the TS value. The detection significance of a source model is estimated using the so called Test Statistic (TS) which is defined as
\begin{equation}
TS=2(lnL(M_s+M_b)−lnL(M_b))
\end{equation}
where $lnL(M_s+M_b)$ is the log-likelihood value obtained when fitting the source and the background together to the data, and $lnL(M_b)$ is the log-likelihood value obtained when fitting only the background model to the data. Under the hypothesis that the model $M_s+M_b$ provides a satisfactory fit of the data, TS follows a $\chi^2_n$ distribution with n degrees of freedom, where n is the number of free parameters in the source model component. Therefore
\begin{equation}
p=\int^{+\infty}_{TS} \chi^2_n(x)dx
\end{equation}

gives the chance probability that the log-likelihood improves by TS/2 when adding the source model $M_s$ due to statistical fluctuations only. For n=1 the significance in Gaussian sigma is given by $\sqrt{TS}$. In figure \ref{map}, under each simulated interacting supernova remnants, is reported the TS values coming from the ctlike likelihood analysis.
Ctlike takes in input the model to be fitted: spectral, spatial and background. For the spatial model I used the same model of the simulated sources with free normalization, for the spatial model I used a <RadialGaussian> model at fixed position but with free $\sigma$. For the background I used only instrumental background with only the normalization free. Ctlike generates an output model XML file that contains the values of the best fitting model parameters. For all free parameters, an error attribute is added providing the statistical uncertainty in the parameter estimate.  
Ctlike as ctobssim write two file, the first is an xml file that containes the best fit parameters resulting from the  maximum likelihood analysis, instead the second is a log file that contains information registered during the analysis. In figure \ref{resu} is reported an example of the analysis output. 

\section{Results}

From the analysis of the 10 realization randomly extracted, I found that CTA detect between 29 and 52 interacting supernova remnants. The detected sources are those having a detection at 5$\sigma$ or TS > 25, (see table \ref{res}). 

\begin{table}[b]
\centering
\begin{tabular}{|ccccccccccc|}
\hline
N$^{\circ}$ simulated sources & 194 & 198 & 191 & 211 & 216 & 181 & 199 & 212 & 214 & 211 \\
N$^{\circ}$ detected sources & 47 & 44 & 40 &45 &50 &29 &39 & 42 & 45 & 52 \\
\hline
\end{tabular}
\caption{Number of simulated and detected sources for each realization considered. }	
	\label{res}
\end{table}

\begin{figure}[ht]
    \centering
    \includegraphics[width=.9\textwidth]{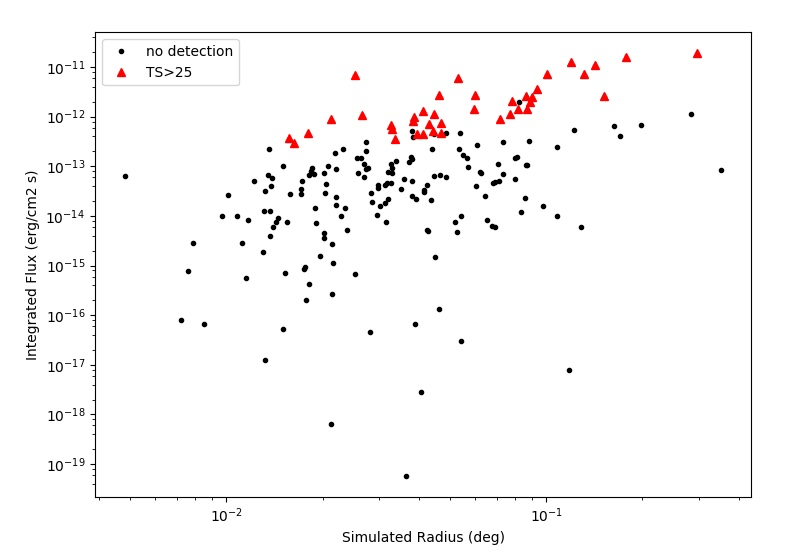}
	\caption{Plot of the simulated radius versus the integrated flux from 1 to 10 TeV of the selected realization. Red symbol marks the detected sources.}
    \label{rea11}
\end{figure}

\begin{figure}[ht]
    \centering
    \includegraphics[width=.85\textwidth]{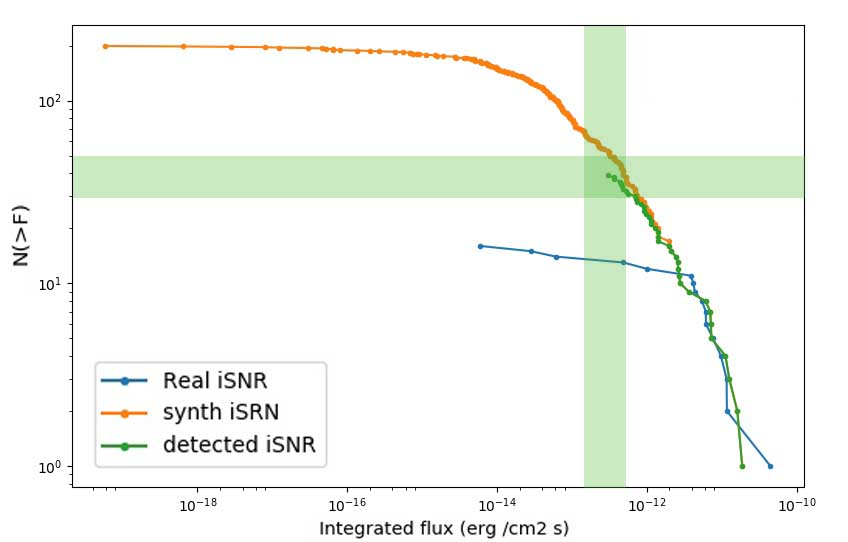}
	\caption{LogN-LogS of the real(blue), simulated (orange) and detected (green) iSNR. On the x-axis flux is integrated from 1 to 10 TeV.}
    \label{log11}
\end{figure}

\begin{sidewaysfigure}[ht]
    \centering
    \includegraphics[width=1\textwidth]{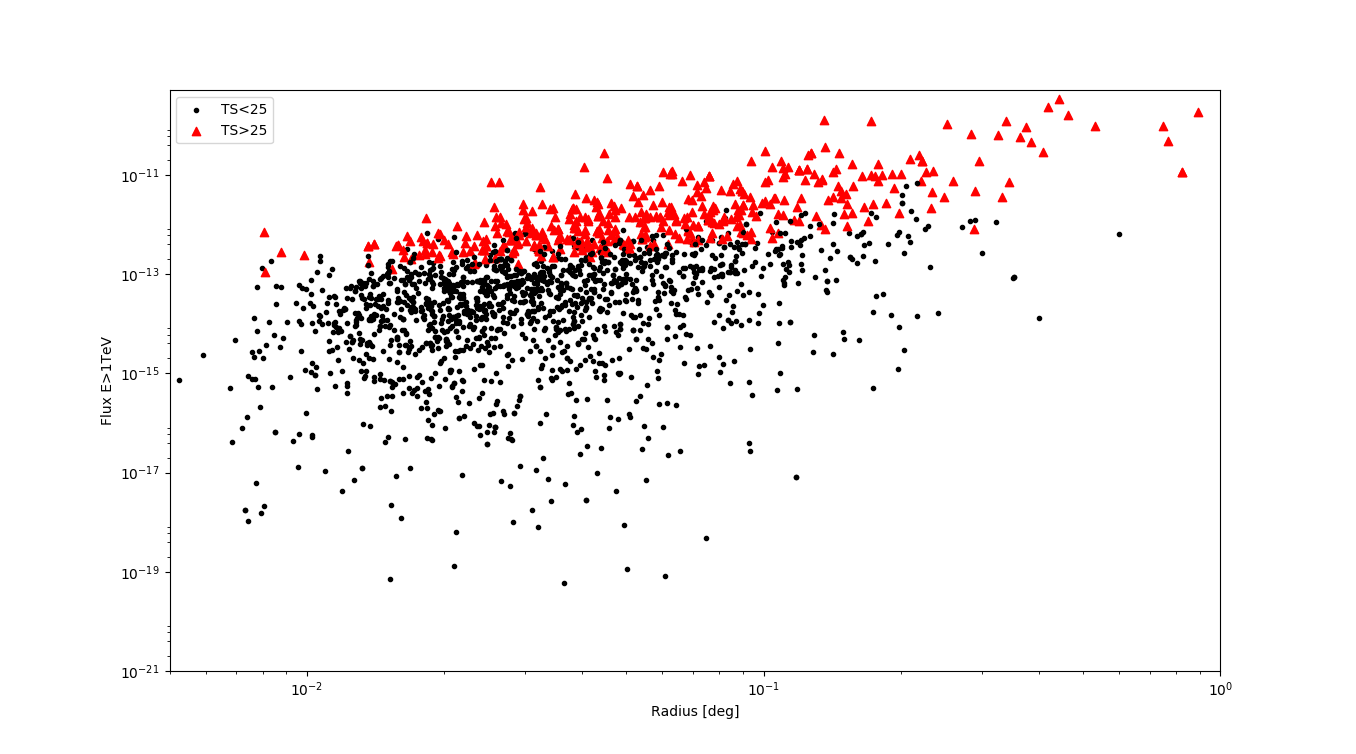}
	\caption{Plot of the simulated radius versus the integrated flux from 1 to 10 TeV of all the 10 realization. Red symbol marks the detected sources.}
    \label{grande}
\end{sidewaysfigure}

For example the results of the analysis of one realization are reported. I simulated 199 interacting systems, 39 of  of which have significance > $5\sigma$.  The figures \ref{log11} show the LogN-LogS curve of the interacting supernova remnants detected in comparison with that of the real iSNRs and with that of the entire realization.  As expected, the logN-logS of the detected iSNR follow the real iSNR curves, for bright sources, but extends to lower fluxes, according with the better sensitivity expected for CTA.
In figure \ref{rea11} is reported the integrated flux between 100 GeV e 10 TeV versus the simulated radius. 
The sources with TS > 25 and the non-detectable ones are separated by an horizontal line at the CTA sensitivity until fino $\sim$ 0.1$^{\circ}$ namely the CTA angular resolution. Figure \ref{grande} show the same behavior in all realizations. 
Depending on the realization CTA detect between $\sim$ 30 and 50 synthetic interacting supernova remnants.
Taking into account that simulations and analysis were done under ideal conditions (only one source for each simulation and only with instrumental background) this means doubling the number of SNR / MC systems already known.

\chapter{Conclusion}

The main goal of this thesis was to create a synthetic population of Galactic interacting supernova remnants that reflects features of the real iSNR population. The strategy was to realize the LogN-LogS of the real interacting supernova remnants and compared it with those obtained from the population model.
Since this sources are composed of a supernova remnants and a molecular cloud, the population of molecular clouds and the protons injection spectra has been reproduced separately. The synthetic population of molecular clouds was derived from the Rice et al. catalog of real molecular clouds. The synthetic MCs have realistic features in terms of mass and distribution in the Galaxy.
I created the synthetic protons distributions that interact within the molecular clouds starting from the 16 spectra of the real interacting supernova remnants. Generate the synthetic spectra requires besides the particle distribution, the density of the target protons. Both the parameters of the injection protons spectra and the target density of the clouds were extracted from the gaussian distribution obtained from the real values that are contained in the $\gamma$-SNR catalog that I've made. 
The $\gamma$-SNR catalog contains all the SNR emitting in $\gamma$-ray. For each SNR considered was carried out an extensive literature research of $\gamma$-ray spectra and physical characteristics.
The spectral fit was performed using a model-independent approach, namely broken power law as protons distribution and $\pi^0$-decay as radiative model.
The key results of the $\gamma$-SNR catalog was the creation of the LogN-LogS curve of the interacting supernova remnants obtained by integrating all the interacting SNRs spectra above 100 MeV.
Since not all the molecular clouds are illuminated by a supernova remnants, a mass dependent probability function was introduced. A uniform probability at 1.5\% was the only function that, reproduce the real iSNR LogN-LogS curve. The number of interacting systems resulting from the synthetic iSNR depend on the realization, for this reason a large number of realizations were made.
Using the CTA simulation softwares, 10 realization were simulated and analysed. For each ralization CTA detected between $\sim$ 30 and 50 synthetic interacting supernova remnants. 
CTA will the be able to extend significantly  the number of the known SNR/MC systems. This will allow to better understand the role of SNRs in the acceleration of Galactic cosmic rays.
\newpage
\section*{Appendix}
\subsubsection*{Naima fitting code}
Naima requires that all data set are written in ipac format, the ".ipac" file must
contain at least the folliwing information: the mean channel energy, differential flux,
upper flux error, lower flux error and upper limit flag. Furthermore the file must
contain the confidence level (cl) of the flux error bar (cl=0.95 in this case).
The first code  is the initialization of the prior function and of the broken power law that is the protons distributions. The second code refers to the definition of the function that returns the $\pi^0$-decay spectra, instead the last code is the procedute to start the NAIMA fitting, namely the file reading and the parameters settings. 		\\

\begin{figure}[ht]
     \centering
     
     \includegraphics[width=.9\textwidth]{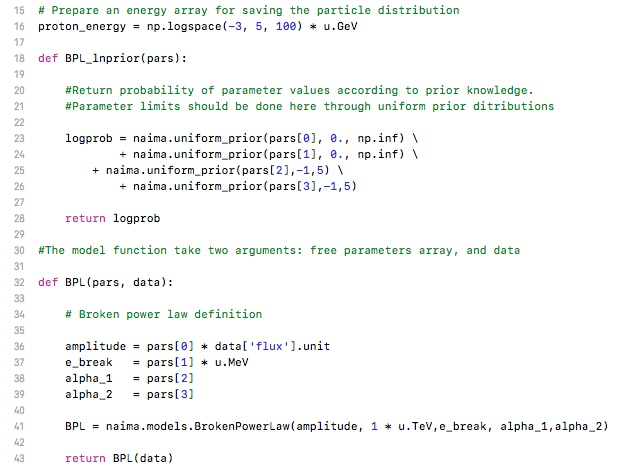}
      \caption*{Broken power law definition}
      \label{fig:a}
\end{figure}
 
     \begin{figure}[ht]
         \centering
         \includegraphics[width=.9\textwidth]{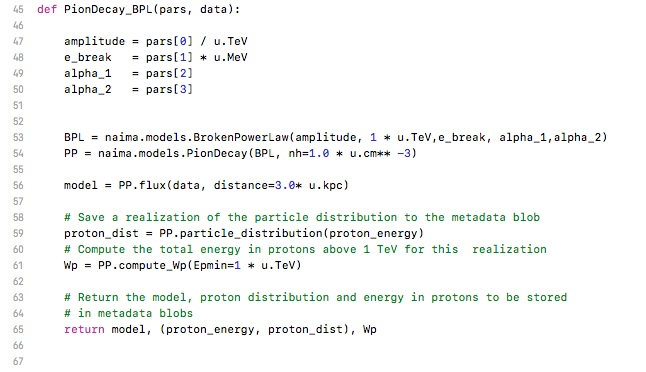}
         \caption*{Pion decay model definition}
         \label{fig:b}
     \end{figure}
   
     \begin{figure}[ht]
         \centering
         \includegraphics[width=.9\textwidth]{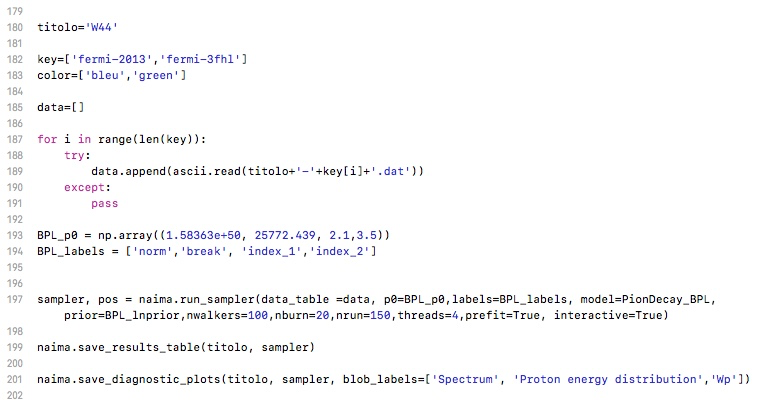}
         \caption*{Starting MCMC procedure}
         \label{fig:c}
     \end{figure}
       
\begin{landscape}

\begin{table}[]
\footnotesize
\begin{tabular}{@{}ccccccccccc@{}}
\toprule
SNR\_Name      & Type  & Other\_Names                & TeVCat         & l                     & b                     & Distance       & Age           & MC\_Mass  & Flux100GeV                      & Flux1TeV                        \\ \midrule
               &       &                             &                & $\mathrm{{}^{\circ}}$ & $\mathrm{{}^{\circ}}$ & $\mathrm{kpc}$ & $\mathrm{yr}$ & $M_{sun}$ & $\mathrm{erg\,s^{-1}\,cm^{-2}}$ & $\mathrm{erg\,s^{-1}\,cm^{-2}}$ \\
               \hline
G359.1-0.5     & INT   & HESSJ1745-303               & TeV J1745-303  & 358.7                 & -0.65                 & 7.6            & 10000.0       & 50000.0   & 4.177e-12                       & 8.425e-13                       \\
HESSJ1731-347  & SHELL & HESSJ1731-347               & TeV J1732-347  & 353.54                & -0.68                 & 3.2            & 14500.0       & 0.0       & 1.6595e-11                      & 6.615e-12                       \\
CTB37B         & SHELL & HESSJ1713-381 G348.7+0.3    & TeV J1713-382  & 348.63                & 0.38                  & 13.2           & 5000.0        & 0.0       & 1.55e-13                        & 4.436e-14                       \\
CTB37A         & INT   & G348.5+0.1                  & TeV J1714-385  & 348.38                & 0.1                   & 9.0            & 16000.0       & 67000.0   & 3.886e-12                       & 8.347e-13                       \\
RXJ1713.7-3946 & SHELL & G347.3-0.5                  & TeV J1713-397  & 347.33                & -0.48                 & 1.0            & 1600.0        & 0.0       & 5.84869e-10                     & 3.1538e-10                      \\
SN1006(NE)     & SHELL &                             & TeV J1504-418  & 327.84                & 14.56                 & 1.6            & 1010.0        & 0.0       & 3.3779e-11                      & 2.2221e-11                      \\
SN1006(SW)     & SHELL &                             & TeV J1504-421  & 327.86                & 15.35                 & 1.6            & 1010.0        & 0.0       & 2.0084e-11                      & 9.8858e-12                      \\
G318.2+0.1     & INT   & HESSJ1457-593               & TeV J1457-594  & 318.36                & -0.44                 & 9.2            & 8000.0        & 0.0       & 4.87e-13                        & 8.7534e-14                      \\
RCW86          & SHELL & G315.4-2.3 MSH14-63         & TeV J1442-624  & 315.41                & -2.31                 & 2.5            & 1800.0        & 0.0       & 1.3462e-11                      & 5.9342e-12                      \\
G298.6-0.0     &       &                             &                & 298.6                 & 0.0                   & 5.0            & 1000.0        & 0.0       & 5.2442e-11                      & 5.1438e-11                      \\
Vela Junior    & SHELL & RXJ0852.0-4622              & TeV J0852-463  & 266.28                & -1.25                 & 1.0            & 3000.0        & 0.0       & 0.0                             & 3.1786e-10                      \\
Puppis A       &       &                             &                & 260.4                 & -3.4                  & 2.2            & 4450.0        & 0.0       & 5.305e-12                       & 7.9818e-13                      \\
IC443          & INT   &                             & TeV J0616+225  & 189.1                 & 3.0                   & 1.5            & 10000.0       & 10000.0   & 1.1617e-11                      & 8.5012e-13                      \\
Tycho          & SHELL &                             & TeV J0025+641  & 120.1                 & 1.4                   & 4.5            & 400.0         & 0.0       & 0.0                             & 9.4669e-14                      \\
Cas A          & SHELL &                             & TeV J2323+588  & 111.7                 & -2.1                  & 3.4            & 330.0         & 0.0       & 3.726e-12                       & 8.1634e-13                      \\
Gamma Cygni    &       &                             &                & 78.2                  & 2.1                   & 1.5            & 6600.0        & 0.0       & 2.3169e-11                      & 7.9406e-12                      \\
Cygnus Loop    &       &                             &                & 74.0                  & -8.5                  & 0.8            & 15000.0       & 0.0       & 1.8e-13                         & 2.7842e-16                      \\
W51C           & INT   & G49.2-0.7 HESSJ1923+141     & TeV J1923+141  & 49.2                  & -0.7                  & 5.6            & 30000.0       & 190000.0  & 1.1725e-11                      & 1.9496e-12                      \\
W49B           & INT   & G43.3-0.2 HESSJ1911+090     & TeV J1911+090  & 43.3                  & -0.2                  & 8.0            & 2000.0        & 1000000.0 & 5.477e-12                       & 3.8876e-13                      \\
W44            & INT   & G34.7-0.4                   & none           & 34.7                  & -0.4                  & 3.0            & 20000.0       & 500000.0  & 1.003e-12                       & 2.8471e-14                      \\
W41            & INT   & G23.3-0.3 HESSJ1834-087     & TeV J1834-087  & 23.24                 & -0.33                 & 4.8            & 100000.0      & 88000.0   & 6.118e-12                       & 1.5368e-12                      \\
W28north       & INT   & HESSJ1801-233 3EGJ1800-2338 & TeV J1801-233  & 6.65                  & -0.27                 & 1.9            & 35000.0       & 50000.0   & 9.776e-12                       & 1.4924e-12                      \\
W28A           & INT   & HESSJ1800-240A              & TeV J1800-240A & 5.88                  & -0.39                 & 1.9            & 35000.0       & 40000.0   & 6.153e-12                       & 2.6322e-12                      \\
W28B           & INT   & HESSJ1800-240B              & TeV J1800-240B & 5.88                  & -0.39                 & 1.9            & 35000.0       & 60000.0   & 7.626e-12                       & 2.8944e-12                      \\
W28C           & INT   & HESSJ1800-240C              & TeV J1800-240C & 5.88                  & -0.39                 & 1.9            & 35000.0       & 20000.0   & 4.376e-12                       & 2.0365e-12                      \\
G349.7+0.2     & INT   &                             &                & 349.7                 & 0.2                   & 11.5           & 1800.0        & 5000.0    & 2.8e-14                         & 4.4379e-15                      \\
HESSJ1912+101  &       &                             & TeV J1912+101  & 44.39                 & -0.07                 & 4.1            & 100000.0      & 12000.0   & 2.6006e-11                      & 1.0625e-11                      \\
HESSJ1534-571  &       &                             & TeV J1534-571  & 323.65                & -0.92                 & 3.5            & 10000.0       & 1000.0    & 2.0676e-11                      & 8.4906e-12                      \\
MSH1739        & INT   &                             &                & 357.7                 & -0.1                  & 11.8           & -1.0          & 35000.0   & 6.1e-14                         & 4.57884e-16                     \\
HB21           & INT   &                             &                & 89.0                  & 4.7                   & 1.7            & 15874.0       & 55000.0   & 6e-15                           & 1.3092e-17                      \\
HESSJ1614-518  & SHELL &                             & TeV J1614-518  & 331.52                & -0.58                 & 1.5            & 30000.0       & 0.0       & 6.4802e-11                      & 2.7512e-11                      \\
W30            & INT   &                             &                & 8.7                   & -0.1                  & 4.6            & 20493.0       & 190000.0  & 4.4596e-11                      & 1.1907e-11                      \\
3C 391         &       & Kes 77                      &                & 31.9                  & 0.0                   & 7.1            & 9000.0        & 0.0       & 1.8e-14                         & 1.4426e-16                      \\
CTB109         &       &                             &                & 109.1                 & 1.0                   & 3.2            & 14000.0       & 0.0       & 2.191e-12                       & 8.8208e-13                      \\
G337.0-0.1     &       & CTB 33                      &                & 337.0                 & -0.1                  & 11.0           & 5000.0        & 0.0       & 7e-15                           & 2.67153e-17                     \\
S147           &       &                             &                & 180.0                 & -1.7                  & 1.39           & 30000.0       & 0.0       & 1.22117e-10                     & 1.0661e-10                      \\
Kes17          &       &                             &                & 304.6                 & 0.1                   & 9.7            & 14000.0       & 0.0       & 0.0                             & 1.8863e-14                      \\ \bottomrule
\end{tabular}
\caption*{Catalog of $\gamma$-SNRs}
\label{cat}
\end{table}
\end{landscape}

\printbibliography

@article{tevcat,
   author = {{Wakely}, S.~P. and {Horan}, D.},
    title = "{TeVCat: An online catalog for Very High Energy Gamma-Ray Astronomy}",
  journal = {International Cosmic Ray Conference},
     year = 2008,
   volume = 3,
    pages = {1341-1344},
   adsurl = {http://adsabs.harvard.edu/abs/2008ICRC....3.1341W}
}

@article{green,
       author = {{Green}, D.~A.},
        title = "{VizieR Online Data Catalog: A Catalogue of Galactic Supernova Remnants (Green 2017)}",
      journal = {VizieR Online Data Catalog},
         year = 2017,
        month = Jun,
          eid = {VII/278},
        pages = {VII/278},
       adsurl = {https://ui.adsabs.harvard.edu/\#abs/2017yCat.7278....0G}
}

@article{unimanitoba,
       author = {{Ferrand}, Gilles and {Safi-Harb}, Samar},
        title = "{A census of high-energy observations of Galactic supernova remnants}",
      journal = {Advances in Space Research},
         year = 2012,
        month = May,
       volume = {49},
        pages = {1313-1319},
          doi = {10.1016/j.asr.2012.02.004},
archivePrefix = {arXiv},
       eprint = {1202.0245},
 primaryClass = {astro-ph.HE},
       adsurl = {https://ui.adsabs.harvard.edu/\#abs/2012AdSpR..49.1313F}
}

@article{3fgl,
       author = {{Acero}, F. and {Ackermann}, M. and {Ajello}, M. and {Albert}, A. and
        {Atwood}, W.~B. and {Axelsson}, M. and {Baldini}, L. and
        {Ballet}, J. and {Barbiellini}, G. and {Bastieri}, D. and
        {Belfiore}, A. and {Bellazzini}, R. and {Bissaldi}, E. and
        {Blandford}, R.~D. and {Bloom}, E.~D. and {Bogart}, J.~R. and
        {Bonino}, R. and {Bottacini}, E. and {Bregeon}, J. and {Britto},
        R.~J. and {Bruel}, P. and {Buehler}, R. and {Burnett}, T.~H. and
        {Buson}, S. and {Caliandro}, G.~A. and {Cameron}, R.~A. and
        {Caputo}, R. and {Caragiulo}, M. and {Caraveo}, P.~A. and
        {Casandjian}, J.~M. and {Cavazzuti}, E. and {Charles}, E. and
        {Chaves}, R.~C.~G. and {Chekhtman}, A. and {Cheung}, C.~C. and
        {Chiang}, J. and {Chiaro}, G. and {Ciprini}, S. and {Claus}, R.
        and {Cohen-Tanugi}, J. and {Cominsky}, L.~R. and {Conrad}, J.
        and {Cutini}, S. and {D'Ammando}, F. and {de Angelis}, A. and
        {DeKlotz}, M. and {de Palma}, F. and {Desiante}, R. and {Digel},
        S.~W. and {Di Venere}, L. and {Drell}, P.~S. and {Dubois}, R.
        and {Dumora}, D. and {Favuzzi}, C. and {Fegan}, S.~J. and
        {Ferrara}, E.~C. and {Finke}, J. and {Franckowiak}, A. and
        {Fukazawa}, Y. and {Funk}, S. and {Fusco}, P. and {Gargano}, F.
        and {Gasparrini}, D. and {Giebels}, B. and {Giglietto}, N. and
        {Giommi}, P. and {Giordano}, F. and {Giroletti}, M. and
        {Glanzman}, T. and {Godfrey}, G. and {Grenier}, I.~A. and
        {Grondin}, M. -H. and {Grove}, J.~E. and {Guillemot}, L. and
        {Guiriec}, S. and {Hadasch}, D. and {Harding}, A.~K. and {Hays},
        E. and {Hewitt}, J.~W. and {Hill}, A.~B. and {Horan}, D. and
        {Iafrate}, G. and {Jogler}, T. and {J{\'o}hannesson}, G. and
        {Johnson}, R.~P. and {Johnson}, A.~S. and {Johnson}, T.~J. and
        {Johnson}, W.~N. and {Kamae}, T. and {Kataoka}, J. and
        {Katsuta}, J. and {Kuss}, M. and {La Mura}, G. and {Landriu}, D.
        and {Larsson}, S. and {Latronico}, L. and {Lemoine-Goumard}, M.
        and {Li}, J. and {Li}, L. and {Longo}, F. and {Loparco}, F. and
        {Lott}, B. and {Lovellette}, M.~N. and {Lubrano}, P. and
        {Madejski}, G.~M. and {Massaro}, F. and {Mayer}, M. and
        {Mazziotta}, M.~N. and {McEnery}, J.~E. and {Michelson}, P.~F.
        and {Mirabal}, N. and {Mizuno}, T. and {Moiseev}, A.~A. and
        {Mongelli}, M. and {Monzani}, M.~E. and {Morselli}, A. and
        {Moskalenko}, I.~V. and {Murgia}, S. and {Nuss}, E. and {Ohno},
        M. and {Ohsugi}, T. and {Omodei}, N. and {Orienti}, M. and
        {Orlando}, E. and {Ormes}, J.~F. and {Paneque}, D. and
        {Panetta}, J.~H. and {Perkins}, J.~S. and {Pesce-Rollins}, M.
        and {Piron}, F. and {Pivato}, G. and {Porter}, T.~A. and
        {Racusin}, J.~L. and {Rando}, R. and {Razzano}, M. and
        {Razzaque}, S. and {Reimer}, A. and {Reimer}, O. and {Reposeur},
        T. and {Rochester}, L.~S. and {Romani}, R.~W. and {Salvetti}, D.
        and {S{\'a}nchez-Conde}, M. and {Saz Parkinson}, P.~M. and
        {Schulz}, A. and {Siskind}, E.~J. and {Smith}, D.~A. and
        {Spada}, F. and {Spandre}, G. and {Spinelli}, P. and {Stephens},
        T.~E. and {Strong}, A.~W. and {Suson}, D.~J. and {Takahashi}, H.
        and {Takahashi}, T. and {Tanaka}, Y. and {Thayer}, J.~G. and
        {Thayer}, J.~B. and {Thompson}, D.~J. and {Tibaldo}, L. and
        {Tibolla}, O. and {Torres}, D.~F. and {Torresi}, E. and {Tosti},
        G. and {Troja}, E. and {Van Klaveren}, B. and {Vianello}, G. and
        {Winer}, B.~L. and {Wood}, K.~S. and {Wood}, M. and {Zimmer}, S.
        and {Fermi-LAT Collaboration}},
        title = "{Fermi Large Area Telescope Third Source Catalog}",
      journal = {The Astrophysical Journal Supplement Series},
         year = 2015,
        month = Jun,
       volume = {218},
          eid = {23},
        pages = {23},
          doi = {10.1088/0067-0049/218/2/23},
archivePrefix = {arXiv},
       eprint = {1501.02003},
 primaryClass = {astro-ph.HE},
       adsurl = {https://ui.adsabs.harvard.edu/\#abs/2015ApJS..218...23A}
}

@article{3fhl,
       author = {{Ajello}, M. and {Atwood}, W.~B. and {Baldini}, L. and {Ballet}, J. and
        {Barbiellini}, G. and {Bastieri}, D. and {Bellazzini}, R. and
        {Bissaldi}, E. and {Blandford}, R.~D. and {Bloom}, E.~D. and
        {Bonino}, R. and {Bregeon}, J. and {Britto}, R.~J. and {Bruel},
        P. and {Buehler}, R. and {Buson}, S. and {Cameron}, R.~A. and
        {Caputo}, R. and {Caragiulo}, M. and {Caraveo}, P.~A. and
        {Cavazzuti}, E. and {Cecchi}, C. and {Charles}, E. and
        {Chekhtman}, A. and {Cheung}, C.~C. and {Chiaro}, G. and
        {Ciprini}, S. and {Cohen}, J.~M. and {Costantin}, D. and
        {Costanza}, F. and {Cuoco}, A. and {Cutini}, S. and {D'Ammando},
        F. and {de Palma}, F. and {Desiante}, R. and {Digel}, S.~W. and
        {Di Lalla}, N. and {Di Mauro}, M. and {Di Venere}, L. and
        {Dom{\'\i}nguez}, A. and {Drell}, P.~S. and {Dumora}, D. and
        {Favuzzi}, C. and {Fegan}, S.~J. and {Ferrara}, E.~C. and
        {Fortin}, P. and {Franckowiak}, A. and {Fukazawa}, Y. and
        {Funk}, S. and {Fusco}, P. and {Gargano}, F. and {Gasparrini},
        D. and {Giglietto}, N. and {Giommi}, P. and {Giordano}, F. and
        {Giroletti}, M. and {Glanzman}, T. and {Green}, D. and
        {Grenier}, I.~A. and {Grondin}, M. -H. and {Grove}, J.~E. and
        {Guillemot}, L. and {Guiriec}, S. and {Harding}, A.~K. and
        {Hays}, E. and {Hewitt}, J.~W. and {Horan}, D. and
        {J{\'o}hannesson}, G. and {Kensei}, S. and {Kuss}, M. and {La
        Mura}, G. and {Larsson}, S. and {Latronico}, L. and {Lemoine-
        Goumard}, M. and {Li}, J. and {Longo}, F. and {Loparco}, F. and
        {Lott}, B. and {Lubrano}, P. and {Magill}, J.~D. and {Maldera},
        S. and {Manfreda}, A. and {Mazziotta}, M.~N. and {McEnery},
        J.~E. and {Meyer}, M. and {Michelson}, P.~F. and {Mirabal}, N.
        and {Mitthumsiri}, W. and {Mizuno}, T. and {Moiseev}, A.~A. and
        {Monzani}, M.~E. and {Morselli}, A. and {Moskalenko}, I.~V. and
        {Negro}, M. and {Nuss}, E. and {Ohsugi}, T. and {Omodei}, N. and
        {Orienti}, M. and {Orlando}, E. and {Palatiello}, M. and
        {Paliya}, V.~S. and {Paneque}, D. and {Perkins}, J.~S. and
        {Persic}, M. and {Pesce-Rollins}, M. and {Piron}, F. and
        {Porter}, T.~A. and {Principe}, G. and {Rain{\`o}}, S. and
        {Rando}, R. and {Razzano}, M. and {Razzaque}, S. and {Reimer},
        A. and {Reimer}, O. and {Reposeur}, T. and {Saz Parkinson},
        P.~M. and {Sgr{\`o}}, C. and {Simone}, D. and {Siskind}, E.~J.
        and {Spada}, F. and {Spandre}, G. and {Spinelli}, P. and
        {Stawarz}, L. and {Suson}, D.~J. and {Takahashi}, M. and {Tak},
        D. and {Thayer}, J.~G. and {Thayer}, J.~B. and {Thompson}, D.~J.
        and {Torres}, D.~F. and {Torresi}, E. and {Troja}, E. and
        {Vianello}, G. and {Wood}, K. and {Wood}, M.},
        title = "{3FHL: The Third Catalog of Hard Fermi-LAT Sources}",
      journal = {The Astrophysical Journal Supplement Series},
         year = 2017,
        month = Oct,
       volume = {232},
          eid = {18},
        pages = {18},
          doi = {10.3847/1538-4365/aa8221},
archivePrefix = {arXiv},
       eprint = {1702.00664},
 primaryClass = {astro-ph.HE},
       adsurl = {https://ui.adsabs.harvard.edu/\#abs/2017ApJS..232...18A}
}

@article{naima,
   author = {{Zabalza}, V.},
    title = {naima: a Python package for inference of relativistic particle
             energy distributions from observed nonthermal spectra},
     year = 2015,
  journal = {Proc.~of International Cosmic Ray Conference 2015},
    pages = "922",
   eprint = {1509.03319},
   adsurl = {http://adsabs.harvard.edu/abs/2015arXiv150903319Z},
}

@article{emcee,
       author = {{Foreman-Mackey}, Daniel and {Hogg}, David W. and {Lang}, Dustin and
        {Goodman}, Jonathan},
        title = "{emcee: The MCMC Hammer}",
      journal = {Publications of the Astronomical Society of the Pacific},
         year = 2013,
        month = Mar,
       volume = {125},
        pages = {306},
          doi = {10.1086/670067},
archivePrefix = {arXiv},
       eprint = {1202.3665},
 primaryClass = {astro-ph.IM},
       adsurl = {https://ui.adsabs.harvard.edu/\#abs/2013PASP..125..306F}
}

@article{w44-fermi,
       author = {{Ackermann}, M. and {Ajello}, M. and {Allafort}, A. and {Baldini}, L.
        and {Ballet}, J. and {Barbiellini}, G. and {Baring}, M.~G. and
        {Bastieri}, D. and {Bechtol}, K. and {Bellazzini}, R. and
        {Blandford}, R.~D. and {Bloom}, E.~D. and {Bonamente}, E. and
        {Borgland}, A.~W. and {Bottacini}, E. and {Brandt}, T.~J. and
        {Bregeon}, J. and {Brigida}, M. and {Bruel}, P. and {Buehler},
        R. and {Busetto}, G. and {Buson}, S. and {Caliandro}, G.~A. and
        {Cameron}, R.~A. and {Caraveo}, P.~A. and {Casandjian}, J.~M.
        and {Cecchi}, C. and {{\c{C}}elik}, {\"O}. and {Charles}, E. and
        {Chaty}, S. and {Chaves}, R.~C.~G. and {Chekhtman}, A. and
        {Cheung}, C.~C. and {Chiang}, J. and {Chiaro}, G. and {Cillis},
        A.~N. and {Ciprini}, S. and {Claus}, R. and {Cohen-Tanugi}, J.
        and {Cominsky}, L.~R. and {Conrad}, J. and {Corbel}, S. and
        {Cutini}, S. and {D'Ammando}, F. and {de Angelis}, A. and {de
        Palma}, F. and {Dermer}, C.~D. and {do Couto e Silva}, E. and
        {Drell}, P.~S. and {Drlica-Wagner}, A. and {Falletti}, L. and
        {Favuzzi}, C. and {Ferrara}, E.~C. and {Franckowiak}, A. and
        {Fukazawa}, Y. and {Funk}, S. and {Fusco}, P. and {Gargano}, F.
        and {Germani}, S. and {Giglietto}, N. and {Giommi}, P. and
        {Giordano}, F. and {Giroletti}, M. and {Glanzman}, T. and
        {Godfrey}, G. and {Grenier}, I.~A. and {Grondin}, M. -H. and
        {Grove}, J.~E. and {Guiriec}, S. and {Hadasch}, D. and
        {Hanabata}, Y. and {Harding}, A.~K. and {Hayashida}, M. and
        {Hayashi}, K. and {Hays}, E. and {Hewitt}, J.~W. and {Hill},
        A.~B. and {Hughes}, R.~E. and {Jackson}, M.~S. and {Jogler}, T.
        and {J{\'o}hannesson}, G. and {Johnson}, A.~S. and {Kamae}, T.
        and {Kataoka}, J. and {Katsuta}, J. and {Kn{\"o}dlseder}, J. and
        {Kuss}, M. and {Lande}, J. and {Larsson}, S. and {Latronico}, L.
        and {Lemoine-Goumard}, M. and {Longo}, F. and {Loparco}, F. and
        {Lovellette}, M.~N. and {Lubrano}, P. and {Madejski}, G.~M. and
        {Massaro}, F. and {Mayer}, M. and {Mazziotta}, M.~N. and
        {McEnery}, J.~E. and {Mehault}, J. and {Michelson}, P.~F. and
        {Mignani}, R.~P. and {Mitthumsiri}, W. and {Mizuno}, T. and
        {Moiseev}, A.~A. and {Monzani}, M.~E. and {Morselli}, A. and
        {Moskalenko}, I.~V. and {Murgia}, S. and {Nakamori}, T. and
        {Nemmen}, R. and {Nuss}, E. and {Ohno}, M. and {Ohsugi}, T. and
        {Omodei}, N. and {Orienti}, M. and {Orlando}, E. and {Ormes},
        J.~F. and {Paneque}, D. and {Perkins}, J.~S. and {Pesce-
        Rollins}, M. and {Piron}, F. and {Pivato}, G. and {Rain{\`o}},
        S. and {Rando}, R. and {Razzano}, M. and {Razzaque}, S. and
        {Reimer}, A. and {Reimer}, O. and {Ritz}, S. and {Romoli}, C.
        and {S{\'a}nchez-Conde}, M. and {Schulz}, A. and {Sgr{\`o}}, C.
        and {Simeon}, P.~E. and {Siskind}, E.~J. and {Smith}, D.~A. and
        {Spandre}, G. and {Spinelli}, P. and {Stecker}, F.~W. and
        {Strong}, A.~W. and {Suson}, D.~J. and {Tajima}, H. and
        {Takahashi}, H. and {Takahashi}, T. and {Tanaka}, T. and
        {Thayer}, J.~G. and {Thayer}, J.~B. and {Thompson}, D.~J. and
        {Thorsett}, S.~E. and {Tibaldo}, L. and {Tibolla}, O. and
        {Tinivella}, M. and {Troja}, E. and {Uchiyama}, Y. and {Usher},
        T.~L. and {Vandenbroucke}, J. and {Vasileiou}, V. and
        {Vianello}, G. and {Vitale}, V. and {Waite}, A.~P. and {Werner},
        M. and {Winer}, B.~L. and {Wood}, K.~S. and {Wood}, M. and
        {Yamazaki}, R. and {Yang}, Z. and {Zimmer}, S.},
        title = "{Detection of the Characteristic Pion-Decay Signature in Supernova Remnants}",
      journal = {Science},
         year = 2013,
        month = Feb,
       volume = {339},
        pages = {807-811},
          doi = {10.1126/science.1231160},
archivePrefix = {arXiv},
       eprint = {1302.3307},
 primaryClass = {astro-ph.HE},
       adsurl = {https://ui.adsabs.harvard.edu/\#abs/2013Sci...339..807A}
}

@article{w44-agile,
       author = {{Cardillo}, M. and {Tavani}, M. and {Giuliani}, A. and {Yoshiike}, S.
        and {Sano}, H. and {Fukuda}, T. and {Fukui}, Y. and
        {Castelletti}, G. and {Dubner}, G.},
        title = "{The supernova remnant W44: Confirmations and challenges for cosmic-ray acceleration}",
      journal = {\aap},
         year = 2014,
        month = May,
       volume = {565},
          eid = {A74},
        pages = {A74},
          doi = {10.1051/0004-6361/201322685},
archivePrefix = {arXiv},
       eprint = {1403.1250},
 primaryClass = {astro-ph.HE},
       adsurl = {https://ui.adsabs.harvard.edu/\#abs/2014A&A...565A..74C}
}

@article{snrcat,
       author = {{Acero}, F. and {Ackermann}, M. and {Ajello}, M. and {Baldini}, L. and
        {Ballet}, J. and {Barbiellini}, G. and {Bastieri}, D. and
        {Bellazzini}, R. and {Bissaldi}, E. and {Blandford}, R.~D. and
        {Bloom}, E.~D. and {Bonino}, R. and {Bottacini}, E. and
        {Brandt}, T.~J. and {Bregeon}, J. and {Bruel}, P. and {Buehler},
        R. and {Buson}, S. and {Caliandro}, G.~A. and {Cameron}, R.~A.
        and {Caputo}, R. and {Caragiulo}, M. and {Caraveo}, P.~A. and
        {Casandjian}, J.~M. and {Cavazzuti}, E. and {Cecchi}, C. and
        {Chekhtman}, A. and {Chiang}, J. and {Chiaro}, G. and {Ciprini},
        S. and {Claus}, R. and {Cohen}, J.~M. and {Cohen-Tanugi}, J. and
        {Cominsky}, L.~R. and {Condon}, B. and {Conrad}, J. and
        {Cutini}, S. and {D'Ammando}, F. and {de Angelis}, A. and {de
        Palma}, F. and {Desiante}, R. and {Digel}, S.~W. and {Di
        Venere}, L. and {Drell}, P.~S. and {Drlica-Wagner}, A. and
        {Favuzzi}, C. and {Ferrara}, E.~C. and {Franckowiak}, A. and
        {Fukazawa}, Y. and {Funk}, S. and {Fusco}, P. and {Gargano}, F.
        and {Gasparrini}, D. and {Giglietto}, N. and {Giommi}, P. and
        {Giordano}, F. and {Giroletti}, M. and {Glanzman}, T. and
        {Godfrey}, G. and {Gomez-Vargas}, G.~A. and {Grenier}, I.~A. and
        {Grondin}, M. -H. and {Guillemot}, L. and {Guiriec}, S. and
        {Gustafsson}, M. and {Hadasch}, D. and {Harding}, A.~K. and
        {Hayashida}, M. and {Hays}, E. and {Hewitt}, J.~W. and {Hill},
        A.~B. and {Horan}, D. and {Hou}, X. and {Iafrate}, G. and
        {Jogler}, T. and {J{\'o}hannesson}, G. and {Johnson}, A.~S. and
        {Kamae}, T. and {Katagiri}, H. and {Kataoka}, J. and {Katsuta},
        J. and {Kerr}, M. and {Kn{\"o}dlseder}, J. and {Kocevski}, D.
        and {Kuss}, M. and {Laffon}, H. and {Lande}, J. and {Larsson},
        S. and {Latronico}, L. and {Lemoine-Goumard}, M. and {Li}, J.
        and {Li}, L. and {Longo}, F. and {Loparco}, F. and {Lovellette},
        M.~N. and {Lubrano}, P. and {Magill}, J. and {Maldera}, S. and
        {Marelli}, M. and {Mayer}, M. and {Mazziotta}, M.~N. and
        {Michelson}, P.~F. and {Mitthumsiri}, W. and {Mizuno}, T. and
        {Moiseev}, A.~A. and {Monzani}, M.~E. and {Moretti}, E. and
        {Morselli}, A. and {Moskalenko}, I.~V. and {Murgia}, S. and
        {Nemmen}, R. and {Nuss}, E. and {Ohsugi}, T. and {Omodei}, N.
        and {Orienti}, M. and {Orlando}, E. and {Ormes}, J.~F. and
        {Paneque}, D. and {Perkins}, J.~S. and {Pesce-Rollins}, M. and
        {Petrosian}, V. and {Piron}, F. and {Pivato}, G. and {Porter},
        T.~A. and {Rain{\`o}}, S. and {Rando}, R. and {Razzano}, M. and
        {Razzaque}, S. and {Reimer}, A. and {Reimer}, O. and {Renaud},
        M. and {Reposeur}, T. and {Rousseau}, R. and {Saz Parkinson},
        P.~M. and {Schmid}, J. and {Schulz}, A. and {Sgr{\`o}}, C. and
        {Siskind}, E.~J. and {Spada}, F. and {Spandre}, G. and
        {Spinelli}, P. and {Strong}, A.~W. and {Suson}, D.~J. and
        {Tajima}, H. and {Takahashi}, H. and {Tanaka}, T. and {Thayer},
        J.~B. and {Thompson}, D.~J. and {Tibaldo}, L. and {Tibolla}, O.
        and {Torres}, D.~F. and {Tosti}, G. and {Troja}, E. and
        {Uchiyama}, Y. and {Vianello}, G. and {Wells}, B. and {Wood},
        K.~S. and {Wood}, M. and {Yassine}, M. and {den Hartog}, P.~R.
        and {Zimmer}, S.},
        title = "{The First Fermi LAT Supernova Remnant Catalog}",
      journal = {The Astrophysical Journal Supplement Series},
         year = 2016,
        month = May,
       volume = {224},
          eid = {8},
        pages = {8},
          doi = {10.3847/0067-0049/224/1/8},
archivePrefix = {arXiv},
       eprint = {1511.06778},
 primaryClass = {astro-ph.HE},
       adsurl = {https://ui.adsabs.harvard.edu/\#abs/2016ApJS..224....8A}
}

@article{skymap,
       author = {{Atwood}, W.~B. and {Abdo}, A.~A. and {Ackermann}, M. and
         {Althouse}, W. and {Anderson}, B. and {Axelsson}, M. and {Baldini}, L. and
         {Ballet}, J. and {Band}, D.~L. and {Barbiellini}, G. and {Bartelt}, J. and
         {Bastieri}, D. and {Baughman}, B.~M. and {Bechtol}, K. and
         {B{\'e}d{\'e}r{\`e}de}, D. and {Bellardi}, F. and {Bellazzini}, R. and
         {Berenji}, B. and {Bignami}, G.~F. and {Bisello}, D. and
         {Bissaldi}, E. and {Blandford}, R.~D. and {Bloom}, E.~D. and
         {Bogart}, J.~R. and {Bonamente}, E. and {Bonnell}, J. and {Borgland
        }, A.~W. and {Bouvier}, A. and {Bregeon}, J. and {Brez}, A. and
         {Brigida}, M. and {Bruel}, P. and {Burnett}, T.~H. and {Busetto}, G. and
         {Caliandro}, G.~A. and {Cameron}, R.~A. and {Caraveo}, P.~A. and
         {Carius}, S. and {Carlson}, P. and {Casandjian}, J.~M. and
         {Cavazzuti}, E. and {Ceccanti}, M. and {Cecchi}, C. and {Charles}, E. and
         {Chekhtman}, A. and {Cheung}, C.~C. and {Chiang}, J. and {Chipaux}, R. and
         {Cillis}, A.~N. and {Ciprini}, S. and {Claus}, R. and
         {Cohen-Tanugi}, J. and {Condamoor}, S. and {Conrad}, J. and
         {Corbet}, R. and {Corucci}, L. and {Costamante}, L. and {Cutini}, S. and
         {Davis}, D.~S. and {Decotigny}, D. and {DeKlotz}, M. and
         {Dermer}, C.~D. and {de Angelis}, A. and {Digel}, S.~W. and
         {do Couto e Silva}, E. and {Drell}, P.~S. and {Dubois}, R. and
         {Dumora}, D. and {Edmonds}, Y. and {Fabiani}, D. and {Farnier}, C. and
         {Favuzzi}, C. and {Flath}, D.~L. and {Fleury}, P. and {Focke}, W.~B. and
         {Funk}, S. and {Fusco}, P. and {Gargano}, F. and {Gasparrini}, D. and
         {Gehrels}, N. and {Gentit}, F. -X. and {Germani}, S. and {Giebels}, B. and
         {Giglietto}, N. and {Giommi}, P. and {Giordano}, F. and {Glanzman}, T. and
         {Godfrey}, G. and {Grenier}, I.~A. and {Grondin}, M. -H. and
         {Grove}, J.~E. and {Guillemot}, L. and {Guiriec}, S. and {Haller}, G. and
         {Harding}, A.~K. and {Hart}, P.~A. and {Hays}, E. and {Healey}, S.~E. and
         {Hirayama}, M. and {Hjalmarsdotter}, L. and {Horn}, R. and
         {Hughes}, R.~E. and {J{\'o}hannesson}, G. and {Johansson}, G. and
         {Johnson}, A.~S. and {Johnson}, R.~P. and {Johnson}, T.~J. and
         {Johnson}, W.~N. and {Kamae}, T. and {Katagiri}, H. and {Kataoka}, J. and
         {Kavelaars}, A. and {Kawai}, N. and {Kelly}, H. and {Kerr}, M. and
         {Klamra}, W. and {Kn{\"o}dlseder}, J. and {Kocian}, M.~L. and
         {Komin}, N. and {Kuehn}, F. and {Kuss}, M. and {Landriu}, D. and
         {Latronico}, L. and {Lee}, B. and {Lee}, S. -H. and
         {Lemoine-Goumard}, M. and {Lionetto}, A.~M. and {Longo}, F. and
         {Loparco}, F. and {Lott}, B. and {Lovellette}, M.~N. and {Lubrano}, P. and
         {Madejski}, G.~M. and {Makeev}, A. and {Marangelli}, B. and
         {Massai}, M.~M. and {Mazziotta}, M.~N. and {McEnery}, J.~E. and
         {Menon}, N. and {Meurer}, C. and {Michelson}, P.~F. and {Minuti}, M. and
         {Mirizzi}, N. and {Mitthumsiri}, W. and {Mizuno}, T. and
         {Moiseev}, A.~A. and {Monte}, C. and {Monzani}, M.~E. and
         {Moretti}, E. and {Morselli}, A. and {Moskalenko}, I.~V. and
         {Murgia}, S. and {Nakamori}, T. and {Nishino}, S. and {Nolan}, P.~L. and
         {Norris}, J.~P. and {Nuss}, E. and {Ohno}, M. and {Ohsugi}, T. and
         {Omodei}, N. and {Orlando}, E. and {Ormes}, J.~F. and
         {Paccagnella}, A. and {Paneque}, D. and {Panetta}, J.~H. and
         {Parent}, D. and {Pearce}, M. and {Pepe}, M. and {Perazzo}, A. and
         {Pesce-Rollins}, M. and {Picozza}, P. and {Pieri}, L. and
         {Pinchera}, M. and {Piron}, F. and {Porter}, T.~A. and {Poupard}, L. and
         {Rain{\`o}}, S. and {Rando}, R. and {Rapposelli}, E. and {Razzano}, M. and
         {Reimer}, A. and {Reimer}, O. and {Reposeur}, T. and {Reyes}, L.~C. and
         {Ritz}, S. and {Rochester}, L.~S. and {Rodriguez}, A.~Y. and
         {Romani}, R.~W. and {Roth}, M. and {Russell}, J.~J. and {Ryde}, F. and
         {Sabatini}, S. and {Sadrozinski}, H.~F. -W. and {Sanchez}, D. and {Sand
        er}, A. and {Sapozhnikov}, L. and {Parkinson}, P.~M. Saz and
         {Scargle}, J.~D. and {Schalk}, T.~L. and {Scolieri}, G. and
         {Sgr{\`o}}, C. and {Share}, G.~H. and {Shaw}, M. and {Shimokawabe}, T. and
         {Shrader}, C. and {Sierpowska-Bartosik}, A. and {Siskind}, E.~J. and
         {Smith}, D.~A. and {Smith}, P.~D. and {Spandre}, G. and {Spinelli}, P. and
         {Starck}, J. -L. and {Stephens}, T.~E. and {Strickman}, M.~S. and
         {Strong}, A.~W. and {Suson}, D.~J. and {Tajima}, H. and
         {Takahashi}, H. and {Takahashi}, T. and {Tanaka}, T. and {Tenze}, A. and
         {Tether}, S. and {Thayer}, J.~B. and {Thayer}, J.~G. and
         {Thompson}, D.~J. and {Tibaldo}, L. and {Tibolla}, O. and
         {Torres}, D.~F. and {Tosti}, G. and {Tramacere}, A. and {Turri}, M. and
         {Usher}, T.~L. and {Vilchez}, N. and {Vitale}, V. and {Wang}, P. and
         {Watters}, K. and {Winer}, B.~L. and {Wood}, K.~S. and {Ylinen}, T. and
         {Ziegler}, M.},
        title = "{The Large Area Telescope on the Fermi Gamma-Ray Space Telescope Mission}",
      journal = {\apj},
         year = "2009",
        month = "Jun",
       volume = {697},
        pages = {1071-1102},
          doi = {10.1088/0004-637X/697/2/1071},
archivePrefix = {arXiv},
       eprint = {0902.1089},
 primaryClass = {astro-ph.IM},
       adsurl = {https://ui.adsabs.harvard.edu/\#abs/2009ApJ...697.1071A}
}

@article{funk,
       author = {{Funk}, Stefan},
        title = "{Ground- and Space-Based Gamma-Ray Astronomy}",
      journal = {Annual Review of Nuclear and Particle Science},
         year = "2015",
        month = "Oct",
       volume = {65},
        pages = {245-277},
          doi = {10.1146/annurev-nucl-102014-022036},
archivePrefix = {arXiv},
       eprint = {1508.05190},
 primaryClass = {astro-ph.HE},
       adsurl = {https://ui.adsabs.harvard.edu/\#abs/2015ARNPS..65..245F}
}

@article{magic,
      author         = "Albert, J. and others",
      title          = "{Discovery of VHE Gamma Radiation from IC443 with the
                        MAGIC Telescope}",
      collaboration  = "MAGIC",
      journal        = "Astrophys. J.",
      volume         = "664",
      year           = "2007",
      pages          = "L87-L90",
      doi            = "10.1086/520957",
      eprint         = "0705.3119",
      archivePrefix  = "arXiv",
      primaryClass   = "astro-ph",
      reportNumber   = "MPP-2007-62",
      SLACcitation   = "%%CITATION = ARXIV:0705.3119;%%"
}

@article{agile,
   author = {{Giuliani}, A. and {Cardillo}, M. and {Tavani}, M. and {Fukui}, Y. and 
	{Yoshiike}, S. and {Torii}, K. and {Dubner}, G. and {Castelletti}, G. and 
	{Barbiellini}, G. and {Bulgarelli}, A. and {Caraveo}, P. and 
	{Costa}, E. and {Cattaneo}, P.~W. and {Chen}, A. and {Contessi}, T. and 
	{Del Monte}, E. and {Donnarumma}, I. and {Evangelista}, Y. and 
	{Feroci}, M. and {Gianotti}, F. and {Lazzarotto}, F. and {Lucarelli}, F. and 
	{Longo}, F. and {Marisaldi}, M. and {Mereghetti}, S. and {Pacciani}, L. and 
	{Pellizzoni}, A. and {Piano}, G. and {Picozza}, P. and {Pittori}, C. and 
	{Pucella}, G. and {Rapisarda}, M. and {Rappoldi}, A. and {Sabatini}, S. and 
	{Soffitta}, P. and {Striani}, E. and {Trifoglio}, M. and {Trois}, A. and 
	{Vercellone}, S. and {Verrecchia}, F. and {Vittorini}, V. and 
	{Colafrancesco}, S. and {Giommi}, P. and {Bignami}, G.},
    title = "{Neutral Pion Emission from Accelerated Protons in the Supernova Remnant W44}",
  journal = {ApJ Let.},
archivePrefix = "arXiv",
   eprint = {1111.4868},
 primaryClass = "astro-ph.HE",
     year = 2011,
    month = dec,
   volume = 742,
      eid = {L30},
    pages = {L30},
      doi = {10.1088/2041-8205/742/2/L30},
   adsurl = {https://ui.adsabs.harvard.edu/abs/2011ApJ...742L..30G}
}

@article{w51c,
   author = {{Jogler}, T. and {Funk}, S.},
    title = "{Revealing W51C as a Cosmic Ray Source Using Fermi-LAT Data}",
  journal = {\apj},
     year = 2016,
    month = jan,
   volume = 816,
      eid = {100},
    pages = {100},
      doi = {10.3847/0004-637X/816/2/100},
   adsurl = {https://ui.adsabs.harvard.edu/abs/2016ApJ...816..100J}
}

@article{g359fermi,
       author = {{Hui}, C.~Y. and {Yeung}, P.~K.~H. and {Ng}, C.~W. and {Lin}, L.~C.~C. and
         {Tam}, P.~H.~T. and {Cheng}, K.~S. and {Kong}, A.~K.~H. and
         {Chernyshov}, D.~O. and {Dogiel}, V.~A.},
        title = "{Observing two dark accelerators around the Galactic Centre with Fermi Large Area Telescope}",
      journal = {Monthly Notices of the Royal Astronomical Society},
         year = "2016",
        month = "Apr",
       volume = {457},
       number = {4},
        pages = {4262-4271},
          doi = {10.1093/mnras/stw209},
archivePrefix = {arXiv},
       eprint = {1601.06500},
 primaryClass = {astro-ph.HE},
       adsurl = {https://ui.adsabs.harvard.edu/abs/2016MNRAS.457.4262H}
}

@article{g359hess,
       author = {{Aharonian}, F. and {Akhperjanian}, A.~G. and {Barres de Almeida}, U. and
         {Bazer-Bachi}, A.~R. and {Behera}, B. and {Beilicke}, M. and
         {Benbow}, W. and {Bernl{\"o}hr}, K. and {Boisson}, C. and {Bolz}, O. and
         {Borrel}, V. and {Braun}, I. and {Brion}, E. and {Brown}, A.~M. and
         {B{\"u}hler}, R. and {Bulik}, T. and {B{\"u}sching}, I. and
         {Boutelier}, T. and {Carrigan}, S. and {Chadwick}, P.~M. and
         {Chounet}, L. -M. and {Clapson}, A.~C. and {Coignet}, G. and
         {Cornils}, R. and {Costamante}, L. and {Dalton}, M. and {Degrange}, B. and
         {Dickinson}, H.~J. and {Djannati-Ata{\"\i}}, A. and {Domainko}, W. and
         {O'C. Drury}, L. and {Dubois}, F. and {Dubus}, G. and {Dyks}, J. and
         {Egberts}, K. and {Emmanoulopoulos}, D. and {Espigat}, P. and
         {Farnier}, C. and {Feinstein}, F. and {Fiasson}, A. and
         {F{\"o}rster}, A. and {Fontaine}, G. and {Funk}, S. and
         {F{\"u}{\ss}ling}, M. and {Gallant}, Y.~A. and {Giebels}, B. and
         {Glicenstein}, J.~F. and {Gl{\"u}ck}, B. and {Goret}, P. and
         {Hadjichristidis}, C. and {Hauser}, D. and {Hauser}, M. and
         {Heinzelmann}, G. and {Henri}, G. and {Hermann}, G. and
         {Hinton}, J.~A. and {Hoffmann}, A. and {Hofmann}, W. and
         {Holleran}, M. and {Hoppe}, S. and {Horns}, D. and {Jacholkowska}, A. and
         {de Jager}, O.~C. and {Jung}, I. and {Katarzy{\'n}ski}, K. and
         {Kendziorra}, E. and {Kerschhaggl}, M. and {Kh{\'e}lifi}, B. and
         {Keogh}, D. and {Komin}, Nu. and {Kosack}, K. and {Lamanna}, G. and
         {Latham}, I.~J. and {Lemoine-Goumard}, M. and {Lenain}, J. -P. and
         {Lohse}, T. and {Martin}, J.~M. and {Martineau-Huynh}, O. and
         {Marcowith}, A. and {Masterson}, C. and {Maurin}, D. and
         {McComb}, T.~J.~L. and {Moderski}, R. and {Moulin}, E. and
         {Naumann-Godo}, M. and {de Naurois}, M. and {Nedbal}, D. and
         {Nekrassov}, D. and {Nolan}, S.~J. and {Ohm}, S. and {Olive}, J. -P. and
         {de O{\~n}a Wilhelmi}, E. and {Orford}, K.~J. and {Osborne}, J.~L. and
         {Ostrowski}, M. and {Panter}, M. and {Pedaletti}, G. and
         {Pelletier}, G. and {Petrucci}, P. -O. and {Pita}, S. and
         {P{\"u}hlhofer}, G. and {Punch}, M. and {Raubenheimer}, B.~C. and
         {Raue}, M. and {Rayner}, S.~M. and {Renaud}, M. and {Ripken}, J. and
         {Rob}, L. and {Rosier-Lees}, S. and {Rowell}, G. and {Rudak}, B. and
         {Ruppel}, J. and {Sahakian}, V. and {Santangelo}, A. and
         {Schlickeiser}, R. and {Sch{\"o}ck}, F.~M. and {Schr{\"o}der}, R. and
         {Schwanke}, U. and {Schwarzburg}, S. and {Schwemmer}, S. and
         {Shalchi}, A. and {Sol}, H. and {Spangler}, D. and {Stawarz}, {\L}. and
         {Steenkamp}, R. and {Stegmann}, C. and {Superina}, G. and {Tam}, P.~H. and
         {Tavernet}, J. -P. and {Terrier}, R. and {van Eldik}, C. and
         {Vasileiadis}, G. and {Venter}, C. and {Vialle}, J.~P. and
         {Vincent}, P. and {Vivier}, M. and {V{\"o}lk}, H.~J. and {Volpe}, F. and
         {Wagner}, S.~J. and {Ward}, M. and {Zdziarski}, A.~A. and {Zech}, A.},
        title = "{Exploring a SNR/molecular cloud association within HESS J1745-303}",
      journal = {Astronomy and Astrophysics},
         year = "2008",
        month = "May",
       volume = {483},
       number = {2},
        pages = {509-517},
          doi = {10.1051/0004-6361:20079230},
archivePrefix = {arXiv},
       eprint = {0803.2844},
 primaryClass = {astro-ph},
       adsurl = {https://ui.adsabs.harvard.edu/abs/2008A&A...483..509A}
}

@article{J1731fermi15,
       author = {{Acero}, F. and {Lemoine-Goumard}, M. and {Renaud}, M. and {Ballet}, J. and
         {Hewitt}, J.~W. and {Rousseau}, R. and {Tanaka}, T.},
        title = "{Study of TeV shell supernova remnants at gamma-ray energies}",
      journal = {Astronomy and Astrophysics},
         year = "2015",
        month = "Aug",
       volume = {580},
          eid = {A74},
        pages = {A74},
          doi = {10.1051/0004-6361/201525932},
archivePrefix = {arXiv},
       eprint = {1506.02307},
 primaryClass = {astro-ph.HE},
       adsurl = {https://ui.adsabs.harvard.edu/abs/2015A&A...580A..74A}
}

@article{j1731fermi18,
       author = {{Guo}, Xiao-Lei and {Xin}, Yu-Liang and {Liao}, Neng-Hui and
         {Yuan}, Qiang and {Gao}, Wei-Hong and {Fan}, Yi-Zhong},
        title = "{Detection of GeV Gamma-Ray Emission in the Direction of HESS J1731-347 with Fermi-LAT}",
      journal = {The Astrophysical Journal},
         year = "2018",
        month = "Jan",
       volume = {853},
       number = {1},
          eid = {2},
        pages = {2},
          doi = {10.3847/1538-4357/aaa3f8},
archivePrefix = {arXiv},
       eprint = {1711.03729},
 primaryClass = {astro-ph.HE},
       adsurl = {https://ui.adsabs.harvard.edu/abs/2018ApJ...853....2G}
}

@article{j1731hess,
       author = {{H.~E.~S.~S. Collaboration} and {Abramowski}, A. and {Acero}, F. and
         {Aharonian}, F. and {Akhperjanian}, A.~G. and {Anton}, G. and
         {Balzer}, A. and {Barnacka}, A. and {Barres de Almeida}, U. and
         {Becherini}, Y. and {Becker}, J. and {Behera}, B. and
         {Bernl{\"o}hr}, K. and {Bochow}, A. and {Boisson}, C. and
         {Bolmont}, J. and {Bordas}, P. and {Brucker}, J. and {Brun}, F. and
         {Brun}, P. and {Bulik}, T. and {B{\"u}sching}, I. and {Carrigan}, S. and
         {Casanova}, S. and {Cerruti}, M. and {Chadwick}, P.~M. and
         {Charbonnier}, A. and {Chaves}, R.~C.~G. and {Cheesebrough}, A. and
         {Chounet}, L. -M. and {Clapson}, A.~C. and {Coignet}, G. and
         {Cologna}, G. and {Conrad}, J. and {Dalton}, M. and {Daniel}, M.~K. and
         {Davids}, I.~D. and {Degrange}, B. and {Deil}, C. and
         {Dickinson}, H.~J. and {Djannati-Ata{\"\i}}, A. and {Domainko}, W. and
         {Drury}, L. O'C. and {Dubois}, F. and {Dubus}, G. and {Dutson}, K. and
         {Dyks}, J. and {Dyrda}, M. and {Egberts}, K. and {Eger}, P. and
         {Espigat}, P. and {Fallon}, L. and {Farnier}, C. and {Fegan}, S. and
         {Feinstein}, F. and {Fernandes}, M.~V. and {Fiasson}, A. and
         {Fontaine}, G. and {F{\"o}rster}, A. and {F{\"u}{\ss}ling}, M. and
         {Gallant}, Y.~A. and {Gast}, H. and {G{\'e}rard}, L. and {Gerbig}, D. and
         {Giebels}, B. and {Glicenstein}, J.~F. and {Gl{\"u}ck}, B. and
         {Goret}, P. and {G{\"o}ring}, D. and {H{\"a}ffner}, S. and
         {Hague}, J.~D. and {Hampf}, D. and {Hauser}, M. and {Heinz}, S. and
         {Heinzelmann}, G. and {Henri}, G. and {Hermann}, G. and
         {Hinton}, J.~A. and {Hoffmann}, A. and {Hofmann}, W. and
         {Hofverberg}, P. and {Holler}, M. and {Horns}, D. and
         {Jacholkowska}, A. and {de Jager}, O.~C. and {Jahn}, C. and
         {Jamrozy}, M. and {Jung}, I. and {Kastendieck}, M.~A. and
         {Katarzy{\'n}ski}, K. and {Katz}, U. and {Kaufmann}, S. and
         {Keogh}, D. and {Khangulyan}, D. and {Kh{\'e}lifi}, B. and
         {Klochkov}, D. and {Klu{\'z}niak}, W. and {Kneiske}, T. and
         {Komin}, Nu. and {Kosack}, K. and {Kossakowski}, R. and {Laffon}, H. and
         {Lamanna}, G. and {Lennarz}, D. and {Lohse}, T. and {Lopatin}, A. and
         {Lu}, C. -C. and {Marandon}, V. and {Marcowith}, A. and {Masbou}, J. and
         {Maurin}, D. and {Maxted}, N. and {McComb}, T.~J.~L. and
         {Medina}, M.~C. and {M{\'e}hault}, J. and {Moderski}, R. and
         {Moulin}, E. and {Naumann}, C.~L. and {Naumann-Godo}, M. and
         {de Naurois}, M. and {Nedbal}, D. and {Nekrassov}, D. and {Nguyen}, N. and
         {Nicholas}, B. and {Niemiec}, J. and {Nolan}, S.~J. and {Ohm}, S. and
         {de O{\~n}a Wilhelmi}, D. and {Opitz}, B. and {Ostrowski}, M. and
         {Oya}, I. and {Panter}, M. and {Paz Arribas}, M. and {Pedaletti}, G. and
         {Pelletier}, G. and {Petrucci}, P. -O. and {Pita}, S. and
         {P{\"u}hlhofer}, G. and {Punch}, M. and {Quirrenbach}, A. and
         {Raue}, M. and {Rayner}, S.~M. and {Reimer}, A. and {Reimer}, O. and
         {Renaud}, M. and {de los Reyes}, R. and {Rieger}, F. and {Ripken}, J. and
         {Rob}, L. and {Rosier-Lees}, S. and {Rowell}, G. and {Rudak}, B. and
         {Rulten}, C.~B. and {Ruppel}, J. and {Ryde}, F. and {Sahakian}, V. and
         {Santangelo}, A. and {Schlickeiser}, R. and {Sch{\"o}ck}, F.~M. and
         {Schulz}, A. and {Schwanke}, U. and {Schwarzburg}, S. and
         {Schwemmer}, S. and {Sikora}, M. and {Skilton}, J.~L. and {Sol}, H. and
         {Spengler}, G. and {Stawarz}, {\L}. and {Steenkamp}, R. and
         {Stegmann}, C. and {Stinzing}, F. and {Stycz}, K. and {Sushch}, I. and
         {Szostek}, A. and {Tavernet}, J. -P. and {Terrier}, R. and
         {Tluczykont}, M. and {Valerius}, K. and {van Eldik}, C. and
         {Vasileiadis}, G. and {Venter}, C. and {Vialle}, J.~P. and {Viana}, A. and
         {Vincent}, P. and {V{\"o}lk}, H.~J. and {Volpe}, F. and {Vorobiov}, S. and
         {Vorster}, M. and {Wagner}, S.~J. and {Ward}, M. and {White}, R. and
         {Wierzcholska}, A. and {Zacharias}, M. and {Zajczyk}, A. and
         {Zdziarski}, A.~A. and {Zech}, A. and {Zechlin}, H. -S.},
        title = "{A new SNR with TeV shell-type morphology: HESS J1731-347}",
      journal = {Astronomy and Astrophysics},
         year = "2011",
        month = "Jul",
       volume = {531},
          eid = {A81},
        pages = {A81},
          doi = {10.1051/0004-6361/201016425},
archivePrefix = {arXiv},
       eprint = {1105.3206},
 primaryClass = {astro-ph.HE},
       adsurl = {https://ui.adsabs.harvard.edu/abs/2011A&A...531A..81H}
}

@article{37bhess,
       author = {{Aharonian}, F. and {Akhperjanian}, A.~G. and {Barres de Almeida}, U. and
         {Bazer-Bachi}, A.~R. and {Behera}, B. and {Beilicke}, M. and
         {Benbow}, W. and {Bernl{\"o}hr}, K. and {Boisson}, C. and {Borrel}, V. and
         {Braun}, I. and {Brion}, E. and {Brucker}, J. and {B{\"u}hler}, R. and
         {Bulik}, T. and {B{\"u}sching}, I. and {Boutelier}, T. and
         {Carrigan}, S. and {Chadwick}, P.~M. and {Chaves}, R.~C.~G. and
         {Chounet}, L. -M. and {Clapson}, A.~C. and {Coignet}, G. and
         {Cornils}, R. and {Costamante}, L. and {Dalton}, M. and {Degrange}, B. and
         {Dickinson}, H.~J. and {Djannati-Ata{\"\i}}, A. and {Domainko}, W. and
         {O'C. Drury}, L. and {Dubois}, F. and {Dubus}, G. and {Dyks}, J. and
         {Egberts}, K. and {Emmanoulopoulos}, D. and {Espigat}, P. and
         {Farnier}, C. and {Feinstein}, F. and {Fiasson}, A. and
         {F{\"o}rster}, A. and {Fontaine}, G. and {Funk}, S. and
         {F{\"u}{\ss}ling}, M. and {Gabici}, S. and {Gallant}, Y.~A. and
         {Giebels}, B. and {Glicenstein}, J.~F. and {Gl{\"u}ck}, B. and
         {Goret}, P. and {Hadjichristidis}, C. and {Hauser}, D. and
         {Hauser}, M. and {Heinzelmann}, G. and {Henri}, G. and {Hermann}, G. and
         {Hinton}, J.~A. and {Hoffmann}, A. and {Hofmann}, W. and
         {Holleran}, M. and {Hoppe}, S. and {Horns}, D. and {Jacholkowska}, A. and
         {de Jager}, O.~C. and {Jung}, I. and {Katarzy{\'n}ski}, K. and
         {Kaufmann}, S. and {Kendziorra}, E. and {Kerschhaggl}, M. and
         {Khangulyan}, D. and {Kh{\'e}lifi}, B. and {Keogh}, D. and
         {Komin}, Nu. and {Kosack}, K. and {Lamanna}, G. and {Latham}, I.~J. and
         {Lemoine-Goumard}, M. and {Lenain}, J. -P. and {Lohse}, T. and
         {Martin}, J.~M. and {Martineau-Huynh}, O. and {Marcowith}, A. and
         {Masterson}, C. and {Maurin}, D. and {McComb}, T.~J.~L. and
         {Moderski}, R. and {Moulin}, E. and {Naumann-Godo}, M. and
         {de Naurois}, M. and {Nedbal}, D. and {Nekrassov}, D. and
         {Nolan}, S.~J. and {Ohm}, S. and {Olive}, J. -P. and
         {de O{\~n}a Wilhelmi}, E. and {Orford}, K.~J. and {Osborne}, J.~L. and
         {Ostrowski}, M. and {Panter}, M. and {Pedaletti}, G. and
         {Pelletier}, G. and {Petrucci}, P. -O. and {Pita}, S. and
         {P{\"u}hlhofer}, G. and {Punch}, M. and {Quirrenbach}, A. and
         {Raubenheimer}, B.~C. and {Raue}, M. and {Rayner}, S.~M. and
         {Renaud}, M. and {Rieger}, F. and {Reimer}, O. and {Ripken}, J. and
         {Rob}, L. and {Rosier-Lees}, S. and {Rowell}, G. and {Rudak}, B. and
         {Ruppel}, J. and {Sahakian}, V. and {Santangelo}, A. and
         {Schlickeiser}, R. and {Sch{\"o}ck}, F.~M. and {Schr{\"o}der}, R. and
         {Schwanke}, U. and {Schwarzburg}, S. and {Schwemmer}, S. and
         {Shalchi}, A. and {Skilton}, J.~L. and {Sol}, H. and {Spangler}, D. and
         {Stawarz}, {\L}. and {Steenkamp}, R. and {Stegmann}, C. and
         {Superina}, G. and {Tam}, P.~H. and {Tavernet}, J. -P. and
         {Terrier}, R. and {van Eldik}, C. and {Vasileiadis}, G. and
         {Venter}, C. and {Vialle}, J.~P. and {Vincent}, P. and {Vivier}, M. and
         {V{\"o}lk}, H.~J. and {Volpe}, F. and {Wagner}, S.~J. and {Ward}, M. and
         {Zdziarski}, A.~A. and {Zech}, A.},
        title = "{Chandra and HESS observations of the supernova remnant CTB 37B}",
      journal = {Astronomy and Astrophysics},
         year = "2008",
        month = "Aug",
       volume = {486},
       number = {3},
        pages = {829-836},
          doi = {10.1051/0004-6361:200809655},
archivePrefix = {arXiv},
       eprint = {0803.0682},
 primaryClass = {astro-ph},
       adsurl = {https://ui.adsabs.harvard.edu/abs/2008A&A...486..829A}
}

@article{37bfermi,
       author = {{Xin}, Yu-Liang and {Liang}, Yun-Feng and {Li}, Xiang and {Yuan}, Qiang and
         {Liu}, Si-Ming and {Wei}, Da-Ming},
        title = "{A GeV Source in the Direction of Supernova Remnant CTB 37B}",
      journal = {The Astrophysical Journal},
         year = "2016",
        month = "Jan",
       volume = {817},
       number = {1},
          eid = {64},
        pages = {64},
          doi = {10.3847/0004-637X/817/1/64},
archivePrefix = {arXiv},
       eprint = {1509.08548},
 primaryClass = {astro-ph.HE},
       adsurl = {https://ui.adsabs.harvard.edu/abs/2016ApJ...817...64X}
}

@article{37ahess,
       author = {{Aharonian}, F. and {Akhperjanian}, A.~G. and {Barres de Almeida}, U. and
         {Bazer-Bachi}, A.~R. and {Behera}, B. and {Beilicke}, M. and
         {Benbow}, W. and {Bernl{\"o}hr}, K. and {Boisson}, C. and {Borrel}, V. and
         {Braun}, I. and {Brion}, E. and {Brucker}, J. and {B{\"u}hler}, R. and
         {Bulik}, T. and {B{\"u}sching}, I. and {Boutelier}, T. and
         {Carrigan}, S. and {Chadwick}, P.~M. and {Chaves}, R. and
         {Chounet}, L. -M. and {Clapson}, A.~C. and {Coignet}, G. and
         {Cornils}, R. and {Costamante}, L. and {Dalton}, M. and {Degrange}, B. and
         {Dickinson}, H.~J. and {Djannati-Ata{\"\i}}, A. and {Domainko}, W. and
         {O'C. Drury}, L. and {Dubois}, F. and {Dubus}, G. and {Dyks}, J. and
         {Egberts}, K. and {Emmanoulopoulos}, D. and {Espigat}, P. and
         {Farnier}, C. and {Feinstein}, F. and {Fiasson}, A. and
         {F{\"o}rster}, A. and {Fontaine}, G. and {Funk}, S. and
         {F{\"u}{\ss}ling}, M. and {Gabici}, S. and {Gallant}, Y.~A. and
         {Giebels}, B. and {Glicenstein}, J.~F. and {Gl{\"u}ck}, B. and
         {Goret}, P. and {Hadjichristidis}, C. and {Hauser}, D. and
         {Hauser}, M. and {Heinzelmann}, G. and {Henri}, G. and {Hermann}, G. and
         {Hinton}, J.~A. and {Hoffmann}, A. and {Hofmann}, W. and
         {Holleran}, M. and {Hoppe}, S. and {Horns}, D. and {Jacholkowska}, A. and
         {de Jager}, O.~C. and {Jung}, I. and {Katarzy{\'n}ski}, K. and
         {Kaufmann}, S. and {Kendziorra}, E. and {Kerschhaggl}, M. and
         {Khangulyan}, D. and {Kh{\'e}lifi}, B. and {Keogh}, D. and
         {Komin}, Nu. and {Kosack}, K. and {Lamanna}, G. and {Latham}, I.~J. and
         {Lemoine-Goumard}, M. and {Lenain}, J. -P. and {Lohse}, T. and
         {Martin}, J.~M. and {Martineau-Huynh}, O. and {Marcowith}, A. and
         {Masterson}, C. and {Maurin}, D. and {McComb}, T.~J.~L. and
         {Moderski}, R. and {Moulin}, E. and {Nakajima}, H. and
         {Naumann-Godo}, M. and {de Naurois}, M. and {Nedbal}, D. and
         {Nekrassov}, D. and {Nolan}, S.~J. and {Ohm}, S. and {Olive}, J. -P. and
         {de O{\~n}a Wilhelmi}, E. and {Orford}, K.~J. and {Osborne}, J.~L. and
         {Ostrowski}, M. and {Panter}, M. and {Pedaletti}, G. and
         {Pelletier}, G. and {Petrucci}, P. -O. and {Pita}, S. and
         {P{\"u}hlhofer}, G. and {Punch}, M. and {Quirrenbach}, A. and
         {Raubenheimer}, B.~C. and {Raue}, M. and {Rayner}, S.~M. and
         {Reimer}, O. and {Renaud}, M. and {Rieger}, F. and {Ripken}, J. and
         {Rob}, L. and {Rosier-Lees}, S. and {Rowell}, G. and {Rudak}, B. and
         {Ruppel}, J. and {Sahakian}, V. and {Santangelo}, A. and
         {Schlickeiser}, R. and {Sch{\"o}ck}, F.~M. and {Schr{\"o}der}, R. and
         {Schwanke}, U. and {Schwarzburg}, S. and {Schwemmer}, S. and
         {Shalchi}, A. and {Skilton}, J.~L. and {Sol}, H. and {Spangler}, D. and
         {Stawarz}, {\L}. and {Steenkamp}, R. and {Stegmann}, C. and
         {Superina}, G. and {Tam}, P.~H. and {Tavernet}, J. -P. and
         {Terrier}, R. and {Tibolla}, O. and {van Eldik}, C. and
         {Vasileiadis}, G. and {Venter}, C. and {Vialle}, J.~P. and
         {Vincent}, P. and {Vivier}, M. and {V{\"o}lk}, H.~J. and {Volpe}, F. and
         {Wagner}, S.~J. and {Ward}, M. and {Zdziarski}, A.~A. and {Zech}, A.},
        title = "{Discovery of a VHE gamma-ray source coincident with the supernova remnant CTB 37A}",
      journal = {Astronomy and Astrophysics},
         year = "2008",
        month = "Nov",
       volume = {490},
       number = {2},
        pages = {685-693},
          doi = {10.1051/0004-6361:200809722},
archivePrefix = {arXiv},
       eprint = {0803.0702},
 primaryClass = {astro-ph},
       adsurl = {https://ui.adsabs.harvard.edu/abs/2008A&A...490..685A}
}

@inproceedings{37afermi08,
       author = {{Castro}, Daniel and {Slane}, Patrick},
        title = "{Fermi LAT Observations of Supernova Remnants Interacting with Molecular Clouds}",
    booktitle = {38th COSPAR Scientific Assembly},
         year = "2010",
       volume = {38},
        month = "Jan",
        pages = {3},
       adsurl = {https://ui.adsabs.harvard.edu/abs/2010cosp...38.2754C}
}

@article{37afermi13,
       author = {{Brandt}, T.~J. and {Fermi-LAT Collaboration}},
        title = "{A view of supernova remnant CTB 37A with the Fermi Gamma-ray Space Telescope}",
      journal = {Advances in Space Research},
         year = "2013",
        month = "Jan",
       volume = {51},
       number = {2},
        pages = {247-252},
          doi = {10.1016/j.asr.2011.07.021},
archivePrefix = {arXiv},
       eprint = {1304.7996},
 primaryClass = {astro-ph.HE},
       adsurl = {https://ui.adsabs.harvard.edu/abs/2013AdSpR..51..247B}
}

@article{1713hess,
       author = {{Aharonian}, F. and {Akhperjanian}, A.~G. and {Bazer-Bachi}, A.~R. and
         {Beilicke}, M. and {Benbow}, W. and {Berge}, D. and {Bernl{\"o}hr}, K. and
         {Boisson}, C. and {Bolz}, O. and {Borrel}, V. and {Braun}, I. and
         {Brion}, E. and {Brown}, A.~M. and {B{\"u}hler}, R. and
         {B{\"u}sching}, I. and {Carrigan}, S. and {Chadwick}, P.~M. and
         {Chounet}, L. -M. and {Coignet}, G. and {Cornils}, R. and
         {Costamante}, L. and {Degrange}, B. and {Dickinson}, H.~J. and
         {Djannati-Ata{\"\i}}, A. and {Drury}, L.~O. 'c. and {Dubus}, G. and
         {Egberts}, K. and {Emmanoulopoulos}, D. and {Espigat}, P. and
         {Feinstein}, F. and {Ferrero}, E. and {Fiasson}, A. and {Fontaine}, G. and
         {Funk}, Seb. and {Funk}, S. and {F{\"u}{\ss}ling}, M. and
         {Gallant}, Y.~A. and {Giebels}, B. and {Glicenstein}, J.~F. and
         {Gl{\"u}ck}, B. and {Goret}, P. and {Hadjichristidis}, C. and
         {Hauser}, D. and {Hauser}, M. and {Heinzelmann}, G. and {Henri}, G. and
         {Hermann}, G. and {Hinton}, J.~A. and {Hoffmann}, A. and {Hofmann}, W. and
         {Holleran}, M. and {Hoppe}, S. and {Horns}, D. and {Jacholkowska}, A. and
         {de Jager}, O.~C. and {Kendziorra}, E. and {Kerschhaggl}, M. and
         {Kh{\'e}lifi}, B. and {Komin}, Nu. and {Konopelko}, A. and
         {Kosack}, K. and {Lamanna}, G. and {Latham}, I.~J. and {Le Gallou}, R. and
         {Lemi{\`e}re}, A. and {Lemoine-Goumard}, M. and {Lohse}, T. and
         {Martin}, J.~M. and {Martineau-Huynh}, O. and {Marcowith}, A. and
         {Masterson}, C. and {Maurin}, G. and {McComb}, T.~J.~L. and
         {Moulin}, E. and {de Naurois}, M. and {Nedbal}, D. and {Nolan}, S.~J. and
         {Noutsos}, A. and {Olive}, J. -P. and {Orford}, K.~J. and
         {Osborne}, J.~L. and {Panter}, M. and {Pelletier}, G. and {Pita}, S. and
         {P{\"u}hlhofer}, G. and {Punch}, M. and {Ranchon}, S. and
         {Raubenheimer}, B.~C. and {Raue}, M. and {Rayner}, S.~M. and
         {Reimer}, A. and {Reimer}, O. and {Ripken}, J. and {Rob}, L. and {Rolland
        }, L. and {Rosier-Lees}, S. and {Rowell}, G. and {Sahakian}, V. and
         {Santangelo}, A. and {Saug{\'e}}, L. and {Schlenker}, S. and
         {Schlickeiser}, R. and {Schr{\"o}der}, R. and {Schwanke}, U. and
         {Schwarzburg}, S. and {Schwemmer}, S. and {Shalchi}, A. and {Sol}, H. and
         {Spangler}, D. and {Spanier}, F. and {Steenkamp}, R. and
         {Stegmann}, C. and {Superina}, G. and {Tam}, P.~H. and
         {Tavernet}, J. -P. and {Terrier}, R. and {Tluczykont}, M. and
         {van Eldik}, C. and {Vasileiadis}, G. and {Venter}, C. and
         {Vialle}, J.~P. and {Vincent}, P. and {V{\"o}lk}, H.~J. and
         {Wagner}, S.~J. and {Ward}, M. and {H.~E.~S.~S. Collaboration}},
        title = "{Primary particle acceleration above 100 TeV in the shell-type supernova remnant RX J1713.7 - 3946 with deep H.E.S.S. observations (Corrigendum)}",
      journal = {Astronomy and Astrophysics},
         year = "2011",
        month = "Jul",
       volume = {531},
          eid = {C1},
        pages = {C1},
          doi = {10.1051/0004-6361/20066381e},
       adsurl = {https://ui.adsabs.harvard.edu/abs/2011A&A...531C...1A}
}

@article{1713fermi,
       author = {{Federici}, S. and {Pohl}, M. and {Telezhinsky}, I. and {Wilhelm}, A. and
         {Dwarkadas}, V.~V.},
        title = "{Analysis of GeV-band {\ensuremath{\gamma}}-ray emission from supernova remnant RX J1713.7-3946}",
      journal = {Astronomy and Astrophysics},
         year = "2015",
        month = "May",
       volume = {577},
          eid = {A12},
        pages = {A12},
          doi = {10.1051/0004-6361/201424947},
archivePrefix = {arXiv},
       eprint = {1502.06355},
 primaryClass = {astro-ph.HE},
       adsurl = {https://ui.adsabs.harvard.edu/abs/2015A&A...577A..12F}
}

@article{sn16,
       author = {{Xing}, Yi and {Wang}, Zhongxiang and {Zhang}, Xiao and {Chen}, Yang},
        title = "{The Likely Fermi Detection of the Supernova Remnant SN 1006}",
      journal = {The Astrophysical Journal},
         year = "2016",
        month = "May",
       volume = {823},
       number = {1},
          eid = {44},
        pages = {44},
          doi = {10.3847/0004-637X/823/1/44},
archivePrefix = {arXiv},
       eprint = {1603.00998},
 primaryClass = {astro-ph.HE},
       adsurl = {https://ui.adsabs.harvard.edu/abs/2016ApJ...823...44X}
}

@article{sn17,
       author = {{Condon}, B. and {Lemoine-Goumard}, M. and {Acero}, F. and
         {Katagiri}, H.},
        title = "{Detection of Two TeV Shell-type Remnants at GeV Energies with FERMI LAT: HESS J1731-347 and SN 1006}",
      journal = {The Astrophysical Journal},
         year = "2017",
        month = "Dec",
       volume = {851},
       number = {2},
          eid = {100},
        pages = {100},
          doi = {10.3847/1538-4357/aa9be8},
archivePrefix = {arXiv},
       eprint = {1711.05499},
 primaryClass = {astro-ph.HE},
       adsurl = {https://ui.adsabs.harvard.edu/abs/2017ApJ...851..100C}
}

@article{sn10,
       author = {{Acero}, F. and {Aharonian}, F. and {Akhperjanian}, A.~G. and
         {Anton}, G. and {Barres de Almeida}, U. and {Bazer-Bachi}, A.~R. and
         {Becherini}, Y. and {Behera}, B. and {Beilicke}, M. and
         {Bernl{\"o}hr}, K. and {Bochow}, A. and {Boisson}, C. and
         {Bolmont}, J. and {Borrel}, V. and {Brucker}, J. and {Brun}, F. and
         {Brun}, P. and {B{\"u}hler}, R. and {Bulik}, T. and {B{\"u}sching}, I. and
         {Boutelier}, T. and {Chadwick}, P.~M. and {Charbonnier}, A. and
         {Chaves}, R.~C.~G. and {Cheesebrough}, A. and {Conrad}, J. and
         {Chounet}, L. -M. and {Clapson}, A.~C. and {Coignet}, G. and
         {Dalton}, M. and {Daniel}, M.~K. and {Davids}, I.~D. and
         {Degrange}, B. and {Deil}, C. and {Dickinson}, H.~J. and
         {Djannati-Ata{\"\i}}, A. and {Domainko}, W. and {O'C. Drury}, L. and
         {Dubois}, F. and {Dubus}, G. and {Dyks}, J. and {Dyrda}, M. and
         {Egberts}, K. and {Eger}, P. and {Espigat}, P. and {Fallon}, L. and
         {Farnier}, C. and {Fegan}, S. and {Feinstein}, F. and {Fiasson}, A. and
         {F{\"o}rster}, A. and {Fontaine}, G. and {F{\"u}{\ss}ling}, M. and
         {Gabici}, S. and {Gallant}, Y.~A. and {G{\'e}rard}, L. and
         {Gerbig}, D. and {Giebels}, B. and {Glicenstein}, J.~F. and
         {Gl{\"u}ck}, B. and {Goret}, P. and {G{\"o}ring}, D. and {Hauser}, D. and
         {Hauser}, M. and {Heinz}, S. and {Heinzelmann}, G. and {Henri}, G. and
         {Hermann}, G. and {Hinton}, J.~A. and {Hoffmann}, A. and {Hofmann}, W. and
         {Hofverberg}, P. and {Holleran}, M. and {Hoppe}, S. and {Horns}, D. and
         {Jacholkowska}, A. and {de Jager}, O.~C. and {Jahn}, C. and {Jung}, I. and
         {Katarzy{\'n}ski}, K. and {Katz}, U. and {Kaufmann}, S. and
         {Kerschhaggl}, M. and {Khangulyan}, D. and {Kh{\'e}lifi}, B. and
         {Keogh}, D. and {Klochkov}, D. and {Klu{\'z}niak}, W. and
         {Kneiske}, T. and {Komin}, Nu. and {Kosack}, K. and {Kossakowski}, R. and
         {Lamanna}, G. and {Lemoine-Goumard}, M. and {Lenain}, J. -P. and
         {Lohse}, T. and {Marandon}, V. and {Marcowith}, A. and {Masbou}, J. and
         {Maurin}, D. and {McComb}, T.~J.~L. and {Medina}, M.~C. and
         {M{\'e}hault}, J. and {Moderski}, R. and {Moulin}, E. and
         {Naumann-Godo}, M. and {de Naurois}, M. and {Nedbal}, D. and
         {Nekrassov}, D. and {Nicholas}, B. and {Niemiec}, J. and
         {Nolan}, S.~J. and {Ohm}, S. and {Olive}, J. -F. and
         {de O{\~n}a Wilhelmi}, E. and {Orford}, K.~J. and {Ostrowski}, M. and
         {Panter}, M. and {Paz Arribas}, M. and {Pedaletti}, G. and
         {Pelletier}, G. and {Petrucci}, P. -O. and {Pita}, S. and
         {P{\"u}hlhofer}, G. and {Punch}, M. and {Quirrenbach}, A. and
         {Raubenheimer}, B.~C. and {Raue}, M. and {Rayner}, S.~M. and
         {Reimer}, O. and {Renaud}, M. and {de Los Reyes}, R. and {Rieger}, F. and
         {Ripken}, J. and {Rob}, L. and {Rosier-Lees}, S. and {Rowell}, G. and
         {Rudak}, B. and {Rulten}, C.~B. and {Ruppel}, J. and {Ryde}, F. and
         {Sahakian}, V. and {Santangelo}, A. and {Schlickeiser}, R. and
         {Sch{\"o}ck}, F.~M. and {Sch{\"o}nwald}, A. and {Schwanke}, U. and
         {Schwarzburg}, S. and {Schwemmer}, S. and {Shalchi}, A. and
         {Sushch}, I. and {Sikora}, M. and {Skilton}, J.~L. and {Sol}, H. and
         {Stawarz}, {\L}. and {Steenkamp}, R. and {Stegmann}, C. and
         {Stinzing}, F. and {Superina}, G. and {Szostek}, A. and {Tam}, P.~H. and
         {Tavernet}, J. -P. and {Terrier}, R. and {Tibolla}, O. and
         {Tluczykont}, M. and {van Eldik}, C. and {Vasileiadis}, G. and
         {Venter}, C. and {Venter}, L. and {Vialle}, J.~P. and {Vincent}, P. and
         {Vink}, J. and {Vivier}, M. and {V{\"o}lk}, H.~J. and {Volpe}, F. and
         {Vorobiov}, S. and {Wagner}, S.~J. and {Ward}, M. and
         {Zdziarski}, A.~A. and {Zech}, A. and {H.~E.~S.~S. Collaboration}},
        title = "{First detection of VHE {\ensuremath{\gamma}}-rays from SN 1006 by HESS}",
      journal = {Astronomy and Astrophysics},
         year = "2010",
        month = "Jun",
       volume = {516},
          eid = {A62},
        pages = {A62},
          doi = {10.1051/0004-6361/200913916},
archivePrefix = {arXiv},
       eprint = {1004.2124},
 primaryClass = {astro-ph.HE},
       adsurl = {https://ui.adsabs.harvard.edu/abs/2010A&A...516A..62A}
}

@article{86fermi,
       author = {{Ajello}, M. and {Baldini}, L. and {Barbiellini}, G. and {Bastieri}, D. and
         {Bellazzini}, R. and {Bissaldi}, E. and {Bloom}, E.~D. and
         {Bonino}, R. and {Bottacini}, E. and {Brandt}, T.~J. and {Bregeon}, J. and
         {Bruel}, P. and {Buehler}, R. and {Caliandro}, G.~A. and
         {Cameron}, R.~A. and {Caragiulo}, M. and {Cavazzuti}, E. and
         {Charles}, E. and {Chekhtman}, A. and {Ciprini}, S. and
         {Cohen-Tanugi}, J. and {Condon}, B. and {Costanza}, F. and
         {Cutini}, S. and {D'Ammando}, F. and {de Palma}, F. and {Desiante}, R. and
         {Di Lalla}, N. and {Di Mauro}, M. and {Di Venere}, L. and
         {Drell}, P.~S. and {Dubner}, G. and {Dumora}, D. and {Duvidovich}, L. and
         {Favuzzi}, C. and {Focke}, W.~B. and {Fusco}, P. and {Gargano}, F. and
         {Gasparrini}, D. and {Giacani}, E. and {Giglietto}, N. and
         {Glanzman}, T. and {Green}, D.~A. and {Grenier}, I.~A. and
         {Guiriec}, S. and {Hays}, E. and {Hewitt}, J.~W. and {Hill}, A.~B. and
         {Horan}, D. and {Jogler}, T. and {J{\'o}hannesson}, G. and
         {Jung-Richardt}, I. and {Kensei}, S. and {Kuss}, M. and {Larsson}, S. and
         {Latronico}, L. and {Lemoine-Goumard}, M. and {Li}, J. and {Li}, L. and
         {Longo}, F. and {Loparco}, F. and {Lovellette}, M.~N. and
         {Lubrano}, P. and {Magill}, J. and {Maldera}, S. and {Manfreda}, A. and
         {Mayer}, M. and {Mazziotta}, M.~N. and {McEnery}, J.~E. and
         {Michelson}, P.~F. and {Mitthumsiri}, W. and {Mizuno}, T. and
         {Monzani}, M.~E. and {Morselli}, A. and {Moskalenko}, I.~V. and
         {Negro}, M. and {Nuss}, E. and {Orienti}, M. and {Orlando}, E. and
         {Ormes}, J.~F. and {Paneque}, D. and {Perkins}, J.~S. and
         {Pesce-Rollins}, M. and {Piron}, F. and {Pivato}, G. and
         {Porter}, T.~A. and {Rain{\`o}}, S. and {Rando}, R. and {Razzano}, M. and
         {Reimer}, A. and {Reimer}, O. and {Reposeur}, T. and {Schmid}, J. and
         {Schulz}, A. and {Sgr{\`o}}, C. and {Simone}, D. and {Siskind}, E.~J. and
         {Spada}, F. and {Spandre}, G. and {Spinelli}, P. and {Thayer}, J.~B. and
         {Tibaldo}, L. and {Torres}, D.~F. and {Tosti}, G. and {Troja}, E. and
         {Uchiyama}, Y. and {Vianello}, G. and {Vink}, J. and {Wood}, K.~S. and
         {Yassine}, M.},
        title = "{Deep Morphological and Spectral Study of the SNR RCW 86 with Fermi-LAT}",
      journal = {The Astrophysical Journal},
         year = "2016",
        month = "Mar",
       volume = {819},
       number = {2},
          eid = {98},
        pages = {98},
          doi = {10.3847/0004-637X/819/2/98},
archivePrefix = {arXiv},
       eprint = {1601.06534},
 primaryClass = {astro-ph.HE},
       adsurl = {https://ui.adsabs.harvard.edu/abs/2016ApJ...819...98A}
}

@article{86hess,
       author = {{H.~E.~S.~S. Collaboration} and {Abramowski}, A. and {Aharonian}, F. and
         {Ait Benkhali}, F. and {Akhperjanian}, A.~G. and {Ang{\"u}ner}, E.~O. and
         {Backes}, M. and {Balzer}, A. and {Becherini}, Y. and
         {Becker Tjus}, J. and {Berge}, D. and {Bernhard}, S. and
         {Bernl{\"o}hr}, K. and {Birsin}, E. and {Blackwell}, R. and
         {B{\"o}ttcher}, M. and {Boisson}, C. and {Bolmont}, J. and
         {Bordas}, P. and {Bregeon}, J. and {Brun}, F. and {Brun}, P. and
         {Bryan}, M. and {Bulik}, T. and {Carr}, J. and {Casanova}, S. and
         {Chakraborty}, N. and {Chalme-Calvet}, R. and {Chaves}, R.~C.~G. and
         {Chen}, A. and {Chevalier}, J. and {Chr{\'e}tien}, M. and
         {Colafrancesco}, S. and {Cologna}, G. and {Condon}, B. and
         {Conrad}, J. and {Couturier}, C. and {Cui}, Y. and {Davids}, I.~D. and
         {Degrange}, B. and {Deil}, C. and {deWilt}, P. and
         {Djannati-Ata{\"\i}}, A. and {Domainko}, W. and {Donath}, A. and
         {Drury}, L.~O. 'C. and {Dubus}, G. and {Dutson}, K. and {Dyks}, J. and
         {Dyrda}, M. and {Edwards}, T. and {Egberts}, K. and {Eger}, P. and
         {Ernenwein}, J. -P. and {Espigat}, P. and {Farnier}, C. and
         {Fegan}, S. and {Feinstein}, F. and {Fernandes}, M.~V. and {Fernand
        ez}, D. and {Fiasson}, A. and {Fontaine}, G. and {F{\"o}rster}, A. and
         {F{\"u}{\ss}ling}, M. and {Gabici}, S. and {Gajdus}, M. and
         {Gallant}, Y.~A. and {Garrigoux}, T. and {Giavitto}, G. and
         {Giebels}, B. and {Glicenstein}, J.~F. and {Gottschall}, D. and
         {Goyal}, A. and {Grondin}, M. -H. and {Grudzi{\'n}ska}, M. and
         {Hadasch}, D. and {H{\"a}ffner}, S. and {Hahn}, J. and {Hawkes}, J. and
         {Heinzelmann}, G. and {Henri}, G. and {Hermann}, G. and {Hervet}, O. and
         {Hillert}, A. and {Hinton}, J.~A. and {Hofmann}, W. and
         {Hofverberg}, P. and {Hoischen}, C. and {Holler}, M. and {Horns}, D. and
         {Ivascenko}, A. and {Jacholkowska}, A. and {Jamrozy}, M. and
         {Janiak}, M. and {Jankowsky}, F. and {Jung-Richardt}, I. and
         {Kastendieck}, M.~A. and {Katarzy{\'n}ski}, K. and {Katz}, U. and
         {Kerszberg}, D. and {Kh{\'e}lifi}, B. and {Kieffer}, M. and
         {Klepser}, S. and {Klochkov}, D. and {Klu{\'z}niak}, W. and
         {Kolitzus}, D. and {Komin}, Nu. and {Kosack}, K. and {Krakau}, S. and
         {Krayzel}, F. and {Kr{\"u}ger}, P.~P. and {Laffon}, H. and
         {Lamanna}, G. and {Lau}, J. and {Lefaucheur}, J. and {Lefranc}, V. and
         {Lemi{\`e}re}, A. and {Lemoine-Goumard}, M. and {Lenain}, J. -P. and
         {Lohse}, T. and {Lopatin}, A. and {Lorentz}, M. and {Lu}, C. -C. and
         {Lui}, R. and {Marandon}, V. and {Marcowith}, A. and {Mariaud}, C. and
         {Marx}, R. and {Maurin}, G. and {Maxted}, N. and {Mayer}, M. and
         {Meintjes}, P.~J. and {Menzler}, U. and {Meyer}, M. and
         {Mitchell}, A.~M.~W. and {Moderski}, R. and {Mohamed}, M. and
         {Mor{\r{a}}}, K. and {Moulin}, E. and {Murach}, T. and
         {de Naurois}, M. and {Niemiec}, J. and {Oakes}, L. and {Odaka}, H. and
         {{\"O}ttl}, S. and {Ohm}, S. and {Opitz}, B. and {Ostrowski}, M. and
         {Oya}, I. and {Panter}, M. and {Parsons}, R.~D. and {Paz Arribas}, M. and
         {Pekeur}, N.~W. and {Pelletier}, G. and {Petrucci}, P. -O. and
         {Peyaud}, B. and {Pita}, S. and {Poon}, H. and {Prokhorov}, D. and
         {Prokoph}, H. and {P{\"u}hlhofer}, G. and {Punch}, M. and
         {Quirrenbach}, A. and {Raab}, S. and {Reichardt}, I. and {Reimer}, A. and
         {Reimer}, O. and {Renaud}, M. and {de los Reyes}, R. and {Rieger}, F. and
         {Romoli}, C. and {Rosier-Lees}, S. and {Rowell}, G. and {Rudak}, B. and
         {Rulten}, C.~B. and {Sahakian}, V. and {Salek}, D. and
         {Sanchez}, D.~A. and {Santangelo}, A. and {Sasaki}, M. and
         {Schlickeiser}, R. and {Sch{\"u}ssler}, F. and {Schulz}, A. and
         {Schwanke}, U. and {Schwemmer}, S. and {Seyffert}, A.~S. and
         {Simoni}, R. and {Sol}, H. and {Spanier}, F. and {Spengler}, G. and
         {Spies}, F. and {Stawarz}, {\L}. and {Steenkamp}, R. and
         {Stegmann}, C. and {Stinzing}, F. and {Stycz}, K. and {Sushch}, I. and
         {Tavernet}, J. -P. and {Tavernier}, T. and {Taylor}, A.~M. and
         {Terrier}, R. and {Tluczykont}, M. and {Trichard}, C. and {Tuffs}, R. and
         {Valerius}, K. and {van der Walt}, J. and {van Eldik}, C. and
         {van Soelen}, B. and {Vasileiadis}, G. and {Veh}, J. and {Venter}, C. and
         {Viana}, A. and {Vincent}, P. and {Vink}, J. and {Voisin}, F. and
         {V{\"o}lk}, H.~J. and {Vuillaume}, T. and {Wagner}, S.~J. and
         {Wagner}, P. and {Wagner}, R.~M. and {Weidinger}, M. and {White}, R. and
         {Wierzcholska}, A. and {Willmann}, P. and {W{\"o}rnlein}, A. and
         {Wouters}, D. and {Yang}, R. and {Zabalza}, V. and {Zaborov}, D. and
         {Zacharias}, M. and {Zdziarski}, A.~A. and {Zech}, A. and {Zefi}, F. and
         {{\.Z}ywucka}, N.},
        title = "{Detailed spectral and morphological analysis of the shell type supernova remnant RCW 86}",
      journal = {Astronomy and Astrophysics},
         year = "2018",
        month = "Apr",
       volume = {612},
          eid = {A4},
        pages = {A4},
          doi = {10.1051/0004-6361/201526545},
archivePrefix = {arXiv},
       eprint = {1601.04461},
 primaryClass = {astro-ph.HE},
       adsurl = {https://ui.adsabs.harvard.edu/abs/2018A&A...612A...4H}
}

@article{velahess,
       author = {{H.~E.~S.~S. Collaboration} and {Abdalla}, H. and {Abramowski}, A. and
         {Aharonian}, F. and {Ait Benkhali}, F. and {Akhperjanian}, A.~G. and
         {Ang{\"u}ner}, E.~O. and {Arakawa}, M. and {Arrieta}, M. and
         {Aubert}, P. and {Backes}, M. and {Balzer}, A. and {Barnard}, M. and
         {Becherini}, Y. and {Becker Tjus}, J. and {Berge}, D. and
         {Bernhard}, S. and {Bernl{\"o}hr}, K. and {Blackwell}, R. and
         {B{\"o}ttcher}, M. and {Boisson}, C. and {Bolmont}, J. and
         {Bordas}, P. and {Bregeon}, J. and {Brun}, F. and {Brun}, P. and
         {Bryan}, M. and {B{\"u}chele}, M. and {Bulik}, T. and {Capasso}, M. and
         {Carr}, J. and {Casanova}, S. and {Cerruti}, M. and {Chakraborty}, N. and
         {Chalme-Calvet}, R. and {Chaves}, R.~C.~G. and {Chen}, A. and
         {Chevalier}, J. and {Chr{\'e}tien}, M. and {Coffaro}, M. and
         {Colafrancesco}, S. and {Cologna}, G. and {Condon}, B. and
         {Conrad}, J. and {Cui}, Y. and {Davids}, I.~D. and {Decock}, J. and
         {Degrange}, B. and {Deil}, C. and {Devin}, J. and {deWilt}, P. and
         {Dirson}, L. and {Djannati-Ata{\"\i}}, A. and {Domainko}, W. and
         {Donath}, A. and {Drury}, L.~O. 'C. and {Dutson}, K. and {Dyks}, J. and
         {Edwards}, T. and {Egberts}, K. and {Eger}, P. and {Ernenwein}, J. -P. and
         {Eschbach}, S. and {Farnier}, C. and {Fegan}, S. and {Fernand
        es}, M.~V. and {Fiasson}, A. and {Fontaine}, G. and {F{\"o}rster}, A. and
         {Funk}, S. and {F{\"u}{\ss}ling}, M. and {Gabici}, S. and {Gajdus}, M. and
         {Gallant}, Y.~A. and {Garrigoux}, T. and {Giavitto}, G. and
         {Giebels}, B. and {Glicenstein}, J.~F. and {Gottschall}, D. and
         {Goyal}, A. and {Grondin}, M. -H. and {Hahn}, J. and {Haupt}, M. and
         {Hawkes}, J. and {Heinzelmann}, G. and {Henri}, G. and {Hermann}, G. and
         {Hervet}, O. and {Hinton}, J.~A. and {Hofmann}, W. and {Hoischen}, C. and
         {Holler}, M. and {Horns}, D. and {Ivascenko}, A. and {Iwasaki}, H. and
         {Jacholkowska}, A. and {Jamrozy}, M. and {Janiak}, M. and
         {Jankowsky}, D. and {Jankowsky}, F. and {Jingo}, M. and {Jogler}, T. and
         {Jouvin}, L. and {Jung-Richardt}, I. and {Kastendieck}, M.~A. and
         {Katarzy{\'n}ski}, K. and {Katsuragawa}, M. and {Katz}, U. and
         {Kerszberg}, D. and {Khangulyan}, D. and {Kh{\'e}lifi}, B. and
         {Kieffer}, M. and {King}, J. and {Klepser}, S. and {Klochkov}, D. and
         {Klu{\'z}niak}, W. and {Kolitzus}, D. and {Komin}, Nu. and
         {Krakau}, S. and {Kraus}, M. and {Kr{\"u}ger}, P.~P. and {Laffon}, H. and
         {Lamanna}, G. and {Lau}, J. and {Lees}, J. -P. and {Lefaucheur}, J. and
         {Lefranc}, V. and {Lemi{\`e}re}, A. and {Lemoine-Goumard}, M. and
         {Lenain}, J. -P. and {Leser}, E. and {Lohse}, T. and {Lorentz}, M. and
         {Liu}, R. and {L{\'o}pez-Coto}, R. and {Lypova}, I. and {Marandon}, V. and
         {Marcowith}, A. and {Mariaud}, C. and {Marx}, R. and {Maurin}, G. and
         {Maxted}, N. and {Mayer}, M. and {Meintjes}, P.~J. and {Meyer}, M. and
         {Mitchell}, A.~M.~W. and {Moderski}, R. and {Mohamed}, M. and
         {Mohrmann}, L. and {Mor{\r{a}}}, K. and {Moulin}, E. and {Murach}, T. and
         {Nakashima}, S. and {de Naurois}, M. and {Niederwanger}, F. and
         {Niemiec}, J. and {Oakes}, L. and {O'Brien}, P. and {Odaka}, H. and
         {{\"O}ttl}, S. and {Ohm}, S. and {Ostrowski}, M. and {Oya}, I. and
         {Padovani}, M. and {Panter}, M. and {Parsons}, R.~D. and
         {Paz Arribas}, M. and {Pekeur}, N.~W. and {Pelletier}, G. and
         {Perennes}, C. and {Petrucci}, P. -O. and {Peyaud}, B. and {Piel}, Q. and
         {Pita}, S. and {Poon}, H. and {Prokhorov}, D. and {Prokoph}, H. and
         {P{\"u}hlhofer}, G. and {Punch}, M. and {Quirrenbach}, A. and
         {Raab}, S. and {Reimer}, A. and {Reimer}, O. and {Renaud}, M. and
         {de los Reyes}, R. and {Richter}, S. and {Rieger}, F. and {Romoli}, C. and
         {Rowell}, G. and {Rudak}, B. and {Rulten}, C.~B. and {Sahakian}, V. and
         {Saito}, S. and {Salek}, D. and {Sanchez}, D.~A. and {Santangelo}, A. and
         {Sasaki}, M. and {Schlickeiser}, R. and {Sch{\"u}ssler}, F. and
         {Schulz}, A. and {Schwanke}, U. and {Schwemmer}, S. and
         {Seglar-Arroyo}, M. and {Settimo}, M. and {Seyffert}, A.~S. and
         {Shafi}, N. and {Shilon}, I. and {Simoni}, R. and {Sol}, H. and
         {Spanier}, F. and {Spengler}, G. and {Spies}, F. and {Stawarz}, {\L}. and
         {Steenkamp}, R. and {Stegmann}, C. and {Stycz}, K. and {Sushch}, I. and
         {Takahashi}, T. and {Tavernet}, J. -P. and {Tavernier}, T. and
         {Taylor}, A.~M. and {Terrier}, R. and {Tibaldo}, L. and {Tiziani}, D. and
         {Tluczykont}, M. and {Trichard}, C. and {Tsuji}, N. and {Tuffs}, R. and
         {Uchiyama}, Y. and {van der Walt}, D.~J. and {van Eldik}, C. and
         {van Rensburg}, C. and {van Soelen}, B. and {Vasileiadis}, G. and
         {Veh}, J. and {Venter}, C. and {Viana}, A. and {Vincent}, P. and
         {Vink}, J. and {Voisin}, F. and {V{\"o}lk}, H.~J. and {Vuillaume}, T. and
         {Wadiasingh}, Z. and {Wagner}, S.~J. and {Wagner}, P. and
         {Wagner}, R.~M. and {White}, R. and {Wierzcholska}, A. and
         {Willmann}, P. and {W{\"o}rnlein}, A. and {Wouters}, D. and {Yang}, R. and
         {Zabalza}, V. and {Zaborov}, D. and {Zacharias}, M. and {Zanin}, R. and
         {Zdziarski}, A.~A. and {Zech}, A. and {Zefi}, F. and {Ziegler}, A. and
         {{\.Z}ywucka}, N.},
        title = "{Deeper H.E.S.S. observations of Vela Junior (RX J0852.0-4622): Morphology studies and resolved spectroscopy}",
      journal = {Astronomy and Astrophysics},
         year = "2018",
        month = "Apr",
       volume = {612},
          eid = {A7},
        pages = {A7},
          doi = {10.1051/0004-6361/201630002},
archivePrefix = {arXiv},
       eprint = {1611.01863},
 primaryClass = {astro-ph.HE},
       adsurl = {https://ui.adsabs.harvard.edu/abs/2018A&A...612A...7H}
}

@article{velefermi,
       author = {{Tanaka}, T. and {Allafort}, A. and {Ballet}, J. and {Funk}, S. and
         {Giordano}, F. and {Hewitt}, J. and {Lemoine-Goumard}, M. and
         {Tajima}, H. and {Tibolla}, O. and {Uchiyama}, Y.},
        title = "{Gamma-Ray Observations of the Supernova Remnant RX J0852.0-4622 with the Fermi Large Area Telescope}",
      journal = {The Astrophysical Journal},
         year = "2011",
        month = "Oct",
       volume = {740},
       number = {2},
          eid = {L51},
        pages = {L51},
          doi = {10.1088/2041-8205/740/2/L51},
archivePrefix = {arXiv},
       eprint = {1109.4658},
 primaryClass = {astro-ph.HE},
       adsurl = {https://ui.adsabs.harvard.edu/abs/2011ApJ...740L..51T}
}

@article{pupphess,
       author = {{H.~E.~S.~S. Collaboration} and {Abramowski}, A. and {Aharonian}, F. and
         {Ait Benkhali}, F. and {Akhperjanian}, A.~G. and {Ang{\"u}ner}, E.~O. and
         {Backes}, M. and {Balenderan}, S. and {Balzer}, A. and {Barnacka}, A. and
         {Becherini}, Y. and {Becker Tjus}, J. and {Berge}, D. and
         {Bernhard}, S. and {Bernl{\"o}hr}, K. and {Birsin}, E. and
         {Biteau}, J. and {B{\"o}ttcher}, M. and {Boisson}, C. and
         {Bolmont}, J. and {Bordas}, P. and {Bregeon}, J. and {Brun}, F. and
         {Brun}, P. and {Bryan}, M. and {Bulik}, T. and {Carrigan}, S. and
         {Casanova}, S. and {Chadwick}, P.~M. and {Chakraborty}, N. and
         {Chalme-Calvet}, R. and {Chaves}, R.~C.~G. and {Chr{\'e}tien}, M. and
         {Colafrancesco}, S. and {Cologna}, G. and {Conrad}, J. and
         {Couturier}, C. and {Cui}, Y. and {Davids}, I.~D. and {Degrange}, B. and
         {Deil}, C. and {deWilt}, P. and {Djannati-Ata{\"\i}}, A. and
         {Domainko}, W. and {Donath}, A. and {O'C. Drury}, L. and {Dubus}, G. and
         {Dutson}, K. and {Dyks}, J. and {Dyrda}, M. and {Edwards}, T. and
         {Egberts}, K. and {Eger}, P. and {Espigat}, P. and {Farnier}, C. and
         {Fegan}, S. and {Feinstein}, F. and {Fernandes}, M.~V. and {Fernand
        ez}, D. and {Fiasson}, A. and {Fontaine}, G. and {F{\"o}rster}, A. and
         {F{\"u}{\ss}ling}, M. and {Gabici}, S. and {Gajdus}, M. and
         {Gallant}, Y.~A. and {Garrigoux}, T. and {Giavitto}, G. and
         {Giebels}, B. and {Glicenstein}, J.~F. and {Gottschall}, D. and
         {Grondin}, M. -H. and {Grudzi{\'n}ska}, M. and {Hadasch}, D. and
         {H{\"a}ffner}, S. and {Hahn}, J. and {Harris}, J. and
         {Heinzelmann}, G. and {Henri}, G. and {Hermann}, G. and {Hervet}, O. and
         {Hillert}, A. and {Hinton}, J.~A. and {Hofmann}, W. and
         {Hofverberg}, P. and {Holler}, M. and {Horns}, D. and {Ivascenko}, A. and
         {Jacholkowska}, A. and {Jahn}, C. and {Jamrozy}, M. and {Janiak}, M. and
         {Jankowsky}, F. and {Jung-Richardt}, I. and {Kastendieck}, M.~A. and
         {Katarzy{\'n}ski}, K. and {Katz}, U. and {Kaufmann}, S. and
         {Kh{\'e}lifi}, B. and {Kieffer}, M. and {Klepser}, S. and
         {Klochkov}, D. and {Klu{\'z}niak}, W. and {Kolitzus}, D. and
         {Komin}, Nu. and {Kosack}, K. and {Krakau}, S. and {Krayzel}, F. and
         {Kr{\"u}ger}, P.~P. and {Laffon}, H. and {Lamanna}, G. and
         {Lefaucheur}, J. and {Lefranc}, V. and {Lemi{\`e}re}, A. and
         {Lemoine-Goumard}, M. and {Lenain}, J. -P. and {Lohse}, T. and
         {Lopatin}, A. and {Lu}, C. -C. and {Marandon}, V. and {Marcowith}, A. and
         {Marx}, R. and {Maurin}, G. and {Maxted}, N. and {Mayer}, M. and
         {McComb}, T.~J.~L. and {M{\'e}hault}, J. and {Meintjes}, P.~J. and
         {Menzler}, U. and {Meyer}, M. and {Mitchell}, A.~M.~W. and
         {Moderski}, R. and {Mohamed}, M. and {Mor{\r{a}}}, K. and {Moulin}, E. and
         {Murach}, T. and {de Naurois}, M. and {Niemiec}, J. and {Nolan}, S.~J. and
         {Oakes}, L. and {Odaka}, H. and {Ohm}, S. and {Opitz}, B. and
         {Ostrowski}, M. and {Oya}, I. and {Panter}, M. and {Parsons}, R.~D. and
         {Arribas}, M. Paz and {Pekeur}, N.~W. and {Pelletier}, G. and
         {Petrucci}, P. -O. and {Peyaud}, B. and {Pita}, S. and {Poon}, H. and
         {P{\"u}hlhofer}, G. and {Punch}, M. and {Quirrenbach}, A. and
         {Raab}, S. and {Reichardt}, I. and {Reimer}, A. and {Reimer}, O. and
         {Renaud}, M. and {de los Reyes}, R. and {Rieger}, F. and {Romoli}, C. and
         {Rosier-Lees}, S. and {Rowell}, G. and {Rudak}, B. and {Rulten}, C.~B. and
         {Sahakian}, V. and {Salek}, D. and {Sanchez}, D.~A. and
         {Santangelo}, A. and {Schlickeiser}, R. and {Sch{\"u}ssler}, F. and
         {Schulz}, A. and {Schwanke}, U. and {Schwarzburg}, S. and
         {Schwemmer}, S. and {Sol}, H. and {Spanier}, F. and {Spengler}, G. and
         {Spies}, F. and {Stawarz}, {\L}. and {Steenkamp}, R. and
         {Stegmann}, C. and {Stinzing}, F. and {Stycz}, K. and {Sushch}, I. and
         {Tavernet}, J. -P. and {Tavernier}, T. and {Taylor}, A.~M. and
         {Terrier}, R. and {Tluczykont}, M. and {Trichard}, C. and
         {Valerius}, K. and {van Eldik}, C. and {van Soelen}, B. and
         {Vasileiadis}, G. and {Veh}, J. and {Venter}, C. and {Viana}, A. and
         {Vincent}, P. and {Vink}, J. and {V{\"o}lk}, H.~J. and {Volpe}, F. and
         {Vorster}, M. and {Vuillaume}, T. and {Wagner}, S.~J. and {Wagner}, P. and
         {Wagner}, R.~M. and {Ward}, M. and {Weidinger}, M. and {Weitzel}, Q. and
         {White}, R. and {Wierzcholska}, A. and {Willmann}, P. and
         {W{\"o}rnlein}, A. and {Wouters}, D. and {Yang}, R. and {Zabalza}, V. and
         {Zaborov}, D. and {Zacharias}, M. and {Zdziarski}, A.~A. and
         {Zech}, A. and {Zechlin}, H. -S.},
        title = "{H.E.S.S. reveals a lack of TeV emission from the supernova remnant Puppis A}",
      journal = {Astronomy and Astrophysics},
         year = "2015",
        month = "Feb",
       volume = {575},
          eid = {A81},
        pages = {A81},
          doi = {10.1051/0004-6361/201424805},
archivePrefix = {arXiv},
       eprint = {1412.6997},
 primaryClass = {astro-ph.HE},
       adsurl = {https://ui.adsabs.harvard.edu/abs/2015A&A...575A..81H}
}

@article{puppfermi,
       author = {{Xin}, Yu-Liang and {Guo}, Xiao-Lei and {Liao}, Neng-Hui and
         {Yuan}, Qiang and {Liu}, Si-Ming and {Wei}, Da-Ming},
        title = "{Revisiting SNR Puppis A with Seven Years of Fermi Large Area Telescope Observations}",
      journal = {The Astrophysical Journal},
         year = "2017",
        month = "Jul",
       volume = {843},
       number = {2},
          eid = {90},
        pages = {90},
          doi = {10.3847/1538-4357/aa74bb},
archivePrefix = {arXiv},
       eprint = {1703.03911},
 primaryClass = {astro-ph.HE},
       adsurl = {https://ui.adsabs.harvard.edu/abs/2017ApJ...843...90X}
}

@article{443ver,
       author = {{Acciari}, V.~A. and {Aliu}, E. and {Arlen}, T. and {Aune}, T. and
         {Bautista}, M. and {Beilicke}, M. and {Benbow}, W. and
         {Bradbury}, S.~M. and {Buckley}, J.~H. and {Bugaev}, V. and {Butt}, Y. and
         {Byrum}, K. and {Cannon}, A. and {Celik}, O. and {Cesarini}, A. and
         {Chow}, Y.~C. and {Ciupik}, L. and {Cogan}, P. and {Colin}, P. and
         {Cui}, W. and {Daniel}, M.~K. and {Dickherber}, R. and {Duke}, C. and
         {Dwarkadas}, V.~V. and {Ergin}, T. and {Fegan}, S.~J. and
         {Finley}, J.~P. and {Finnegan}, G. and {Fortin}, P. and {Fortson}, L. and
         {Furniss}, A. and {Gall}, D. and {Gibbs}, K. and {Gillanders}, G.~H. and
         {Godambe}, S. and {Grube}, J. and {Guenette}, R. and {Gyuk}, G. and
         {Hanna}, D. and {Hays}, E. and {Holder}, J. and {Horan}, D. and
         {Hui}, C.~M. and {Humensky}, T.~B. and {Imran}, A. and {Kaaret}, P. and
         {Karlsson}, N. and {Kertzman}, M. and {Kieda}, D. and {Kildea}, J. and
         {Konopelko}, A. and {Krawczynski}, H. and {Krennrich}, F. and
         {Lang}, M.~J. and {LeBohec}, S. and {Maier}, G. and {McCann}, A. and
         {McCutcheon}, M. and {Millis}, J. and {Moriarty}, P. and {Ong}, R.~A. and
         {Otte}, A.~N. and {Pandel}, D. and {Perkins}, J.~S. and {Pohl}, M. and
         {Quinn}, J. and {Ragan}, K. and {Reyes}, L.~C. and {Reynolds}, P.~T. and
         {Roache}, E. and {Rose}, H.~J. and {Schroedter}, M. and
         {Sembroski}, G.~H. and {Smith}, A.~W. and {Steele}, D. and
         {Swordy}, S.~P. and {Theiling}, M. and {Toner}, J.~A. and
         {Valcarcel}, L. and {Varlotta}, A. and {Vassiliev}, V.~V. and
         {Vincent}, S. and {Wagner}, R.~G. and {Wakely}, S.~P. and
         {Ward}, J.~E. and {Weekes}, T.~C. and {Weinstein}, A. and
         {Weisgarber}, T. and {Williams}, D.~A. and {Wissel}, S. and {Wood}, M. and
         {Zitzer}, B.},
        title = "{Observation of Extended Very High Energy Emission from the Supernova Remnant IC 443 with VERITAS}",
      journal = {The Astrophysical Journal},
         year = "2009",
        month = "Jun",
       volume = {698},
       number = {2},
        pages = {L133-L137},
          doi = {10.1088/0004-637X/698/2/L133},
archivePrefix = {arXiv},
       eprint = {0905.3291},
 primaryClass = {astro-ph.HE},
       adsurl = {https://ui.adsabs.harvard.edu/abs/2009ApJ...698L.133A}
}

@article{443mag,
       author = {{Albert}, J. and {Aliu}, E. and {Anderhub}, H. and {Antoranz}, P. and
         {Armada}, A. and {Baixeras}, C. and {Barrio}, J.~A. and {Bartko}, H. and
         {Bastieri}, D. and {Becker}, J.~K. and {Bednarek}, W. and {Berger}, K. and
         {Bigongiari}, C. and {Biland}, A. and {Bock}, R.~K. and {Bordas}, P. and
         {Bosch-Ramon}, V. and {Bretz}, T. and {Britvitch}, I. and {Camara}, M. and
         {Carmona}, E. and {Chilingarian}, A. and {Coarasa}, J.~A. and
         {Commichau}, S. and {Contreras}, J.~L. and {Cortina}, J. and
         {Costado}, M.~T. and {Curtef}, V. and {Danielyan}, V. and {Dazzi}, F. and
         {De Angelis}, A. and {Delgado}, C. and {de los Reyes}, R. and
         {De Lotto}, B. and {Domingo-Santamar{\'\i}a}, E. and {Dorner}, D. and
         {Doro}, M. and {Errando}, M. and {Fagiolini}, M. and {Ferenc}, D. and
         {Fern{\'a}ndez}, E. and {Firpo}, R. and {Flix}, J. and
         {Fonseca}, M.~V. and {Font}, L. and {Fuchs}, M. and {Galante}, N. and
         {Garc{\'\i}a-L{\'o}pez}, R.~J. and {Garczarczyk}, M. and {Gaug}, M. and
         {Giller}, M. and {Goebel}, F. and {Hakobyan}, D. and {Hayashida}, M. and
         {Hengstebeck}, T. and {Herrero}, A. and {H{\"o}hne}, D. and {Hose}, J. and
         {Hsu}, C.~C. and {Jacon}, P. and {Jogler}, T. and {Kosyra}, R. and
         {Kranich}, D. and {Kritzer}, R. and {Laille}, A. and {Lindfors}, E. and
         {Lombardi}, S. and {Longo}, F. and {L{\'o}pez}, J. and {L{\'o}pez}, M. and
         {Lorenz}, E. and {Majumdar}, P. and {Maneva}, G. and {Mannheim}, K. and
         {Mansutti}, O. and {Mariotti}, M. and {Mart{\'\i}nez}, M. and
         {Mazin}, D. and {Merck}, C. and {Meucci}, M. and {Meyer}, M. and {Mirand
        a}, J.~M. and {Mirzoyan}, R. and {Mizobuchi}, S. and {Moralejo}, A. and
         {Nieto}, D. and {Nilsson}, K. and {Ninkovic}, J. and
         {O{\~n}a-Wilhelmi}, E. and {Otte}, N. and {Oya}, I. and {Paneque}, D. and
         {Panniello}, M. and {Paoletti}, R. and {Paredes}, J.~M. and
         {Pasanen}, M. and {Pascoli}, D. and {Pauss}, F. and {Pegna}, R. and
         {Persic}, M. and {Peruzzo}, L. and {Piccioli}, A. and {Prandini}, E. and
         {Puchades}, N. and {Raymers}, A. and {Rhode}, W. and {Rib{\'o}}, M. and
         {Rico}, J. and {Rissi}, M. and {Robert}, A. and {R{\"u}gamer}, S. and
         {Saggion}, A. and {Saito}, T. and {S{\'a}nchez}, A. and {Sartori}, P. and
         {Scalzotto}, V. and {Scapin}, V. and {Schmitt}, R. and {Schweizer}, T. and
         {Shayduk}, M. and {Shinozaki}, K. and {Shore}, S.~N. and {Sidro}, N. and
         {Sillanp{\"a}{\"a}}, A. and {Sobczynska}, D. and {Stamerra}, A. and
         {Stark}, L.~S. and {Takalo}, L. and {Temnikov}, P. and {Tescaro}, D. and
         {Teshima}, M. and {Torres}, D.~F. and {Turini}, N. and {Vankov}, H. and
         {Vitale}, V. and {Wagner}, R.~M. and {Wibig}, T. and {Wittek}, W. and
         {Zandanel}, F. and {Zanin}, R. and {Zapatero}, J.},
        title = "{Discovery of Very High Energy Gamma Radiation from IC 443 with the MAGIC Telescope}",
      journal = {The Astrophysical Journal},
         year = "2007",
        month = "Aug",
       volume = {664},
       number = {2},
        pages = {L87-L90},
          doi = {10.1086/520957},
archivePrefix = {arXiv},
       eprint = {0705.3119},
 primaryClass = {astro-ph},
       adsurl = {https://ui.adsabs.harvard.edu/abs/2007ApJ...664L..87A}
}

@article{443fer,
       author = {{Ackermann}, M. and {Ajello}, M. and {Allafort}, A. and {Baldini}, L. and
         {Ballet}, J. and {Barbiellini}, G. and {Baring}, M.~G. and
         {Bastieri}, D. and {Bechtol}, K. and {Bellazzini}, R. and {Bland
        ford}, R.~D. and {Bloom}, E.~D. and {Bonamente}, E. and {Borgland
        }, A.~W. and {Bottacini}, E. and {Brandt}, T.~J. and {Bregeon}, J. and
         {Brigida}, M. and {Bruel}, P. and {Buehler}, R. and {Busetto}, G. and
         {Buson}, S. and {Caliandro}, G.~A. and {Cameron}, R.~A. and
         {Caraveo}, P.~A. and {Casandjian}, J.~M. and {Cecchi}, C. and
         {{\c{C}}elik}, {\"O}. and {Charles}, E. and {Chaty}, S. and
         {Chaves}, R.~C.~G. and {Chekhtman}, A. and {Cheung}, C.~C. and
         {Chiang}, J. and {Chiaro}, G. and {Cillis}, A.~N. and {Ciprini}, S. and
         {Claus}, R. and {Cohen-Tanugi}, J. and {Cominsky}, L.~R. and
         {Conrad}, J. and {Corbel}, S. and {Cutini}, S. and {D'Ammando}, F. and
         {de Angelis}, A. and {de Palma}, F. and {Dermer}, C.~D. and
         {do Couto e Silva}, E. and {Drell}, P.~S. and {Drlica-Wagner}, A. and
         {Falletti}, L. and {Favuzzi}, C. and {Ferrara}, E.~C. and
         {Franckowiak}, A. and {Fukazawa}, Y. and {Funk}, S. and {Fusco}, P. and
         {Gargano}, F. and {Germani}, S. and {Giglietto}, N. and {Giommi}, P. and
         {Giordano}, F. and {Giroletti}, M. and {Glanzman}, T. and
         {Godfrey}, G. and {Grenier}, I.~A. and {Grondin}, M. -H. and
         {Grove}, J.~E. and {Guiriec}, S. and {Hadasch}, D. and {Hanabata}, Y. and
         {Harding}, A.~K. and {Hayashida}, M. and {Hayashi}, K. and {Hays}, E. and
         {Hewitt}, J.~W. and {Hill}, A.~B. and {Hughes}, R.~E. and
         {Jackson}, M.~S. and {Jogler}, T. and {J{\'o}hannesson}, G. and
         {Johnson}, A.~S. and {Kamae}, T. and {Kataoka}, J. and {Katsuta}, J. and
         {Kn{\"o}dlseder}, J. and {Kuss}, M. and {Lande}, J. and {Larsson}, S. and
         {Latronico}, L. and {Lemoine-Goumard}, M. and {Longo}, F. and
         {Loparco}, F. and {Lovellette}, M.~N. and {Lubrano}, P. and
         {Madejski}, G.~M. and {Massaro}, F. and {Mayer}, M. and
         {Mazziotta}, M.~N. and {McEnery}, J.~E. and {Mehault}, J. and
         {Michelson}, P.~F. and {Mignani}, R.~P. and {Mitthumsiri}, W. and
         {Mizuno}, T. and {Moiseev}, A.~A. and {Monzani}, M.~E. and
         {Morselli}, A. and {Moskalenko}, I.~V. and {Murgia}, S. and
         {Nakamori}, T. and {Nemmen}, R. and {Nuss}, E. and {Ohno}, M. and
         {Ohsugi}, T. and {Omodei}, N. and {Orienti}, M. and {Orlando}, E. and
         {Ormes}, J.~F. and {Paneque}, D. and {Perkins}, J.~S. and
         {Pesce-Rollins}, M. and {Piron}, F. and {Pivato}, G. and
         {Rain{\`o}}, S. and {Rando}, R. and {Razzano}, M. and {Razzaque}, S. and
         {Reimer}, A. and {Reimer}, O. and {Ritz}, S. and {Romoli}, C. and
         {S{\'a}nchez-Conde}, M. and {Schulz}, A. and {Sgr{\`o}}, C. and
         {Simeon}, P.~E. and {Siskind}, E.~J. and {Smith}, D.~A. and {Spand
        re}, G. and {Spinelli}, P. and {Stecker}, F.~W. and {Strong}, A.~W. and
         {Suson}, D.~J. and {Tajima}, H. and {Takahashi}, H. and
         {Takahashi}, T. and {Tanaka}, T. and {Thayer}, J.~G. and
         {Thayer}, J.~B. and {Thompson}, D.~J. and {Thorsett}, S.~E. and
         {Tibaldo}, L. and {Tibolla}, O. and {Tinivella}, M. and {Troja}, E. and
         {Uchiyama}, Y. and {Usher}, T.~L. and {Vandenbroucke}, J. and
         {Vasileiou}, V. and {Vianello}, G. and {Vitale}, V. and {Waite}, A.~P. and
         {Werner}, M. and {Winer}, B.~L. and {Wood}, K.~S. and {Wood}, M. and
         {Yamazaki}, R. and {Yang}, Z. and {Zimmer}, S.},
        title = "{Detection of the Characteristic Pion-Decay Signature in Supernova Remnants}",
      journal = {Science},
         year = "2013",
        month = "Feb",
       volume = {339},
       number = {6121},
        pages = {807-811},
          doi = {10.1126/science.1231160},
archivePrefix = {arXiv},
       eprint = {1302.3307},
 primaryClass = {astro-ph.HE},
       adsurl = {https://ui.adsabs.harvard.edu/abs/2013Sci...339..807A}
}

@article{443ag,
       author = {{Tavani}, M. and {Giuliani}, A. and {Chen}, A.~W. and {Argan}, A. and
         {Barbiellini}, G. and {Bulgarelli}, A. and {Caraveo}, P. and
         {Cattaneo}, P.~W. and {Cocco}, V. and {Contessi}, T. and {D'Ammand
        o}, F. and {Costa}, E. and {De Paris}, G. and {Del Monte}, E. and
         {Di Cocco}, G. and {Donnarumma}, I. and {Evangelista}, Y. and
         {Ferrari}, A. and {Feroci}, M. and {Fuschino}, F. and {Galli}, M. and
         {Gianotti}, F. and {Labanti}, C. and {Lapshov}, I. and
         {Lazzarotto}, F. and {Lipari}, P. and {Longo}, F. and {Marisaldi}, M. and
         {Mastropietro}, M. and {Mereghetti}, S. and {Morelli}, E. and
         {Moretti}, E. and {Morselli}, A. and {Pacciani}, L. and
         {Pellizzoni}, A. and {Perotti}, F. and {Piano}, G. and {Picozza}, P. and
         {Pilia}, M. and {Pucella}, G. and {Prest}, M. and {Rapisarda}, M. and
         {Rappoldi}, A. and {Scalise}, E. and {Rubini}, A. and {Sabatini}, S. and
         {Striani}, E. and {Soffitta}, P. and {Trifoglio}, M. and {Trois}, A. and
         {Vallazza}, E. and {Vercellone}, S. and {Vittorini}, V. and
         {Zambra}, A. and {Zanello}, D. and {Pittori}, C. and {Verrecchia}, F. and
         {Santolamazza}, P. and {Giommi}, P. and {Colafrancesco}, S. and
         {Antonelli}, L.~A. and {Salotti}, L.},
        title = "{Direct Evidence for Hadronic Cosmic-Ray Acceleration in the Supernova Remnant IC 443}",
      journal = {The Astrophysical Journal},
         year = "2010",
        month = "Feb",
       volume = {710},
       number = {2},
        pages = {L151-L155},
          doi = {10.1088/2041-8205/710/2/L151},
archivePrefix = {arXiv},
       eprint = {1001.5150},
 primaryClass = {astro-ph.HE},
       adsurl = {https://ui.adsabs.harvard.edu/abs/2010ApJ...710L.151T}
}

@article{ty,
       author = {{Archambault}, S. and {Archer}, A. and {Benbow}, W. and {Bird}, R. and
         {Bourbeau}, E. and {Buchovecky}, M. and {Buckley}, J.~H. and
         {Bugaev}, V. and {Cerruti}, M. and {Connolly}, M.~P. and {Cui}, W. and
         {Dwarkadas}, V.~V. and {Errando}, M. and {Falcone}, A. and {Feng}, Q. and
         {Finley}, J.~P. and {Fleischhack}, H. and {Fortson}, L. and
         {Furniss}, A. and {Griffin}, S. and {H{\"u}tten}, M. and {Hanna}, D. and
         {Holder}, J. and {Johnson}, C.~A. and {Kaaret}, P. and {Kar}, P. and
         {Kelley-Hoskins}, N. and {Kertzman}, M. and {Kieda}, D. and
         {Krause}, M. and {Kumar}, S. and {Lang}, M.~J. and {Maier}, G. and
         {McArthur}, S. and {McCann}, A. and {Moriarty}, P. and {Mukherjee}, R. and
         {Nieto}, D. and {O'Brien}, S. and {Ong}, R.~A. and {Otte}, A.~N. and
         {Park}, N. and {Pohl}, M. and {Popkow}, A. and {Pueschel}, E. and
         {Quinn}, J. and {Ragan}, K. and {Reynolds}, P.~T. and
         {Richards}, G.~T. and {Roache}, E. and {Sadeh}, I. and {Santander}, M. and
         {Sembroski}, G.~H. and {Shahinyan}, K. and {Slane}, P. and
         {Staszak}, D. and {Telezhinsky}, I. and {Trepanier}, S. and
         {Tyler}, J. and {Wakely}, S.~P. and {Weinstein}, A. and
         {Weisgarber}, T. and {Wilcox}, P. and {Wilhelm}, A. and
         {Williams}, D.~A. and {Zitzer}, B.},
        title = "{Gamma-Ray Observations of Tycho{\textquoteright}s Supernova Remnant with VERITAS and Fermi}",
      journal = {The Astrophysical Journal},
         year = "2017",
        month = "Feb",
       volume = {836},
       number = {1},
          eid = {23},
        pages = {23},
          doi = {10.3847/1538-4357/836/1/23},
archivePrefix = {arXiv},
       eprint = {1701.06740},
 primaryClass = {astro-ph.HE},
       adsurl = {https://ui.adsabs.harvard.edu/abs/2017ApJ...836...23A}
}

@inproceedings{casaver,
       author = {{Kumar}, S. and {VERITAS Collaboration}},
        title = "{A detailed study of gamma-ray emission from Cassiopeia A using VERITAS}",
    booktitle = {34th International Cosmic Ray Conference (ICRC2015)},
         year = "2015",
       series = {International Cosmic Ray Conference},
       volume = {34},
        month = "Jul",
          eid = {760},
        pages = {760},
archivePrefix = {arXiv},
       eprint = {1508.07453},
 primaryClass = {astro-ph.HE},
       adsurl = {https://ui.adsabs.harvard.edu/abs/2015ICRC...34..760K}
}

@article{casamag,
       author = {{Ahnen}, M.~L. and {Ansoldi}, S. and {Antonelli}, L.~A. and
         {Arcaro}, C. and {Babi{\'c}}, A. and {Banerjee}, B. and {Bangale}, P. and
         {Barres de Almeida}, U. and {Barrio}, J.~A. and
         {Becerra Gonz{\'a}lez}, J. and {Bednarek}, W. and {Bernardini}, E. and
         {Berti}, A. and {Bhattacharyya}, W. and {Biasuzzi}, B. and {Biland
        }, A. and {Blanch}, O. and {Bonnefoy}, S. and {Bonnoli}, G. and
         {Carosi}, R. and {Carosi}, A. and {Chatterjee}, A. and {Colak}, M. and
         {Colin}, P. and {Colombo}, E. and {Contreras}, J.~L. and {Cortina}, J. and
         {Covino}, S. and {Cumani}, P. and {Da Vela}, P. and {Dazzi}, F. and
         {De Angelis}, A. and {De Lotto}, B. and {de O{\~n}a Wilhelmi}, E. and
         {Di Pierro}, F. and {Doert}, M. and {Dom{\'\i}nguez}, A. and
         {Dominis Prester}, D. and {Dorner}, D. and {Doro}, M. and
         {Einecke}, S. and {Eisenacher Glawion}, D. and {Elsaesser}, D. and
         {Engelkemeier}, M. and {Fallah Ramazani}, V. and
         {Fern{\'a}ndez-Barral}, A. and {Fidalgo}, D. and {Fonseca}, M.~V. and
         {Font}, L. and {Fruck}, C. and {Galindo}, D. and
         {Garc{\'\i}a L{\'o}pez}, R.~J. and {Garczarczyk}, M. and {Gaug}, M. and
         {Giammaria}, P. and {Godinovi{\'c}}, N. and {Gora}, D. and
         {Guberman}, D. and {Hadasch}, D. and {Hahn}, A. and {Hassan}, T. and
         {Hayashida}, M. and {Herrera}, J. and {Hose}, J. and {Hrupec}, D. and
         {Inada}, T. and {Ishio}, K. and {Konno}, Y. and {Kubo}, H. and
         {Kushida}, J. and {Kuve{\v{z}}di{\'c}}, D. and {Lelas}, D. and
         {Lindfors}, E. and {Lombardi}, S. and {Longo}, F. and {L{\'o}pez}, M. and
         {Maggio}, C. and {Majumdar}, P. and {Makariev}, M. and {Maneva}, G. and
         {Manganaro}, M. and {Mannheim}, K. and {Maraschi}, L. and
         {Mariotti}, M. and {Mart{\'\i}nez}, M. and {Mazin}, D. and
         {Menzel}, U. and {Minev}, M. and {Mirzoyan}, R. and {Moralejo}, A. and
         {Moreno}, V. and {Moretti}, E. and {Neustroev}, V. and
         {Niedzwiecki}, A. and {Nievas Rosillo}, M. and {Nilsson}, K. and
         {Ninci}, D. and {Nishijima}, K. and {Noda}, K. and {Nogu{\'e}s}, L. and
         {Paiano}, S. and {Palacio}, J. and {Paneque}, D. and {Paoletti}, R. and
         {Paredes}, J.~M. and {Pedaletti}, G. and {Peresano}, M. and
         {Perri}, L. and {Persic}, M. and {Prada Moroni}, P.~G. and {Prand
        ini}, E. and {Puljak}, I. and {Garcia}, J.~R. and {Reichardt}, I. and
         {Rhode}, W. and {Rib{\'o}}, M. and {Rico}, J. and {Righi}, C. and
         {Saito}, T. and {Satalecka}, K. and {Schroeder}, S. and
         {Schweizer}, T. and {Shore}, S.~N. and {Sitarek}, J. and
         {{\v{S}}nidari{\'c}}, I. and {Sobczynska}, D. and {Stamerra}, A. and
         {Strzys}, M. and {Suri{\'c}}, T. and {Takalo}, L. and {Tavecchio}, F. and
         {Temnikov}, P. and {Terzi{\'c}}, T. and {Tescaro}, D. and
         {Teshima}, M. and {Torres-Alb{\`a}}, N. and {Treves}, A. and
         {Vanzo}, G. and {Vazquez Acosta}, M. and {Vovk}, I. and {Ward}, J.~E. and
         {Will}, M. and {Zari{\'c}}, D.},
        title = "{A cut-off in the TeV gamma-ray spectrum of the SNR Cassiopeia A}",
      journal = {Monthly Notices of the Royal Astronomical Society},
         year = "2017",
        month = "Dec",
       volume = {472},
       number = {3},
        pages = {2956-2962},
          doi = {10.1093/mnras/stx2079},
archivePrefix = {arXiv},
       eprint = {1707.01583},
 primaryClass = {astro-ph.HE},
       adsurl = {https://ui.adsabs.harvard.edu/abs/2017MNRAS.472.2956A}
}

@article{gcfermi,
       author = {{Fraija}, N. and {Araya}, M.},
        title = "{The Gigaelectronvolt Counterpart of VER J2019+407 in the Northern Shell of the Supernova Remnant G78.2+2.1 ({\ensuremath{\gamma}} Cygni)}",
      journal = {The Astrophysical Journal},
         year = "2016",
        month = "Jul",
       volume = {826},
       number = {1},
          eid = {31},
        pages = {31},
          doi = {10.3847/0004-637X/826/1/31},
archivePrefix = {arXiv},
       eprint = {1605.00571},
 primaryClass = {astro-ph.HE},
       adsurl = {https://ui.adsabs.harvard.edu/abs/2016ApJ...826...31F}
}

@article{gcver,
       author = {{Aliu}, E. and {Archambault}, S. and {Arlen}, T. and {Aune}, T. and
         {Beilicke}, M. and {Benbow}, W. and {Bird}, R. and {Bouvier}, A. and
         {Bradbury}, S.~M. and {Buckley}, J.~H. and {Bugaev}, V. and
         {Byrum}, K. and {Cannon}, A. and {Cesarini}, A. and {Ciupik}, L. and
         {Collins-Hughes}, E. and {Connolly}, M.~P. and {Cui}, W. and
         {Dickherber}, R. and {Duke}, C. and {Dumm}, J. and {Dwarkadas}, V.~V. and
         {Errando}, M. and {Falcone}, A. and {Federici}, S. and {Feng}, Q. and
         {Finley}, J.~P. and {Finnegan}, G. and {Fortson}, L. and {Furniss}, A. and
         {Galante}, N. and {Gall}, D. and {Gillanders}, G.~H. and {Godambe}, S. and
         {Gotthelf}, E.~V. and {Griffin}, S. and {Grube}, J. and {Gyuk}, G. and
         {Hanna}, D. and {Holder}, J. and {Huan}, H. and {Hughes}, G. and
         {Humensky}, T.~B. and {Kaaret}, P. and {Karlsson}, N. and
         {Kertzman}, M. and {Khassen}, Y. and {Kieda}, D. and {Krawczynski}, H. and
         {Krennrich}, F. and {Lang}, M.~J. and {Lee}, K. and {Madhavan}, A.~S. and
         {Maier}, G. and {Majumdar}, P. and {McArthur}, S. and {McCann}, A. and
         {Millis}, J. and {Moriarty}, P. and {Mukherjee}, R. and {Nelson}, T. and
         {O'Faol{\'a}in de Bhr{\'o}ithe}, A. and {Ong}, R.~A. and {Orr}, M. and
         {Otte}, A.~N. and {Pandel}, D. and {Park}, N. and {Perkins}, J.~S. and
         {Pohl}, M. and {Popkow}, A. and {Prokoph}, H. and {Quinn}, J. and
         {Ragan}, K. and {Reyes}, L.~C. and {Reynolds}, P.~T. and {Roache}, E. and
         {Rose}, H.~J. and {Ruppel}, J. and {Saxon}, D.~B. and {Schroedter}, M. and
         {Sembroski}, G.~H. and {{\c{S}}ent{\"u}rk}, G.~D. and {Skole}, C. and
         {Telezhinsky}, I. and {Te{\v{s}}i{\'c}}, G. and {Theiling}, M. and
         {Thibadeau}, S. and {Tsurusaki}, K. and {Tyler}, J. and {Varlotta}, A. and
         {Vassiliev}, V.~V. and {Vincent}, S. and {Wakely}, S.~P. and
         {Ward}, J.~E. and {Weekes}, T.~C. and {Weinstein}, A. and
         {Weisgarber}, T. and {Welsing}, R. and {Williams}, D.~A. and
         {Zitzer}, B.},
        title = "{Discovery of TeV Gamma-Ray Emission toward Supernova Remnant SNR G78.2+2.1}",
      journal = {The Astrophysical Journal},
         year = "2013",
        month = "Jun",
       volume = {770},
       number = {2},
          eid = {93},
        pages = {93},
          doi = {10.1088/0004-637X/770/2/93},
archivePrefix = {arXiv},
       eprint = {1305.6508},
 primaryClass = {astro-ph.HE},
       adsurl = {https://ui.adsabs.harvard.edu/abs/2013ApJ...770...93A}
}

@article{cl,
       author = {{Reichardt}, I. and {Terrier}, R. and {West}, J. and {Safi-Harb}, S. and
         {de O{\~n}a-Wilhelmi}, E. and {Rico}, J.},
        title = "{Fermi/LAT Study of the Cygnus Loop Supernova Remnant: Discovery of a Point-like Source and of Spectral Differences in its gamma-ray emission}",
      journal = {arXiv e-prints},
         year = "2015",
        month = "Feb",
          eid = {arXiv:1502.03053},
        pages = {arXiv:1502.03053},
archivePrefix = {arXiv},
       eprint = {1502.03053},
 primaryClass = {astro-ph.HE},
       adsurl = {https://ui.adsabs.harvard.edu/abs/2015arXiv150203053R}
}

@article{51m,
       author = {{Aleksi{\'c}}, J. and {Alvarez}, E.~A. and {Antonelli}, L.~A. and
         {Antoranz}, P. and {Asensio}, M. and {Backes}, M. and
         {Barres de Almeida}, U. and {Barrio}, J.~A. and {Bastieri}, D. and
         {Becerra Gonz{\'a}lez}, J. and {Bednarek}, W. and {Berger}, K. and
         {Bernardini}, E. and {Biland}, A. and {Blanch}, O. and {Bock}, R.~K. and
         {Boller}, A. and {Bonnoli}, G. and {Borla Tridon}, D. and {Bretz}, T. and
         {Ca{\~n}ellas}, A. and {Carmona}, E. and {Carosi}, A. and {Colin}, P. and
         {Colombo}, E. and {Contreras}, J.~L. and {Cortina}, J. and
         {Cossio}, L. and {Covino}, S. and {Da Vela}, P. and {Dazzi}, F. and
         {De Angelis}, A. and {De Caneva}, G. and {De Cea del Pozo}, E. and
         {De Lotto}, B. and {Delgado Mendez}, C. and {Diago Ortega}, A. and
         {Doert}, M. and {Dom{\'\i}nguez}, A. and {Dominis Prester}, D. and
         {Dorner}, D. and {Doro}, M. and {Eisenacher}, D. and {Elsaesser}, D. and
         {Ferenc}, D. and {Fonseca}, M.~V. and {Font}, L. and {Fruck}, C. and
         {Garc{\'\i}a L{\'o}pez}, R.~J. and {Garczarczyk}, M. and {Garrido}, D. and
         {Giavitto}, G. and {Godinovi{\'c}}, N. and
         {Gonz{\'a}lez Mu{\~n}oz}, A. and {Gozzini}, S.~R. and {Hadasch}, D. and
         {H{\"a}fner}, D. and {Herrero}, A. and {Hildebrand}, D. and {Hose}, J. and
         {Hrupec}, D. and {Huber}, B. and {Jankowski}, F. and {Jogler}, T. and
         {Kadenius}, V. and {Kellermann}, H. and {Klepser}, S. and
         {Kr{\"a}henb{\"u}hl}, T. and {Krause}, J. and {La Barbera}, A. and
         {Lelas}, D. and {Leonardo}, E. and {Lewandowska}, N. and
         {Lindfors}, E. and {Lombardi}, S. and {L{\'o}pez}, M. and
         {L{\'o}pez-Coto}, R. and {L{\'o}pez-Oramas}, A. and {Lorenz}, E. and
         {Makariev}, M. and {Maneva}, G. and {Mankuzhiyil}, N. and
         {Mannheim}, K. and {Maraschi}, L. and {Mariotti}, M. and
         {Mart{\'\i}nez}, M. and {Mazin}, D. and {Meucci}, M. and {Mirand
        a}, J.~M. and {Mirzoyan}, R. and {Mold{\'o}n}, J. and {Moralejo}, A. and
         {Munar-Adrover}, P. and {Niedzwiecki}, A. and {Nieto}, D. and
         {Nilsson}, K. and {Nowak}, N. and {Orito}, R. and {Paiano}, S. and
         {Paneque}, D. and {Paoletti}, R. and {Pardo}, S. and {Paredes}, J.~M. and
         {Partini}, S. and {Perez-Torres}, M.~A. and {Persic}, M. and
         {Pilia}, M. and {Pochon}, J. and {Prada}, F. and {Prada Moroni}, P.~G. and
         {Prandini}, E. and {Puerto Gimenez}, I. and {Puljak}, I. and
         {Reichardt}, I. and {Reinthal}, R. and {Rhode}, W. and {Rib{\'o}}, M. and
         {Rico}, J. and {R{\"u}gamer}, S. and {Saggion}, A. and {Saito}, K. and
         {Saito}, T.~Y. and {Salvati}, M. and {Satalecka}, K. and
         {Scalzotto}, V. and {Scapin}, V. and {Schultz}, C. and {Schweizer}, T. and
         {Shore}, S.~N. and {Sillanp{\"a}{\"a}}, A. and {Sitarek}, J. and
         {Snidaric}, I. and {Sobczynska}, D. and {Spanier}, F. and {Spiro}, S. and
         {Stamatescu}, V. and {Stamerra}, A. and {Steinke}, B. and {Storz}, J. and
         {Strah}, N. and {Sun}, S. and {Suri{\'c}}, T. and {Takalo}, L. and
         {Takami}, H. and {Tavecchio}, F. and {Temnikov}, P. and
         {Terzi{\'c}}, T. and {Tescaro}, D. and {Teshima}, M. and {Tibolla}, O. and
         {Torres}, D.~F. and {Treves}, A. and {Uellenbeck}, M. and {Vogler}, P. and
         {Wagner}, R.~M. and {Weitzel}, Q. and {Zabalza}, V. and {Zandanel}, F. and
         {Zanin}, R.},
        title = "{Morphological and spectral properties of the W51 region measured with the MAGIC telescopes}",
      journal = {Astronomy and Astrophysics},
         year = "2012",
        month = "May",
       volume = {541},
          eid = {A13},
        pages = {A13},
          doi = {10.1051/0004-6361/201218846},
archivePrefix = {arXiv},
       eprint = {1201.4074},
 primaryClass = {astro-ph.HE},
       adsurl = {https://ui.adsabs.harvard.edu/abs/2012A&A...541A..13A}
}

@article{51f,
       author = {{Jogler}, T. and {Funk}, S.},
        title = "{Revealing W51C as a Cosmic Ray Source Using Fermi-LAT Data}",
      journal = {The Astrophysical Journal},
         year = "2016",
        month = "Jan",
       volume = {816},
       number = {2},
          eid = {100},
        pages = {100},
          doi = {10.3847/0004-637X/816/2/100},
       adsurl = {https://ui.adsabs.harvard.edu/abs/2016ApJ...816..100J}
}

@article{49,
       author = {{H.~E.~S.~S. Collaboration} and {Abdalla}, H. and {Abramowski}, A. and
         {Aharonian}, F. and {Ait Benkhali}, F. and {Akhperjanian}, A.~G. and
         {Andersson}, T. and {Ang{\"u}ner}, E.~O. and {Arrieta}, M. and
         {Aubert}, P. and {Backes}, M. and {Balzer}, A. and {Barnard}, M. and
         {Becherini}, Y. and {Becker Tjus}, J. and {Berge}, D. and
         {Bernhard}, S. and {Bernl{\"o}hr}, K. and {Blackwell}, R. and
         {B{\"o}ttcher}, M. and {Boisson}, C. and {Bolmont}, J. and
         {Bordas}, P. and {Bregeon}, J. and {Brun}, F. and {Brun}, P. and
         {Bryan}, M. and {Bulik}, T. and {Capasso}, M. and {Carr}, J. and
         {Casanova}, S. and {Cerruti}, M. and {Chakraborty}, N. and
         {Chalme-Calvet}, R. and {Chaves}, R.~C.~G. and {Chen}, A. and
         {Chevalier}, J. and {Chr{\'e}tien}, M. and {Colafrancesco}, S. and
         {Cologna}, G. and {Condon}, B. and {Conrad}, J. and {Cui}, Y. and
         {Davids}, I.~D. and {Decock}, J. and {Degrange}, B. and {Deil}, C. and
         {Devin}, J. and {deWilt}, P. and {Dirson}, L. and
         {Djannati-Ata{\"\i}}, A. and {Domainko}, W. and {Donath}, A. and
         {Drury}, L.~O. 'C. and {Dubus}, G. and {Dutson}, K. and {Dyks}, J. and
         {Edwards}, T. and {Egberts}, K. and {Eger}, P. and {Ernenwein}, J. -P. and
         {Eschbach}, S. and {Farnier}, C. and {Fegan}, S. and {Fernand
        es}, M.~V. and {Fiasson}, A. and {Fontaine}, G. and {F{\"o}rster}, A. and
         {Funk}, S. and {F{\"u}{\ss}ling}, M. and {Gabici}, S. and {Gajdus}, M. and
         {Gallant}, Y.~A. and {Garrigoux}, T. and {Giavitto}, G. and
         {Giebels}, B. and {Glicenstein}, J.~F. and {Gottschall}, D. and
         {Goyal}, A. and {Grondin}, M. -H. and {Hadasch}, D. and {Hahn}, J. and
         {Haupt}, M. and {Hawkes}, J. and {Heinzelmann}, G. and {Henri}, G. and
         {Hermann}, G. and {Hervet}, O. and {Hinton}, J.~A. and {Hofmann}, W. and
         {Hoischen}, C. and {Holler}, M. and {Horns}, D. and {Ivascenko}, A. and
         {Jacholkowska}, A. and {Jamrozy}, M. and {Janiak}, M. and
         {Jankowsky}, D. and {Jankowsky}, F. and {Jingo}, M. and {Jogler}, T. and
         {Jouvin}, L. and {Jung-Richardt}, I. and {Kastendieck}, M.~A. and
         {Katarzy{\'n}ski}, K. and {Katz}, U. and {Kerszberg}, D. and
         {Kh{\'e}lifi}, B. and {Kieffer}, M. and {King}, J. and {Klepser}, S. and
         {Klochkov}, D. and {Klu{\'z}niak}, W. and {Kolitzus}, D. and
         {Komin}, Nu. and {Kosack}, K. and {Krakau}, S. and {Kraus}, M. and
         {Krayzel}, F. and {Kr{\"u}ger}, P.~P. and {Laffon}, H. and
         {Lamanna}, G. and {Lau}, J. and {Lees}, J. -P. and {Lefaucheur}, J. and
         {Lefranc}, V. and {Lemi{\`e}re}, A. and {Lemoine-Goumard}, M. and
         {Lenain}, J. -P. and {Leser}, E. and {Lohse}, T. and {Lorentz}, M. and
         {Liu}, R. and {L{\'o}pez-Coto}, R. and {Lypova}, I. and {Marandon}, V. and
         {Marcowith}, A. and {Mariaud}, C. and {Marx}, R. and {Maurin}, G. and
         {Maxted}, N. and {Mayer}, M. and {Meintjes}, P.~J. and {Meyer}, M. and
         {Mitchell}, A.~M.~W. and {Moderski}, R. and {Mohamed}, M. and
         {Mohrmann}, L. and {Mor{\r{a}}}, K. and {Moulin}, E. and {Murach}, T. and
         {de Naurois}, M. and {Niederwanger}, F. and {Niemiec}, J. and
         {Oakes}, L. and {O'Brien}, P. and {Odaka}, H. and {{\"O}ttl}, S. and
         {Ohm}, S. and {Ostrowski}, M. and {Oya}, I. and {Padovani}, M. and
         {Panter}, M. and {Parsons}, R.~D. and {Pekeur}, N.~W. and
         {Pelletier}, G. and {Perennes}, C. and {Petrucci}, P. -O. and
         {Peyaud}, B. and {Piel}, Q. and {Pita}, S. and {Poon}, H. and
         {Prokhorov}, D. and {Prokoph}, H. and {P{\"u}hlhofer}, G. and
         {Punch}, M. and {Quirrenbach}, A. and {Raab}, S. and {Reimer}, A. and
         {Reimer}, O. and {Renaud}, M. and {de los Reyes}, R. and {Rieger}, F. and
         {Romoli}, C. and {Rosier-Lees}, S. and {Rowell}, G. and {Rudak}, B. and
         {Rulten}, C.~B. and {Sahakian}, V. and {Salek}, D. and
         {Sanchez}, D.~A. and {Santangelo}, A. and {Sasaki}, M. and
         {Schlickeiser}, R. and {Sch{\"u}ssler}, F. and {Schulz}, A. and
         {Schwanke}, U. and {Schwemmer}, S. and {Settimo}, M. and
         {Seyffert}, A.~S. and {Shafi}, N. and {Shilon}, I. and {Simoni}, R. and
         {Sol}, H. and {Spanier}, F. and {Spengler}, G. and {Spies}, F. and
         {Stawarz}, {\L}. and {Steenkamp}, R. and {Stegmann}, C. and
         {Stinzing}, F. and {Stycz}, K. and {Sushch}, I. and {Tavernet}, J. -P. and
         {Tavernier}, T. and {Taylor}, A.~M. and {Terrier}, R. and
         {Tibaldo}, L. and {Tiziani}, D. and {Tluczykont}, M. and
         {Trichard}, C. and {Tuffs}, R. and {Uchiyama}, Y. and
         {van der Walt}, D.~J. and {van Eldik}, C. and {van Rensburg}, C. and
         {van Soelen}, B. and {Vasileiadis}, G. and {Veh}, J. and {Venter}, C. and
         {Viana}, A. and {Vincent}, P. and {Vink}, J. and {Voisin}, F. and
         {V{\"o}lk}, H.~J. and {Vuillaume}, T. and {Wadiasingh}, Z. and
         {Wagner}, S.~J. and {Wagner}, P. and {Wagner}, R.~M. and {White}, R. and
         {Wierzcholska}, A. and {Willmann}, P. and {W{\"o}rnlein}, A. and
         {Wouters}, D. and {Yang}, R. and {Zabalza}, V. and {Zaborov}, D. and
         {Zacharias}, M. and {Zdziarski}, A.~A. and {Zech}, A. and {Zefi}, F. and
         {Ziegler}, A. and {{\.Z}ywucka}, N. and {Fermi-LAT Collaboration} and
         {Katsuta}, J.},
        title = "{The supernova remnant W49B as seen with H.E.S.S. and Fermi-LAT}",
      journal = {Astronomy and Astrophysics},
         year = "2018",
        month = "Apr",
       volume = {612},
          eid = {A5},
        pages = {A5},
          doi = {10.1051/0004-6361/201527843},
archivePrefix = {arXiv},
       eprint = {1609.00600},
 primaryClass = {astro-ph.HE},
       adsurl = {https://ui.adsabs.harvard.edu/abs/2018A&A...612A...5H}
}

@article{41mag,
       author = {{Albert}, J. and {Aliu}, E. and {Anderhub}, H. and {Antoranz}, P. and
         {Armada}, A. and {Asensio}, M. and {Baixeras}, C. and {Barrio}, J.~A. and
         {Bartelt}, M. and {Bartko}, H. and {Bastieri}, D. and
         {Bavikadi}, S.~R. and {Bednarek}, W. and {Berger}, K. and
         {Bigongiari}, C. and {Biland}, A. and {Bisesi}, E. and {Bock}, R.~K. and
         {Bordas}, P. and {Bosch-Ramon}, V. and {Bretz}, T. and {Britvitch}, I. and
         {Camara}, M. and {Carmona}, E. and {Chilingarian}, A. and
         {Ciprini}, S. and {Coarasa}, J.~A. and {Commichau}, S. and
         {Contreras}, J.~L. and {Cortina}, J. and {Curtef}, V. and
         {Dame}, T.~M. and {Danielyan}, V. and {Dazzi}, F. and {De Angelis}, A. and
         {de los Reyes}, R. and {De Lotto}, B. and {Domingo-Santamar{\'\i}}, E. and
         {Dorner}, D. and {Doro}, M. and {Errando}, M. and {Fagiolini}, M. and
         {Ferenc}, D. and {Fern{\'a}ndez}, E. and {Firpo}, R. and {Flix}, J. and
         {Fonseca}, M.~V. and {Font}, L. and {Fuchs}, M. and {Galante}, N. and
         {Garczarczyk}, M. and {Gaug}, M. and {Giller}, M. and {Goebel}, F. and
         {Hakobyan}, D. and {Hayashida}, M. and {Hengstebeck}, T. and
         {H{\"o}hne}, D. and {Hose}, J. and {Hsu}, C.~C. and {Isar}, P.~G. and
         {Jacon}, P. and {Kalekin}, O. and {Kasyra}, R. and {Kranich}, D. and
         {Laatiaoui}, M. and {Laille}, A. and {Lenisa}, T. and {Liebing}, P. and
         {Lindfors}, E. and {Lombardi}, S. and {Longo}, F. and {L{\'o}pez}, J. and
         {L{\'o}pez}, M. and {Lorenz}, E. and {Lucarelli}, F. and
         {Majumdar}, P. and {Maneva}, G. and {Mannheim}, K. and {Mansutti}, O. and
         {Mariotti}, M. and {Mart{\'\i}nez}, M. and {Mase}, K. and {Mazin}, D. and
         {Merck}, C. and {Meucci}, M. and {Meyer}, M. and {Miranda}, J.~M. and
         {Mirzoyan}, R. and {Mizobuchi}, S. and {Moralejo}, A. and
         {Nilsson}, K. and {O{\~n}a-Wilhelmi}, E. and {Ordu{\~n}a}, R. and
         {Otte}, N. and {Oya}, I. and {Paneque}, D. and {Paoletti}, R. and
         {Paredes}, J.~M. and {Pasanen}, M. and {Pascoli}, D. and {Pauss}, F. and
         {Pavel}, N. and {Pegna}, R. and {Persic}, M. and {Peruzzo}, L. and
         {Piccioli}, A. and {Poller}, M. and {Prandini}, E. and {Raymers}, A. and
         {Rico}, J. and {Rhode}, W. and {Rib{\'o}}, M. and {Riegel}, B. and
         {Rissi}, M. and {Robert}, A. and {R{\"u}gamer}, S. and {Saggion}, A. and
         {S{\'a}nchez}, A. and {Sartori}, P. and {Scalzotto}, V. and
         {Scapin}, V. and {Schmitt}, R. and {Schweizer}, T. and {Shayduk}, M. and
         {Shinozaki}, K. and {Shore}, S.~N. and {Sidro}, N. and
         {Sillanp{\"a}{\"a}}, A. and {Sobczynska}, D. and {Stamerra}, A. and
         {Stark}, L.~S. and {Takalo}, L. and {Temnikov}, P. and {Tescaro}, D. and
         {Teshima}, M. and {Tonello}, N. and {Torres}, A. and {Torres}, D.~F. and
         {Turini}, N. and {Vankov}, H. and {Vitale}, V. and {Wagner}, R.~M. and
         {Wibig}, T. and {Wittek}, W. and {Zanin}, R. and {Zapatero}, J.},
        title = "{Observation of VHE Gamma Radiation from HESS J1834-087/W41 with the MAGIC Telescope}",
      journal = {The Astrophysical Journal},
         year = "2006",
        month = "May",
       volume = {643},
       number = {1},
        pages = {L53-L56},
          doi = {10.1086/504917},
archivePrefix = {arXiv},
       eprint = {astro-ph/0604197},
 primaryClass = {astro-ph},
       adsurl = {https://ui.adsabs.harvard.edu/abs/2006ApJ...643L..53A}
}

@article{41,
       author = {{H.~E.~S.~S. Collaboration} and {Abramowski}, A. and {Aharonian}, F. and
         {Ait Benkhali}, F. and {Akhperjanian}, A.~G. and {Ang{\"u}ner}, E. and
         {Anton}, G. and {Backes}, M. and {Balenderan}, S. and {Balzer}, A. and
         {Barnacka}, A. and {Becherini}, Y. and {Becker Tjus}, J. and
         {Bernl{\"o}hr}, K. and {Birsin}, E. and {Bissaldi}, E. and
         {Biteau}, J. and {B{\"o}ttcher}, M. and {Boisson}, C. and
         {Bolmont}, J. and {Bordas}, P. and {Brucker}, J. and {Brun}, F. and
         {Brun}, P. and {Bulik}, T. and {Carrigan}, S. and {Casanova}, S. and
         {Chadwick}, P.~M. and {Chalme-Calvet}, R. and {Chaves}, R.~C.~G. and
         {Cheesebrough}, A. and {Chr{\'e}tien}, M. and {Colafrancesco}, S. and
         {Cologna}, G. and {Conrad}, J. and {Couturier}, C. and {Cui}, Y. and
         {Dalton}, M. and {Daniel}, M.~K. and {Davids}, I.~D. and
         {Degrange}, B. and {Deil}, C. and {deWilt}, P. and {Dickinson}, H.~J. and
         {Djannati-Ata{\"\i}}, A. and {Domainko}, W. and {O'C. Drury}, L. and
         {Dubus}, G. and {Dutson}, K. and {Dyks}, J. and {Dyrda}, M. and
         {Edwards}, T. and {Egberts}, K. and {Eger}, P. and {Espigat}, P. and
         {Farnier}, C. and {Fegan}, S. and {Feinstein}, F. and {Fernand
        es}, M.~V. and {Fernandez}, D. and {Fiasson}, A. and {Fontaine}, G. and
         {F{\"o}rster}, A. and {F{\"u}{\ss}ling}, M. and {Gajdus}, M. and
         {Gallant}, Y.~A. and {Garrigoux}, T. and {Giavitto}, G. and
         {Giebels}, B. and {Glicenstein}, J.~F. and {Grondin}, M. -H. and
         {Grudzi{\'n}ska}, M. and {H{\"a}ffner}, S. and {Hahn}, J. and
         {Harris}, J. and {Heinzelmann}, G. and {Henri}, G. and {Hermann}, G. and
         {Hervet}, O. and {Hillert}, A. and {Hinton}, J.~A. and {Hofmann}, W. and
         {Hofverberg}, P. and {Holler}, M. and {Horns}, D. and
         {Jacholkowska}, A. and {Jahn}, C. and {Jamrozy}, M. and {Janiak}, M. and
         {Jankowsky}, F. and {Jung}, I. and {Kastendieck}, M.~A. and
         {Katarzy{\'n}ski}, K. and {Katz}, U. and {Kaufmann}, S. and
         {Kh{\'e}lifi}, B. and {Kieffer}, M. and {Klepser}, S. and
         {Klochkov}, D. and {Klu{\'z}niak}, W. and {Kneiske}, T. and
         {Kolitzus}, D. and {Komin}, Nu. and {Kosack}, K. and {Krakau}, S. and
         {Krayzel}, F. and {Kr{\"u}ger}, P.~P. and {Laffon}, H. and
         {Lamanna}, G. and {Lefaucheur}, J. and {Lemi{\`e}re}, A. and
         {Lemoine-Goumard}, M. and {Lenain}, J. -P. and {Lohse}, T. and
         {Lopatin}, A. and {Lu}, C. -C. and {Marandon}, V. and {Marcowith}, A. and
         {Marx}, R. and {Maurin}, G. and {Maxted}, N. and {Mayer}, M. and
         {McComb}, T.~J.~L. and {M{\'e}hault}, J. and {Meintjes}, P.~J. and
         {Menzler}, U. and {Meyer}, M. and {Moderski}, R. and {Mohamed}, M. and
         {Moulin}, E. and {Murach}, T. and {Naumann}, C.~L. and
         {de Naurois}, M. and {Niemiec}, J. and {Nolan}, S.~J. and {Oakes}, L. and
         {Odaka}, H. and {Ohm}, S. and {de O{\~n}a Wilhelmi}, E. and
         {Opitz}, B. and {Ostrowski}, M. and {Oya}, I. and {Panter}, M. and
         {Parsons}, R.~D. and {Paz Arribas}, M. and {Pekeur}, N.~W. and
         {Pelletier}, G. and {Perez}, J. and {Petrucci}, P. -O. and
         {Peyaud}, B. and {Pita}, S. and {Poon}, H. and {P{\"u}hlhofer}, G. and
         {Punch}, M. and {Quirrenbach}, A. and {Raab}, S. and {Raue}, M. and
         {Reichardt}, I. and {Reimer}, A. and {Reimer}, O. and {Renaud}, M. and
         {de los Reyes}, R. and {Rieger}, F. and {Rob}, L. and {Romoli}, C. and
         {Rosier-Lees}, S. and {Rowell}, G. and {Rudak}, B. and {Rulten}, C.~B. and
         {Sahakian}, V. and {Sanchez}, D.~A. and {Santangelo}, A. and
         {Schlickeiser}, R. and {Sch{\"u}ssler}, F. and {Schulz}, A. and
         {Schwanke}, U. and {Schwarzburg}, S. and {Schwemmer}, S. and {Sol}, H. and
         {Spengler}, G. and {Spies}, F. and {Stawarz}, {\L}. and
         {Steenkamp}, R. and {Stegmann}, C. and {Stinzing}, F. and {Stycz}, K. and
         {Sushch}, I. and {Tavernet}, J. -P. and {Tavernier}, T. and
         {Taylor}, A.~M. and {Terrier}, R. and {Tluczykont}, M. and
         {Trichard}, C. and {Valerius}, K. and {van Eldik}, C. and
         {van Soelen}, B. and {Vasileiadis}, G. and {Venter}, C. and
         {Viana}, A. and {Vincent}, P. and {V{\"o}lk}, H.~J. and {Volpe}, F. and
         {Vorster}, M. and {Vuillaume}, T. and {Wagner}, S.~J. and {Wagner}, P. and
         {Wagner}, R.~M. and {Ward}, M. and {Weidinger}, M. and {Weitzel}, Q. and
         {White}, R. and {Wierzcholska}, A. and {Willmann}, P. and
         {W{\"o}rnlein}, A. and {Wouters}, D. and {Yang}, R. and {Zabalza}, V. and
         {Zacharias}, M. and {Zdziarski}, A.~A. and {Zech}, A. and
         {Zechlin}, H. -S.},
        title = "{Probing the gamma-ray emission from HESS J1834-087 using H.E.S.S. and Fermi LAT observations}",
      journal = {Astronomy and Astrophysics},
         year = "2015",
        month = "Feb",
       volume = {574},
          eid = {A27},
        pages = {A27},
          doi = {10.1051/0004-6361/201322694},
archivePrefix = {arXiv},
       eprint = {1407.0862},
 primaryClass = {astro-ph.HE},
       adsurl = {https://ui.adsabs.harvard.edu/abs/2015A&A...574A..27H}
}

@article{28,
       author = {{Cui}, Yudong and {Yeung}, Paul K.~H. and {Tam}, P.~H. Thomas and
         {P{\"u}hlhofer}, Gerd},
        title = "{Leaked GeV CRs from a Broken Shell: Explaining 9 Years of Fermi-LAT Data of SNR W28}",
      journal = {The Astrophysical Journal},
         year = "2018",
        month = "Jun",
       volume = {860},
       number = {1},
          eid = {69},
        pages = {69},
          doi = {10.3847/1538-4357/aac37b},
archivePrefix = {arXiv},
       eprint = {1805.03372},
 primaryClass = {astro-ph.HE},
       adsurl = {https://ui.adsabs.harvard.edu/abs/2018ApJ...860...69C}
}

@article{28ag,
       author = {{Giuliani}, A. and {Tavani}, M. and {Bulgarelli}, A. and {Striani}, E. and
         {Sabatini}, S. and {Cardillo}, M. and {Fukui}, Y. and {Kawamura}, A. and
         {Ohama}, A. and {Furukawa}, N. and {Torii}, K. and {Sano}, H. and
         {Aharonian}, F.~A. and {Verrecchia}, F. and {Argan}, A. and
         {Barbiellini}, G. and {Caraveo}, P.~A. and {Cattaneo}, P.~W. and
         {Chen}, A.~W. and {Cocco}, V. and {Costa}, E. and {D'Ammando}, F. and
         {Del Monte}, E. and {de Paris}, G. and {Di Cocco}, G. and
         {Donnarumma}, I. and {Evangelista}, Y. and {Feroci}, M. and
         {Fiorini}, M. and {Froysland}, T. and {Fuschino}, F. and {Galli}, M. and
         {Gianotti}, F. and {Labanti}, C. and {Lapshov}, Y. and
         {Lazzarotto}, F. and {Lipari}, P. and {Longo}, F. and {Marisaldi}, M. and
         {Mereghetti}, S. and {Morselli}, A. and {Moretti}, E. and
         {Pacciani}, L. and {Pellizzoni}, A. and {Perotti}, F. and
         {Picozza}, P. and {Pilia}, M. and {Prest}, M. and {Pucella}, G. and
         {Rapisarda}, M. and {Rappoldi}, A. and {Soffitta}, P. and
         {Trifoglio}, M. and {Trois}, A. and {Vallazza}, E. and
         {Vercellone}, S. and {Vittorini}, V. and {Zambra}, A. and
         {Zanello}, D. and {Pittori}, C. and {Santolamazza}, P. and
         {Giommi}, P. and {Colafrancesco}, S. and {Salotti}, L.},
        title = "{AGILE detection of GeV {\ensuremath{\gamma}}-ray emission from the SNR W28}",
      journal = {Astronomy and Astrophysics},
         year = "2010",
        month = "Jun",
       volume = {516},
          eid = {L11},
        pages = {L11},
          doi = {10.1051/0004-6361/201014256},
archivePrefix = {arXiv},
       eprint = {1005.0784},
 primaryClass = {astro-ph.HE},
       adsurl = {https://ui.adsabs.harvard.edu/abs/2010A&A...516L..11G}
}

@article{349,
       author = {{H.~E.~S.~S. Collaboration} and {Abramowski}, A. and {Aharonian}, F. and
         {Ait Benkhali}, F. and {Akhperjanian}, A.~G. and {Ang{\"u}ner}, E.~O. and
         {Backes}, M. and {Balenderan}, S. and {Balzer}, A. and {Barnacka}, A. and
         {Becherini}, Y. and {Becker Tjus}, J. and {Berge}, D. and
         {Bernhard}, S. and {Bernl{\"o}hr}, K. and {Birsin}, E. and
         {Biteau}, J. and {B{\"o}ttcher}, M. and {Boisson}, C. and
         {Bolmont}, J. and {Bordas}, P. and {Bregeon}, J. and {Brun}, F. and
         {Brun}, P. and {Bryan}, M. and {Bulik}, T. and {Carrigan}, S. and
         {Casanova}, S. and {Chadwick}, P.~M. and {Chakraborty}, N. and
         {Chalme-Calvet}, R. and {Chaves}, R.~C.~G. and {Chr{\'e}tien}, M. and
         {Colafrancesco}, S. and {Cologna}, G. and {Conrad}, J. and
         {Couturier}, C. and {Cui}, Y. and {Davids}, I.~D. and {Degrange}, B. and
         {Deil}, C. and {deWilt}, P. and {Djannati-Ata{\"\i}}, A. and
         {Domainko}, W. and {Donath}, A. and {O'C. Drury}, L. and {Dubus}, G. and
         {Dutson}, K. and {Dyks}, J. and {Dyrda}, M. and {Edwards}, T. and
         {Egberts}, K. and {Eger}, P. and {Espigat}, P. and {Farnier}, C. and
         {Fegan}, S. and {Feinstein}, F. and {Fernandes}, M.~V. and {Fernand
        ez}, D. and {Fiasson}, A. and {Fontaine}, G. and {F{\"o}rster}, A. and
         {F{\"u}{\ss}ling}, M. and {Gabici}, S. and {Gajdus}, M. and
         {Gallant}, Y.~A. and {Garrigoux}, T. and {Giavitto}, G. and
         {Giebels}, B. and {Glicenstein}, J.~F. and {Gottschall}, D. and
         {Grondin}, M. -H. and {Grudzi{\'n}ska}, M. and {Hadasch}, D. and
         {H{\"a}ffner}, S. and {Hahn}, J. and {Harris}, J. and
         {Heinzelmann}, G. and {Henri}, G. and {Hermann}, G. and {Hervet}, O. and
         {Hillert}, A. and {Hinton}, J.~A. and {Hofmann}, W. and
         {Hofverberg}, P. and {Holler}, M. and {Horns}, D. and {Ivascenko}, A. and
         {Jacholkowska}, A. and {Jahn}, C. and {Jamrozy}, M. and {Janiak}, M. and
         {Jankowsky}, F. and {Jung-Richardt}, I. and {Kastendieck}, M.~A. and
         {Katarzy{\'n}ski}, K. and {Katz}, U. and {Kaufmann}, S. and
         {Kh{\'e}lifi}, B. and {Kieffer}, M. and {Klepser}, S. and
         {Klochkov}, D. and {Klu{\'z}niak}, W. and {Kolitzus}, D. and
         {Komin}, Nu. and {Kosack}, K. and {Krakau}, S. and {Krayzel}, F. and
         {Kr{\"u}ger}, P.~P. and {Laffon}, H. and {Lamanna}, G. and
         {Lefaucheur}, J. and {Lefranc}, V. and {Lemi{\`e}re}, A. and
         {Lemoine-Goumard}, M. and {Lenain}, J. -P. and {Lohse}, T. and
         {Lopatin}, A. and {Lu}, C. -C. and {Marandon}, V. and {Marcowith}, A. and
         {Marx}, R. and {Maurin}, G. and {Maxted}, N. and {Mayer}, M. and
         {McComb}, T.~J.~L. and {M{\'e}hault}, J. and {Meintjes}, P.~J. and
         {Menzler}, U. and {Meyer}, M. and {Mitchell}, A.~M.~W. and
         {Moderski}, R. and {Mohamed}, M. and {Mor{\r{a}}}, K. and {Moulin}, E. and
         {Murach}, T. and {de Naurois}, M. and {Niemiec}, J. and {Nolan}, S.~J. and
         {Oakes}, L. and {Odaka}, H. and {Ohm}, S. and {Opitz}, B. and
         {Ostrowski}, M. and {Oya}, I. and {Panter}, M. and {Parsons}, R.~D. and
         {Arribas}, M. Paz and {Pekeur}, N.~W. and {Pelletier}, G. and
         {Petrucci}, P. -O. and {Peyaud}, B. and {Pita}, S. and {Poon}, H. and
         {P{\"u}hlhofer}, G. and {Punch}, M. and {Quirrenbach}, A. and
         {Raab}, S. and {Reichardt}, I. and {Reimer}, A. and {Reimer}, O. and
         {Renaud}, M. and {de los Reyes}, R. and {Rieger}, F. and {Romoli}, C. and
         {Rosier-Lees}, S. and {Rowell}, G. and {Rudak}, B. and {Rulten}, C.~B. and
         {Sahakian}, V. and {Salek}, D. and {Sanchez}, D.~A. and
         {Santangelo}, A. and {Schlickeiser}, R. and {Sch{\"u}ssler}, F. and
         {Schulz}, A. and {Schwanke}, U. and {Schwarzburg}, S. and
         {Schwemmer}, S. and {Sol}, H. and {Spanier}, F. and {Spengler}, G. and
         {Spies}, F. and {Stawarz}, {\L}. and {Steenkamp}, R. and
         {Stegmann}, C. and {Stinzing}, F. and {Stycz}, K. and {Sushch}, I. and
         {Tavernet}, J. -P. and {Tavernier}, T. and {Taylor}, A.~M. and
         {Terrier}, R. and {Tluczykont}, M. and {Trichard}, C. and
         {Valerius}, K. and {van Eldik}, C. and {van Soelen}, B. and
         {Vasileiadis}, G. and {Veh}, J. and {Venter}, C. and {Viana}, A. and
         {Vincent}, P. and {Vink}, J. and {V{\"o}lk}, H.~J. and {Volpe}, F. and
         {Vorster}, M. and {Vuillaume}, T. and {Wagner}, S.~J. and {Wagner}, P. and
         {Wagner}, R.~M. and {Ward}, M. and {Weidinger}, M. and {Weitzel}, Q. and
         {White}, R. and {Wierzcholska}, A. and {Willmann}, P. and
         {W{\"o}rnlein}, A. and {Wouters}, D. and {Yang}, R. and {Zabalza}, V. and
         {Zaborov}, D. and {Zacharias}, M. and {Zdziarski}, A.~A. and
         {Zech}, A. and {Zechlin}, H. -S.},
        title = "{H.E.S.S. detection of TeV emission from the interaction region between the supernova remnant G349.7+0.2 and a molecular cloud}",
      journal = {Astronomy and Astrophysics},
         year = "2015",
        month = "Jan",
       volume = {574},
          eid = {A100},
        pages = {A100},
          doi = {10.1051/0004-6361/201425070},
archivePrefix = {arXiv},
       eprint = {1412.2251},
 primaryClass = {astro-ph.HE},
       adsurl = {https://ui.adsabs.harvard.edu/abs/2015A&A...574A.100H}
}

@inproceedings{j,
       author = {{Gottschall}, D. and {Capasso}, M. and {Deil}, C. and
         {Djannati-Atai}, A. and {Donath}, A. and {Eger}, P. and {Marandon}, V. and
         {Maxted}, N. and {P{\"u}hlhofer}, G. and {Renaud}, M. and {Sasaki}, M. and
         {Terrier}, R. and {Vink}, J. and {H.~E.~S.~S. Collaboration}},
        title = "{Discovery of new TeV supernova remnant shells in the Galactic plane with H.E.S.S.}",
    booktitle = {6th International Symposium on High Energy Gamma-Ray Astronomy},
         year = "2017",
       series = {American Institute of Physics Conference Series},
       volume = {1792},
        month = "Jan",
          eid = {040030},
        pages = {040030},
          doi = {10.1063/1.4968934},
archivePrefix = {arXiv},
       eprint = {1612.00261},
 primaryClass = {astro-ph.HE},
       adsurl = {https://ui.adsabs.harvard.edu/abs/2017AIPC.1792d0030G}
}

@article{1534,
       author = {{Araya}, Miguel},
        title = "{Detection of GeV Gamma-Rays from HESS J1534-571 and Multiwavelength Implications for the Origin of the Nonthermal Emission}",
      journal = {The Astrophysical Journal},
         year = "2017",
        month = "Jul",
       volume = {843},
       number = {1},
          eid = {12},
        pages = {12},
          doi = {10.3847/1538-4357/aa7261},
archivePrefix = {arXiv},
       eprint = {1705.03902},
 primaryClass = {astro-ph.HE},
       adsurl = {https://ui.adsabs.harvard.edu/abs/2017ApJ...843...12A}
}

@article{1739,
       author = {{Castro}, Daniel and {Slane}, Patrick and {Carlton}, Ashley and
         {Figueroa-Feliciano}, Enectali},
        title = "{Fermi-LAT Observations of Supernova Remnants Interacting with Molecular Clouds: W41, MSH 17-39, and G337.7-0.1}",
      journal = {The Astrophysical Journal},
         year = "2013",
        month = "Sep",
       volume = {774},
       number = {1},
          eid = {36},
        pages = {36},
          doi = {10.1088/0004-637X/774/1/36},
archivePrefix = {arXiv},
       eprint = {1305.3623},
 primaryClass = {astro-ph.HE},
       adsurl = {https://ui.adsabs.harvard.edu/abs/2013ApJ...774...36C}
}

@article{21,
       author = {{Pivato}, G. and {Hewitt}, J.~W. and {Tibaldo}, L. and {Acero}, F. and
         {Ballet}, J. and {Brandt}, T.~J. and {de Palma}, F. and {Giordano}, F. and
         {Janssen}, G.~H. and {J{\'o}hannesson}, G. and {Smith}, D.~A.},
        title = "{Fermi LAT and WMAP Observations of the Supernova Remnant HB 21}",
      journal = {The Astrophysical Journal},
         year = "2013",
        month = "Dec",
       volume = {779},
       number = {2},
          eid = {179},
        pages = {179},
          doi = {10.1088/0004-637X/779/2/179},
archivePrefix = {arXiv},
       eprint = {1311.0393},
 primaryClass = {astro-ph.HE},
       adsurl = {https://ui.adsabs.harvard.edu/abs/2013ApJ...779..179P}
}

@article{30f,
       author = {{Ajello}, M. and {Allafort}, A. and {Baldini}, L. and {Ballet}, J. and
         {Barbiellini}, G. and {Bastieri}, D. and {Bechtol}, K. and
         {Bellazzini}, R. and {Berenji}, B. and {Blandford}, R.~D. and
         {Bloom}, E.~D. and {Bonamente}, E. and {Borgland}, A.~W. and
         {Bregeon}, J. and {Brigida}, M. and {Bruel}, P. and {Buehler}, R. and
         {Buson}, S. and {Caliandro}, G.~A. and {Cameron}, R.~A. and
         {Caraveo}, P.~A. and {Casandjian}, J.~M. and {Cecchi}, C. and
         {Charles}, E. and {Chekhtman}, A. and {Ciprini}, S. and {Claus}, R. and
         {Cohen-Tanugi}, J. and {Cutini}, S. and {de Angelis}, A. and
         {de Palma}, F. and {Dermer}, C.~D. and {Silva}, E. do Couto e. and
         {Drell}, P.~S. and {Drlica-Wagner}, A. and {Dubois}, R. and
         {Favuzzi}, C. and {Fegan}, S.~J. and {Ferrara}, E.~C. and
         {Focke}, W.~B. and {Frailis}, M. and {Fukazawa}, Y. and {Fukui}, Y. and
         {Fusco}, P. and {Gargano}, F. and {Gasparrini}, D. and {Germani}, S. and
         {Giglietto}, N. and {Giommi}, P. and {Giordano}, F. and
         {Giroletti}, M. and {Glanzman}, T. and {Godfrey}, G. and
         {Grove}, J.~E. and {Guiriec}, S. and {Hadasch}, D. and {Hanabata}, Y. and
         {Harding}, A.~K. and {Hayashi}, K. and {Hays}, E. and {Itoh}, R. and
         {J{\'o}hannesson}, G. and {Johnson}, A.~S. and {Kamae}, T. and
         {Katagiri}, H. and {Kataoka}, J. and {Kn{\"o}dlseder}, J. and
         {Kubo}, H. and {Kuss}, M. and {Lande}, J. and {Latronico}, L. and
         {Lee}, S. -H. and {Lionetto}, A.~M. and {Longo}, F. and {Loparco}, F. and
         {Lovellette}, M.~N. and {Lubrano}, P. and {Mazziotta}, M.~N. and
         {Mehault}, J. and {Michelson}, P.~F. and {Mizuno}, T. and
         {Moiseev}, A.~A. and {Monte}, C. and {Monzani}, M.~E. and
         {Morselli}, A. and {Moskalenko}, I.~V. and {Murgia}, S. and
         {Nakamori}, T. and {Naumann-Godo}, M. and {Nishino}, S. and
         {Nolan}, P.~L. and {Norris}, J.~P. and {Nuss}, E. and {Ohno}, M. and
         {Ohsugi}, T. and {Okumura}, A. and {Omodei}, N. and {Orlando}, E. and
         {Ormes}, J.~F. and {Paneque}, D. and {Parent}, D. and {Pelassa}, V. and
         {Pesce-Rollins}, M. and {Pierbattista}, M. and {Piron}, F. and
         {Porter}, T.~A. and {Rain{\`o}}, S. and {Rando}, R. and {Reimer}, A. and
         {Reimer}, O. and {Reposeur}, T. and {Roth}, M. and
         {Sadrozinski}, H.~F. -W. and {Sgr{\`o}}, C. and {Siskind}, E.~J. and
         {Smith}, P.~D. and {Spandre}, G. and {Spinelli}, P. and {Suson}, D.~J. and
         {Tajima}, H. and {Takahashi}, H. and {Tanaka}, T. and {Thayer}, J.~G. and
         {Thayer}, J.~B. and {Tibaldo}, L. and {Tibolla}, O. and
         {Torres}, D.~F. and {Tosti}, G. and {Tramacere}, A. and {Troja}, E. and
         {Uchiyama}, Y. and {Uehara}, T. and {Usher}, T.~L. and {Vand
        enbroucke}, J. and {Van Etten}, A. and {Vasileiou}, V. and
         {Vianello}, G. and {Vilchez}, N. and {Vitale}, V. and {Waite}, A.~P. and
         {Wang}, P. and {Winer}, B.~L. and {Wood}, K.~S. and {Yamamoto}, H. and
         {Yamazaki}, R. and {Yang}, Z. and {Yasuda}, H. and {Ziegler}, M. and
         {Zimmer}, S.},
        title = "{Fermi Large Area Telescope Observations of the Supernova Remnant G8.7-0.1}",
      journal = {The Astrophysical Journal},
         year = "2012",
        month = "Jan",
       volume = {744},
       number = {1},
          eid = {80},
        pages = {80},
          doi = {10.1088/0004-637X/744/1/80},
archivePrefix = {arXiv},
       eprint = {1109.3017},
 primaryClass = {astro-ph.HE},
       adsurl = {https://ui.adsabs.harvard.edu/abs/2012ApJ...744...80A}
}

@article{30h,
       author = {{Aharonian}, F. and {Akhperjanian}, A.~G. and {Bazer-Bachi}, A.~R. and
         {Beilicke}, M. and {Benbow}, W. and {Berge}, D. and {Bernl{\"o}hr}, K. and
         {Boisson}, C. and {Bolz}, O. and {Borrel}, V. and {Braun}, I. and
         {Breitling}, F. and {Brown}, A.~M. and {Chadwick}, P.~M. and
         {Chounet}, L. -M. and {Cornils}, R. and {Costamante}, L. and
         {Degrange}, B. and {Dickinson}, H.~J. and {Djannati-Ata{\"\i}}, A. and
         {Drury}, L. O'C. and {Dubus}, G. and {Emmanoulopoulos}, D. and
         {Espigat}, P. and {Feinstein}, F. and {Fontaine}, G. and {Fuchs}, Y. and
         {Funk}, S. and {Gallant}, Y.~A. and {Giebels}, B. and {Gillessen}, S. and
         {Glicenstein}, J.~F. and {Goret}, P. and {Hadjichristidis}, C. and
         {Hauser}, M. and {Heinzelmann}, G. and {Henri}, G. and {Hermann}, G. and
         {Hinton}, J.~A. and {Hofmann}, W. and {Holleran}, M. and {Horns}, D. and
         {Jacholkowska}, A. and {de Jager}, O.~C. and {Kh{\'e}lifi}, B. and
         {Komin}, Nu. and {Konopelko}, A. and {Latham}, I.~J. and
         {Le Gallou}, R. and {Lemi{\`e}re}, A. and {Lemoine-Goumard}, M. and
         {Leroy}, N. and {Lohse}, T. and {Martin}, J.~M. and
         {Martineau-Huynh}, O. and {Marcowith}, A. and {Masterson}, C. and
         {McComb}, T.~J.~L. and {de Naurois}, M. and {Nolan}, S.~J. and
         {Noutsos}, A. and {Orford}, K.~J. and {Osborne}, J.~L. and
         {Ouchrif}, M. and {Panter}, M. and {Pelletier}, G. and {Pita}, S. and
         {P{\"u}hlhofer}, G. and {Punch}, M. and {Raubenheimer}, B.~C. and
         {Raue}, M. and {Raux}, J. and {Rayner}, S.~M. and {Reimer}, A. and
         {Reimer}, O. and {Ripken}, J. and {Rob}, L. and {Rolland}, L. and
         {Rowell}, G. and {Sahakian}, V. and {Saug{\'e}}, L. and
         {Schlenker}, S. and {Schlickeiser}, R. and {Schuster}, C. and
         {Schwanke}, U. and {Siewert}, M. and {Sol}, H. and {Spangler}, D. and
         {Steenkamp}, R. and {Stegmann}, C. and {Tavernet}, J. -P. and
         {Terrier}, R. and {Th{\'e}oret}, C.~G. and {Tluczykont}, M. and
         {Vasileiadis}, G. and {Venter}, C. and {Vincent}, P. and
         {V{\"o}lk}, H.~J. and {Wagner}, S.~J.},
        title = "{The H.E.S.S. Survey of the Inner Galaxy in Very High Energy Gamma Rays}",
      journal = {The Astrophysical Journal},
         year = "2006",
        month = "Jan",
       volume = {636},
       number = {2},
        pages = {777-797},
          doi = {10.1086/498013},
archivePrefix = {arXiv},
       eprint = {astro-ph/0510397},
 primaryClass = {astro-ph},
       adsurl = {https://ui.adsabs.harvard.edu/abs/2006ApJ...636..777A}
}

@inproceedings{391f,
       author = {{Castro}, Daniel and {Slane}, Patrick},
        title = "{Fermi LAT Observations of Supernova Remnants Interacting with Molecular Clouds}",
    booktitle = {38th COSPAR Scientific Assembly},
         year = "2010",
       volume = {38},
        month = "Jan",
        pages = {3},
       adsurl = {https://ui.adsabs.harvard.edu/abs/2010cosp...38.2754C}
}

@article{109,
       author = {{Castro}, Daniel and {Slane}, Patrick and {Ellison}, Donald C. and
         {Patnaude}, Daniel J.},
        title = "{Fermi-LAT Observations and a Broadband Study of Supernova Remnant CTB 109}",
      journal = {The Astrophysical Journal},
         year = "2012",
        month = "Sep",
       volume = {756},
       number = {1},
          eid = {88},
        pages = {88},
          doi = {10.1088/0004-637X/756/1/88},
archivePrefix = {arXiv},
       eprint = {1207.1432},
 primaryClass = {astro-ph.HE},
       adsurl = {https://ui.adsabs.harvard.edu/abs/2012ApJ...756...88C}
}

@article{147,
       author = {{Katsuta}, J. and {Uchiyama}, Y. and {Tanaka}, T. and {Tajima}, H. and
         {Bechtol}, K. and {Funk}, S. and {Lande}, J. and {Ballet}, J. and
         {Hanabata}, Y. and {Lemoine-Goumard}, M. and {Takahashi}, T.},
        title = "{Fermi Large Area Telescope Observation of Supernova Remnant S147}",
      journal = {The Astrophysical Journal},
         year = "2012",
        month = "Jun",
       volume = {752},
       number = {2},
          eid = {135},
        pages = {135},
          doi = {10.1088/0004-637X/752/2/135},
archivePrefix = {arXiv},
       eprint = {1204.4703},
 primaryClass = {astro-ph.HE},
       adsurl = {https://ui.adsabs.harvard.edu/abs/2012ApJ...752..135K}
}

@article{k17,
       author = {{Gelfand}, Joseph D. and {Castro}, Daniel and {Slane}, Patrick O. and
         {Temim}, Tea and {Hughes}, John P. and {Rakowski}, Cara},
        title = "{Supernova Remnant Kes 17: An Efficient Cosmic Ray Accelerator inside a Molecular Cloud}",
      journal = {The Astrophysical Journal},
         year = "2013",
        month = "Nov",
       volume = {777},
       number = {2},
          eid = {148},
        pages = {148},
          doi = {10.1088/0004-637X/777/2/148},
archivePrefix = {arXiv},
       eprint = {1311.6894},
 primaryClass = {astro-ph.GA},
       adsurl = {https://ui.adsabs.harvard.edu/abs/2013ApJ...777..148G}
}

@article{ev,
   author = {{Woltjer}, L.},
    title = "{Supernova Remnants}",
  journal = {\araa},
     year = 1972,
   volume = 10,
    pages = {129},
      doi = {10.1146/annurev.aa.10.090172.001021},
   adsurl = {https://ui.adsabs.harvard.edu/abs/1972ARA%26A..10..129W}
}

@book{book,
   author = {{Padmanabhan}, T.},
    title = "{Theoretical Astrophysics - Volume 2, Stars and Stellar Systems}",
booktitle = {Theoretical Astrophysics - Volume 2, Stars and Stellar Systems, by T.~Padmanabhan, pp.~594.~Cambridge University Press, July 2001.~ISBN-10: 0521562414.~ISBN-13: 9780521562416.~LCCN: QB801 .P23 2001},
     year = 2001,
    month = jul,
    pages = {594},
      doi = {10.2277/0521562414},
   adsurl = {https://ui.adsabs.harvard.edu/abs/2001thas.book.....P}
}

@book{long,
   author = {{Longair}, M.~S.},
    title = "{High Energy Astrophysics}",
booktitle = {High Energy Astrophysics, by Malcolm S.~Longair, Cambridge, UK: Cambridge University Press, 2011},
     year = 2011,
    month = feb,
   adsurl = {https://ui.adsabs.harvard.edu/abs/2011hea..book.....L}
}

@article{acc77,
   author = {{Krymskii}, G.~F.},
    title = "{A regular mechanism for the acceleration of charged particles on the front of a shock wave}",
  journal = {Akademiia Nauk SSSR Doklady},
     year = 1977,
    month = jun,
   volume = 234,
    pages = {1306-1308},
   adsurl = {https://ui.adsabs.harvard.edu/abs/1977DoSSR.234.1306K}
}

@inproceedings{ax,
   author = {{Axford}, W.~I. and {Leer}, E. and {Skadron}, G.},
    title = "{The acceleration of cosmic rays by shock waves}",
booktitle = {International Cosmic Ray Conference},
     year = 1977,
   series = {International Cosmic Ray Conference},
   volume = 11,
    pages = {132-137},
   adsurl = {https://ui.adsabs.harvard.edu/abs/1977ICRC...11..132A}
}

@article{bell1,
   author = {{Bell}, A.~R.},
    title = "{The acceleration of cosmic rays in shock fronts. I}",
  journal = {\mnras},
     year = 1978,
    month = jan,
   volume = 182,
    pages = {147-156},
      doi = {10.1093/mnras/182.2.147},
   adsurl = {https://ui.adsabs.harvard.edu/abs/1978MNRAS.182..147B}
}

@ARTICLE{bell2,
   author = {{Bell}, A.~R.},
    title = "{The acceleration of cosmic rays in shock fronts. II}",
  journal = {\mnras},
     year = 1978,
    month = feb,
   volume = 182,
    pages = {443-455},
      doi = {10.1093/mnras/182.3.443},
   adsurl = {https://ui.adsabs.harvard.edu/abs/1978MNRAS.182..443B}
}

@ARTICLE{bo,
   author = {{Blandford}, R.~D. and {Ostriker}, J.~P.},
    title = "{Particle acceleration by astrophysical shocks}",
  journal = {\apjl},
     year = 1978,
    month = apr,
   volume = 221,
    pages = {L29-L32},
      doi = {10.1086/182658},
   adsurl = {https://ui.adsabs.harvard.edu/abs/1978ApJ...221L..29B}
}

@ARTICLE{hh,
   author = {{Hinton}, J.~A. and {Hofmann}, W.},
    title = "{Teraelectronvolt Astronomy}",
  journal = {\araa},
archivePrefix = "arXiv",
   eprint = {1006.5210},
 primaryClass = "astro-ph.HE",
     year = 2009,
    month = sep,
   volume = 47,
    pages = {523-565},
      doi = {10.1146/annurev-astro-082708-101816},
   adsurl = {https://ui.adsabs.harvard.edu/abs/2009ARA%26A..47..523H}
}

@ARTICLE{kelner,
       author = {{Kelner}, S.~R. and {Aharonian}, F.~A. and {Bugayov}, V.~V.},
        title = "{Energy spectra of gamma rays, electrons, and neutrinos produced at proton-proton interactions in the very high energy regime}",
      journal = {\prd},
         year = "2006",
        month = "Aug",
       volume = {74},
       number = {3},
          eid = {034018},
        pages = {034018},
          doi = {10.1103/PhysRevD.74.034018},
archivePrefix = {arXiv},
       eprint = {astro-ph/0606058},
 primaryClass = {astro-ph},
       adsurl = {https://ui.adsabs.harvard.edu/abs/2006PhRvD..74c4018K}
}

@ARTICLE{brems,
   author = {{Blumenthal}, G.~R. and {Gould}, R.~J.},
    title = "{Bremsstrahlung, Synchrotron Radiation, and Compton Scattering of High-Energy Electrons Traversing Dilute Gases}",
  journal = {Reviews of Modern Physics},
     year = 1970,
   volume = 42,
    pages = {237-271},
      doi = {10.1103/RevModPhys.42.237},
   adsurl = {https://ui.adsabs.harvard.edu/abs/1970RvMP...42..237B}
}

@article{rey,
author = {Reynolds, Stephen P.},
title = {Supernova Remnants at High Energy},
journal = {Annual Review of Astronomy and Astrophysics},
volume = {46},
number = {1},
pages = {89-126},
year = {2008},
doi = {10.1146/annurev.astro.46.060407.145237},
URL = { 
        https://doi.org/10.1146/annurev.astro.46.060407.145237},
eprint = { 
        https://doi.org/10.1146/annurev.astro.46.060407.145237}
 }

@ARTICLE{kawa,
   author = {{Fukui}, Y. and {Kawamura}, A.},
    title = "{Molecular Clouds in Nearby Galaxies}",
  journal = {\araa},
     year = 2010,
    month = sep,
   volume = 48,
    pages = {547-580},
      doi = {10.1146/annurev-astro-081309-130854},
   adsurl = {https://ui.adsabs.harvard.edu/abs/2010ARA%26A..48..547F}
}

@ARTICLE{bolatto,
       author = {{Bolatto}, Alberto D. and {Wolfire}, Mark and {Leroy}, Adam K.},
        title = "{The CO-to-H$_{2}$ Conversion Factor}",
      journal = {\araa},
         year = "2013",
        month = "Aug",
       volume = {51},
       number = {1},
        pages = {207-268},
          doi = {10.1146/annurev-astro-082812-140944},
archivePrefix = {arXiv},
       eprint = {1301.3498},
 primaryClass = {astro-ph.GA},
       adsurl = {https://ui.adsabs.harvard.edu/abs/2013ARA&A..51..207B}
}

@ARTICLE{wilson,
   author = {{Wilson}, R.~W. and {Jefferts}, K.~B. and {Penzias}, A.~A.},
    title = "{Carbon Monoxide in the Orion Nebula}",
  journal = {\apjl},
     year = 1970,
    month = jul,
   volume = 161,
    pages = {L43},
      doi = {10.1086/180567},
   adsurl = {https://ui.adsabs.harvard.edu/abs/1970ApJ...161L..43W}
}

@ARTICLE{comb,
   author = {{Combes}, F. and {Encrenaz}, P.~J. and {Lucas}, R. and {Weliachew}, L.
	},
    title = "{Observation of the CO Molecule in the Spiral Arms of External Galaxies}",
  journal = {\aap},
     year = 1977,
    month = mar,
   volume = 55,
    pages = {311},
   adsurl = {https://ui.adsabs.harvard.edu/abs/1977A%26A....55..311C}
}

@ARTICLE{dame,
   author = {{Dame}, T.~M. and {Ungerechts}, H. and {Cohen}, R.~S. and {de Geus}, E.~J. and 
	{Grenier}, I.~A. and {May}, J. and {Murphy}, D.~C. and {Nyman}, L.-A. and 
	{Thaddeus}, P.},
    title = "{A composite CO survey of the entire Milky Way}",
  journal = {\apj},
     year = 1987,
    month = nov,
   volume = 322,
    pages = {706-720},
      doi = {10.1086/165766},
   adsurl = {https://ui.adsabs.harvard.edu/abs/1987ApJ...322..706D}
}

@ARTICLE{tadd,
   author = {{Blitz}, L. and {Thaddeus}, P.},
    title = "{Giant molecular complexes and OB associations. I - The Rosette molecular complex}",
  journal = {\apj},
     year = 1980,
    month = oct,
   volume = 241,
    pages = {676-696},
      doi = {10.1086/158379},
   adsurl = {https://ui.adsabs.harvard.edu/abs/1980ApJ...241..676B}
}

@ARTICLE{jack,
       author = {{Jackson}, J.~M. and {Rathborne}, J.~M. and {Shah}, R.~Y. and
         {Simon}, R. and {Bania}, T.~M. and {Clemens}, D.~P. and
         {Chambers}, E.~T. and {Johnson}, A.~M. and {Dormody}, M. and
         {Lavoie}, R. and {Heyer}, M.~H.},
        title = "{The Boston University-Five College Radio Astronomy Observatory Galactic Ring Survey}",
      journal = {\apjs},
         year = "2006",
        month = "Mar",
       volume = {163},
       number = {1},
        pages = {145-159},
          doi = {10.1086/500091},
archivePrefix = {arXiv},
       eprint = {astro-ph/0602160},
 primaryClass = {astro-ph},
       adsurl = {https://ui.adsabs.harvard.edu/abs/2006ApJS..163..145J}
}

@ARTICLE{mf,
       author = {{Crutcher}, Richard M.},
        title = "{Magnetic Fields in Molecular Clouds}",
      journal = {\araa},
         year = "2012",
        month = "Sep",
       volume = {50},
        pages = {29-63},
          doi = {10.1146/annurev-astro-081811-125514},
       adsurl = {https://ui.adsabs.harvard.edu/abs/2012ARA&A..50...29C}
}

@ARTICLE{brand,
   author = {{Brand}, J. and {Blitz}, L.},
    title = "{The Velocity Field of the Outer Galaxy}",
  journal = {\aap},
     year = 1993,
    month = aug,
   volume = 275,
    pages = {67},
   adsurl = {https://ui.adsabs.harvard.edu/abs/1993A%26A...275...67B}
}

@ARTICLE{rom,
       author = {{Roman-Duval}, Julia and {Jackson}, James M. and {Heyer}, Mark and
         {Johnson}, Alexis and {Rathborne}, Jill and {Shah}, Ronak and
         {Simon}, Robert},
        title = "{Kinematic Distances to Molecular Clouds Identified in the Galactic Ring Survey}",
      journal = {\apj},
         year = "2009",
        month = "Jul",
       volume = {699},
       number = {2},
        pages = {1153-1170},
          doi = {10.1088/0004-637X/699/2/1153},
archivePrefix = {arXiv},
       eprint = {0905.0723},
 primaryClass = {astro-ph.GA},
       adsurl = {https://ui.adsabs.harvard.edu/abs/2009ApJ...699.1153R}
}

@ARTICLE{ptu,
       author = {{Ptuskin}, Vladimir},
        title = "{Propagation of galactic cosmic rays}",
      journal = {Astroparticle Physics},
         year = "2012",
        month = "Dec",
       volume = {39},
        pages = {44-51},
          doi = {10.1016/j.astropartphys.2011.11.004},
       adsurl = {https://ui.adsabs.harvard.edu/abs/2012APh....39...44P}
}

@ARTICLE{a,
       author = {{Aharonian}, F.~A.},
        title = "{Vary High and Ultra High Energy Gamma-Rays from Giant Molecular Clouds}",
      journal = {\apss},
         year = "1991",
        month = "Jun",
       volume = {180},
       number = {2},
        pages = {305-320},
          doi = {10.1007/BF00648185},
       adsurl = {https://ui.adsabs.harvard.edu/abs/1991Ap&SS.180..305A}
}

@ARTICLE{peda,
       author = {{Pedaletti}, G. and {Torres}, D.~F. and {Gabici}, S. and
         {de O{\~n}a Wilhelmi}, E. and {Mazin}, D. and {Stamatescu}, V.},
        title = "{On the potential of the Cherenkov Telescope Array for the study of cosmic-ray diffusion in molecular clouds}",
      journal = {\aap},
         year = "2013",
        month = "Feb",
       volume = {550},
          eid = {A123},
        pages = {A123},
          doi = {10.1051/0004-6361/201220583},
archivePrefix = {arXiv},
       eprint = {1301.5240},
 primaryClass = {astro-ph.HE},
       adsurl = {https://ui.adsabs.harvard.edu/abs/2013A&A...550A.123P}
}

@ARTICLE{issa,
       author = {{Issa}, M.~R. and {Wolfendale}, A.~W.},
        title = "{{\ensuremath{\gamma}} Rays from the cosmic ray irradiation of local molecular clouds}",
      journal = {\nat},
         year = "1981",
        month = "Aug",
       volume = {292},
       number = {5822},
        pages = {430-433},
          doi = {10.1038/292430a0},
       adsurl = {https://ui.adsabs.harvard.edu/abs/1981Natur.292..430I}
}

@ARTICLE{mont,
       author = {{Montmerle}, T.},
        title = "{On gamma-ray sources, supernova remnants, OB associations, and the origin of cosmic rays.}",
      journal = {\apj},
         year = "1979",
        month = "Jul",
       volume = {231},
        pages = {95-110},
          doi = {10.1086/157166},
       adsurl = {https://ui.adsabs.harvard.edu/abs/1979ApJ...231...95M}
}

@ARTICLE{aha94,
       author = {{Aharonian}, F.~A. and {Drury}, L. O'C. and {Voelk}, H.~J.},
        title = "{GeV/TeV gamma-ray emission from dense molecular clouds overtaken by supernova shells}",
      journal = {\aap},
         year = "1994",
        month = "May",
       volume = {285},
        pages = {645-647},
       adsurl = {https://ui.adsabs.harvard.edu/abs/1994A&A...285..645A}
}

@ARTICLE{frail,
       author = {{Frail}, D.~A. and {Goss}, W.~M. and {Reynoso}, E.~M. and
         {Giacani}, E.~B. and {Green}, A.~J. and {Otrupcek}, R.},
        title = "{A Survey for OH (1720 MHz) Maser Emission Toward Supernova Remnants}",
      journal = {\aj},
         year = "1996",
        month = "Apr",
       volume = {111},
        pages = {1651},
          doi = {10.1086/117904},
       adsurl = {https://ui.adsabs.harvard.edu/abs/1996AJ....111.1651F}
}

@ARTICLE{cav,
       author = {{Jiang}, Bing and {Chen}, Yang and {Wang}, Junzhi and {Su}, Yang and
         {Zhou}, Xin and {Safi-Harb}, Samar and {DeLaney}, Tracey},
        title = "{Cavity of Molecular Gas Associated with Supernova Remnant 3C 397}",
      journal = {\apj},
         year = "2010",
        month = "Apr",
       volume = {712},
       number = {2},
        pages = {1147-1156},
          doi = {10.1088/0004-637X/712/2/1147},
archivePrefix = {arXiv},
       eprint = {1001.2204},
 primaryClass = {astro-ph.GA},
       adsurl = {https://ui.adsabs.harvard.edu/abs/2010ApJ...712.1147J}
}

@ARTICLE{w28radio,
       author = {{Aharonian}, F. and {Akhperjanian}, A.~G. and {Bazer-Bachi}, A.~R. and
         {Behera}, B. and {Beilicke}, M. and {Benbow}, W. and {Berge}, D. and
         {Bernl{\"o}hr}, K. and {Boisson}, C. and {Bolz}, O. and {Borrel}, V. and
         {Braun}, I. and {Brion}, E. and {Brown}, A.~M. and {B{\"u}hler}, R. and
         {Bulik}, T. and {B{\"u}sching}, I. and {Boutelier}, T. and
         {Carrigan}, S. and {Chadwick}, P.~M. and {Chounet}, L. -M. and
         {Clapson}, A.~C. and {Coignet}, G. and {Cornils}, R. and
         {Costamante}, L. and {Degrange}, B. and {Dickinson}, H.~J. and
         {Djannati-Ata{\"\i}}, A. and {Domainko}, W. and {O'C. Drury}, L. and
         {Dubus}, G. and {Dyks}, J. and {Egberts}, K. and {Emmanoulopoulos}, D. and
         {Espigat}, P. and {Farnier}, C. and {Feinstein}, F. and {Fiasson}, A. and
         {F{\"o}rster}, A. and {Fontaine}, G. and {Fukui}, Y. and {Funk}, Seb. and
         {Funk}, S. and {F{\"u}{\ss}ling}, M. and {Gallant}, Y.~A. and
         {Giebels}, B. and {Glicenstein}, J.~F. and {Gl{\"u}ck}, B. and
         {Goret}, P. and {Hadjichristidis}, C. and {Hauser}, D. and
         {Hauser}, M. and {Heinzelmann}, G. and {Henri}, G. and {Hermann}, G. and
         {Hinton}, J.~A. and {Hoffmann}, A. and {Hofmann}, W. and
         {Holleran}, M. and {Hoppe}, S. and {Horns}, D. and {Jacholkowska}, A. and
         {de Jager}, O.~C. and {Kendziorra}, E. and {Kerschhaggl}, M. and
         {Kh{\'e}lifi}, B. and {Komin}, Nu. and {Kosack}, K. and {Lamanna}, G. and
         {Latham}, I.~J. and {Le Gallou}, R. and {Lemi{\`e}re}, A. and
         {Lemoine-Goumard}, M. and {Lenain}, J. -P. and {Lohse}, T. and
         {Martin}, J.~M. and {Martineau-Huynh}, O. and {Marcowith}, A. and
         {Masterson}, C. and {Maurin}, G. and {McComb}, T.~J.~L. and
         {Moderski}, R. and {Moriguchi}, Y. and {Moulin}, E. and
         {de Naurois}, M. and {Nedbal}, D. and {Nolan}, S.~J. and
         {Olive}, J. -P. and {Orford}, K.~J. and {Osborne}, J.~L. and
         {Ostrowski}, M. and {Panter}, M. and {Pedaletti}, G. and
         {Pelletier}, G. and {Petrucci}, P. -O. and {Pita}, S. and
         {P{\"u}hlhofer}, G. and {Punch}, M. and {Ranchon}, S. and
         {Raubenheimer}, B.~C. and {Raue}, M. and {Rayner}, S.~M. and
         {Reimer}, O. and {Renaud}, M. and {Ripken}, J. and {Rob}, L. and {Rolland
        }, L. and {Rosier-Lees}, S. and {Rowell}, G. and {Rudak}, B. and
         {Ruppel}, J. and {Sahakian}, V. and {Santangelo}, A. and
         {Saug{\'e}}, L. and {Schlenker}, S. and {Schlickeiser}, R. and
         {Schr{\"o}der}, R. and {Schwanke}, U. and {Schwarzburg}, S. and
         {Schwemmer}, S. and {Shalchi}, A. and {Sol}, H. and {Spangler}, D. and
         {Stawarz}, {\L}. and {Steenkamp}, R. and {Stegmann}, C. and
         {Superina}, G. and {Takeuchi}, T. and {Tam}, P.~H. and
         {Tavernet}, J. -P. and {Terrier}, R. and {van Eldik}, C. and
         {Vasileiadis}, G. and {Venter}, C. and {Vialle}, J.~P. and
         {Vincent}, P. and {Vivier}, M. and {V{\"o}lk}, H.~J. and {Volpe}, F. and
         {Wagner}, S.~J. and {Ward}, M.},
        title = "{Discovery of very high energy gamma-ray emission coincident with molecular clouds in the W 28 (G6.4-0.1) field}",
      journal = {\aap},
         year = "2008",
        month = "Apr",
       volume = {481},
       number = {2},
        pages = {401-410},
          doi = {10.1051/0004-6361:20077765},
archivePrefix = {arXiv},
       eprint = {0801.3555},
 primaryClass = {astro-ph},
       adsurl = {https://ui.adsabs.harvard.edu/abs/2008A&A...481..401A}
}

@ARTICLE{lmd,
   author = {{Dopita}, M.~A. and {Vogt}, F.~P.~A. and {Sutherland}, R.~S. and 
	{Seitenzahl}, I.~R. and {Ruiter}, A.~J. and {Ghavamian}, P.},
    title = "{Shocked Interstellar Clouds and Dust Grain Destruction in the LMC Supernova Remnant N132D}",
  journal = {\apjs},
archivePrefix = "arXiv",
   eprint = {1806.04276},
     year = 2018,
    month = jul,
   volume = 237,
      eid = {10},
    pages = {10},
      doi = {10.3847/1538-4365/aac837},
   adsurl = {https://ui.adsabs.harvard.edu/abs/2018ApJS..237...10D}
}

@ARTICLE{slane,
       author = {{Slane}, Patrick and {Bykov}, Andrei and {Ellison}, Donald C. and
         {Dubner}, Gloria and {Castro}, Daniel},
        title = "{Supernova Remnants Interacting with Molecular Clouds: X-Ray and Gamma-Ray Signatures}",
      journal = {\ssr},
         year = "2015",
        month = "May",
       volume = {188},
       number = {1-4},
        pages = {187-210},
          doi = {10.1007/s11214-014-0062-6},
archivePrefix = {arXiv},
       eprint = {1406.4364},
 primaryClass = {astro-ph.HE},
       adsurl = {https://ui.adsabs.harvard.edu/abs/2015SSRv..188..187S}
}

@ARTICLE{brogan,
       author = {{Brogan}, C.~L. and {Goss}, W.~M. and {Hunter}, T.~R. and
         {Richards}, A.~M.~S. and {Chandler}, C.~J. and {Lazendic}, J.~S. and
         {Koo}, B. -C. and {Hoffman}, I.~M. and {Claussen}, M.~J.},
        title = "{OH (1720 MHz) Masers: A Multiwavelength Study of the Interaction between the W51C Supernova Remnant and the W51B Star Forming Region}",
      journal = {\apj},
         year = "2013",
        month = "Jul",
       volume = {771},
       number = {2},
          eid = {91},
        pages = {91},
          doi = {10.1088/0004-637X/771/2/91},
archivePrefix = {arXiv},
       eprint = {1305.2793},
 primaryClass = {astro-ph.GA},
       adsurl = {https://ui.adsabs.harvard.edu/abs/2013ApJ...771...91B}
}

@ARTICLE{443co,
       author = {{Su}, Yang and {Fang}, Min and {Yang}, Ji and {Zhou}, Ping and
         {Chen}, Yang},
        title = "{Molecular Environment of the Supernova Remnant IC 443: Discovery of the Molecular Shells Surrounding the Remnant}",
      journal = {\apj},
         year = "2014",
        month = "Jun",
       volume = {788},
       number = {2},
          eid = {122},
        pages = {122},
          doi = {10.1088/0004-637X/788/2/122},
archivePrefix = {arXiv},
       eprint = {1405.7098},
 primaryClass = {astro-ph.SR},
       adsurl = {https://ui.adsabs.harvard.edu/abs/2014ApJ...788..122S}
}

@ARTICLE{radiosnr,
       author = {{Dubner}, Gloria and {Giacani}, Elsa},
        title = "{Radio emission from supernova remnants}",
      journal = {\aapr},
         year = "2015",
        month = "Sep",
       volume = {23},
          eid = {3},
        pages = {3},
          doi = {10.1007/s00159-015-0083-5},
archivePrefix = {arXiv},
       eprint = {1508.07294},
 primaryClass = {astro-ph.HE},
       adsurl = {https://ui.adsabs.harvard.edu/abs/2015A&ARv..23....3D}
}

@ARTICLE{case,
   author = {{Case}, G.~L. and {Bhattacharya}, D.},
    title = "{A New {$\Sigma$}-D Relation and Its Application to the Galactic Supernova Remnant Distribution}",
  journal = {\apj},
   eprint = {astro-ph/9807162},
     year = 1998,
    month = sep,
   volume = 504,
    pages = {761-772},
      doi = {10.1086/306089},
   adsurl = {https://ui.adsabs.harvard.edu/abs/1998ApJ...504..761C}
}

@ARTICLE{pablo,
       author = {{Pavlovi{\'c}}, M.~Z. and {Uro{\v{s}}evi{\'c}}, D. and
         {Vukoti{\'c}}, B. and {Arbutina}, B. and {G{\"o}ker}, {\"U}. D.},
        title = "{The Radio Surface-brightness-to-Diameter Relation for Galactic Supernova Remnants: Sample Selection and Robust Analysis with Various Fitting Offsets}",
      journal = {\apjs},
         year = "2013",
        month = "Jan",
       volume = {204},
       number = {1},
          eid = {4},
        pages = {4},
          doi = {10.1088/0067-0049/204/1/4},
archivePrefix = {arXiv},
       eprint = {1210.4602},
 primaryClass = {astro-ph.GA},
       adsurl = {https://ui.adsabs.harvard.edu/abs/2013ApJS..204....4P}
}

@ARTICLE{kassim,
       author = {{Kassim}, Namir E. and {Hertz}, Paul and {van Dyk}, Schuyler D. and
         {Weiler}, Kurt W.},
        title = "{X-Ray Observations of Supernova Remnants as Distance Indicators}",
      journal = {\apjl},
         year = "1994",
        month = "Jun",
       volume = {427},
        pages = {L95},
          doi = {10.1086/187373},
       adsurl = {https://ui.adsabs.harvard.edu/abs/1994ApJ...427L..95K}
}

@INPROCEEDINGS{chere,
   author = {{Blackett}, P.~M.~S.},
    title = "{A possible contribution to the night sky from the Cerenkov radiation emitted by cosmic rays}",
booktitle = {The Emission Spectra of the Night Sky and Aurorae},
     year = 1948,
    pages = {34},
   adsurl = {https://ui.adsabs.harvard.edu/abs/1948esns.conf...34B}
}

@INPROCEEDINGS{hillas,
   author = {{Hillas}, A.~M.},
    title = "{Cerenkov light images of EAS produced by primary gamma}",
booktitle = {19th International Cosmic Ray Conference (ICRC19), Volume 3},
     year = 1985,
   series = {International Cosmic Ray Conference},
   volume = 3,
   editor = {{Jones}, F.~C.},
    month = aug,
   adsurl = {https://ui.adsabs.harvard.edu/abs/1985ICRC....3..445H}
}

@ARTICLE{crab,
   author = {{Weekes}, T.~C. and {Cawley}, M.~F. and {Fegan}, D.~J. and {Gibbs}, K.~G. and 
	{Hillas}, A.~M. and {Kowk}, P.~W. and {Lamb}, R.~C. and {Lewis}, D.~A. and 
	{Macomb}, D. and {Porter}, N.~A. and {Reynolds}, P.~T. and {Vacanti}, G.
	},
    title = "{Observation of TeV gamma rays from the Crab nebula using the atmospheric Cerenkov imaging technique}",
  journal = {\apj},
     year = 1989,
    month = jul,
   volume = 342,
    pages = {379-395},
      doi = {10.1086/167599},
   adsurl = {https://ui.adsabs.harvard.edu/abs/1989ApJ...342..379W}
}

@ARTICLE{rice,
       author = {{Rice}, Thomas S. and {Goodman}, Alyssa A. and {Bergin}, Edwin A. and
         {Beaumont}, Christopher and {Dame}, T.~M.},
        title = "{A Uniform Catalog of Molecular Clouds in the Milky Way}",
      journal = {\apj},
         year = "2016",
        month = "May",
       volume = {822},
       number = {1},
          eid = {52},
        pages = {52},
          doi = {10.3847/0004-637X/822/1/52},
archivePrefix = {arXiv},
       eprint = {1602.02791},
 primaryClass = {astro-ph.GA},
       adsurl = {https://ui.adsabs.harvard.edu/abs/2016ApJ...822...52R}
}

@ARTICLE{yus,
       author = {{Yusef-Zadeh}, Farhad and {Bally}, John},
        title = "{A non-thermal axially symmetric radio wake towards the galactic centre}",
      journal = {\nat},
         year = "1987",
        month = "Dec",
       volume = {330},
       number = {6147},
        pages = {455-458},
          doi = {10.1038/330455a0},
       adsurl = {https://ui.adsabs.harvard.edu/abs/1987Natur.330..455Y}
}

@ARTICLE{aha08,
       author = {{Aharonian}, F. and {Akhperjanian}, A.~G. and {Barres de Almeida}, U. and
         {Bazer-Bachi}, A.~R. and {Behera}, B. and {Beilicke}, M. and
         {Benbow}, W. and {Bernl{\"o}hr}, K. and {Boisson}, C. and {Bolz}, O. and
         {Borrel}, V. and {Braun}, I. and {Brion}, E. and {Brown}, A.~M. and
         {B{\"u}hler}, R. and {Bulik}, T. and {B{\"u}sching}, I. and
         {Boutelier}, T. and {Carrigan}, S. and {Chadwick}, P.~M. and
         {Chounet}, L. -M. and {Clapson}, A.~C. and {Coignet}, G. and
         {Cornils}, R. and {Costamante}, L. and {Dalton}, M. and {Degrange}, B. and
         {Dickinson}, H.~J. and {Djannati-Ata{\"\i}}, A. and {Domainko}, W. and
         {O'C. Drury}, L. and {Dubois}, F. and {Dubus}, G. and {Dyks}, J. and
         {Egberts}, K. and {Emmanoulopoulos}, D. and {Espigat}, P. and
         {Farnier}, C. and {Feinstein}, F. and {Fiasson}, A. and
         {F{\"o}rster}, A. and {Fontaine}, G. and {Funk}, S. and
         {F{\"u}{\ss}ling}, M. and {Gallant}, Y.~A. and {Giebels}, B. and
         {Glicenstein}, J.~F. and {Gl{\"u}ck}, B. and {Goret}, P. and
         {Hadjichristidis}, C. and {Hauser}, D. and {Hauser}, M. and
         {Heinzelmann}, G. and {Henri}, G. and {Hermann}, G. and
         {Hinton}, J.~A. and {Hoffmann}, A. and {Hofmann}, W. and
         {Holleran}, M. and {Hoppe}, S. and {Horns}, D. and {Jacholkowska}, A. and
         {de Jager}, O.~C. and {Jung}, I. and {Katarzy{\'n}ski}, K. and
         {Kendziorra}, E. and {Kerschhaggl}, M. and {Kh{\'e}lifi}, B. and
         {Keogh}, D. and {Komin}, Nu. and {Kosack}, K. and {Lamanna}, G. and
         {Latham}, I.~J. and {Lemoine-Goumard}, M. and {Lenain}, J. -P. and
         {Lohse}, T. and {Martin}, J.~M. and {Martineau-Huynh}, O. and
         {Marcowith}, A. and {Masterson}, C. and {Maurin}, D. and
         {McComb}, T.~J.~L. and {Moderski}, R. and {Moulin}, E. and
         {Naumann-Godo}, M. and {de Naurois}, M. and {Nedbal}, D. and
         {Nekrassov}, D. and {Nolan}, S.~J. and {Ohm}, S. and {Olive}, J. -P. and
         {de O{\~n}a Wilhelmi}, E. and {Orford}, K.~J. and {Osborne}, J.~L. and
         {Ostrowski}, M. and {Panter}, M. and {Pedaletti}, G. and
         {Pelletier}, G. and {Petrucci}, P. -O. and {Pita}, S. and
         {P{\"u}hlhofer}, G. and {Punch}, M. and {Raubenheimer}, B.~C. and
         {Raue}, M. and {Rayner}, S.~M. and {Renaud}, M. and {Ripken}, J. and
         {Rob}, L. and {Rosier-Lees}, S. and {Rowell}, G. and {Rudak}, B. and
         {Ruppel}, J. and {Sahakian}, V. and {Santangelo}, A. and
         {Schlickeiser}, R. and {Sch{\"o}ck}, F.~M. and {Schr{\"o}der}, R. and
         {Schwanke}, U. and {Schwarzburg}, S. and {Schwemmer}, S. and
         {Shalchi}, A. and {Sol}, H. and {Spangler}, D. and {Stawarz}, {\L}. and
         {Steenkamp}, R. and {Stegmann}, C. and {Superina}, G. and {Tam}, P.~H. and
         {Tavernet}, J. -P. and {Terrier}, R. and {van Eldik}, C. and
         {Vasileiadis}, G. and {Venter}, C. and {Vialle}, J.~P. and
         {Vincent}, P. and {Vivier}, M. and {V{\"o}lk}, H.~J. and {Volpe}, F. and
         {Wagner}, S.~J. and {Ward}, M. and {Zdziarski}, A.~A. and {Zech}, A.},
        title = "{Exploring a SNR/molecular cloud association within HESS J1745-303}",
      journal = {\aap},
         year = "2008",
        month = "May",
       volume = {483},
       number = {2},
        pages = {509-517},
          doi = {10.1051/0004-6361:20079230},
archivePrefix = {arXiv},
       eprint = {0803.2844},
 primaryClass = {astro-ph},
       adsurl = {https://ui.adsabs.harvard.edu/abs/2008A&A...483..509A}
}

@ARTICLE{fuku,
       author = {{Fukuda}, T. and {Yoshiike}, S. and {Sano}, H. and {Torii}, K. and
         {Yamamoto}, H. and {Acero}, F. and {Fukui}, Y.},
        title = "{Interstellar Protons in the TeV {\ensuremath{\gamma}}-Ray SNR HESS J1731-347: Possible Evidence for the Coexistence of Hadronic and Leptonic {\ensuremath{\gamma}}-Rays}",
      journal = {\apj},
         year = "2014",
        month = "Jun",
       volume = {788},
       number = {1},
          eid = {94},
        pages = {94},
          doi = {10.1088/0004-637X/788/1/94},
archivePrefix = {arXiv},
       eprint = {1405.2599},
 primaryClass = {astro-ph.HE},
       adsurl = {https://ui.adsabs.harvard.edu/abs/2014ApJ...788...94F}
}

@ARTICLE{guo,
       author = {{Guo}, Xiao-Lei and {Xin}, Yu-Liang and {Liao}, Neng-Hui and
         {Yuan}, Qiang and {Gao}, Wei-Hong and {Fan}, Yi-Zhong},
        title = "{Detection of GeV Gamma-Ray Emission in the Direction of HESS J1731-347 with Fermi-LAT}",
      journal = {\apj},
         year = "2018",
        month = "Jan",
       volume = {853},
       number = {1},
          eid = {2},
        pages = {2},
          doi = {10.3847/1538-4357/aaa3f8},
archivePrefix = {arXiv},
       eprint = {1711.03729},
 primaryClass = {astro-ph.HE},
       adsurl = {https://ui.adsabs.harvard.edu/abs/2018ApJ...853....2G}
}

@ARTICLE{blu,
       author = {{Blumer}, Harsha and {Safi-Harb}, Samar and {Kothes}, Roland and
         {Rogers}, Adam and {Gotthelf}, Eric V.},
        title = "{X-ray and radio studies of SNR CTB 37B hosting the magnetar CXOU J171405.7-381031}",
      journal = {\mnras},
         year = "2019",
        month = "Aug",
       volume = {487},
       number = {4},
        pages = {5019-5028},
          doi = {10.1093/mnras/stz1656},
archivePrefix = {arXiv},
       eprint = {1906.07249},
 primaryClass = {astro-ph.HE},
       adsurl = {https://ui.adsabs.harvard.edu/abs/2019MNRAS.487.5019B}
}

@INPROCEEDINGS{fukui,
   author = {{Fukui}, Y.},
    title = "{Molecular and Atomic Gas in the Young TeV {$\gamma$}-Ray SNRs RX J1713.7-3946 and RX J0852.0-4622; Evidence for the Hadronic Production of {$\gamma$}-Rays}",
booktitle = {Cosmic Rays in Star-Forming Environments},
     year = 2013,
   series = {Astrophysics and Space Science Proceedings},
   volume = 34,
archivePrefix = "arXiv",
   eprint = {1304.1261},
   editor = {{Torres}, D.~F. and {Reimer}, O.},
    pages = {249},
      doi = {10.1007/978-3-642-35410-6_17},
   adsurl = {https://ui.adsabs.harvard.edu/abs/2013ASSP...34..249F}
}

@ARTICLE{tsuji,
   author = {{Tsuji}, N. and {Uchiyama}, Y.},
    title = "{Expansion measurements of supernova remnant RX J1713.7-3946}",
  journal = {\pasj},
archivePrefix = "arXiv",
   eprint = {1609.07886},
 primaryClass = "astro-ph.HE",
     year = 2016,
    month = dec,
   volume = 68,
      eid = {108},
    pages = {108},
      doi = {10.1093/pasj/psw102},
   adsurl = {https://ui.adsabs.harvard.edu/abs/2016PASJ...68..108T}
}

@ARTICLE{wink,
       author = {{Winkler}, P. Frank and {Gupta}, Gaurav and {Long}, Knox S.},
        title = "{The SN 1006 Remnant: Optical Proper Motions, Deep Imaging, Distance, and Brightness at Maximum}",
      journal = {\apj},
         year = "2003",
        month = "Mar",
       volume = {585},
       number = {1},
        pages = {324-335},
          doi = {10.1086/345985},
archivePrefix = {arXiv},
       eprint = {astro-ph/0208415},
 primaryClass = {astro-ph},
       adsurl = {https://ui.adsabs.harvard.edu/abs/2003ApJ...585..324W}
}

@INPROCEEDINGS{hofv,
   author = {{Hofverberg}, P. and {Chaves}, R.~C.~G. and {Fiasson}, A. and 
	{Kosack}, K. and {M{\'e}hault}, J. and {de On{\~a} Wilhelmi}, E. and 
	{H.E.S.S.~Collaboration}},
    title = "{Discovery of VHE gamma-rays from the vicinity of the shell-type SNR G318.2+0.1 with H.E.S.S.}",
booktitle = {25th Texas Symposium on Relativistic Astrophysics},
     year = 2010,
archivePrefix = "arXiv",
   eprint = {1104.5119},
      eid = {196},
    pages = {196},
   adsurl = {https://ui.adsabs.harvard.edu/abs/2010tsra.confE.196H}
}

@ARTICLE{bocc,
       author = {{Bocchino}, F. and {Vink}, J. and {Favata}, F. and {Maggio}, A. and
         {Sciortino}, S.},
        title = "{A BeppoSAX and ROSAT view of the RCW86 supernova remnant}",
      journal = {\aap},
         year = "2000",
        month = "Aug",
       volume = {360},
        pages = {671-682},
archivePrefix = {arXiv},
       eprint = {astro-ph/0005369},
 primaryClass = {astro-ph},
       adsurl = {https://ui.adsabs.harvard.edu/abs/2000A&A...360..671B}
}

@ARTICLE{ajello,
   author = {{Ajello}, M. and {Baldini}, L. and {Barbiellini}, G. and {Bastieri}, D. and 
	{Bellazzini}, R. and {Bissaldi}, E. and {Bloom}, E.~D. and {Bonino}, R. and 
	{Bottacini}, E. and {Brandt}, T.~J. and {Bregeon}, J. and {Bruel}, P. and 
	{Buehler}, R. and {Caliandro}, G.~A. and {Cameron}, R.~A. and 
	{Caragiulo}, M. and {Cavazzuti}, E. and {Charles}, E. and {Chekhtman}, A. and 
	{Ciprini}, S. and {Cohen-Tanugi}, J. and {Condon}, B. and {Costanza}, F. and 
	{Cutini}, S. and {D'Ammando}, F. and {de Palma}, F. and {Desiante}, R. and 
	{Di Lalla}, N. and {Di Mauro}, M. and {Di Venere}, L. and {Drell}, P.~S. and 
	{Dubner}, G. and {Dumora}, D. and {Duvidovich}, L. and {Favuzzi}, C. and 
	{Focke}, W.~B. and {Fusco}, P. and {Gargano}, F. and {Gasparrini}, D. and 
	{Giacani}, E. and {Giglietto}, N. and {Glanzman}, T. and {Green}, D.~A. and 
	{Grenier}, I.~A. and {Guiriec}, S. and {Hays}, E. and {Hewitt}, J.~W. and 
	{Hill}, A.~B. and {Horan}, D. and {Jogler}, T. and {J{\'o}hannesson}, G. and 
	{Jung-Richardt}, I. and {Kensei}, S. and {Kuss}, M. and {Larsson}, S. and 
	{Latronico}, L. and {Lemoine-Goumard}, M. and {Li}, J. and {Li}, L. and 
	{Longo}, F. and {Loparco}, F. and {Lovellette}, M.~N. and {Lubrano}, P. and 
	{Magill}, J. and {Maldera}, S. and {Manfreda}, A. and {Mayer}, M. and 
	{Mazziotta}, M.~N. and {McEnery}, J.~E. and {Michelson}, P.~F. and 
	{Mitthumsiri}, W. and {Mizuno}, T. and {Monzani}, M.~E. and 
	{Morselli}, A. and {Moskalenko}, I.~V. and {Negro}, M. and {Nuss}, E. and 
	{Orienti}, M. and {Orlando}, E. and {Ormes}, J.~F. and {Paneque}, D. and 
	{Perkins}, J.~S. and {Pesce-Rollins}, M. and {Piron}, F. and 
	{Pivato}, G. and {Porter}, T.~A. and {Rain{\`o}}, S. and {Rando}, R. and 
	{Razzano}, M. and {Reimer}, A. and {Reimer}, O. and {Reposeur}, T. and 
	{Schmid}, J. and {Schulz}, A. and {Sgr{\`o}}, C. and {Simone}, D. and 
	{Siskind}, E.~J. and {Spada}, F. and {Spandre}, G. and {Spinelli}, P. and 
	{Thayer}, J.~B. and {Tibaldo}, L. and {Torres}, D.~F. and {Tosti}, G. and 
	{Troja}, E. and {Uchiyama}, Y. and {Vianello}, G. and {Vink}, J. and 
	{Wood}, K.~S. and {Yassine}, M.},
    title = "{Deep Morphological and Spectral Study of the SNR RCW 86 with Fermi-LAT}",
  journal = {\apj},
archivePrefix = "arXiv",
   eprint = {1601.06534},
 primaryClass = "astro-ph.HE",
     year = 2016,
    month = mar,
   volume = 819,
      eid = {98},
    pages = {98},
      doi = {10.3847/0004-637X/819/2/98},
   adsurl = {https://ui.adsabs.harvard.edu/abs/2016ApJ...819...98A}
}

@ARTICLE{shan,
       author = {{Shan}, Su-Su and {Zhu}, Hui and {Tian}, Wen-Wu and {Zhang}, Hai-Yan and
         {Yang}, Ai-Yuan and {Zhang}, Meng-Fei},
        title = "{The distance measurements of supernova remnants in the fourth Galactic quadrant}",
      journal = {Research in Astronomy and Astrophysics},
         year = "2019",
        month = "Jul",
       volume = {19},
       number = {7},
          eid = {092},
        pages = {092},
          doi = {10.1088/1674-4527/19/7/92},
archivePrefix = {arXiv},
       eprint = {1901.02882},
 primaryClass = {astro-ph.GA},
       adsurl = {https://ui.adsabs.harvard.edu/abs/2019RAA....19...92S}
}

@ARTICLE{bamba,
       author = {{Bamba}, Aya and {Sawada}, Makoto and {Nakano}, Yuto and
         {Terada}, Yukikatsu and {Hewitt}, John and {Petre}, Robert and
         {Angelini}, Lorella},
        title = "{New identification of the mixed-morphology supernova remnant G298.6-0.0 with possible gamma-ray association}",
      journal = {\pasj},
         year = "2016",
        month = "Jun",
       volume = {68},
          eid = {S5},
        pages = {S5},
          doi = {10.1093/pasj/psv096},
archivePrefix = {arXiv},
       eprint = {1509.00214},
 primaryClass = {astro-ph.HE},
       adsurl = {https://ui.adsabs.harvard.edu/abs/2016PASJ...68S...5B}
}

@ARTICLE{allen,
   author = {{Allen}, G.~E. and {Chow}, K. and {DeLaney}, T. and {Filipovi{\'c}}, M.~D. and 
	{Houck}, J.~C. and {Pannuti}, T.~G. and {Stage}, M.~D.},
    title = "{On the Expansion Rate, Age, and Distance of the Supernova Remnant G266.2-1.2 (Vela Jr.)}",
  journal = {\apj},
archivePrefix = "arXiv",
   eprint = {1410.7435},
 primaryClass = "astro-ph.HE",
     year = 2015,
    month = jan,
   volume = 798,
      eid = {82},
    pages = {82},
      doi = {10.1088/0004-637X/798/2/82},
   adsurl = {https://ui.adsabs.harvard.edu/abs/2015ApJ...798...82A}
}

@ARTICLE{becker,
       author = {{Becker}, Werner and {Prinz}, Tobias and {Winkler}, P. Frank and
         {Petre}, Robert},
        title = "{The Proper Motion of the Central Compact Object RX J0822-4300 in the Supernova Remnant Puppis A}",
      journal = {\apj},
         year = "2012",
        month = "Aug",
       volume = {755},
       number = {2},
          eid = {141},
        pages = {141},
          doi = {10.1088/0004-637X/755/2/141},
archivePrefix = {arXiv},
       eprint = {1204.3510},
 primaryClass = {astro-ph.HE},
       adsurl = {https://ui.adsabs.harvard.edu/abs/2012ApJ...755..141B}
}

@ARTICLE{lee,
   author = {{Lee}, J.-J. and {Koo}, B.-C. and {Yun}, M.~S. and {Stanimirovi{\'c}}, S. and 
	{Heiles}, C. and {Heyer}, M.},
    title = "{A 21 cm Spectral and Continuum Study of IC 443 Using the Very Large Array and the Arecibo Telescope}",
  journal = {\aj},
     year = 2008,
    month = mar,
   volume = 135,
    pages = {796-808},
      doi = {10.1088/0004-6256/135/3/796},
   adsurl = {https://ui.adsabs.harvard.edu/abs/2008AJ....135..796L}
}

@ARTICLE{yu,
       author = {{Yu}, Bin and {Chen}, B.~Q. and {Jiang}, B.~W. and {Zijlstra}, A.},
        title = "{Three-dimensional dust mapping of 12 supernovae remnants in the Galactic anticentre}",
      journal = {\mnras},
         year = "2019",
        month = "Sep",
       volume = {488},
       number = {3},
        pages = {3129-3142},
          doi = {10.1093/mnras/stz1940},
archivePrefix = {arXiv},
       eprint = {1907.04326},
 primaryClass = {astro-ph.GA},
       adsurl = {https://ui.adsabs.harvard.edu/abs/2019MNRAS.488.3129Y}
}

@ARTICLE{yato,
       author = {{Hayato}, Asami and {Yamaguchi}, Hiroya and {Tamagawa}, Toru and
         {Katsuda}, Satoru and {Hwang}, Una and {Hughes}, John P. and
         {Ozawa}, Midori and {Bamba}, Aya and {Kinugasa}, Kenzo and
         {Terada}, Yukikatsu and {Furuzawa}, Akihiro and {Kunieda}, Hideyo and
         {Makishima}, Kazuo},
        title = "{Expansion Velocity of Ejecta in Tycho's Supernova Remnant Measured by Doppler Broadened X-ray Line Emission}",
      journal = {\apj},
         year = "2010",
        month = "Dec",
       volume = {725},
       number = {1},
        pages = {894-903},
          doi = {10.1088/0004-637X/725/1/894},
archivePrefix = {arXiv},
       eprint = {1009.6031},
 primaryClass = {astro-ph.HE},
       adsurl = {https://ui.adsabs.harvard.edu/abs/2010ApJ...725..894H}
}

@ARTICLE{ala,
       author = {{Alarie}, Alexandre and {Bilodeau}, Antoine and {Drissen}, Laurent},
        title = "{A hyperspectral view of Cassiopeia A}",
      journal = {\mnras},
         year = "2014",
        month = "Jul",
       volume = {441},
       number = {4},
        pages = {2996-3008},
          doi = {10.1093/mnras/stu774},
       adsurl = {https://ui.adsabs.harvard.edu/abs/2014MNRAS.441.2996A}
}

@ARTICLE{fe06,
       author = {{Fesen}, Robert A. and {Hammell}, Molly C. and {Morse}, Jon and
         {Chevalier}, Roger A. and {Borkowski}, Kazimierz J. and
         {Dopita}, Michael A. and {Gerardy}, Christopher L. and
         {Lawrence}, Stephen S. and {Raymond}, John C. and
         {van den Bergh}, Sidney},
        title = "{The Expansion Asymmetry and Age of the Cassiopeia A Supernova Remnant}",
      journal = {\apj},
         year = "2006",
        month = "Jul",
       volume = {645},
       number = {1},
        pages = {283-292},
          doi = {10.1086/504254},
archivePrefix = {arXiv},
       eprint = {astro-ph/0603371},
 primaryClass = {astro-ph},
       adsurl = {https://ui.adsabs.harvard.edu/abs/2006ApJ...645..283F}
}

@ARTICLE{clump,
       author = {{Shan}, S.~S. and {Zhu}, H. and {Tian}, W.~W. and {Zhang}, M.~F. and
         {Zhang}, H.~Y. and {Wu}, D. and {Yang}, A.~Y.},
        title = "{Distances of Galactic Supernova Remnants Using Red Clump Stars}",
      journal = {\apjs},
         year = "2018",
        month = "Oct",
       volume = {238},
       number = {2},
          eid = {35},
        pages = {35},
          doi = {10.3847/1538-4365/aae07a},
archivePrefix = {arXiv},
       eprint = {1810.06014},
 primaryClass = {astro-ph.IM},
       adsurl = {https://ui.adsabs.harvard.edu/abs/2018ApJS..238...35S}
}

@article{Uchiyama,
      author         = "Uchiyama, Yasunobu and Takahashi, Tadayuki and Aharonian,
                        Felix and Mattox, John",
      title          = "{ASCA View of the supernova remnant gamma Cygni
                        (G78.2+2.1): Bremsstrahlung x-ray spectrum from
                        loss-flattened electron distribution}",
      journal        = "Astrophys. J.",
      volume         = "571",
      year           = "2002",
      pages          = "866",
      doi            = "10.1086/340121",
      eprint         = "astro-ph/0202414",
      archivePrefix  = "arXiv",
      primaryClass   = "astro-ph",
      reportNumber   = "ISAS-RN-743",
      SLACcitation   = "%%CITATION = ASTRO-PH/0202414;%%"
}

@ARTICLE{Floop,
       author = {{Fesen}, Robert A. and {Weil}, Kathryn E. and {Cisneros}, Ignacio A. and
         {Blair}, William P. and {Raymond}, John C.},
        title = "{The Cygnus Loop's distance, properties, and environment driven morphology}",
      journal = {\mnras},
         year = "2018",
        month = "Dec",
       volume = {481},
       number = {2},
        pages = {1786-1798},
          doi = {10.1093/mnras/sty2370},
archivePrefix = {arXiv},
       eprint = {1809.01713},
 primaryClass = {astro-ph.HE},
       adsurl = {https://ui.adsabs.harvard.edu/abs/2018MNRAS.481.1786F}
}

\end{document}